\documentclass{elsart}
\usepackage{graphicx}
\usepackage{amssymb,amsmath}

\newcommand{\dd}{\textrm{d}}

\newcommand{\bv}{{\bf v}}
\newcommand{\bx}{{\bf x}}
\newcommand{\bP}{{\bf P}}
\newcommand{\bQ}{{\bf Q}}
\newcommand{\bp}{{\bf p}}
\newcommand{\bq}{{\bf q}}
\newcommand{\ofp}{{\mathcal{L}}^{0}_{FP}}
\newcommand{\fxt}{{f({\bf x},t)}}
\newcommand{\fsta}{{f^{st}({\bf x})}}
\newcommand{\Lex}{{\mathcal{L}}^{ext}_{FP}}
\newcommand{\be}{\begin{equation}}
\newcommand{\ee}{\end{equation}}
\newcommand{\bea}{\begin{eqnarray}}
\newcommand{\eea}{\end{eqnarray}}

\newcommand{\RE}{\tt{Re}}
\newcommand{\bm}{\mathbf}

\newcommand{\ftilde}{\tilde{f}}
\newcommand{\bmu}{{\boldsymbol{\mu}}}
\newcommand{\bnu}{{\boldsymbol{\nu}}}
\newcommand{\bpt}{{\tilde{\bf p}}}
\newcommand{\tB}{{\tilde{B}}}
\newcommand{\pt}{{\tilde{p}}}
\newcommand{\obp}{{\bf p}'}
\newcommand{\opt}{{\tilde{p}}'}
\newcommand{\oW}{W'}
\newcommand{\of}{f'}
\newcommand{\oA}{A'}

\newcommand{\obpt}{{\tilde{\bf p}}'}
\newcommand{\oft}{{\tilde{f}}'}

\newcommand{\hcL}{{{\mathcal L}}}

\newcommand{\bu}{\text{\bf u}}

\newcommand{\bsigma}{\boldsymbol{\sigma}}
\newcommand{\mH}{\mathcal{H}}

\newcommand{\deriv}[2]{{\frac{d#1}{d#2}}}

\newcommand{\p}{{\bf p}}
\newcommand{\q}{{\bf q}}

\newcommand{\x}{{\bf x}}

\newcommand{\za}{\alpha}
\newcommand{\zb}{\beta}
\newcommand{\zd}{\delta}
\newcommand{\ze}{\epsilon}
\newcommand{\zg}{\gamma}
\newcommand{\zl}{\lambda}
\newcommand{\zL}{\Lambda}

\newcommand{\zs}{\sigma}
\newcommand{\zt}{\tau}

\newcommand{\zF}{\Phi}

\newcommand{\zR}{I\hskip-3.4pt R}

\newcommand{\zW}{\Omega}

\newcommand{\zD}{{\Delta}}
\newcommand{\zG}{{\Gamma}}

\newcommand{\WFR}{$\Omega$-FR }
\newcommand{\WFRs}{$\Omega$-FR's }
\newcommand{\LFR}{$\Lambda$-FR }
\newcommand{\Wt}{\overline{\Omega}_{t_0,t_0+\tau}}
\newcommand{\Ft}{\overline{\phi}_{t_0,t_0+\tau}}

\newcommand{\Wz}{\overline{\Omega}_{0,\tau}}

\newcommand{\cc}{{\bf c}}
\newcommand{\noi}{\noindent}
\newcommand {\bdm} {\begin{displaymath}}
\newcommand {\edm} {\end{displaymath}}
\newcommand {\ba}  {\begin{array}}
\newcommand {\ea}  {\end{array}}

\newcommand{\td}{T_r}
\newcommand{\ts}{T_\ell}
\newcommand{\xt}{X_t}

\newcommand{\kT}{{\cal T}}

\newcommand{\se}{{(s)}}

\newcommand{\ie}{i.e.\ }

\numberwithin{equation}{section}

\graphicspath{{Figures/}}

\begin{document}

\begin{frontmatter}

\title{Fluctuation-Dissipation: Response Theory in Statistical Physics}

\author[cam]{Umberto Marini Bettolo Marconi},
\author[dfr]{Andrea Puglisi},
\author[dft]{Lamberto Rondoni}
\author[dfir]{and Angelo Vulpiani}

\address[cam]{Universit\`a di Camerino, Dipartimento di Matematica e
Fisica, Via Madonna delle Carceri, I-62032 Camerino, Italy}

\address[dfr]{Universit\`a di Roma ``La Sapienza'', Dipartimento di
Fisica, p.le Aldo Moro 2, I-00185 Roma, Italy}

\address[dft]{Dipartimento di Matematica and INFM, Politecnico di Torino,
Corso Duca degli Abruzzi 24, I-10129, Torino, Italy}

\address[dfir]{Universit\`a di Roma ``La Sapienza'', Dipartimento di
Fisica and INFN, p.le Aldo Moro 2, I-00185 Roma, Italy}


\begin{abstract}
General aspects of the Fluctuation-Dissipation Relation (FDR), and
Response Theory are considered.  After analyzing the conceptual and
historical relevance of fluctuations in statistical mechanics, we
illustrate the relation between the relaxation of spontaneous
fluctuations, and the response to an external perturbation.  These
studies date back to Einstein's work on Brownian Motion, were
continued by Nyquist and Onsager and culminated in  Kubo's linear response theory.

The FDR has been originally developed in the framework of statistical
mechanics of Hamiltonian systems, nevertheless a generalized FDR holds
under rather general hypotheses, regardless of the Hamiltonian, or
equilibrium nature of the system.  In the last decade, this subject
was revived by the works on Fluctuation Relations
(FR) concerning far from equilibrium systems. The connection of
these works with large deviation theory is analyzed.

Some examples, beyond the standard applications of statistical
mechanics, where fluctuations play a major role are discussed: fluids,
granular media, nano-systems and biological systems.
\end{abstract}

\begin{keyword}
Correlation functions \sep Non-equilibrium phenomena \sep Large deviations \sep Fluctuation Relations

\PACS 05.40.-a, 05.70.Ln, 45.70.-n, 47.52.+j, 02.50.Ga, 05.60.-k, 47.27.-i, 47.27.eb, 51.20.+d
\end{keyword}
\end{frontmatter}

\clearpage
\tableofcontents
\newpage

\vspace{3cm}

\hspace{4cm}\parbox{10cm}{
\noindent To the memory of Robert H. Kraichnan (1928-2008), whose conspicuous
and remarkable achievements include fundamental contributions to the 
subject of our review.}

\vspace{1cm}

\section{Introduction}
\label{sec:intro}

\subsection{Introductory remarks and plan of the paper}

Recent developments in non equilibrium statistical physics have
convinced us that times are ripe for a review of the vast subject
concerning the fluctuations of systems described by statistical
mechanics.  This issue is important even beyond the ``traditional'' applications of
statistical mechanics, e.g.\ in a wide range of disciplines
ranging from the study of small biological systems to turbulence, from
climate studies to granular media etc.  Moreover, the improved 
resolution in real experiments and the computational capability reached in 
numerical simulations has led to an
increased ability to unveil the detailed nature of fluctuations,
posing new questions and challenges to the theorists. 

One of the most important and general results concerning systems
described by statistical mechanics is the existence of a relation
between the spontaneous fluctuations and the response to external fields 
of physical observables.  This result has
applications both in equilibrium statistical mechanics, where it is
used to relate the correlation functions to macroscopically measurable
quantities such as specific heats, susceptibilities and
compressibilities, and in non equilibrium systems, where it offers the
possibility of studying the response to time dependent external
fields, by analyzing time-dependent correlations.

The idea of relating the amplitude of the dissipation to that of the
fluctuations dates back to Einstein's work on Brownian motion.  Later,
Onsager put forward his regression hypothesis stating that the
relaxation of a macroscopic nonequilibrium perturbation follows the
same laws which govern the dynamics of fluctuations in
equilibrium systems. This principle is the basis of the FDR theorem of
Callen and Welton, and of Kubo's theory of time dependent correlation
functions.  This result represents a fundamental tool in
non-equilibrium statistical mechanics since it allows one to predict
the average response to external perturbations, without applying any
perturbation. In fact, via an equilibrium molecular dynamics simulation
one can compute correlation functions at equilibrium and then, using the
Green-Kubo formula, obtain the transport coefficients of model liquids
without resorting to approximation schemes.

Although the FDR theory was originally applied to Hamiltonian systems
near thermodynamic equilibrium, it has been 
realized that a generalized FDR holds for a vast class of systems 
with chaotic dynamics of special interest in the study of natural 
systems, such as geophysics and climate.

A renewed interest toward the theory of fluctuations has been
motivated by the study of nonequilibrium processes.  In 1993 Evans,
Cohen and Morriss considered the fluctuations of the entropy
production rate in a shearing fluid, and proposed the so called {\em Fluctuation Relation}
(FR). This represents a general result concerning systems arbitrarily
far from equilibrium. Moreover it is consistent with the Green-Kubo
and Onsager relations, when equilibrium is approached.  Starting with
the developments of the FR proposed by Evans and Searles and, in
different conditions, by Gallavotti and Cohen, continuing with the
Crooks and Jarzynski relations, and the recent works by Derrida and coworkers and Jona-Lasinio
and coworkers on dynamical path probabilities, the last 10-15 years have
produced a whole new theoretical framework which encompasses the
previous linear response theory and goes beyond that, to include far from
equilibrium phenomena, such as turbulence and the dynamics of granular materials.

In this paper, we only consider systems which are ergodic and have an
invariant phase space distribution, which is reached in physically
relevant time scales.  Therefore, we do not discuss aging
and glassy behaviours.  For such interesting issues the reader is
referred to recent reviews~\cite{MPRR98,CR03,LN07}.

This review is organized as follows. In this section we give a brief
historical overview on the origin of FDR, in particular of its
relevance to the ``proof'' of the existence of atoms. In Section 2 we
present the classical fundamental contributions of Onsager and Kubo to
the general theory of FDR. The third Section is devoted to a discussion of
FDR in chaotic systems. It is shown that a generalized FDR holds under
rather general conditions, even in non-Hamiltonian systems, and in
non-equilibrium situations. In addition, we discuss the connection
between this FDR and the foundations of statistical mechanics, and
anomalous diffusion. Section 4 treats non-standard applications of the
FDR: fluid dynamics, climate, granular materials and biophysical
systems. The different approaches to the FR are
discussed in Section 5. In Section 6 we discuss the numerical and experimental investigations
of the FR in stochastic systems, granular gases and conducting systems. Section
7 concerns the seminal work of Onsager and Machlup, and the
recent results in extended non-equilibrium systems, in terms of large
deviations theory. Four Appendices 
conclude the paper.

\subsection{Historical overview}

The study of the fluctuations, and of their relation with the response
to external perturbations, has an important conceptual and historical
relevance, apart from its technical interest which is
now universally recognized.  Therefore we give a brief overview of Statistical Mechanics at the beginning
of the $20$-th century.

\subsubsection{Fluctuation Phenomena and Atoms}

In spite of the great successes of the atomistic hypothesis in
chemistry and kinetic theory, towards the end of the $19$-th century,
some influential scientists such as Ostwald and Mach still claimed that
it was possible to avoid the atomistic perspective in the descriptions
of nature.  Indeed, there was no unquestionable evidence of the
existence of atoms at that time.  {\it The atomic theory plays a role
in physics similar to that of certain auxiliary concepts in
mathematics; it is a mathematical model for facilitating the mental
reproduction of facts}~\cite{M83}, was stated by Mach.

Even Boltzmann and Gibbs, who already knew the expression for the mean
square energy fluctuation, correctly thought that 
fluctuations would be too small to be observed in macroscopic systems:
{\it In the molecular theory we assume that the laws of the phenomena
found in nature do not essentially deviate from the limits that they
would approach in the case of an infinite number of infinitely small
molecules}~\cite{B96}.

{\it ... [the fluctuations] would be in general vanishing quantities,
since such experience would not be wide enough to embrace the more
considerable divergences from the mean values}~\cite{G02}.

One of the reasons of Einstein's and Smoluchowski's interest for
statistical mechanics was to find conclusive evidence for the atomic
hypothesis.  At variance with the ideas of Boltzmann and Gibbs, their
efforts attributed a central role to the fluctuations.  {\it The
equation we finally obtained {\tt (eq. 1.11)}  would yield an
exact determination of the universal constant {\tt (i.e. the Avogadro
number)}, if it were possible to determine the average of the square
of the energy fluctuation of the system; this is however not possible
according our present knowledge}~\cite{E04}.

\subsubsection{Brownian motion: a macroscopic window on the microscopic world}

Einstein's hope came true in the
understanding of a ``strange'' phenomenon: the Brownian motion (BM).
In 1827 the Scottish botanist Robert Brown noticed that pollen grains
suspended in water jiggled about under the lens of the microscope,
following a zigzag path.  Initially, he believed that such an activity
was peculiar to the male sexual cells of plants;
further study revealed that the same motion could be
observed  even
with chips of glass or granite or particles of smoke.
For some decades such a phenomenon was considered as a sort
of curiosity, although the BM attracted the interest of important
scientists, such as Poincar\'e~\cite{M01}.

The breakthrough in the understanding of the BM is due to the
independent works by Einstein~\cite{E05,E06} and
Smoluchowski~\cite{s06} at the beginning of the $20$th century.  A
few years later, Langevin proposed an approach in terms of a
stochastic differential equation, which takes into account the effect
of molecular collisions by means of an average force, given by the
fluid friction, and of a random fluctuating term~\cite{l08}.  

The experimental works by Perrin and Svendborg on the BM can be
considered as the  first ``proof'' of the existence of atoms, while
Langevin's equation, which has been the first example of stochastic
differential equation, inspired the mathematical theory of continuous
time stochastic processes.

The basic physical assumptions in both  Einstein's
and Langevin's approaches are:

a) Stokes' law for the frictional force exerted on a body moving in a liquid;

b) the  equipartition of kinetic energy among the degrees of
freedom of the system, i.e. between the particles of the fluid and the
grain performing BM.

A colloidal particle suspended in a liquid at temperature $T$ is thus
assimilated to a particle of the liquid, so that it possesses an
average kinetic energy $RT/(2N_{A})$, in each spatial direction, where
$R$ is the perfect gas constant and $N_{A}$ is the Avogadro
number. Accordingly, one obtains:
\begin{equation}
\label{BM.1}
\frac{1}{2}m\langle v_x^2 \rangle=\frac{RT}{2N_{A}} \,\,.
\end{equation}

According to Stokes' law, a spherical particle of radius $a$, moving
in a liquid with speed $V$ in the $x$ direction, experiences a
viscous drag force:
$$
f_{S}= -\alpha V= -6\pi \eta a V \, ,
$$
where $\eta$ is the  viscosity.  This law holds if $a$
is much larger than the average distance between the liquid molecules,
in which case  Stokes' force ($f_S$) represents the average macroscopic effect of the
irregular impacts of the molecules of the fluid.

Therefore, isolating the average force, the dynamical equation of the
particle in the $x$ direction  can be written as
\begin{equation}
\label{BM.2}
m { dV \over dt} =f_{S}+ f_R(t)=  -\alpha V+f_R(t)
\end{equation}
where $f_R(t)$ is a random fluctuating force mimicking the effects of
the impact of the molecules on the particle.  Of course $\langle
f_R(t)\rangle=0$, and its characteristic time, being
determined by the collisions of the molecules of the liquid with the
colloidal particle, is much smaller than $\tau= m/ \alpha$.  One
can then assume that $f_R$ is a Gaussian stochastic process with
$\langle f_R(t) f_R(t')\rangle = c \delta(t-t')$, where $c$ can be
determined by the energy equipartition.  Elementary computations
give~\cite{G90}
\begin{equation}
\label{BM.3}
 \langle [x(t)-x(0)]^2 \rangle= \frac{2RT}{\alpha \, N_{A}}
\left[t-\frac{m}{\alpha}\left(1
-e^{-\frac{\alpha}{m}t}\right)\right] \, .
\end{equation}

For a grain of size $O(1 \mu)$, in a common liquid (such as water) at
room temperature, the characteristic time
$\tau=m/\alpha=m/6 \pi \eta a$ is $O(10^{-7} s)$.
 For $t \gg \tau $, we get
\begin{equation}
\label{BM.4}
 \langle [x(t)-x(0)]^2 \rangle \simeq
\frac{2RT}{\alpha \, N_{A}} t =\frac{RT}{3 N_{A} \pi \eta a}t \,\, .
\end{equation}
This leads to the celebrated Einstein relation
for the diffusion coefficient $D$, in terms of macroscopic
variables and of the Avogadro number:
\begin{equation}
  D= \lim_{t \to \infty}  {  {\langle [x(t)-x(0)]^2 \rangle}
\over {2 t}}= \frac{RT}{6 N_{A} \pi \eta a} \,\,.
\end{equation}
Let us stress that (1.5) allows the determination of
the Avogadro number (a microscopic quantity) from experimentally
accessible macroscopic quantities, thus providing a non ambiguous
link between the microscopic and macroscopic levels of description, in
physics.

The theoretical work by Einstein and the experiments by Perrin gave a
clear and conclusive evidence of the relationship between the
diffusion coefficient (which is  measurable at the macroscopic
level) and the Avogadro number (which is related to the microscopic
description)~\cite{P13}.  Such a result could be considered as a
``proof'' of the existence of atoms: after that, even two champions of
the energetic point of view, like Helm and Ostwald, accepted atomism
as a physical fact and not as a mere useful hypothesis.  In a lecture
in Paris, in 1911, Arrhenius, summarizing the works of Einstein and
Perrin, declared {\it after this, it does not seem possible to doubt
that the molecular theory entertained by the philosophers of
antiquity, Leucippus and Democritos, has attained the truth at least
in essentials}~\cite{M01}.

Although Langevin's approach, like Einstein's, is, from a
mathematical point of view, rather simple, there is a very subtle
conceptual point at the basis of his theory of the Brownian motion.
The ingenious idea is the assumption of the validity of Stokes law
(which has a macroscopic nature), together with the assumption that
the Brownian particle is in statistical equilibrium with the molecules
in the liquid.  In other words, in spite of the fact that the mass of
the colloidal particle is exceedingly larger than the mass of the
molecules, energy equipartition is assumed to hold.

\subsection{Fluctuations in equilibrium statistical mechanics}

Equilibrium statistical mechanics predicts both the average value of
the thermodynamic observables and the size of their fluctuations about
their equilibrium values. In this section, we briefly report the salient
features of the theory of fluctuations of extensive thermodynamic
variables.  

Let us consider a large isolated system in thermodynamic
equilibrium, and single out a small region representing the
subsystem of interest (here named S), while the rest of the system
constitutes a reservoir (R).  Suppose that, due to the coupling with
R, the extensive variable $X$ of the system S can
fluctuate about its equilibrium value, while the remaining extensive
variables are held fixed.
The probability  of observing a fluctuation of size
smaller than a value $X_0$ is given by the canonical distribution
function:
\begin{equation}
P(X<X_0)=e^{-\psi(h)}\int_{-\infty}^{X_0} e^{- h X}d N(X)
\label{f1}
\end{equation}
where $h$ is the intensive thermodynamic field conjugated to $X$,
 $N(X)$ is the cumulative number of microstates with $X<X_0$, and the
 integral is to be interpreted in the Riemann-Stieltjes sense if the
 spectrum has also a discrete part. The function $\psi(h)$, which normalizes the
 distribution function, is determined by the condition
\begin{equation}
e^{\psi(h)}=\int_{-\infty}^{\infty} e^{- h X}d N(X)
\label{f2}
\end{equation}
and represents a thermodynamic potential.  The function $\psi(h)$ is a
Legendre transform of the entropy, or generalized Massieu
function~\cite{m70}. Note that, even if this argument is formal and
therefore of general validity, when $X$ is an extensive observable of
a macroscopic system (i.e. in the thermodynamic limit) the function
$\psi(h)$, or better its Legendre transform, can be seen in terms of
large deviations theory, see Appendix B.

If we consider the characteristic function:
\begin{equation}
\langle e^{\alpha X} \rangle =e^{-\psi(h)}\int_{-\infty}^{\infty} 
e^{-(h-\alpha)X} d N(X)=e^{\psi(h-\alpha)} e^{-\psi(h)}
\label{f3}
\end{equation}
and expand $\psi(h-\alpha)$ with respect to $\alpha$, we obtain the
cumulants, $\langle X^n \rangle_c$, of the distribution
\begin{equation}
\ln \langle e^{\alpha X} \rangle =
\sum_{n=1}^{\infty}\frac{\alpha^n}{n!}
\langle X^n \rangle_c=
\sum_{n=1}^{\infty}\frac{(-\alpha)^n}{n!}\frac{d^n}{d h^n}\psi(h) \,\,.
\label{f4}
\end{equation}
As an important example, we consider the energy fluctuations in a
system S, at constant temperature. In this case $X=E$,
$h=\beta^{-1}$ and $\psi(h)=-\beta A(\beta)$, where $A$ is the
Helmholtz free energy.  By employing eq. (\ref{f4}) and using standard
thermodynamic relations, we obtain the following results: \bea \langle
E \rangle_c &\equiv& \langle E \rangle= -\frac{d}{d \beta} (\beta A(\beta))=U
\label{f6a}\\
\langle E^2 \rangle_c&\equiv&\langle E^2 \rangle - \langle E \rangle^2 =
\frac{d^2 (\beta A(\beta))}{d \beta^2} =-\frac{d U}{d \beta}=k_B T^2 C_V\\
\label{f6b}
\langle E^3 \rangle_c&\equiv&\langle E^3 \rangle -3\langle E \rangle \langle E^2\rangle+2\langle E \rangle^3=
-\frac{d^3 (\beta A(\beta))}{d \beta^3} \\&=&k_B^2 T^3 [2 C_V+T \frac{d C_V}{d T}] \,\,.
\label{f6c}
\eea Formulae (\ref{f6a})-(\ref{f6c}) are a manifestation of the
relation between the fluctuations of the quantity $X$ and the
variations of the thermodynamic potential $\psi$, with respect to the
conjugate field $h$.  The generalization to several extensive
variables $X_1,X_2..,X_r$ is straightforward and involves the
generalized potential defined by $\phi(h_1,h_2,..,h_r)$ \be
e^{\phi(h_1,h_2,..,h_r)}=\int..\int dX_1..dX_r e^{(h_1 X_1+ ..+h_r
X_r)} dN(X_1,..,X_r) \ee

\be
\langle\exp{ \sum_j^r \alpha_j X_j} \rangle=
\e^{-\phi(h_1,h_2,..,h_r)}\int..\int dX_1..dX_r 
\exp \left(\sum_j^r (\alpha_j-h_j) X_j  \right) dN(X_1,.,X_r)
\ee

\bea
\langle X_m \rangle_c=
-\frac{\partial}{\partial h_m}\psi(h_1,..,h_r)\\
\langle X_m X_n \rangle_c=
\frac{\partial^2}{\partial h_m\partial h_n}\psi(h_1,..,h_r) \,\,.
\label{f11}
\eea 


Let us, now, consider the problem of relating the properties of 
a system whose Hamiltonian is 
$\mathcal{H}(\bP,\bQ)=\mathcal{H}_0(\bP,\bQ)+\lambda A(\bP,\bQ)$ to those
of a similar system characterized by $\mathcal{H}_0(\bP,\bQ)$ only, where  
$\bP=(\bp_1,...,\bp_N),\bQ=(\bq_1,...,\bq_N)$.
If the perturbation induced by  $\lambda A(\bP,\bQ)$ is small, the
canonical distribution of the system with $\lambda\neq 0$ is 
to linear order given by:
\be
f(\bP,\bQ)=\frac{\exp (-\beta\mathcal{H})}
{\int d\bP d\bQ\exp (-\beta\mathcal{H})}
\simeq \frac{\exp (-\beta\mathcal{H}_0)}
{\int d\bP d\bQ\exp (-\beta\mathcal{H}_0)}\
\frac{1-\lambda \beta A(\bP,\bQ)}
{1-\lambda \beta \langle A(\bP,\bQ)\rangle}_0
\label{fe1}
\ee
or
\be
f(\bP,\bQ)=f_0(\bP,\bQ)\bigl(1-\lambda \beta \left[A(\bP,\bQ)-\langle A(\bP,\bQ)\rangle_0 \right]\bigl)
\label{fe2}
\ee
where $\langle  \rangle_0$ stands for the average relative to
the system with $\lambda=0$.
The corresponding average change of a generic observable $B(\bP,\bQ)$,
induced by the perturbation, is to the same order:
\bea
\langle \Delta B\rangle_0&\simeq& 
\int d\bP d\bQ B(\bP,\bQ)\bigl(f(\bP,\bQ)-f_0(\bP,\bQ)\bigl) \nonumber\\
&\simeq& -\lambda \beta
\int d\bP d\bQ B(\bP,\bQ) \bigl(A(\bP,\bQ)-\langle A(\bP,\bQ)\rangle_0 ) \bigl) f_0(\bP,\bQ))
\nonumber\\
&=&-\lambda \beta\bigl(\langle B A\rangle_0-\langle B 
\rangle_0\langle A\rangle_0\bigl) \,\,.
\label{fe3}
\eea 
The last formula shows that the change of the observable $B$ is related
to the equilibrium pair correlation of $B$ with the perturbation $A$.
With the choice $B(\bP,\bQ)=A(\bP,\bQ)={\cal H}_0$  we obtain:
\be
\frac{\partial \langle {\cal H}_0 \rangle_0}{\partial\beta}=
\lim_{\lambda\to 0}\frac{\langle \Delta {\cal H}_0 \rangle_0}{\lambda \beta}=
-\bigl(\langle{\cal H}_0^2\rangle_0-\langle {\cal H}_0  \rangle_0^2\bigl)=
-k_B T^2 C_V
\label{fe4}
\ee
hence we connect the heat capacity, i.e. the response to an energy perturbation,
to the energy fluctuations.

Another relevant instance of the equilibrium fluctuation-response
theorem is the relation between the correlation function
$\Gamma(r,r')$ (defined below) and the isothermal susceptibility
$\chi_T(r,r')$.  In this case one considers the Hamiltonian to be a
functional of the magnetization density $m(r)$: \be {\cal
H}(m(r))={\cal H}_0(m(r))-\int dr h(r)m(r) \ee By using formula
\eqref{fe3} with $B=m(r)$ and $\lambda A=\int dr' h(r') m(r')$ we
find: 
\be \chi_T(r,r') = \beta [\langle m(r)m(r') \rangle_0- \langle
m(r) \rangle_0\langle m(r') \rangle_0] =\beta \Gamma(r,r') \,\,.\ee

\subsection{Fluctuations in non-equilibrium statistical mechanics:
the Einstein relation between diffusion and mobility}

We already discussed the BM and its historical relevance for the
atomic hypothesis and the development of the modern theory of 
stochastic processes.  In addition, in the celebrated paper of
Einstein one can find the first example of FDR.  Let us write the Langevin equation for the
colloidal particle in ``modern'' terms
\begin{equation}
\label{BM.5}
{dV \over dt}=
 -\gamma V +\sqrt{ {{2 \gamma k_B T} \over m}  } \eta \,\,
\end{equation}
where $\gamma=\alpha/m$ and  $\eta$ is a white noise, i.e. a
Gaussian stochastic process
with $<\eta(t)>=0$ and $<\eta(t)\eta(t')>=\delta(t-t')$.
The time correlation of $V$ is
$$
C_{VV}(t)=<V(t)V(0)>=<V^2>e^{-\gamma t} \,\, ,
$$
an easy computation gives
$$
D=\int_0^{\infty}C_{VV}(t) dt \,\, .
$$
Consider now a small perturbing force $f(t)=F\Theta(t)$,
where $\Theta(t)$ is the Heaviside step function.
The average  response of the
velocity  after a long time (i.e. the drift) to such a perturbation is given by:
\begin{equation}
\label{BM.6}
<\delta V>= {F \over  \gamma}\,\,.
\end{equation}
Defining the mobility $\mu$ as:
$$
<\delta V>= \mu F \,\,
$$
one easily obtains the  celebrated Einstein relation
$$
\mu={D \over {k_BT}}
$$ which gives a link between the diffusion coefficient (a property of
the unperturbed system) and the mobility, which measures the system's
reaction to a small perturbation. This constitutes the first example
of FDR.


\section{The classical linear response theory}

\label{sec:classical}

\subsection{Hamiltonian formulation of the FDR}
\label{hamiltonian} 
In this section we recall the derivation of Kubo's formula, which
holds in the case of equilibrium dynamics~\cite{T07}.  Let us consider a system
described by the Hamiltonian
$\mathcal{H}(\bP,\bQ,t)=\mathcal{H}_0(\bP,\bQ)-\mathcal{F}(t)A(\bP,\bQ)
$, where $\mathcal{H}_0(\bP,\bQ)$ is the time-independent part.  The evolution is
governed by the Hamilton equations: \bea \frac{\partial q_j}{\partial t}&=&
\frac{\partial\mathcal{H}_0}{\partial p_j}-\mathcal{F}(t)
\frac{\partial A}{\partial p_j}=\frac{\partial\mathcal{H}_0}{\partial
p_j} -\mathcal{F}(t)K_j^q \\ \frac{\partial p_j}{\partial t}&=&
-\frac{\partial\mathcal{H}_0}{\partial q_j}+\mathcal{F}(t)
\frac{\partial A}{\partial q_j}=
-\frac{\partial\mathcal{H}_0}{\partial q_j}-\mathcal{F}(t)K_j^p.
\label{hf1}
\eea
where we have introduced the ``generalized forces'' 
\begin{equation} 
K_j^q=\frac{\partial A}{\partial p_j},\qquad
K_j^p=-\frac{\partial A}{\partial q_j}.
\end{equation}

The associated $\Gamma$-phase space probability distribution 
evolves according to the Liouville equation:
\begin{equation} 
\frac{\partial}{\partial t}
f(\bP,\bQ,t)+i[\mathcal{L}_0+\mathcal{L}_{ext}(t)]f(\bP,\bQ,t)=0
\label{hf2}
\end{equation}
where $\mathcal{L}_0$ and $\mathcal{L}_{ext}$ are the Liouville
operators relative to the unperturbed Hamiltonian and to its
perturbation respectively. These can be expressed by means of 
Poisson brackets as:
\begin{equation}
i\mathcal{L}_0 f=\{f,\mathcal{H}_0\}
=\sum_j \left( \frac{\partial
\mathcal{H}_0}{\partial p_j}\frac{\partial}{\partial q_j}-\frac{\partial
\mathcal{H}_0}{\partial q_j}\frac{\partial}{\partial p_j} \right) f
\label{hf3}
\end{equation}
\begin{equation}
i\mathcal{L}_{ext}(t)f=-\mathcal{F}(t)\{f,A\}=-\mathcal{F}(t)\sum_j 
\left( \frac{\partial A}{\partial p_j}\frac{\partial}{\partial q_j}-
\frac{\partial A}{\partial q_j}\frac{\partial}{\partial p_j} \right) f \,\,.
\label{hf4}
\end{equation}

We assume that the system is in thermal equilibrium at $t=t_0$, when 
$\mathcal{H}=\mathcal{H}_0$, and its statistical properties are 
described by  the canonical
equilibrium distribution, $f(\bP,\bQ,t_0)=f_{eq}(\bP,\bQ)$, satisfying the
equation $i\mathcal{L}_{0} f_{eq}(\bP,\bQ)=0$. 
Switching on the perturbation, we  obtain an 
approximate solution of~\eqref{hf2}, in the form:
\begin{equation}
f(\bP,\bQ,t)=f_{eq}(\bP,\bQ)-i\int_{t_0}^t dt' 
e^{-i(t-t')\mathcal{L}_0}\mathcal{L}_{ext}(t')f_{eq}(\bP,\bQ)+ \dotsc
\label{hf6}
\end{equation}
which is valid to first order in $i\mathcal{L}_{ext}(t)$.
Using eq.~\eqref{hf6}, we can compute, to linear order in
$\mathcal{F}(t)$, the ensemble average of an arbitrary 
function $B(\bP,\bQ)$ of the phase space variables $(\bP,\bQ)$, at time $t$,
as follows:
\be
\langle B(t)\rangle=\int d\bP d\bQ B(\bP,\bQ)f(\bP,\bQ,t)
\label{hf7}
\ee
so that its change with respect to the equilibrium value is:
\begin{multline} 
\langle \Delta B(t)\rangle = \langle B(t) \rangle - \langle B(t=t_0)
\rangle =\\ -i  \int d\bP d\bQ B(\bP,\bQ)\int_{t_0}^t dt' 
e^{-i(t-t')\mathcal{L}_0}\mathcal{L}_{ext}(t')f_{eq}(\bP,\bQ)\label{hf8} \,\,.
\end{multline}
Substituting, we find 
\be
\langle \Delta B(t)\rangle= \int d\bP d\bQ B(\bP,\bQ)\int_{t_0}^t dt'
e^{-i(t-t')\mathcal{L}_0}
\mathcal{F}(t')\{f_{eq},A\}.
\label{hf9}
\ee
Using \eqref{hf4} and the fact that $f_{eq}(\bP,\bQ)$ depends on $(\bP,\bQ)$ only through $\mathcal{H}_0$,
we write:
\be
\langle \Delta B(t)\rangle= \int_{t_0}^t dt'\mathcal{F}(t')
\int d\bP d\bQ B(\bP,\bQ) e^{-i(t-t')\mathcal{L}_0}
\{\mathcal{H}_0,A\}\frac{\partial}{\partial
\mathcal{H}_0 }
f_{eq}(\mathcal{H}_0(\bP,\bQ)).
\label{hf10}
\ee
Since $\frac{\partial}{\partial
\mathcal{H}_0 } f_{eq}(\mathcal{H}_0(\bP,\bQ))=-\beta f_{eq}(\mathcal{H}_0(\bP,\bQ))$
and $\{\mathcal{H}_0,A\}=-\frac{d A}{d t}$
we can write
\be
\langle \Delta B(t)\rangle= \beta\int_{t_0}^t dt'\mathcal{F}(t')
\int d\bP d\bQ B(\bP,\bQ) e^{-i(t-t')\mathcal{L}_0}
\frac{d }{d t}A(\bP,\bQ)_{t=0} f_{eq}(\bP,\bQ)  \,\,.
\label{hf11}
\ee
Finally, using the unitarity of the Liouvillian operator,
we find:
\be
\langle \Delta B(t)\rangle= \beta\int_{t_0}^t dt'\mathcal{F}(t')
\int d\bP d\bQ  \frac{d }{d t}A(\bP,\bQ)_{t=t_0} f_{eq}(\bP,\bQ) 
e^{i(t-t')\mathcal{L}_0} B(\bP,\bQ).
\label{hf12}
\ee
If we introduce the response function, $R(t)$, through the relation
\be
\langle \Delta B(t)\rangle= \int_{t_0}^t dt'
R(t-t')\mathcal{F}(t'),
\label{hf13}
\ee
we may write
\be
R(t)=\beta \int d\bP d\bQ  \frac{d }{d t}A(\bP,\bQ)_{t=0} f_{eq}(\bP,\bQ) 
e^{i(t-t')\mathcal{L}_0} B(\bP,\bQ)=\beta\langle \dot A(t_0) B(t)\rangle 
\label{hf14}
\ee
which can also be rewritten as 
\be
R(t)=-\beta\langle A(t_0) \dot B(t)\rangle \,\,.
\label{hf15}
\ee
For later comparison, we can express the response function in a different
guise, by defining the dissipative flux $J(\bP,\bQ)$
\bea
&&J(\bP,\bQ)f_{eq}(\bP,\bQ)=-\{\mathcal{H}_0,A\}f_{eq}(\bP,\bQ)
=-\sum_j \left(\frac{\partial\mathcal{H}_0}{\partial q_j}  
K_j^q+\frac{\partial\mathcal{H}_0}{\partial p_j}
K_j^p \right)f_{eq}(\bP,\bQ)\nonumber\\&=&
\sum_j\left(K_j^q F_j^{(0)} -K_j^p \frac{p_j}{m} \right)f_{eq}(\bP,\bQ)
\label{hf16} 
\eea
where we have introduced the force acting on the j-th particle,
due to the Hamiltonian $\mathcal{H}_0$, as
$F^{(0)}_j= -\frac{\partial\mathcal{H}_0}{\partial q_j}$.
Using \eqref{hf16} we may write:
\be
R(t)=-\beta \int d\bP d\bQ  J(\bP,\bQ) f_{eq}(\bP,\bQ) 
e^{i(t-t')\mathcal{L}_0} B(\bP,\bQ)=-\beta\langle  J(t_0) B(t)\rangle \,\,.
\label{hf17}
\ee


\subsection{Green-Kubo relations}

\subsubsection{Diffusion}

In order to probe the diffusive properties of a system of particles, we apply
a spatially uniform force $h$ along an arbitrary direction, say
the $x$-direction, and consider how the induced change of the average
velocity depends on the external field.  Due to the mutual interactions
among the particles, this average velocity 
reaches a stationary value, because the drift current caused by
the field is balanced by the diffusion current.

In order to achieve such a situation, we add
the following  perturbation to the Hamiltonian:
\be
A(\bP,\bQ)=-h\sum_j q^x_j
\label{appd1}
\ee 
so that the dissipative flux, introduced above, turns out to be:
\be
J(\bP,\bQ)=-\frac{h}{m}\sum_j p^x_j.
\label{appd2}
\ee 
We, now, compute the change of the average velocity along the x-direction,
i.e. we set $B(\bP,\bQ)=U_x(\bP,\bQ)=\frac{1}{m}\sum_j p^x_j$.
By applying formula \eqref{hf17} we have
\be
\langle \Delta U_x(t)\rangle=-\beta\int_{t_0}^t dt'\mathcal{F}(t')
\int d\bP d\bQ J(\bP,\bQ) e^{-i(t-t')\mathcal{L}_0}U_x(\bP,\bQ)f_{eq}(\bP,\bQ) 
\label{appd3}
\ee 
or 
\bea
\langle \Delta U_x(t)\rangle&=&\beta\frac{h}{m^2}
\int_{t_0}^t dt'\mathcal{F}(t')
\int d\bP d\bQ \sum_j p^x_j e^{-i(t-t')\mathcal{L}_0}
\sum_k p^x_k f_{eq}(\bP,\bQ)\nonumber\\
&=&\beta\frac{h}{m^2}
\int_{t_0}^t dt'\mathcal{F}(t')\sum_{jk} \left\langle 
p^x_j(t_0) p^x_k(t-t')\right\rangle_{eq}.
\label{appd4}
\eea 
We, now, set $\mathcal{F}(t')=1$ for $t'\geq 0$, $t_0=0$ and 
take into account the fact that
the momenta of different particles are not correlated in the equilibrium
system:
\be
\langle \Delta U_x(t)\rangle=
\beta\frac{h}{m^2}
\int_{0}^t dt'\sum_{j} \left\langle 
p^x_j(t_0) p^x_j(t-t')\right\rangle_{eq}.
\label{appd5}
\ee
At this stage, we connect the self-correlation function
to the mobility, $\mu$, through  the relation
$\langle \Delta U_x(t)\rangle=\mu h$,
to conclude that
\be
\mu=\frac{\beta}{m^2}
\int_{0}^{\infty} dt'\sum_{j} \left\langle 
p^x_j(0) p^x_j(t')\right\rangle_{eq}.
\label{appd6}
\ee
Finally we obtain the diffusion constant from the 
Einstein relation $D=\mu/\beta$.
\subsubsection{Shear flow}
Let us now discuss how to calculate the coefficient of the shear viscosity.
Assume the following form of the perturbation operator:
\be
A(\bP,\bQ)=\frac{\gamma}{2}\sum_j (q^y_j p^x_j+q^x_j p^y_j)
\label{apps1}
\ee 
where $\gamma$ is the shear rate. The Hamiltonian with the above
perturbation generates the following equations of motion:
\begin{align}
\frac{\partial {q}^x_j}{\partial t}&=\frac{p^x_j}{m}+\gamma q^y_j; \;\;\;\;\; \frac{\partial {q}^y_j}{\partial t}=\frac{p^y_j}{m}+\gamma q^x_j\\
\frac{\partial {p}^x_j}{\partial t}&=F^x_j-\gamma p^y_j; \;\;\;\;\; \frac{\partial {p}^y_j}{\partial t}=F^y_j-\gamma p^x_j
\end{align}
which are known as unthermostatted SLLOD equations describing a
sheared fluid~\cite{EM90}. 

In this case the dissipative flux is:
\be
J(\bP,\bQ)=\frac{\gamma}{2}\sum_j \left(\frac{2}{m}p^x_j p^y_j  +q^y_j F^x_j+q^x_j F^y_j\right)
\label{apps2}
\ee 
and turns out to be proportional to the 
xy component of the pressure tensor
\be
P^{xy}=\frac{1}{2V}\sum_j
\left( \frac{2}{m}p^{x}_j p^{y}_j+ q^y_j F^x_j+ q^x_j F^y_j\right)
\label{apps3}
\ee
via the relation $J(\bP,\bQ)=\gamma V P^{xy}$, where $V$ is the
volume of the system.
By following steps similar to those leading to \eqref{appd6}
we obtain
\bea
\langle \Delta P^{xy}(t)\rangle&=&-\beta
\int_{0}^t dt'
\int d\bP d\bQ J(\bP,\bQ)  e^{-i(t-t')\mathcal{L}_0} P^{xy}(\bP,\bQ) 
f_{eq}(\bP,\bQ)\nonumber\\
&=&-\beta\gamma V
\int_{0}^t dt' \left\langle P^{xy}(0) P^{xy}(t')
\right\rangle_{eq} \,\,,
\label{apps4}
\eea 
where again we have used \eqref{hf17} and  
$\mathcal{F}(t')=1$ for $t'\geq 0$.

Finally using the definition of the shear viscosity $\eta$
\be
\eta=-\lim_{t\to \infty}\frac{\langle\Delta P^{xy}(t)\rangle}{\gamma}
\ee
we obtain:
\be
\eta=\beta V
\int_{0}^{\infty} dt' \left\langle 
P^{xy}(0) P^{xy}(t')\right\rangle_{eq} \,\,.
\label{apps5}
\ee
Eqs. \eqref{appd6} and \eqref{apps5} are just two examples
of  more general relations known as Green-Kubo relations,
connecting the transport coefficients to the equilibrium
two-time correlations of a system.

A similar method is employed in order to derive relations
for the thermal conductivity, $\lambda$, and the bulk  viscosity coefficient,
$\kappa$.
The first of these relations reads:
\be
\lambda=\frac{1}{3 V k_B T^2}\int_{0}^{\infty} dt' \left\langle 
{\bf S}(0)\cdot {\bf S}(t')\right\rangle_{eq},
\label{apps6}
\ee
where 
\be
{\bf S}= \sum_j \left(  \frac{1}{2m}p^2_j -<\epsilon_j> \right) 
\frac{{\bf p}_j}{m}
+\frac{1}{2}\sum_i \sum_{j\neq i}
\left( {\bf q}_{ij}\cdot{\bf F}_{ij}+U({\bf q}_{ij})\right)  
\frac{{\bf p}_j}{m} 
\label{apps7}
\ee
where $<\epsilon_j>$ is the enthalpy per particle, $U({\bf q}_{ij})$
is the pair potential and
$ {\bf q}_{ij}= {\bf q}_{i}- {\bf q}_{j}$. For details see~\cite{MQ00}.

Similarly, the bulk  viscosity reads:
\be
\kappa=\frac{1}{9 V k_B T}\int_{0}^{\infty} dt' 
\sum_{ab} \left\langle \delta \mathcal{J}^{aa}(0) \delta \mathcal{J}^{bb}(t')\right\rangle_{eq}\,\,,
\ee
where $a$ and $b$ run over the spatial components,
\be
\mathcal{J}^{ab}=\sum_j \frac{p_{i}^a p_{i}^b}{m}+\sum_j q_{i}^a F_{i}^b
\label{apps8}
\ee
and
\be
\delta \mathcal{J}^{ab}=\mathcal{J}^{ab}-\langle \mathcal{J}^{ab} \rangle.
\ee

\subsection{Linear Response for stochastic dynamics}

\subsubsection{Langevin Dynamics}

Previously we discussed only Hamiltonian systems perturbed by a weak
time-dependent external force. In the present section, we derive in a
different fashion the FDR in systems subject to dissipative forces and
random forces.  Let us consider the following stochastic differential
equation for the scalar variable $A(t)$
\begin{equation}
\frac{d A(t)}{dt}=-\gamma
\frac{\partial H[A(t)]}{\partial A(t)}+\sqrt{2\gamma T} \eta(t)
\label{u1}
\end{equation}
where  $H[A(t)]$ is a function of  $A(t)$ and $\eta$ is a white noise.


We define the two-time correlation function as:
\begin{equation}
C(t,t')=\langle A(t) A(t')\rangle
\label{u3}
\end{equation}
and assume without loss of generality $t>t'$.
In order to calculate the response of the
system  to a time dependent external perturbation, we consider the 
following perturbation
\begin{equation}
H_h[ A(t)]= H[ A(t)]-h(t)A(t) \,\,.
\label{u4}
\end{equation}
The variation of the quantity $A(t)$, induced by the presence
of $h(t)$, is measured by the two-time response function, $R(t,t')$, defined
as
\begin{equation}
R(t,t')=\frac{\langle\delta  A(t)\rangle}{\delta h(t')}
\label{u5}
\end{equation}
The Fluctuation Dissipation theorem relates the two-time correlation function
with the response function.
Let us compute the following derivative of $C(t,t')$:
\begin{equation}
\frac{\partial C(t,t') }{\partial t}=
\left\langle \left[-\gamma
\frac{\partial H[A(t)]}{\partial A(t)}+\eta(t)\right]A(t')\right\rangle
\label{u6}
\end{equation}
and subtract the derivative with respect to the argument $t'$:
\begin{equation}
\frac{\partial C(t,t') }{\partial t}-\frac{\partial C(t,t') }{\partial t'} =
-\gamma\langle
\frac{\partial H[A(t)]}{\partial A(t)}A(t')\rangle
+\gamma\langle\frac{\partial H[A(t')]}{\partial A(t')}A(t)\rangle
-\langle A(t)\eta(t')\rangle
\label{u7}
\end{equation}  
where we dropped a noise term because of  causality.
We use now the property, which is proved by
 Onsager's argument (cf.~\cite{K87}, pages 
49-50):
\begin{equation}
\langle A(t) B(t+\tau)\rangle=\langle A(t) B(t-\tau)\rangle
=\langle A(t+\tau) B(t)\rangle
\label{ons1}
\end{equation} 
the first equality derives from time reversal,
the last equality follows from the invariance of the equilibrium 
state under a time translation $t\to t+\tau$.
Therefore, the first two terms in the r.h.s. of eq. \eqref{u7}
cancel each other. Finally,
the last term can be evaluated by means of Novikov's theorem~\cite{n65}
which states that if $\eta$ is a Gaussian process and $g$ is a function of $A(t)$ then
\begin{equation}
\langle g(A(t))\eta(t')\rangle=\int dt''
\frac{\delta \langle g(A(t))\rangle}{\delta \eta(t'')}
\langle \eta(t'')\eta(t')\rangle=\int dt''
\frac{\delta \langle g(A(t))\rangle}{\delta h(t'')}
\langle \eta(t'')\eta(t')\rangle \,\,.
\label{u8}
\end{equation}  
Setting $g(A(t))=A(t)$ we obtain, using eq.(\ref{u5}) and the property
$\langle \eta(t') \eta(t'')\rangle=\delta(t'-t'')$,
\begin{equation}
\langle A(t)\eta(t')\rangle=\int dt''
R(t,t'')\langle \eta(t'')\eta(t')\rangle
=2\gamma T R(t,t')
\label{u9}
\end{equation}
and we find:
\begin{equation}
\frac{\partial C(t,t') }{\partial t}-\frac{\partial C(t,t') }{\partial t'} 
=-2\gamma T R(t,t') \,\,.
\label{u10}
\end{equation}
At equilibrium, the two-time average depends only on the time difference, 
hence, setting $t'=t+s$, we must have: 
\begin{equation}
\frac{\partial C(t,t+s) }{\partial t}=0
\label{u11}
\end{equation}
or
\begin{equation}
\frac{\partial}{\partial t}\langle A(t)A(t+s)\rangle=
-\frac{\partial}{\partial s}\langle A(t)A(t+s)\rangle
\label{u12}
\end{equation}
so that we have
\begin{equation}
\frac{\partial C(t,t') }{\partial t}
=-\frac{\partial C(t,t') }{\partial t'}
\label{u13}
\end{equation}
\begin{equation}
\frac{\partial C(t,t') }{\partial t'} =
\gamma T R(t,t') \,\,.
\label{u14}
\end{equation}
We conclude that the response function has the form:
\begin{equation}
R(t,t')=
\frac{1}{\gamma T}\theta(t-t')\frac{\partial C(t,t') }{\partial t'}  \,\,.
\label{u15}
\end{equation}

In order to connect this result with the static susceptibility, we switch on a constant perturbation $\epsilon$
at time $t=0$, and compute
the change induced on the variable $A$, which is given,
using relation~\eqref{u5}, by the formula \be \langle
A(t)\rangle-\langle A(0)\rangle=\epsilon \int_0^t dt' R(t,t')=
\epsilon \frac{1}{\gamma T}[C(t,t)-C(t,0)] \,\,.\ee 

After defining  the integrated response $\chi$ by the formula:
\begin{equation}
\chi(t,0)=\frac{1}{\gamma T}[C(t,t)-C(t,0)],
\label{ex2}
\end{equation}
we can verify that its $t\to\infty$  limit coincides with  the static response function
\be
\chi_{st}=\frac{1}{\gamma T}[\langle A^2\rangle-\langle A\rangle^2] \,\,.
\ee
Here we have used the fact that the values of $A$ at time $t$ and at time $0$ become uncorrelated as
$t\to\infty$ and that the average factorizes.

\subsubsection{Linear Response in Fokker-Planck equation}
The probability distribution function of the process
described by a stochastic differential equation evolves according to a
Fokker-Planck equation. Within the Fokker-Planck formalism, we can
generalize the method of section \ref{hamiltonian} to systems subjected
to an average drift and a fluctuating term.

In such a case, the distribution $\fxt$ evolves according
to:
\begin{equation} \label{liouville}
\frac{\partial}{\partial t}
\fxt=\left(\mathcal{L}^0_{FP}+\mathcal{L}^{ext}_{FP}(t)\right)\fxt
\end{equation}
where $\mathcal{L}^0_{FP}$ represents the Fokker-Planck operator
of the unperturbed system, defined by
\begin{equation}
\mathcal{L}^0_{FP}\fxt =-\sum_j \partial_j \left[ D^{(1)}_j({\bf x})\fxt \right ]+
\sum_{jk}\partial_j\partial_k D^{(2)}_{jk}({\bf x})\fxt
\end{equation}
where the functions $D^{(1)}_j$ and $D^{(2)}_{jk}$
are related to the underlying Ito stochastic differential equations
by
\begin{equation} 
\frac{d x_j(t)}{dt}=D^{(1)}_j({\bf x})+\sum_k M_{jk}({\bf x})\eta_k(t),
\end{equation}
and 
\be
D^{(2)}_{jk}({\bf x})=\frac{1}{2}M_{jl}({\bf x})M^{T}_{lk}({\bf x})
\ee
where the white noise, $\eta_k$, is a Gaussian process with 
\be
\langle \eta_k(t)\rangle=0, \;\;\;\; \langle \eta_j(t)\eta_k(t')\rangle =\delta_{jk} \delta(t-t') \,\,.
\ee
We probe the response of the system, by adding an infinitesimal
time-dependent perturbation of the form 
\be
\delta D_j^{(1)}({\bf x},t)=\mathcal{F}(t) K_j({\bf x}) \,\,.
\label{fperturbation}
\ee
Such a perturbation induces the following 
change in the Fokker-Planck operator:
\begin{equation} 
\mathcal{L}^{ext}_{FP}(t)\fxt =- \mathcal{F}(t)
\sum_j\partial_j (K_j({\bf x})\fxt) \,\,.
\label{fpl}
\end{equation}

We first determine the perturbed distribution, 
assuming that the system was stationary
in the infinite past, i.e.: $f({\bf x},-\infty)=f_{st}({\bf x})$ and
\begin{equation}
\mathcal{L}^{0}_{FP}\fsta=0 \,\,.
\label{fpst}
\end{equation}
An approximate solution of the perturbed problem
\begin{equation} 
\frac{\partial}{\partial t}
\fxt=[\mathcal{L}_{FP}^0+\mathcal{L}^{ext}_{FP}(t)]\fxt
\label{fpl1}
\end{equation}
to  first order in the
perturbation $\mathcal{L}^{ext}_{FP}$ can be represented as:
$f({\bf x},t)\simeq \fsta+\Delta f({\bf x},t)$, where $\Delta f$
is formally given by:
\begin{equation}
\Delta f({\bf x},t)=
\int_{- \infty}^t dt' e^{(t-t')\ofp}\Lex(t')\fsta \,\,.
\label{solution}
\end{equation}

With this approximation, the deviation of the noise average of a physical
quantity $B({\bf x})$, due to the
perturbation~\eqref{fperturbation}, can be written as:
\begin{equation} 
\langle \Delta B(t)\rangle  
= \langle B({\bf x},t)\rangle - \langle B({\bf x},t=-\infty)
\rangle = \int d^N x B({\bf x})\Delta f({\bf x},t)
\label{fpdeviation2}
\end{equation}
or, substituting eq.~\eqref{solution},
\begin{equation} 
\langle\Delta B(t)\rangle=\int_{- \infty}^t dt' 
\int d^N x  B({\bf x})e^{\ofp(t-t')}\Lex(t')\fsta 
\label{fpdeviation3}
\end{equation}
compactly written as
\begin{equation} 
\Delta B(t)  = \int_{- \infty}^t dt' 
R(t-t')\mathcal{F}(t')
\label{fpdeviation4}
\end{equation}
where the response function $R(t-t')$, for $t\geq t'$, reads:
\begin{equation} 
R(t)  =-
\int d^N x B({\bf x}) e^{\ofp t}\sum_j\partial_j (  K_j({\bf x})\fsta) 
\label{fpdeviation5}
\end{equation}
and $R(t)=0$ for $t < 0$, where~\eqref{fpl} has been used. 

In order to relate the response function 
to the correlation in the unperturbed case, between two operators  
$B({\bf x},t)$ and $\tB({\bf x'},t')$, defined as:
\be
\langle B(t) \tB(t')\rangle= \int\int d^N xd^N x' B({\bf x})\tB({\bf x'})
W_2({\bf x},t;{\bf x'},t')
\label{cor1}
\ee
where $W_2({\bf x},t;{\bf x'},t')$ is the joint 
probability distribution for the unperturbed case, we use the result:
\bea
W_2({\bf x},t;{\bf x'},t')=P({\bf x},t,{\bf x'},t')
f^{st}({\bf x'})
=e^{\ofp (t-t')}\delta({\bf x}-{\bf x'})f^{st}({\bf x'})
\eea
where $P({\bf x},t,{\bf x'},t')$ is the transition probability from 
configuration ${\bf x'}$ to ${\bf x}$. Substituting
such an expression in \eqref{cor1} we obtain
\bea
\langle B(t) \tB(t')\rangle= \int && d^N x \bigr[
\theta(t-t') B({\bf x})e^{\ofp (t-t')} \tB({\bf x})\nonumber\\
&+&\theta(t'-t) \tB({\bf x})e^{-\ofp (t-t')} B({\bf x})\bigr]\fsta \,\,.
\label{corrba}
\eea
Hence, provided we make the identification
\be
\tB(t')=\beta \frac {d A(t')}{dt'}= -\frac{1}{\fsta}\sum_j \partial_j (K_j({\bf x})\fsta ) \,\,,
\ee
we can rewrite the response function as:
\be
R(t)  =\beta
\int d^N x B({\bf x}) e^{\ofp t} \frac{d A({\bf x})}{d t}\fsta=\beta
\left \langle B(t)\left(\frac{ d A(t)}{dt} \right)_{t=0}\right\rangle ,
\label{fpdeviation7}
\end{equation}
so that the analogy with the Hamiltonian formulation of the FDR is complete.
\subsubsection{Kramers equation}
As an important special case, we consider the motion
of a colloidal particle in an external field.
To obtain a more useful expression and make contact with the FDR
in Hamiltonian systems, 
we assume the Fokker-Planck equation to be of the form of a Kramers
equation. 
Such stochastic dynamics can be considered as a Hamiltonian dynamical
system, in which the momentum equation is augmented by a friction and a noise
term. Denoting ${\bf x}=(p_j,q_j)$, we have
\bea
\frac{d q_j}{dt}&=& \frac{\partial\mathcal{H}_0}{\partial p_j}
+\mathcal{F}(t)K_j^q \\
\frac{d p_j}{dt} &=& 
-\frac{\partial\mathcal{H}_0}{\partial q_j}
-\gamma p_j+\mathcal{F}(t)K_j^p+\sqrt{2\gamma m k_B T}\eta_j \,\,.
\label{kr1}
\eea

The associated linear evolution operators are
\be
\mathcal{L}_{FP}^0=-\sum_j\left(\frac{\partial\mathcal{H}_0}{\partial p_j}
\frac{\partial }{\partial q_j}-\frac{\partial\mathcal{H}_0}{\partial q_j}
\frac{\partial }{\partial p_j}\right)
+\gamma\sum_j\left(\frac{\partial }{\partial p_j} p_j
-\beta^{-1}\frac{\partial^2 }{\partial p_j^2} \right)
\label{kr2}
\ee
and
\be
\mathcal{L}_{FP}^{ext}(t)=-\mathcal{F}(t)\sum_j
\left(K_j^q\frac{\partial }{\partial q_j}
+K_j^p\frac{\partial }{\partial p_j}\right) \,\,.
\label{kr3}
\ee
When $\mathcal{F}(t)=0$ the system has an equilibrium distribution
$f^{st}(\bP,\bQ)=\mathcal{N}\exp(-\beta\mathcal{H}_0)$, so that we may explicitly
compute the expression:
\bea
\sum_j K_j(\bP,\bQ)\partial_j f^{st}(\bP,\bQ) &=&-\beta 
(K_j^q \frac{\partial\mathcal{H}_0}{\partial q_j}+
K_j^p\frac{\partial\mathcal{H}_0}{\partial p_j})f^{st}(\bP,\bQ) \nonumber\\
&=&\beta (K_j^q F_j  -K_j^p\frac{p_j}{m})f^{st}(\bP,\bQ)\\ &=&J(\bP,\bQ)f^{st}(\bP,\bQ) 
\label{kr4}
\eea
where the dissipative flux $J(\bP,\bQ)$
is identical to the one in Eq.~\eqref{hf16}.

By comparing eq.~\eqref{kr4} and \eqref{fpdeviation5} one sees 
that, for $t>0$, $R(t)$ can be expressed as the two-time correlation:
\be
R(t)=-\beta\int d^N x B(\bP,\bQ) e^{\ofp t}  J(\bP,\bQ) f^{st}(\bP,\bQ) 
\ee
which is similar to eq. \eqref{hf17}, provided $\mathcal{L}^{0}_{FP}$ is replaced by $-i\mathcal{L}^{0}$.

\subsubsection{Smoluchowski equation}

\label{overdamped}
The Smoluchowski equation can be considered the Fokker-Planck (FP) equation
which arises in the
overdamped
limit, $\gamma\to\infty$ , of Langevin's equation \eqref{kr1}:
\be
\frac{d q_j}{dt} =-\Gamma\frac{\partial\mathcal{H}_0}{\partial q_j}
+\sqrt{2\Gamma k_B T}\eta_j
\label{smol0}
\ee
with $\Gamma=1/m\gamma$. The associated FP equation reads:
\be 
\frac{\partial}{\partial t}f(\{q\},t)
=\left(\mathcal{L}^0_{S}+\mathcal{L}^{ext}_{S}(t)\right)f(\{q\},t)
\label{smol1}
\ee
where $\mathcal{L}^0_{S}=\Gamma
\sum_j\partial_j(\partial_j+\beta\partial_j  \mathcal{H}_0)$
and $\mathcal{L}^{ext}_{S}(t)=-\Gamma\mathcal{F}(t)\sum_j K_j\partial_j$.
In this case we find that the current is determined by the following
condition:
\bea
\sum_j K_j(\{q\})\partial_j f^{st}(\{q\})&=&-\beta 
K_j \frac{\partial\mathcal{H}_0}{\partial q_j}f^{st}(\{q\}) \nonumber\\
&=&\beta K_j F_j f^{st}(\{q\}) =J(\{q\})f^{st}(\{q\})
\label{smol2}
\eea
with $f^{st}(\{q\})=\mathcal{N}\exp(-\beta\mathcal{H}_0^{config}(\{q\}))$.
We conclude that in this case the response function is:
\be
R(t)=-\beta\int d^N q B(\{q\}) e^{\mathcal{L}^0_{S} t}  J(\{q\})
f^{st}(\{q\}) \,\,.
\label{smol3}
\ee

\subsection{Johnson-Nyquist spectrum and Onsager regression hypothesis}

Even before Onsager's and Kubo's great synthesis, important results 
had been obtained in some specific cases or under special
conditions. Such contributions played a significant role in the development
of statistical physics. In the following we illustrate in detail  some of them.

\subsubsection{Frequency response and Johnson-Nyquist spectrum}

Equations~\eqref{hf14}-\eqref{hf15}, relating the response to a given
perturbation and a suitable equilibrium correlation, is often useful
in its frequency representation.
If the external perturbation is periodic in time, 
\begin{equation} \label{periodic}
\mathcal{F}(t)=\RE [K_0\e^{i \omega t}],
\end{equation}
because of linearity, the response reads
\begin{equation}
\langle \Delta B(t) \rangle = \RE [\chi(\omega) K_0 \e^{i \omega t}],
\end{equation}
where $\chi(\omega)$ is the admittance  defined as the Fourier
transform of the response function $R(t)$:
\begin{equation}
\chi(\omega)=\int_{-\infty}^{\infty} R(t)\e^{-i\omega t} dt=\beta \int_{t_0}^{\infty} \langle \dot{A}(t_0)B(t) \rangle\e^{-i\omega t} dt \,\,.
\end{equation}

As an interesting example, we discuss the  generalized
Langevin equation, with a retarded frictional force $\gamma(t)$:
\begin{equation} \label{genelange}
\frac{du(t)}{dt}=-\int_{t_0}^t \gamma(t-t')u(t')dt'+\frac{1}{m}\eta(t)+\frac{1}{m}\mathcal{F}(t), \;\;\;t>t_0
\end{equation}
where the external force $\mathcal{F}(t)$ is also included in this
equation. The usual linear Langevin equation is the limiting case with
$\gamma(t)=\gamma \delta(t)$.  It is immediate to see that the
response of $u(t)$ to a periodic perturbation of the
form~\eqref{periodic} reads
\begin{equation}
\langle u(t) \rangle=\RE[\chi(\omega)K_0 \e^{i \omega t}],
\end{equation}
where the admittance is given by 
\begin{equation}
\chi(\omega)=\frac{1}{m}\frac{1}{i\omega+\tilde{\gamma}(\omega)},
\end{equation}
with $\tilde{\gamma}(\omega)$ the Fourier transform of $\gamma(t)$. In
other words, for a periodic perturbation, the admittance is 
analogous to the mobility. At equilibrium, $\mathcal{F} \equiv 0$, the
self-correlation of $u(t)$ can be calculated, and its Fourier
transform reads
\begin{equation} \label{halfspectrum}
\int_0^\infty \e^{-i \omega t} \langle u(t_0)u(t_0+t) \rangle dt=\langle u^2 \rangle \frac{1}{i \omega + \tilde{\gamma}(\omega)}
\end{equation}
and therefore the FDR 
\begin{equation}
\chi(\omega)=\frac{1}{m\langle u^2 \rangle}\int_0^\infty \langle u(t_0)u(t_0+t)\rangle \e^{-i \omega t} dt
\end{equation}
is recovered. Note also that Eq.~\eqref{halfspectrum} allows us to
calculate the power spectrum of $u(t)$ at equilibrium, i.e.
\begin{equation}
S_u(\omega)=\int_{-\infty}^\infty \e^{-i \omega t} \langle u(t_0)u(t_0+t) \rangle dt=\langle u^2 \rangle \frac{2\tilde{\gamma}(\omega)}{[\tilde{\gamma}(\omega)]^2+\omega^2}.
\end{equation}

A particular case of Eq.~\eqref{genelange} is the equation governing
the charge $Q(t)$ contained in a condenser of capacity $C$, in a simple
$RC$ circuit without any externally applied  electromotive force:
\begin{equation}
R\frac{dQ}{dt}=-\frac{Q(t)}{C}+\eta(t),
\end{equation}
being $\eta(t)$ a noisy internal electromotive force due to the
interactions with a thermostat at temperature $T$.  A series of
experiments by Johnson~\cite{J28}, supported by the theory of
Nyquist~\cite{N28}, demonstrated the existence of such spontaneous
fluctuations of electric tension at the edges of the resistor $u=Q/C$,
due to the thermal motion of charge carriers. The measured quantity
in the experiment was the power spectrum of these tension
fluctuations, i.e. $S_u(\omega)$. The above equation for the $RC$ circuit can be written in the form
\begin{equation}
\frac{du}{dt}=-\frac{u(t)}{RC}+\eta'(t),
\end{equation}
with $\eta'(t)=\eta/RC$. This is equivalent to Eq.~\eqref{genelange}
with $\gamma(t)=\frac{1}{RC}\delta(t)$,
i.e. $\tilde{\gamma}(\omega)=\frac{1}{RC}$, resulting in
\begin{equation} \label{spettro}
S_u(\omega)=\langle u^2 \rangle \frac{2RC}{1+(RC\omega)^2}.
\end{equation}
The energy of the capacitor is $E=Q^2/2C$, and the
equipartition of energy, required at equilibrium, imposes that
$\langle E \rangle=\frac{k_B T}{2}$, which yields $\langle u^2 \rangle = \langle
Q^2 \rangle/C^2=2\langle E \rangle/C= k_BT/C$. This result, put into
Eq.~\eqref{spettro}, gives the spectrum of the Johnson-Nyquist noise
\begin{equation}
S_u(\omega)=\frac{2Rk_BT}{1+(\omega RC)^2}.
\end{equation}
At low frequencies, $\omega \ll 1/RC$, the power spectrum becomes
constant and does not depend on the capacity of the circuit:
\begin{equation}
S_u(\omega) \simeq 2Rk_BT.
\end{equation}

For a modern linear response study of chaotic systems in frequency
domain we refer the reader to~\cite{TKP01}.

\subsubsection{The Onsager regression hypothesis}

In his two seminal papers of 1931~\cite{O31,O31b}, Onsager presented his {\em
regression hypothesis}: ``{\em ...  the average regression of fluctuations
will obey the same laws as the corresponding macroscopic irreversible
process}''~\cite{O31b}. This principle states the equivalence between the law
governing the relaxation of spontaneous deviations from equilibrium
and the law obeyed by the corresponding irreversible process, which is
triggered by  an initial small deviation from equilibrium. The
fluctuation-dissipation theorem~(2.16), justifies this hypothesis
 in a natural way. 

Onsager referred to the relaxation of a macroscopic variable
$\alpha(t)$, an observable function of
the microscopic degrees of freedom, $\alpha(\bP(t),\bQ(t))$, whose
equilibrium average, $\alpha_0$ is given by
\begin{equation}
\alpha_0=\langle \alpha(\bP,\bQ) \rangle=\frac{\int d\bP d\bQ e^{-\beta \mH(\bP,\bQ)} \alpha(\bP,\bQ)}{\int d\bP d\bQ e^{-\beta \mH(\bP,\bQ)}}
\end{equation}
with the unperturbed Hamiltonian $\mH(\bP,\bQ)$.
Fluctuations  around this value are denoted by $\delta
\alpha(t)=\alpha(t)-\alpha_0$. The average regression at time $t_1$ of
a fluctuation $\delta \alpha(t_0)$, at time $t_0$, is
given by the auto-correlation function $\langle \delta \alpha(t_0)
\delta \alpha(t_1) \rangle$. The self-correlation function depends
only on the difference $t_1-t_0$, since we are considering a
time-translation invariant (equilibrium) state, therefore in the
following we take $t_0=0$ and $t_1=t$.

The irreversible processes considered by Onsager, on the other hand,
consist of the average evolution of the observable $\alpha(t)$, with
$t>0$, when a small perturbation $\Delta \mH$ is applied to the
Hamiltonian from time $-\infty$ up to time $0$, and then is switched
off. This corresponds to a perturbation of the Hamiltonian  $\Delta \mH=-\mathcal{F}
\alpha(\bP,\bQ)$, so that the conjugated field of the
perturbation is $\alpha(\bP,\bQ)$ itself.  The average evolution for
times $t>0$ (the so-called ``after-effect function'') is given by
\begin{equation}
\overline{\alpha}(t)=\frac{\int d\bP d\bQ e^{-\beta \mH(\bP,\bQ)+\Delta \mH(\bP,\bQ) } \alpha(\bP,\bQ)}{\int d\bP d\bQ e^{-\beta \mH(\bP,\bQ) +\Delta \mH(\bP,\bQ)}}.
\end{equation}
Furthermore, the perturbation $\mathcal{F}(t)=F \Theta(-t)$ is such
that $\overline{\alpha}(0)-\alpha_0=\delta \alpha(0)$.
The regression hypothesis can therefore be formulated in the following
way:
\begin{equation} \label{regression}
\frac{\overline{\alpha}(t)-\alpha_0}{\overline{\alpha}(0)-\alpha_0}=\frac{\langle \delta \alpha(t) \delta \alpha(0) \rangle}{\langle (\delta \alpha(0))^2\rangle}=\frac{\langle \alpha(t)\alpha(0) \rangle - \alpha_0^2}{\langle \alpha^2 \rangle - \alpha_0^2}.
\end{equation}

By definition, the irreversible evolution in the l.h.s. is governed by
macroscopic {\em linear} laws, as discussed in Onsager's two 1931
papers, so that the fluctuation-dissipation theorem~(2.16) can
be used to compute $\overline{\alpha}(t)$. For this purpose, it is
convenient to consider $\mH+\Delta \mH$ as the unperturbed Hamiltonian,
with ensemble averages given by $\langle ... \rangle_{\Delta \mH}$,
and $-\Delta \mH$ as the perturbation switched {\em on} at time
$0$. From~(2.16) one gets:
\begin{multline} \label{relaxation}
\overline{\alpha}(t)=\langle \alpha \rangle_{\Delta \mH}+\beta F\int_0^t dt'\langle \alpha(0) \dot{\alpha}(t-t')\rangle_{\Delta \mH} =\\\langle \alpha \rangle_{\Delta \mH}+\beta F [\langle \alpha(t) \alpha(0)\rangle_{\Delta \mH}-\langle \alpha^2 \rangle_{\Delta \mH}].
\end{multline}
At $t=0$ one has $\langle \alpha \rangle_{\Delta
\mH}=\alpha_0+\delta\alpha(0)$. Asymptotically, for $t \to +\infty$, one
 has $\overline{\alpha}(t) \to \alpha_0$, and $\langle \alpha(t)
\alpha(0) \rangle_{\Delta \mH} \to (\langle \alpha \rangle_{\Delta
\mH})^2$, which, substituting in~\eqref{relaxation}, gives $\beta F \langle
\alpha^2 \rangle_{\Delta \mH}=\beta F (\langle \alpha \rangle_{\Delta
\mH})^2+\delta \alpha(0)$. With these relations, Eq.~\eqref{relaxation} becomes
\begin{equation} \label{negligenza}
\overline{\alpha}(t)=\alpha_0+\beta F[\langle \alpha(t)\alpha(0) \rangle_{\Delta \mH}-(\langle \alpha \rangle_{\Delta \mH})^2].
\end{equation}
It is now plain that, to be consistent with the linear
approximation, all the effects of the perturbation $\Delta \mH$ can be
neglected in the functions appearing in the r.h.s, which are of second
order in $\alpha$. This means that all the subscripts in the perturbed
averages $\langle ... \rangle_{\Delta \mH}$ can be dropped and
formula~\eqref{negligenza}, i.e.  the regression hypothesis, turns out to be
 a particular case of the Fluctuation-Dissipation theorem.

\subsubsection{The Onsager reciprocal relations}
\label{sec:reciprocal}
In~\cite{O31,O31b}, Onsager derived a relation among off-diagonal
elements of the evolution matrix for linear irreversible processes. It
was, in fact, the central result of that work. Again, it is possible,
{\em a posteriori}, to offer a simpler derivation of that relation,
starting from the Green-Kubo formulas. As we have seen, a direct
consequence of FDR is the possibility to
write the transport coefficients as time-integrals of suitable
correlation functions. In general, if one has $N$  conserved
quantities, $\alpha_i(\bx,t)$, $i=(1,..N)$, and
small spatial and temporal variations of the fields, a linear
 system of evolution equations can be written as
\begin{subequations}
\begin{align}
\frac{\partial \alpha_i(\bx,t)}{\partial t} &=-\nabla \cdot \mathbf{J}_i(\bx,t)\\
\mathbf{J}_i(\bx,t) &=L_{ij}\pmb{\chi}_j,
\end{align}
\end{subequations}
where the fluxes $\mathbf{J}_i$, the transport coefficients $L_{ij}$
and the thermodynamic forces $\pmb{\chi}_j$ have been introduced. The
latter are the forces conjugated to the fluxes in the dissipation
function (entropy production) of non-equilibrium thermodynamics~\cite{DEGM}. A
thermodynamic force $\pmb{\chi}_j$ is in general a gradient of the
same quantity $\alpha_i(\bx,t)$: this is the application of linear
response theory to {\em internal} forces, in fact the fluxes are
responses to internal gradients. For instance, the heat flux $q_x$
along the $\hat{x}$ direction is expressed as $q=-\kappa \partial_x T$
with $\kappa$ the thermal conductivity. Note that for each flux-force
pair $(i,j)$, the coefficient $L_{ij}$ is indeed a matrix of indexes
$(k,l)$ for all possible pairs of the Cartesian components, i.e.
$L_{ij}=(L_{ij}^{kl})$ with $k,l = x,y,z$. We recall the Green-Kubo
relations for these coefficients:
\begin{equation}
L_{ij}^{kl}=\beta V \int_0^\infty ds \langle J_i^k(t) J_j^l(0)\rangle.
\end{equation}
Onsager's reciprocal relations between $L_{ij}^{kl}$ and $L_{ji}^{lk}$ appears as a consequence of microscopic time
reversibility. The equations of motion for the evolution of a phase
space point (describing a system of $N$ particles)
are invariant for the time reversal operation  $I$, which
transforms $\bQ \to \bQ$, $\bP \to -\bP$ and $t \to -t$. This
invariance  implies the invariance of the phase space probability
measure. Denoting by $S^t$ the time evolution operator, which
satisfies $I S^t=S^{-t}I$, the following chain of identities
holds for a generic time correlation function:
\begin{multline}
\langle [S^t J_i^k] J_j^l  \rangle= \langle I \{[S^t J_i^k]  J_j^l\} \rangle= \langle [S^{-t} I J_i^k] I J_j^l \rangle \\= \langle [S^t I J_j^l] I J_i^k \rangle=I_iI_j \langle [S^t J_j^l ] J_i^k \rangle,
\end{multline}
where the invariance to a translation of time has also been used. In
the last passage, we have exploited the fact that the fluxes have the
opposite parity of their conjugated density under time reversal, so
that $I J_i^k = I_i J_i^k$ where $I_i= \pm 1$, for example heat and
mass fluxes are odd, while the momentum flux is even. From the above
identities, the reciprocal relations follow:
\begin{equation}
L_{ij}^{kl}=I_i I_j L_{ji}^{lk}.
\end{equation}


\section{Fluctuation dissipation in chaotic systems and statistical physics}
\label{sec:chaos}

The Fluctuation-Response theory was originally developed in the
context of equilibrium statistical mechanics of Hamiltonian
systems\cite{K66,K86}.  In this Section we will show that a
generalized FDR   holds under rather general
hypotheses\cite{dh75,ht77,ht82,FIV90}: hence also for non-Hamiltonian systems.
For instance, it holds for a vast class of systems with chaotic
dynamics, something that is particularly relevant in
hydrodynamics\cite{Kr59,Kr00}.

Usually the Fluctuation-Response problem has been studied for infinitesimal
perturbations; in many problems of  statistical mechanics this is
not a serious limitation.  However, it is possible to see that a FD
relation holds also for finite perturbations.  This is important
in different contexts, \textit{e.g.} in geophysical or climate
investigations, where the study of small perturbations is rather
academic, while the relaxation of large fluctuations due to fast
changes of the parameters is an issue of practical relevance
\cite{BLMV03}. In the following we derive a FDR which is valid
independently of the number of degrees of freedom.

\subsection{Chaos and the FDR: van Kampen's objection}

Van Kampen argued that the usual derivation of the FDR,
based  on a first-order truncation of the time-dependent
perturbation theory for the evolution of probability density, is not
fully justified \cite{vK71}.  In a nutshell, using the dynamical
systems terminology, van Kampen's argument can be summarized as
follows.  Given a perturbation $\delta {\bf x}(0)$, on the state of
the system ${\bf x}(0)$ at time $0$, one can write a Taylor expansion
for  the difference between the perturbed and the unperturbed
trajectories at time $t$, $\delta {\bf x}(t)$, whose components are:
\begin{equation}
\label{3.1}
\delta  x_i(t)=
\sum_{j} \frac{\partial x_i(t)}{\partial x_j(0)} \delta  x_j(0)
+O(|\delta {\bf x}(0)|^2) \,\,\, .
\end{equation}
After averaging over the initial conditions, one obtains the mean response function:
\begin{equation}
\label{3.2}
R_{i,j}(t)= \Biggl \langle \frac{\partial x_i(t)} {\partial x_j(0)}
\Biggr \rangle =\int \frac{\partial x_i(t)} {\partial x_j(0)}
\rho({\bf x}(0)) d{\bf x}(0 ) \,\,.
\end{equation}
In  a Hamiltonian system one has $\rho({\bf x}) \propto exp\Bigl( -\beta H(\bf{x}) \Bigr)$, so that,
integrating by parts,  we obtain
\begin{equation}
\label{3.3}
R_{i,j}(t)= \beta \Biggl \langle x_i(t)
 \frac {\partial  H({\bf x}(0))} 
{ \partial x_j(0)} \Biggr \rangle 
\end{equation}
which is nothing but the usual FDR \cite{K66,K86}.

In the presence of chaos the terms ${\partial x_i(t)}/{\partial
x_j(0)}$ grow exponentially as $e^{\lambda t}$, where $\lambda$ is the
Lyapunov exponent, therefore it is not possible to use the linear
expansion (\ref{3.1}) for a time larger than $(1/ \lambda) \ln
(L/|\delta{\bf x}(0)|)$, where $L$ is the typical fluctuation size of
the variable ${\bf x}$.  Such an argument suggests  that the linear
response theory is valid only for extremely small and unphysical
perturbations or for short times.  Indeed, according to van Kampen's
argument, the FDR holds up to $1 s$ if a perturbing electric
field smaller than $10^{-20} V/m$ is applied to the electrons of a
conductor, in clear disagreement with experience.

The possible repercussions of van Kampen's arguments are summarized by the following two sentences~ \cite{vK71}:
{\it The basic
linearity assumption of linear theory is shown to be completely
unrealistic and incompatible with basic ideas of statistical mechanics
of irreversible processes}; and {\it Linear
response theory does provide expressions for the phenomenological
coefficients, but I assert that it arrives at these expressions by a
mathematical exercise, rather than by describing the actual mechanism
which is responsible for the response.} These observations
stimulated further research which resulted in a deeper understanding of
the physical mechanisms underlying the validity of the FDR
and of its range of applicability.

As a matter of fact, the success of the linear theory for the
computation of the transport coefficients (e.g. electric
conductivity), in terms of correlation functions of the unperturbed
systems, is transparent, and its validity has been, directly and
indirectly, tested in a huge variety of cases. To reconcile van
Kampen's argument with the well established linear response theory, note that van Kampen considered single
trajectories, while the FDR deals with observables quantities, which
are averages over ensembles of trajectories, and have a much more
regular behavior than the single trajectories.  Indeed, the reason of
the seemingly inexplicable effectiveness of the Linear-Response theory
 lies in the ``constructive role of chaos'' because, as Kubo
observed, ``{\it instability [of the trajectories] instead favors the
stability of distribution functions, working as the cause of the
mixing}''\cite{K86}.

\subsection{Generalized FDR for non Hamiltonian systems}

In the following we will give a derivation, under rather general
hypotheses, of a general FDR, which avoids  van Kampen's
critique.  Consider a dynamical system $ {\bf x}(0) \to {\bf x}(t)=S^t
{\bf x}(0)$ with states ${\bf x}$ belonging to a $N$-dimensional
vector space.  For the sake of generality, we consider the possibility
that the time evolution be not completely deterministic (\textit{e.g.}
it could be described by stochastic differential equations).  We
assume that the system is mixing and that the invariant probability
distribution $\rho({\bf x})$ enjoys some regularity property (see
later), while no assumption is made on $N$.

Our aim is to express the average response of a generic observable $A$
to a perturbation, in terms of suitable correlation functions,
computed according to the invariant measure of the unperturbed system.
The first step is to study the behavior of a single component of ${\bf  x}$,
 say $x_i$, when the system, described by $\rho({\bf x})$, is
subjected to an initial (non-random) perturbation 
${\bf  x}(0) \to {\bf x}(0) + \Delta {\bf x}_{0}$~\footnote{The study of an
  ``impulsive'' perturbation is not a severe limitation, for instance in
  the linear regime the (differential) linear response describes the effect of a generic perturbation.}.  This
instantaneous kick modifies the density of the system into
$\rho'({\bf x})$, which is related to the invariant
distribution by $\rho' ({\bf x}) = \rho ({\bf x} - \Delta {\bf x}_0)$.
We introduce the probability of transition from ${\bf x}_0$ at time
$0$ to ${\bf x}$ at time $t$, $W ({\bf x}_0,0 \to {\bf x},t)$. For a
deterministic system, with evolution law $ {\bf x}(t)=S^{t}{\bf
  x}(0)$, the probability of transition reduces to $W ({\bf x}_0,0 \to
{\bf x},t)=\delta({\bf x}-S^{t}{\bf x}_{0})$, where $\delta(\cdot)$ is the
Dirac delta function.  Then we can write an expression for the mean value of
the variable $x_i$, computed with the density of the perturbed system:
\begin{equation}
\label{3.4}
\Bigl \langle x_i(t) \Bigr \rangle ' = 
\int\!\int x_i \rho' ({\bf x}_0) 
W ({\bf x}_0,0 \to {\bf x},t) \, d{\bf x} \, d{\bf x}_0  \; .
\end{equation}
The mean value of $x_i$ during the unperturbed evolution can be written in
a similar way:
\begin{equation}
\label{3.5}
\Bigl \langle x_i(t) \Bigr \rangle = 
\int\!\int x_i \rho ({\bf x}_0) 
W ({\bf x}_0,0 \to {\bf x},t) \, d{\bf x} \, d{\bf x}_0  \; .
\end{equation}
Therefore, defining $\overline{\delta x_i} =  \langle x_i \rangle' -
\langle x_i \rangle$, we have:
\begin{eqnarray}
\label{3.6}
\overline{\delta x_i} \, (t)  &=&
\int  \int x_i \;
\frac{\rho ({\bf x}_0 - \Delta {\bf x}_0) - \rho ({\bf x}_0)}
{\rho ({\bf x}_0) } \;
\rho ({\bf x}_0) W ({\bf x_0},0 \to {\bf x},t) 
\, d{\bf x} \, d{\bf x}_0 \nonumber \\
&=& \Bigl \langle x_i(t) \;  F({\bf x}_0,\Delta {\bf x}_0) \Bigr \rangle
\end{eqnarray}
where
\begin{equation}
\label{3.7}
F({\bf x}_0,\Delta {\bf x}_0) =
\left[ \frac{\rho ({\bf x}_0 - \Delta {\bf x}_0) - \rho ({\bf x}_0)}
{\rho ({\bf x}_0)} \right] \; .
\end{equation}
Note that the system  is assumed to be mixing, so that the decay to zero of the
time-correlation functions prevents any departure from equilibrium.
 
For an infinitesimal perturbation $\delta {\bf x}(0) = (\delta x_1(0)
\cdots \delta x_N(0))$, the function in (\ref{3.7}) can be expanded to
first order, if $\rho({\bf x})$ is non-vanishing and differentiable,
and one obtains:
\begin{eqnarray}
\label{3.8}
\overline{\delta x_i} \, (t)  &=&
- \sum_j \Biggl
\langle x_i(t) \left. \frac{\partial \ln \rho({\bf x})}{\partial x_j} 
\right|_{t=0}  \Biggr \rangle \delta x_j(0) \nonumber \\
&\equiv&
\sum_j R_{i,j}(t) \delta x_j(0)
\end{eqnarray}
which defines the linear response 
\begin{equation}
\label{3.9}
R_{i,j}(t) = - \Biggl \langle x_i(t) \left.
 \frac{\partial \ln \rho({\bf x})} {\partial x_j} \right|_{t=0}
\Biggr  \rangle 
\end{equation} 
of the variable $x_i$ with respect to a perturbation of $x_j$.
One can easily repeat the computation for a generic observable
$A({\bf x})$, obtaining:
\begin{equation}
\label{3.10}
\overline{\delta A} \, (t)  = -\sum_j
\Biggl \langle A({\bf x}(t)) \left.
\frac{\partial \ln \rho({\bf x})} {\partial x_j} \right|_{t=0} 
\Biggr \rangle \delta x_j(0) \,\, .
\end{equation}

In the above derivation of the FDR, we did not use any approximation
on the evolution of $\delta {\bf x}(t)$.  Starting with the exact
expression (\ref{3.6}) for the response, only a linearization of the
initial perturbed density is needed, and this implies nothing but the
smallness of the initial perturbation.  From the evolution of the
trajectories difference, one can indeed define the leading Lyapunov
exponent $\lambda$, by considering only the positive quantities $|\delta {\bf
x}(t)|$, so that at small $|\delta {\bf x}(0)|$ and large enough $t$
one can write
\begin{equation}
\label{3.11}
\Bigl \langle \ln |\delta {\bf x}(t)|\Bigr \rangle \simeq 
\ln |\delta {\bf x}(0)| + \lambda t \,\, .
\end{equation}
Differently, in the derivation of the FDR, one deals with averages of quantities
with sign, such as $\overline{\delta {\bf x}(t)}$.  This apparently
marginal difference is very important and underlies the possibility of deriving
 the FDR without incurring in  van Kampen's objection.
 
At this point one could object that in  chaotic deterministic
dissipative systems the above machinery cannot be applied, because the
invariant measure is not smooth. 
\footnote{Typically the invariant measure of a chaotic attractor has a
multifractal character and its Renyi dimensions $d_q$ do not
depend on $q$\cite{PV87}.}  In chaotic dissipative systems the invariant
measure is singular, however the previous derivation of the FDR is
still valid if one considers perturbations along the expanding
directions. In addition, one is often interested in some specific
variables, so that a projection is performed, making irrelevant the
singular character of the invariant measure. For a mathematically
oriented presentation of these ideas, related to a class of
particularly well behaved dynamical systems, see~\cite{R98}.  In these
cases, a general response function has two contributions,
corresponding respectively to the expanding (unstable) and the
contracting (stable) directions of the dynamics. The first
contribution can be associated to some correlation function of the
dynamics on the attractor (i.e. the unperturbed system). This is not
true for the second contribution which is very difficult to extract
numerically~\cite{CS07}.  Nevertheless, a small amount of noise, that
is always present in a physical system, smoothens the $\rho({\bf x})$
and the FDR can be derived.  We recall that this ``beneficial'' noise
has the important role of selecting the natural measure, and, in the
numerical experiments, it is provided by the round-off errors of the
computer. Then, although the introduction of noise in a deterministic
framework may look unjustified, the assumption on the smoothness of
the invariant measure along the unstable manifold still allows us to
avoid subtle technical difficulties, 
as confirmed by Ruelle's work~\cite{R98}.

We further observe that these dynamical
mechanisms, which work equally well in low and high dimensional
systems, do not need to constitute the main reason for which $\rho$
can be assumed to be smooth, in systems of physical interest. Indeed,
in systems of many particles, the projections from high dimensional
phase spaces to a low  dimensional space, is the main reason for
dealing with smooth distributions.

In Hamiltonian systems, taking the canonical ensemble as the
equilibrium distribution, one has $ \ln \rho= -\beta H({\bf Q},{\bf
P})\, + constant.$ If $x_i$ denotes the component $q_k$ of the
position vector ${\bf Q}$ and $x_j$ the corresponding component $p_k$
of the momentum ${\bf P}$, Hamilton's equations ($dq_k/dt=\partial
H/\partial p_k$) lead to the usual FDR \cite{K66,K86}
\begin{equation}
\label{3.12}
\frac{ \overline{\delta q_k} \, (t)} {\delta p_k(0)}
=\beta \Biggl \langle q_k(t) \left . \frac {dq_k} {dt} \right|_{t=0} \Biggr \rangle =
- \beta \frac{d}{dt} \Bigl \langle q_k(t) q_k(0) \Bigr \rangle \, .
\end{equation}

In non Hamiltonian systems, where usually the shape of $\rho({\bf x})$
is not known, relation (\ref{3.9}) does not give very detailed
information. 
However it shows  the existence of a connection between
the mean response function $R_{i,j}$ and a suitable correlation
function, computed in the non perturbed systems:
\begin{equation}
\label{3.13}
\Bigl \langle x_i(t)f_j({\bf x}(0)) \Bigr \rangle \; , 
\quad \textrm{with} \quad 
f_j({\bf x})=- \frac{\partial \ln \rho}{\partial x_j} \,\, ,
\end{equation}
where, in general, the function $f_j$ is unknown.  Let us
stress that in spite of the technical difficulty for the determination
of the function $f_j$, which depends on the invariant measure, a FDR
 always holds in mixing systems, if the invariant measure can be considered
``smooth enough''.  We note that the nature of the statistical steady
state (either equilibrium, or non equilibrium) plays no role at all for
the validity of the FDR.

In the case of a multivariate Gaussian distribution, 
$\ln \rho({\bf x})= -{1 \over 2} \sum_{i,j}\alpha_{i,j}x_i x_j + const.$
where $\{ \alpha_{i,j} \}$ is a positive matrix, and 
the elements of the linear response matrix can be written in terms
of the usual  correlation functions:
\begin{equation}
\label{3.14}
R_{i,j} (t) =\sum_k \alpha_{i,k}
{\Bigl \langle x_i(t) x_k(0) \Bigr \rangle }  \; .
\end{equation}
An important nontrivial class of systems with a Gaussian invariant
measure is the inviscid hydrodynamics, where the Liouville theorem
holds, and a quadratic invariant exists, see Section 4.1.

\subsection{FDR, chaos  and molecular dynamics}

It is not necessary to stress further the relevance of the FDR for the
numerical investigation of the transport properties: via a molecular
dynamics simulation one can compute the equilibrium correlation
functions and therefore, using the Green- Kubo formula, the transport
coefficients of liquids models without the introduction of any
simplifying assumption.  This approach has been successfully used in
seminal works~\cite{AGW70,LVK73,EM90}.

Another method which is more fundamental, from a conceptual point of
view, but rather simple is to use
directly the  definitions of transport coefficients: one perturbs the
system with an external force or imposes driving boundary
conditions (e.g. a shear)~\cite{CJ75,J83,CJM79,MECV89} and observes
the relaxation process.  
For such a problem, van Kampen's objection is relevant, to some extent.
For example in numerical simulations, $R_{i,i}(t)$ is obtained by
perturbing the variable $x_i$ at time $t=t_0$ with a small
perturbation of amplitude $\delta x_i(0)$ and then evaluating the
separation $\delta x_i(t|t_0)$ between the two trajectories ${\bf
x}(t)$ and ${\bf x}'(t)$ which are integrated up to a prescribed time
$t_1=t_0+\Delta t$.  At time $t=t_1$ the variable $x_i$ of the
reference trajectory is again perturbed with the same $\delta x_i(0)$,
and a new sample $\delta {\bf x}(t|t_1)$ is computed and so forth.
The procedure is repeated $M \gg 1$ times and the mean response is
then evaluated:
\begin{equation} 
\label{3.15}
R_{i,i}(\tau)= {1 \over M} \sum_{k=1}^M 
{ {\delta  x_i(t_k+\tau|t_k)} \over  {\delta  x_i(0)}}  \,\, .
\end{equation}
In the presence of chaos, the two trajectories ${\bf x}(t)$
and ${\bf x}'(t)$ typically separate exponentially in time, therefore
the mean response is the result of a delicate balance of terms which
grow in time in different directions.  A naive estimate of the  error in the
computation of $R_{i,i}(\tau)$ suggests an increase in time as 
\begin{equation} 
\label{3.16}
e_M(\tau)=\Bigl[ {1 \over M} \sum_{k=1}^M 
\Bigr( { {\delta  x_i(t_k+\tau|t_k)} \over  {\delta  x_i(0)}}\Bigl)^2 -
\Bigr( R_{i,i}(\tau)\Bigl)^2 \Bigl]^{1/2}
\sim 
{ {e^{{L(2) \over 2} \tau}} \over  {\sqrt{M}} } \,\, ,
\end{equation}
 where $L(2)$ is the generalized Lyapunov exponent of order $2$:
\begin{equation} 
\label{3.17}
L(2)= \lim_{\tau \to \infty} {1 \over \tau}
\ln \Biggl \langle \Bigr(
{ {|\delta {\bf x}(\tau)|} \over  {|\delta {\bf x}(0)|} } \Bigl)^2 
\Biggr \rangle \,\, .
\end{equation}
In the above equation  $\delta {\bf x}(\tau)$ is assumed infinitesimal,
i.e. evolving according the linearized dynamics 
(in mathematical terms,  $\delta {\bf x}(\tau)$
is a tangent vector of the system)~\cite{BCFV02}, and
a lower bound for $L(2)$ is given by $2 \lambda$.
Thus very  large $M$ seems to be  necessary,
in order to properly estimate this balance and to compute
$R_{i,i}(\tau)$ for large $\tau$.

On the other hand, the numerical results obtained with molecular
dynamics simulations show  rather clearly that the presence of chaos does not introduce severe limitations
in the range of validity of the FDR, in agreement with the
intuition of Kubo and the derivation of the FDR in the previous
section.  For instance in \cite{CJ75,J83},
for an ion in liquid argon, a perturbing force of order $10^5 eV/cm$
has been used, and a rather good agreement has been obtained between
the results produced perturbing the system and those deduced from the
Green-Kubo formula.  These results are not restricted to systems with a 
large number of degrees of freedom \cite{CFIPV91,FIV90}.  As a final technical remark we note that the
error estimate given by (\ref{3.16}) is rather conservative and
the uncertainty is actually smaller (at least for $\tau$ not too large).  In
\cite{CFIPV91} one can see that for $\tau <{\cal T}$ (where ${\cal T}$
is order of the correlation time) $e_M(\tau)$ first decreases and
later increases exponentially, but with a rate different from $L(2)$.
Indeed it is not completely correct in this context to use at finite time the
Lyapunov exponents (or analogous quantities), which are well defined
only at very large times.
\footnote{The same subtle problem appears in 
the connection between the Kolmogorov-Sinai entropy, 
$h_{KS}$ (which is linked to the Lyapunov exponents via the Pesin 
formula\cite{O93}),
and the variation rate of the coarse grained Gibbs entropy,
 $r_G$.  Detailed numerical computations show that typically the 
(often accepted) identification of the two quantities does not
 hold.
The basic reason of this fact is in the asymptotic 
(with respect to time) nature of $h_{KS}$, 
while $r_G$ is a quantity related to the short time behavior of a given system
\cite{FPV05}.}

\subsection{Linear response theory, foundations of statistical
mechanics and anomalous diffusion}

This section is devoted to a brief discussion of interesting,
and still partly open issues involving the relation
among linear response theory, the foundations of statistical mechanics
and systems with anomalous diffusion.

\subsubsection{Linear response theory and the foundations of statistical
mechanics}

In his celebrated book {\it Mathematical Foundations of Statistical
Mechanics}, Khinchin obtained  important results concerning the ergodic
and mixing problem in statistical mechanics, without recourse to the
notion of metrical transitivity, required by the Birkhoff theorem~\cite{K49}.
The general idea of this approach, which will be considered also in
Chapter 5, is that the systems of interest in statistical mechanics
have a very large number of degrees of freedom and the important
observables are not generic functions.

In a system with many degrees of freedom and weak interactions
among the particles
\footnote{Khinchin considers a separable Hamiltonian system i.e.:
$$
H=\sum_{n=1}^N H_n({\bf q}_n,{\bf p}_n) \,\, .
$$
It is reasonable to think that the result holds also in systems of
particles interacting through a short range potential, as those
studied in~\cite{ML63} for equilibrium statistical mechanics.},
consider an observable $A$ which
contains the coordinates of only one (or a few) particles.
Khinchin has shown that the ergodic hypothesis for a quantity $A$, i.e. the validity of
\begin{equation}
\label{L.1}
\lim_{T \to \infty} \frac{1}{T}\int_0^{\infty} A(t) dt=\langle A \rangle \,\,,
\end{equation}
holds if the auto-correlation of $A$ satisfies the relation
\begin{equation}
\label{L.2}
\lim_{t \to \infty} \langle A(t)A(0) \rangle =\langle A \rangle^2 \,\,\, .
\end{equation}
This result can be reversed, i.e. the auto-correlation
$C_A(t)=\langle A(t)A(0) \rangle$ satisfies (\ref{L.2}) if
Eq. (\ref{L.1}) holds uniformly~\cite{K57}\footnote{Note that this result concerns the ergodic hypothesis for only the given observable $A$, which is much weaker than ergodicity in its strict
technical meaning, indeed it  holds only for specific observables,
and in systems with many degrees of freedom.}.

On the other hand, we know that correlation functions are
closely related to the response function, so it is quite natural
to investigate the connection between ergodicity and linear response theory.
In a series of papers~\cite{L01,L02,L06,L07a,L07b} this problem has been
discussed in detail.
The main result is the possibility to connect ergodic or non ergodic
behaviour
to the response of a system to an external perturbation. One finds
that, in a generic system, irreversibility, i.e. the validity of (\ref{L.2}),
is a necessary but not sufficient condition for ergodicity.

Let us indicate, in a symbolic way, the variables which determine
the macroscopic state and the control parameters with $\{ A \}$
and $\{ P \}$, respectively.
If we perturb the system changing  $\{ P \}$ at time $t=0$, i.e.
$\{ P \} \to \{ P \} + \{ \delta P \} \Theta(t)$ one has
\begin{equation}
\label{L.3}
\overline{\delta A (t)}=\delta P \int_0^t R(t') dt'
\end{equation}
where $\{ R \}$ indicates the (differential) response function.
For the sake of simplicity,  we assume $\langle A \rangle =0$.
If ergodicity holds for the variable $A$, one has~\cite{L01,L02}:
\begin{equation}
\label{L.4}
\lim_{T \to \infty} {1 \over T} \int_0^T \int_0^t R_A(t-t') dt dt'=\chi_A
\end{equation}
where $\chi_A$ is the static susceptibility. The advantage
of~\eqref{L.4} over the usual condition for ergodicity,
Eq.~\eqref{L.1}, is that both sides of~\eqref{L.4} may be evaluated,
for specific models, knowing the unperturbed measure (or equivalently the
Hamiltonian).  In the linear response theory, for Hamiltonian systems,
one has
\begin{equation}
\label{L.5}
R_A(t)=-{d \over dt}C_A(t)=-{d \over dt} \langle A(t)A(0) \rangle \,\,\, .
\end{equation}
Denoting with $\tilde{R}_A(z)$ and $\tilde{C}_A(z)$ the Laplace transform
of $R_A(t)$ and $C_A(t)$, respectively, from (\ref{L.4}-\ref{L.5}),
 and using the condition (valid for differentiable variables)
$dC_A(t)/dt=0$ for $t=0$, the limit $T \to \infty$ produces
\begin{equation}
\label{L.6}
\tilde{R}_A(0) + C_A(T)-{\tilde{C}_A(0) \over T} \to \chi_A,
\end{equation}
or in different form:
\begin{equation}
\label{L.7}
\tilde{R}_A(0)+z\tilde{C}_A(z)|_{z=0}= \chi_A \,\,\, .
\end{equation}
In general one does not know if (\ref{L.4}), or its equivalent
(\ref{L.7}),
holds.
It is possible to show that the two equations hold and coincide if
$C_A(t)$ approaches zero for large $t$   in a fast way, so that
\begin{equation}
\label{L.8}
\int_0^{\infty} C_A(t) dt < \infty \,\, .
\end{equation}
In other words, the condition that $C_A(t) \to 0$ for $t \to \infty$
is only  necessary for ergodicity, while~\eqref{L.8} is
sufficient. Note that (\ref{L.8}) is implied by the Green-Kubo
formulae, when the $\{ \chi \}$ are finite quantities. The Khinchin
theorem, on the other side, requires weak interactions among the
components of the system: this is no more guaranteed, in the general
case, when $C_A(t)$ can have a slow decay.

In~\cite{L01,L02,L06} one can find a detailed discussion on different systems
(including monoatomic chains, oscillator models, chains with impurity, etc.)
where the  powerful recurrence relations method  can be successfully
applied and it is possible to determine whether the condition
(\ref{L.8}) is valid or not. Consider, for instance,  the case of a nearest-neighbor coupled harmonic
oscillator chain of $2N+1$ atoms, with periodic boundary conditions,
containing a tagged particle of mass $M$, while the remaining
have mass $m$. In the $N \to \infty$ limit, it is possible to see that
the validity of (\ref{L.8}) depends on the value of $\lambda=m/M$.
If $\lambda$ is finite, the momentum of the tagged particle
is an ergodic observable.

\subsubsection{The fluctuation- dissipation relation in the case of
anomalous diffusion}

In recent years, many authors have shown that one can have anomalous
diffusion in certain systems~\cite{ZSW93,CMMV99}, i.e. one can have
\begin{equation}
\label{A.2}
\langle x^2(t)\rangle \sim t^{2 \nu} \,\,\, \mbox{with} \,\,\,  \nu \ne
1/2.
\end{equation}
Formally this corresponds to $D=\infty$ if $\nu > 1/2$
(superdiffusion)
and $D=0$ if $\nu < 1/2$ (subdiffusion).

Here we shall not discuss the different mechanisms at the origin of
anomalous diffusion~\cite{CMMV99}.  Our aim is to provide a short
discussion of the following problem, which comes out in a rather
natural way in presence of anomalous diffusion.  We know, from Chapter
2, that, if in an unperturbed system $\langle x(t)\rangle=0$, in the
cases with standard diffusion, i.e. with a finite positive diffusion
coefficient, the response to a small external force is a
linear drift
\begin{equation}
\label{A.3}
\overline{x(t)} \sim t \,\,,
\end{equation}
 where the over-bar indicates the average over the
perturbed system.
One may ask  how  this scenario changes  in presence of anomalous
diffusion.

As a consequence of the simple identity
\begin{equation}
\label{A.4}
\langle x^2(t) \rangle =\int_0^t\int_0^t
C_v(t_1-t_2)dt_1 dt_2  \,\,\,\, , \,\,\,\,
C_v(t_1-t_2)=\langle v(t_1)v(t_2) \rangle \,\, ,
\end{equation}
and the fluctuation dissipation relation,
one has the formal relation
\begin{equation}
\label{A.5}
{d \over dt} \overline{x(t)} =K \int_0^t
 C_v(t')  dt' \,\, ,
\end{equation}
where $K$ is a constant depending on the physical details,
whose value is not relevant for our purpose.
In the  superdiffusive case, one has
\begin{equation}
\label{A.6}
C_v(t) \sim t^{-\alpha} \,\, \mbox{with} \,\, \alpha < 1,
\end{equation}
and $\nu=1-\alpha/2$.

A straightforward consequence of~\eqref{A.5}  seems to be
\begin{equation}
\label{A.7}
\overline{x(t)} \sim t^{2\nu} \,\, .
\end{equation}

On the other hand it is not difficult to realize that such a formal
argument is not reliable.
Let us consider again the  previous discussion, i.e.
take a system whose   state is   $\{ A \}$  and the control
 parameters are $\{ P\}$.
Consider a special   value $\{ P^* \}$ of $\{ P \}$ such that
(\ref{L.8}) does not hold.
Now   perturb the system changing  $\{ P \}$ at time $t=0$, so that
$\{ P^* \} \to \{ P^* \} + \{ \delta P \} \Theta(t)$.
In our case $R_v(t)=C_v(t)$,
taking $t \to \infty$ in  (\ref{L.3}),
we have  $\overline{\delta A (\infty)}= \infty$, if $\nu > 1/2$.
The validity of such a result is highly questionable:
for instance it is easy to obtain a finite result:
\begin{equation}
\label{A.8}
\overline{\delta A (\infty)}= \langle A \rangle_{P^* + \delta P}
- \langle A \rangle_{P^*} \,\, .
\end{equation}

Let us discuss separately the subdiffusive case ($\nu < 1/2$)
and the superdiffusive case ($\nu > 1/2$).\\

\paragraph{The subdiffusive case}

Some authors  show that in the subdiffusive case,
the result (\ref{A.7}) holds~\cite{MBK99}.
This has been explicitly proved in systems described by a
fractional-Fokker-Planck equation.
In the case of an overdamped particle, see Sect.~\ref{overdamped},  the usual
Fokker-Planck equation is
\begin{equation}
\label{A.9}
{\partial \over {\partial t}} P(x,t)={\cal L}_{FP} P(x,t)=
{\partial \over {\partial x}} \left( { V'(x) \over {m\eta_1}}  P(x,t)
\right)
+K_1{\partial^2 \over {\partial x^2}}    P(x,t) \,\, ,
\end{equation}
with $K_1=k_BT/m\eta_1$. For $V=0$ one has
\begin{equation}
\label{A.10}
\langle x^2(t) \rangle \simeq 2 K_1 t \,\,
\end{equation}
while for $V=-Fx$,
\begin{equation}
\label{A.11}
\overline{x(t)} \simeq {F \over m\eta_1} t \,\, ,
\end{equation}
which is nothing but the Einstein relation for such a system.
The  fractional-Fokker-Planck equation is given by
\begin{equation}
\label{A.12}
{\partial \over {\partial t}} P(x,t)=
{\cal D}_T^{1-\gamma}\Bigl [ {\cal L}_{FP} P(x,t) \Bigr ]
\end{equation}
where now ${\cal L}_{FP}$ is defined as in (\ref{A.9})
with the changes $\eta_1 \to \eta_{\gamma}$ and
$K_1 \to K_{\gamma}$, and
${\cal D}_T^{1-\gamma}$ is the so-called Riemann-Liouville fractional
operator defined through:
\begin{equation}
\label{A.13}
{\cal D}_T^{1-\gamma}\Bigl [ f(x,t) \Bigr ]=
{1 \over \Gamma(\gamma)} {\partial \over {\partial t}}
\int_0^t {f(x,t') \over {(t-t')^{1-\gamma}}} dt' \,\,\,
\mbox{with} \,\,\, 0< \gamma \le 1 \,\,,
\end{equation}
being $\Gamma(\gamma)$ the Euler gamma function. For $V=0$ one has a generalization of (\ref{A.10}):
\begin{equation}
\label{A.14}
\langle x^2(t) \rangle \simeq 2 {{K_1 t^\gamma} \over {\Gamma(1+\gamma)}} \,\,
\end{equation}
and for $V=-Fx$,
\begin{equation}
\label{A.15}
\overline{x(t)} \simeq {F \over m\eta_{\gamma} }
{t^{\gamma} \over \Gamma(1+\gamma)}  \,\, ,
\end{equation}
in other words, if we assume the relation $K_{\gamma}=k_B T/m\eta_{\gamma}
$,
one has:
\begin{equation}
\label{A.16}
\overline{x(t)}={1 \over 2} {F\langle x^2(t) \rangle \over k_B T} \,\, ,
\end{equation}
which generalizes the Einstein relation for the subdiffusive case.
The investigation of charge carrier transport in semiconductors~\cite{GSGWS96}
showed that, up to a prefactor, which could not be determined exactly,
eq. (\ref{A.16}) is indeed valid.\\

\paragraph{The superdiffusive case}

This situation is much more delicate, and, as far as we know, the
problem is still open.  The difficulties are due to the fact that the
mechanism for the anomalous diffusion is rather subtle and can be
obtained only in systems with rather peculiar dynamics.  For instance
in deterministic chaotic systems with a fine tuning of the
parameters~\cite{ZSW93,L98}.  Typically the following scenario holds: if
for a certain value $P_0=P^*$ of the control parameters one has
$C_v(t) \sim t^{-\alpha}$ with $\alpha <1$ and therefore $\nu >1/2$, a
small perturbation $P_0 \to P_p=P^* +\delta P$ restores the standard
diffusion with $\nu=1/2$. More complex is the case of non-chaotic
systems, which also show anomalous diffusion~\cite{JR06}, but it is
not completely clear which perturbations restore  normal diffusion.

At first, the difference between $P^*$ and $P^*+\delta P$ can appear a paradox, on the other
hand a precise analysis of the problem shows that  the
anomalous behaviour for $P=P_0=P^*$ induces a sort of ``ghost'' for
$P=P_p=P^* +\delta P$, which is reminiscent of the unperturbed anomalous behavior.  For $P=P_p$ the correlation function $C_v(t)$
is integrable, however for $t<t_c(\delta P)$ (where $t_c(\delta P) \to
\infty $ as $\delta P \to 0$) one has $C_v(t) \sim t^{-\alpha}$,
therefore the difference between the anomalous case (for $P_0$) and
the standard diffusion (for $P_p$) appears only at very large time,
leading to a large diffusion coefficient (diverging for $\delta P \to
0$)~\cite{BCVV95}.

With the above scenario one can try to give a solution to the
problem.
Let us denote by $C_v^{(0)}(t)$ and  $C_v^{(p)}(t)$
the correlation function for $P_0=P^*$ and $P_p=P^* +\delta P$
respectively.
In the presence of anomalous behavior (i.e. when (\ref{L.8}) is not valid)
it is not possible to exchange the limits $\delta P \to 0$ and
$t \to \infty$.
Since the  relaxation time of the perturbed system is $t_c$,
Eq. (\ref{A.5}) is valid only  for $t<t_c$, while for
$t>t_c$, Eq.~(\ref{A.5}) must be replaced by:
\begin{equation}
\label{A.17}
{d \over dt} \overline{x(t)} =K \int_0^{t_c}  C_v^{(0)}(t')  dt'
+K \int_{t_c}^t  C_v^{(p)}(t')  dt'
 \,\, .
\end{equation}
As a consequence Eq. (\ref{A.7}) holds in the range $t<t_c$, while for
larger times one has the standard result (\ref{A.3}), with a different
prefactor depending on $t_c$.  One can say that also in anomalous
diffusion, a ``ghost'' for $P^*$ remains up to large but finite times
$t_c$.\\ A detailed discussion of this problem for a dynamical model
has been given in~\cite{TFWG94}. Both possibilities, that $\delta P
\to 0$ before $t \to \infty$, and that $t \to \infty$ before $\delta P
\to 0$, have been considered. In the first case, one obtains
$\overline{x(t)} \sim t^{2\nu}$, for the response at large times; on
the other hand, for any finite value of $\delta P$,
$d\overline{x(t)}/dt$ saturates to a finite value, that depends on the
perturbation. The general scenario suggested in Eq.~\eqref{A.17} is
therefore verified.

\section{Some applications}
\label{sec:appl}

The FDR has been extensively used in equilibrium statistical
mechanics. Applications in near-equilibrium contexts are discussed in
any advanced book dealing with transport phenomena.  In this section
we want to discuss how to use the FDR in less standard cases, namely
non-equilibrium and non-Hamiltonian systems.

\subsection{Fluid dynamics}

The Fluctuation-Response theory was originally developed within the
 framework of equilibrium statistical mechanics of Hamiltonian
 systems, giving rise to some confusion and misleading ideas about its
 validity.  For instance, some authors claimed that in fully developed
 turbulence (which is a non Hamiltonian and non equilibrium system)
 there is no relation between spontaneous fluctuations and relaxation
 to the statistical steady state \cite{RS78}.

The confusion is due,  at least in part, to 
terminology; for instance, at times  the term FDR 
is used just for Eq. (\ref{3.14})
which is valid in the Gaussian case.
Nevertheless the  FDR  plays a role in
analytical approaches to the statistical description of hydrodynamics,
where Green functions are naturally involved, both in perturbative
theories and in closure schemes \cite{Kr61,MK05,F95}, as well as  in
other  fields, like in geophysical
or climate studies\cite{L75,L78,B80}.

\subsubsection{Interlude on Statistical Mechanics of Fluids}

The Navier-Stokes equations for the evolution of the velocity field
${\bf v}({\bf x},t)$ of an incompressible fluid are:
\begin{equation} 
\label{4.1}
 {\partial  {\bf v}({\bf x},t)   \over \partial t}  + 
  ({\bf v} ({\bf x},t)\cdot \nabla) {\bf v}({\bf x},t) =
  -{ 1 \over \rho} \nabla p({\bf x},t)   +\nu \Delta
  {\bf v}({\bf x},t)  +{\bf F}({\bf x},t)  \, ,
\end{equation}
\begin{equation} 
\label{4.2}
\qquad  \nabla \cdot {\bf v}({\bf x},t)   =0 , 
\end{equation}
where $p({\bf x},t) $ is the pressure in the point 
${\bf x}$ at time $t$;   $\rho$ and $\nu$ are the density and the kinematic
viscosity of the fluid respectively.

For a perfect fluid (i.e. $\nu=0$) and  in the absence of external forces
(${\bf F}=0$),
the evolution of the  velocity field is given by the Euler equations,
which conserve the kinetic energy.
Such a case  may be easily treated within an equilibrium 
 statistical mechanics framework, simply following the approach used for the
 Hamiltonian statistical mechanics \cite{BJPV98}.
Consider a  fluid in a cube of edge $L$ and assume
periodic boundary conditions,  so that
the velocity field can be expressed in terms of Fourier series as follows:
\begin{equation} 
\label{4.3}
{\bf v}({\bf x},t)={ 1 \over L^{d/2} } \sum_{\bf k} {\bf u}({\bf k},t)
 e^{ i {\bf k} {\bf x}}
\end{equation}
with  ${\bf k}=2\pi {\bf n}/L$ and ${\bf n}=(n_1,..., n_d)$, 
$n_j$ being integers and  $d$  the spatial dimension.
By introducing~\eqref{4.3} in the 
Euler equations, and imposing an ultraviolet cutoff,
 ${\bf u}({\bf k})=0$ for $k>K_{max}$,
one obtains a set of ordinary differential equations of the form:
\begin{equation} 
\label{4.4}
{{d Y_a} \over { dt}}= \sum_{b,c}^N  A_{abc} Y_b Y_c
\end{equation}
where $N \sim K_{max}^d$ is the number of degrees of freedom, and the variables $\{
Y_a \}$ are a subset of the full Fourier spectrum $ \{ {\bf u}({\bf
k}) \}$.  The incompressibility condition yields  ${\bf k}\cdot{\bf u}({\bf k})=0$, and the
fact that $ {\bf u}({\bf k})$ is real yields  ${\bf u}({\bf k})={\bf
u}^*({-\bf k})$.\\ As a consequence of the energy conservation one has
\begin{equation} 
\label{4.5}
{1 \over 2}  \sum_a Y_a^2=E=constant \,\, .
\end{equation}
In addition, the coefficients  $A_{abc}$ have the properties 
$ A_{abc}=A_{acb} \,\,\, ,\,\,\, A_{abc}+ A_{bca}+A_{cab}=0$,
and incompressibility makes the  Liouville theorem hold:
\begin{equation} 
\label{4.6}
\sum_a  { {\partial} \over {\partial Y_a}}  
{ {d Y_a} \over { dt} }=0 \,\, ,
\end{equation}
so that the necessary ingredients for  an
equilibrium statistical mechanical approach are all present.
Assuming, in analogy with the treatment of Hamiltonian systems,
a microcanonical distribution on the constant energy surface
${1 \over 2} \sum_a Y_a^2=E$,  for large  $N$
it is easy to obtain the invariant probability density of $Y_a$:
\begin{equation} 
\label{4.7}
P_{inv}(\{Y_a \}) \propto exp \left (- {\beta \over 2} \sum_a Y_a^2 \right )
\end{equation}
and the energy equipartition  $<Y_a^2>=2 {E \over N}=\beta^{-1}$.

In  two dimensions another quantity is conserved: the enstrophy  
${1 \over 2} \int | \nabla \times \, {\bf u}({\bf x}, t)|^2 d {\bf x}$ which,
in terms of the $\{Y_a\}$, can be written as
\begin{equation} 
\label{4.8}
{1 \over 2} \sum_a  k_a^2 Y_a^2=\Omega= const. 
\end{equation}
Assuming a microcanonical distribution on the surface where
both the energy and the enstrophy are constant,
and taking  the large  $N$  limit, one obtains
\begin{equation} 
\label{4.9}
P_{inv}(\{Y_a \}) \propto exp - {1 \over 2}\Bigl( \beta_1 \sum_a Y_a^2
+ \beta_2 \sum_a k_a^2  Y_a^2 \Bigr)
\end{equation}
where the Lagrange multipliers $\beta_1$ and $\beta_2$  are determined
by $E$ and $\Omega$.

Other systems, like magnetohydrodynamics and geostrophic
fluids,  have quadratic invariants
and obey the Liouville theorem in the inviscid limit.
\footnote{Numerical results in $2D$ and $3D$ inviscid hydrodynamics, as well
as in other inviscid cases (e. g. magnetohydrodynamics) show that the
systems described by the inviscid truncated Eq. (\ref{4.4}) with quadratic 
invariant, are ergodic and mixing\cite{OP72,SSMK75}:
 arbitrary initial distributions of $\{ Y_a \}$ evolve towards
the multivariate Gaussian (\ref{4.10}).}
In such situations it is straightforward to generalize the
previous results as
\begin{equation} 
\label{4.10}
P_{inv}(\{Y_a \}) \propto exp \left (- {1 \over 2} \sum_{a,b} \alpha_{a,b}Y_a Y_b \right )
\end{equation}
where $\{ \alpha_{a,b} \}$ is a positive matrix  dependent 
on the  specific form of the
invariants and on  the values of the Lagrange multipliers.

A statistical steady state requires a forcing in the case of a viscous
fluid, which results in:
\begin{equation} 
\label{4.11}
\Bigr({{d} \over { dt}} +\nu k_a^2\Bigl) Y_a= 
\sum_{b,c}^N  A_{abc} Y_b Y_c + f_a \,\, ,
\end{equation}
instead of eq. (\ref{4.4}).
It is well known\cite{F95} that in
 $3D$ the limit $\nu \to 0$ is singular and cannot be interchanged with the
limit $K_{max} \to \infty$,  therefore the statistical mechanics  of an 
inviscid fluid has a rather limited relevance for the Navier-Stokes 
equations at very high Reynolds numbers.
\footnote{In $2D$ the use of conservative statistical mechanics
has been proposed to justify some behaviors of real fluids, e.g. Jupiter's
red spot and the formation of coherent structures \cite{R91,P94}.
This approach, although interesting, is limited to some specific
two-dimensional situations.}
In the $\nu \to 0$ limit
in presence of forcing at large scale (small $k$) one has 
an intermediate range, called the inertial
range, which has neither  pumping nor dissipation and shows   
a strong departure from  equipartition\cite{F95,BJPV98}.
Instead of  $<Y_a^2>=const.$ one has $<Y_a^2> \sim k_a^{-\gamma}$
where $\gamma \simeq 11/3$, the value $\gamma=11/3$ corresponding to the
Kolmogorov spectrum\footnote{$E(k)dk$ is the contribution to the kinetic energy
of  the wave numbers in the range $[k, k+dk]$.}
 $E(k)\sim k^{-5/3}$.

\subsubsection{A general remark on the decay of correlation functions}

Using general arguments,  one could naively conclude that all typical correlation
functions, at large time delay, have to relax to zero with the same
characteristic time related to spectral properties of the  operator
${\cal L}$, governing the time evolution of the probability distribution $P(\{Y\},t)$:
\begin{equation} 
\label{4.13}
{ \partial \over {\partial t}} P(\{Y\},t)={\cal L} P(\{Y\},t).
\end{equation}
The argument goes as follows: let the time evolution of the system be described by
ordinary differential equations $dY_a/dt=X_a(\{Y\})$ so that ${\cal L}
P(\{Y\},t)= -\sum_a {\partial \over {\partial Y_a}} \Bigl(
X_a(\{Y\})P(\{Y\},t)\Bigr)$, for the system (\ref{4.11}), where
$X_a=\sum_{a,b}A_{abc}Y_bY_b-\nu k_a^2Y_a+f_a.$
\footnote{In  the presence of white noise terms $\{ {\eta}_a \}$ in eq. (\ref{4.11}),
i.e. $X_a \to X_a+{\eta}_a$ where $\{ {\eta}_a \}$ are Gaussian processes
with $<\eta_a(t)>=0$ and  
$<{\eta}_a(t){\eta}_b(t')>=2\Lambda_{ab}\delta(t-t')$, 
 one has to add to ${\cal L}$ the term\cite{G90}
 $$\sum_{ab} \Lambda_{ab} {\partial^2 \over {\partial Y_a \partial Y_b} }\, .$$
}
Let us introduce the eigenvalues $\{\alpha_k\}$ and the eigenfunctions
$\{\psi_k\}$ of ${\cal L}$:
\begin{equation} 
\label{4.14}
{\cal L}\psi_k=\alpha_k \psi_k \, .
\end{equation}
Of course $\psi_0=P_{inv}$ and $\alpha_0=0$; moreover the mixing
condition requires that $Re\, \alpha_k <0$ for $k=1,2,...$.
Furthermore assuming that the coefficients $\{g_1,g_2,...\}$ and 
$\{h_1,h_2,...\}$ exist, such that the expansion of the functions $g(\{Y\})$ and 
$h(\{Y\})$ are unique:
\begin{equation} 
\label{4.15}
g(\{Y\})=\sum_{k=0}g_k \psi_k(\{Y\}) \,\,\, , \,\,\,
h(\{Y\})=\sum_{k=0}h_k \psi_k(\{Y\}) \,\, ,
\end{equation}
one has
\begin{equation} 
\label{4.16}
C_{g,f}(t)= \sum_{k=1} g_k h_k <\psi_k^2> e^{\alpha_k t} \, ,
\end{equation}
where $C_{g,f}(t)= <g(\{Y(t)\}) h(\{Y(0)\})> - <g(\{Y\})>< h(\{Y\})>$.
For ``generic'', i.e. not orthogonal to $\psi_1$, so that $g_1\neq 0$
and $h_1\neq 0$, functions $g$ and $f$, the quantity $C_{g,f}(t)$
approaches zero at large time:
\begin{equation} 
\label{4.17}
C_{g,f}(t) \sim e^{-t/\tau_c} \,\,\, , \,\,\, 
\tau_c= {1 \over {|Re \, \alpha_1|}} \, .
\end{equation}
Naively  this argument seems to imply that in hydrodynamic systems
all the  correlation functions,
 $C_{a,a}(t)=<Y_a(t)Y_a(0)>$, relax 
to zero with the  same characteristic time for all $k_a$,
while one expects a whole hierarchy of characteristic
times distinguishing the behavior of the correlation
functions at different scales \cite{F95}. 
The paradox is, of course, only apparent
since the above  argument is valid only at very long times,
i.e. much longer than the longest characteristic time, hence it tells  nothing 
about systems with many different time scales.

\subsubsection{FDR in perfect fluids and in turbulence}

Although the inviscid fluid mechanics is not a Hamiltonian system,
Eqs. (\ref{3.3})
and (\ref{4.7}) (or  (\ref{4.9}) in $2D$) make the FDR take 
the simple form
\begin{equation} 
\label{4.12}
R_{a,b}={C_{a,a}(t) \over C_{a,a}(0)} \delta_{ab} \, 
\end{equation}
because of the Gaussian statistics.  To the best of our knowledge,
Kraichnan was the first to realize that the FDR may hold in a non
Hamiltonian context\cite{Kr59,Kr00}.  As a matter of fact, in the case
of turbulence the scenario is not too simple for two reasons:\\ a) the
(unknown) invariant measure is not Gaussian \\ b) the degrees of
freedom $\{Y_a\}$ are not independent.\\ Nevertheless the general
result discussed in Section 3 suggests that a relation between
response and a suitable correlation function must exist. Indeed, also
in turbulence there are many degrees of freedom, which are likely to
yield a smooth distribution, when projecting on lower dimensional
spaces, as required for the FDR.

The FDR plays an important role in the statistical closure
problem, which has to be tackled by any non linear theory. In our case the evolution equations for
$C_{aa}(t)=<Y_a(t)Y_a(0)>$ have the form
\begin{equation} 
\label{4.18}
\Bigr({{d} \over { dt}} +\nu k_a^2\Bigl) C_{aa}=F_a(C^{(3)}) 
\end{equation}
where $F_a$ is a function of the third-order correlations $C^{(3)}$ i.e.
of terms like $\langle Y_a(t)Y_b(0)Y_c(0) \rangle$.
In the equations for the $C^{(3)}$, fourth-order correlations $C^{(4)}$
appear, and so on.
An analogous situation occurs  in  the kinetic theory of gases, with 
the  Bogolyubov-Born-Green-Kirkwood-Yvon hierarchy\cite{B75}.
Experience has shown that it is quite efficient to truncate the hierarchy
at a certain level, making some closure hypothesis.
The simplest assumption in our case is the quasi-normal approximation,
which assumes that the fourth-order cumulants vanish, i.e that
\begin{multline} 
\label{4.19}
<Y_a\,Y_b\,Y_c\,Y_d>=<Y_a\,Y_b> <Y_c\,Y_d>\\+<Y_a\,Y_c> <Y_b\,Y_d>+<Y_a\,Y_d> <Y_b\,Y_c> \, .
\end{multline}
Unfortunately the resulting set of equations gives 
unphysical results, such as negative $E(k)$ for some  $k$. 
Therefore it is necessary to introduce more precise hypotheses.
One approximation which has been successfully used is the so called
eddy-damped quasi-normal Markovian approximation, which takes
into account the physical mechanism of the energy transfer\cite{L90}.

An alternative approach is the direct-interaction-approximation (DIA)
developed by Kraichnan\cite{Kr61,Kr66} which is one of the few self-consistent
analytical turbulence theory so far discovered.
In the DIA one writes separately equations for responses $\{ R \}$
and for correlations  $\{ C \}$
\begin{equation} 
\label{4.20}
\Bigr({{\partial} \over {\partial t}} +\nu k_a^2\Bigl) C_{a,a}(t,t')=
{\cal F}_a(\{C\}, \{R\}) 
\end{equation}
\begin{equation} 
\label{4.21}
\Bigr({{\partial} \over {\partial t}} +\nu k_a^2\Bigl) R_{a,a}(t,t')=
{\cal G}_a(\{C\}, \{R\}) 
\end{equation}
where  $C_{a,a}(t,t')=<Y_a(t)Y_a(t')>$,
$R_{a,a}(t,t')= {\overline{\delta Y_a(t)}}/\delta Y_a(t')$, 
${\cal F}_a$ and ${\cal G}_a$ are (rather complicated) nonlinear 
integral expressions involving $\{C\}$ and $\{R\}$.
The DIA equations give the correct results (including the FDR in the form 
(\ref{4.12})) in the inviscid limit. 
In addition,  for the ``random coupling model'', i.e. 
\begin{equation} 
\label{4.22}
\Bigr({{d} \over { dt}} +\nu k_a^2\Bigl)Y_a=
\sum_{b,c}A'_{abc}Y_bY_c \, ,
\end{equation}
where $A'_{abc}=\pm A_{abc}$ with the sign varying randomly
from triad  to triad $(a,b,c)$, Kraichnan\cite{Kr61}
showed that DIA equations are exact.
\footnote{The DIA gives exact results also in other cases, e.g.  in
the large-$N$ limit of the spherical model of turbulence~\cite{E94}
where the velocity field in (\ref{4.1}) is now a vector with $N$
components, i.e. ${\bf v} \in {\bf R}^N$, while ${\bf x} \in {\bf
R}^3$.}  Although the DIA equations (\ref{4.20},\ref{4.21}) are closed,
they are very complicated, so simplifications are usually introduced,
like the so called Simple-Pole Model \cite{KL87} where basically one
assumes the FDR in its simplest form:
\begin{equation} 
\label{4.23}
C_{a,a}(t,t')= R_{a,a}(t,t') C_{a,a}(t',t')   \,\,\,\,\, t\ge t' \, . 
\end{equation}
It has been shown that renormalization group methods and the
so called time-ordering approach \cite{KM04,MK05} give the same
results with an exponential form:
\begin{equation} 
\label{4.24}
C_{a,a}(t,t')=  C_{a,a}(t',t') e^{-\omega(k_a)|t-t'|}
  \,\,\,\,\, t\ge t' \, ,
\end{equation}
\begin{equation} 
\label{4.25}
R_{a,a}(t,t')= e^{-\omega(k_a)(t-t')}
  \,\,\,\,\, t\ge t' \, ,
\end{equation}
with $\omega(k) \sim k^{-2/3}$ in agreement with simple dimensional
arguments.

Note that Eq. (\ref{3.13}) indicates that assumption (\ref{4.23})
is equivalent to assume Gaussian statistics for $\{Y\}$ without
imposing a specific covariance matrix.  Since it is well known, from
both numerical and experimental results, that for very
high Reynolds numbers, the statistics of the velocity fields is
rather far from Gaussian, e.g.  one has long tails in the PdF and
intermittent behavior\cite{F95,BJPV98}, eq.(\ref{4.23}), or eqs. (\ref{4.24}, \ref{4.25}), can not be completely
correct.  However on intuitive grounds one expects the times
characterizing the responses $\{R\}$ to approximate the characteristic
correlation times of $\{C\}$.  Numerical investigations \cite{Kr66}
at moderate Reynolds number of the DIA equations, show that
$R_{a,a}(t,t')$ is not exactly proportional to $C_{a,a}(t,t')$,
however comparing the correlation times $\tau_C(k_a)$ (e.g. the time
after which the correlation function becomes lower than $1/2$) and the
response time $\tau_R(k_a)$ (e.g. the time after which the response
function becomes lower than $1/2$), one has that the ratio
$\tau_C(k_a)/\tau_R(k_a)$ stays constant through the inertial range.

\subsubsection{Numerical results for simplified models}

Let us now discuss some numerical tests of the FDR in
simplified models of fluid dynamics.  Such systems are far from being
realistic, but are nontrivial and share the main features of the
Navier- Stokes, or the Euler equations.  Consider the
model \cite{OM80}:
\begin{equation}
\label{4.26}
{dY_n \over dt}= Y_{n+1} Y_{n+2} + Y_{n-1} Y_{n-2} - 2 Y_{n+1} Y_{n-1}
\end{equation}
with $n=(1,2,...,N)$  and the periodic condition $Y_{n+N}=Y_n$.
This system, originally introduced  as a toy model for chaos in
fluid mechanics\cite{OM80}, 
contains some of the main features of inviscid hydrodynamics:
a) quadratic interactions; b) a quadratic invariant, $E={1 \over 2}\sum_{n=1}^{N} Y_n^2$;
c) the Liouville theorem holds.
For sufficiently large $N$
the distribution of each variable $Y_n$ is Gaussian. 
In this situation,
the Gaussian-like FDR holds for each of the $n$ variables, 
i.e. self-response functions to infinitesimal perturbations
are indistinguishable from the corresponding self-correlation functions
\cite{FIV90}.

\begin{figure}[htbp]
\includegraphics[angle=-90,width=15cm,clip=true]{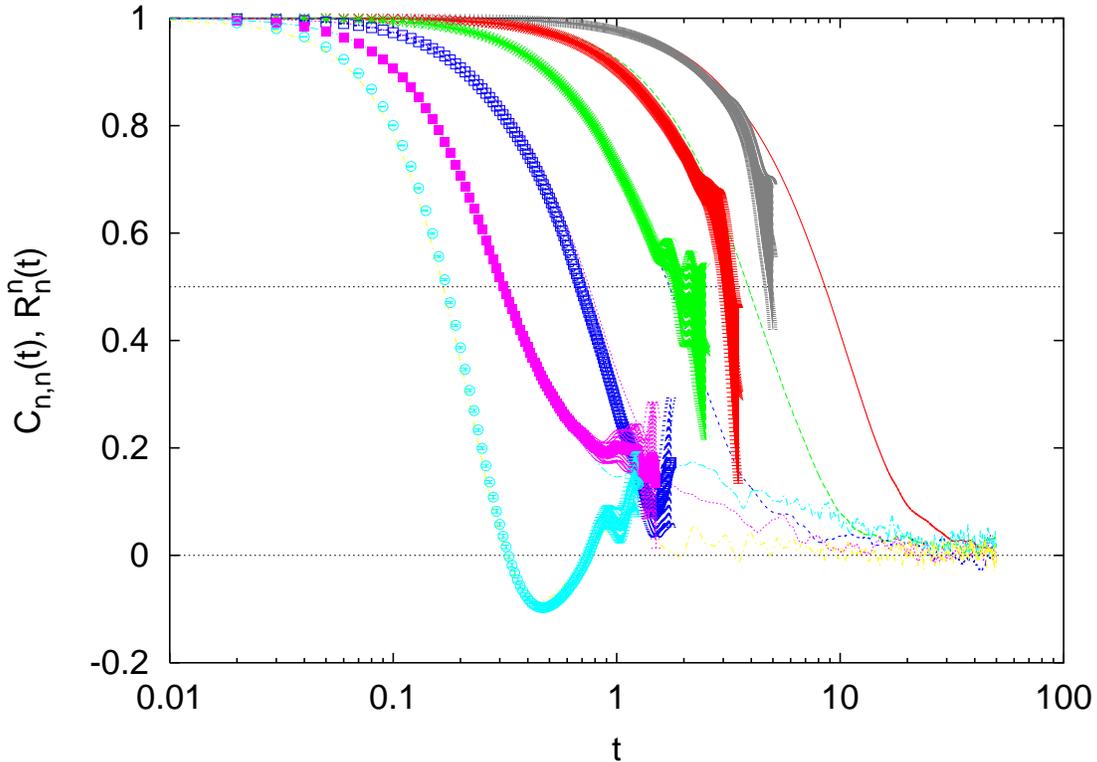}
\caption{
FDRs for the six fastest variables of 
the  model (\ref{4.27}), with $N=20$, $\alpha=0.005$ and $\beta=1.7$,
 $n=5,6,7,8,9,10$. 
Thin lines represent the normalized correlation functions 
$C_{n,n}(t)/C_{n,n}(0)$. 
Self-response functions $R_{n,n}$ 
are plotted with statistically computed error bars. 
\label{fig:4.1}}
\end{figure}

We can slightly modify the system (\ref{4.26}) in order to
have variables with different characteristic times;  for
instance, by rescaling the evolution time of each variable\cite{BDLV02}:
\begin{equation}
\label{4.27}
\frac{dY_n}{dt} = k_n \cdot (Y_{n+1} Y_{n+2} + 
Y_{n-1} Y_{n-2} - 2 Y_{n+1} Y_{n-1}) \, ,
\end{equation}
where the factor $k_n$ is defined as $k_n = \alpha \cdot \beta^n$, with 
$\beta > 1$,
for $n=1,2,...,N/2$, with the "mirror" property $k_{n+N/2}=k_{N/2+1-n}$.
For  the system (\ref{4.27})  a  quadratic integral of motion exists:
\begin{equation}
\label{4.28}
I = \sum_{n=1}^{N} { Y_n^2 \over k_n}
\end{equation}
Moreover, if $N \gg 1$  the  $\{ Y_n \}$ variables  preserve 
the Gaussian statistics with $<Y_nY_m>=<Y_n^2>\delta_{nm} \propto k_n$.
As a consequence  the simplest  FDR, i.e. Eq.~(\ref{4.12}),
holds for each of the variables. 
On the other hand, at variance with  the original  system (\ref{4.26}),
 each $Y_n$  has here its own characteristic time.
Figure \ref{fig:4.1} shows the (expected) validity of
(\ref{4.12}), however the shape of  $C_{nn}(t)$
changes with $n$. 
The correlation time of a
variable $Y_n$ behaves as
\begin{equation}
\label{4.29}
\tau_{C}(k_n) \sim k_n^{-3/2} \, .
\end{equation}
The exponent of this scaling law can be easily explained with a 
dimensional argument, by noticing that 
$<Y_n^2> \sim k_n$, and from (\ref{4.27}) the characteristic 
time results to be just $\tau_C(k_n) \sim \sqrt{<Y_n^2>}/(k_n<Y_n^2>)
 \sim k_n^{-3/2}$.   

\begin{figure}[htbp]
\includegraphics[angle=-0,width=15cm,clip=true]{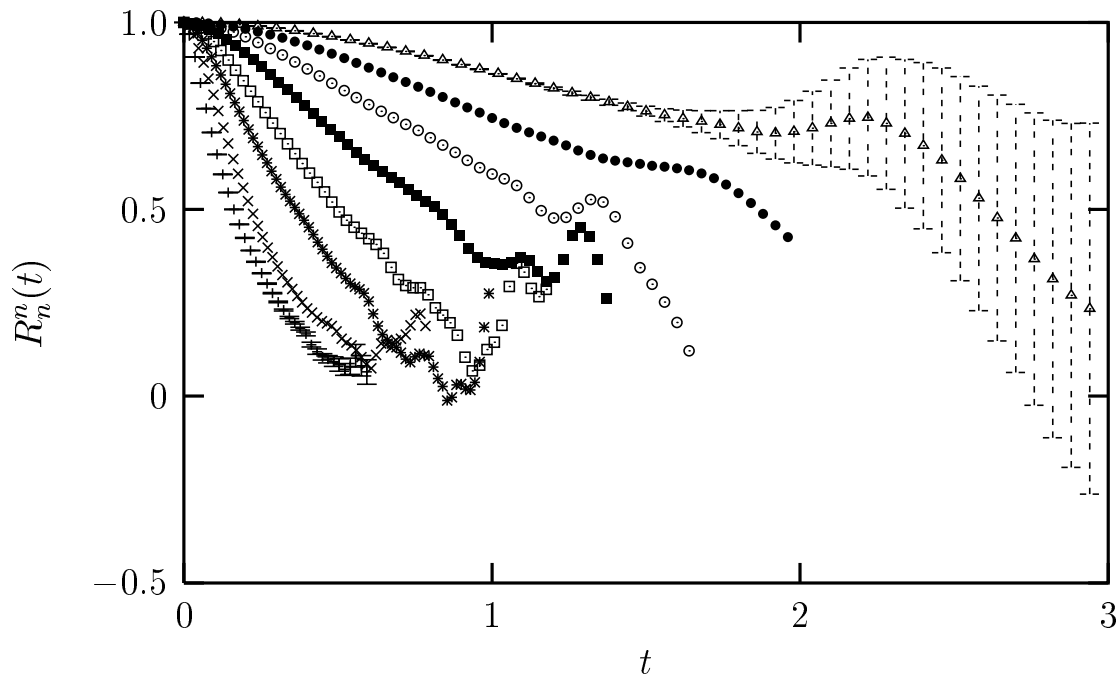}
\caption{
Response functions, $ R_{n,n}(t)$,
 for  shells $n=7,8,\dots,13,14$ (from top  to bottom).
Error bars are shown only for the smallest and the largest scales.
The parameters in eqs. (\ref{4.30}) are $b=0.4$,
$\nu=5.10^{-7}$, and $N=25$.
\label{fig:4.2}}
\end{figure}

\begin{figure}[htbp]
\includegraphics[angle=-0,width=15cm,clip=true]{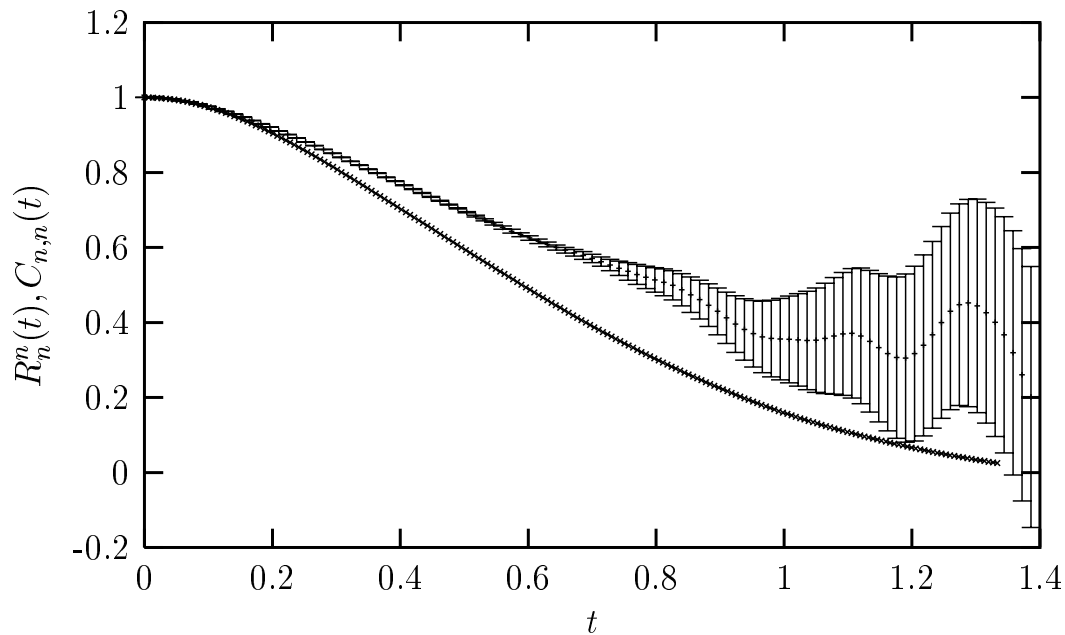}
\caption{
Comparison  between the averaged Response Function, $R_{nn}(t)$, (top)
 and the (normalized) self-correlation, $C_{n,n}(t)/C_{nn}(0)$
 (bottom) for the shell $n=10$. 
\label{fig:4.3}}
\end{figure}

Let us now discuss more interesting cases, i.e. the chaotic
dissipative systems with  non Gaussian statistics, called ``shell models'' which have been introduced to
study the  turbulent energy cascade and 
 share many statistical properties
with  turbulent three dimensional velocity fields \cite{BJPV98,B03}. 
Let us consider   a set of wave-numbers $k_n = 2^n k_0$ with
$n = 0,\dots,N$, and the  shell-velocity (complex) variables $u_n(t)$ 
which must be understood as
the velocity fluctuation over a distance $l_n=k_n^{-1}$.
In the following we present numerical
results for a particular choice, the so-called Sabra model
\cite{LPPPV98,B03}:
\begin{equation}
\label{4.30}
\left(\frac{d}{dt}+\nu k_n^2\right) u_n =
i\left[k_nu_{n+1}^*u_{n+2} + b k_{n-1}
u_{n+1}u_{n-1}^* + (1+b)k_{n-2}u_{n-2}u_{n-1}\right] +f_n
\end{equation}
where $b$ is a free parameter, $\nu$ is the molecular viscosity and
$f_n$ is an external forcing acting only at large scales, necessary to
maintain a stationary temporal evolution.  We note that the shell
models have the same type of quadratic nonlinearities as the
Navier-Stokes equations in the Fourier space. In addition in the
inviscid and unforced limit ($\nu= f_n=0$) the energy (${1 \over 2}\sum_n |u_n|^2$) is conserved and
the volumes in phase space are preserved.  Because of the above properties, the set
of coupled ordinary differential equations (\ref{4.31}) possesses the
features necessary to mimic the Navier-Stokes non-linear evolution.
The main, strong, difference with the previous inviscid models
(\ref{4.26}, \ref{4.27}) is the existence of a mean energy flux from
large to small scales, which drives the system towards a strongly
non-Gaussian stationary state~\cite{BJPV98}.  The shell
models here discussed present exactly the same qualitative
difficulties as the original Navier-Stokes eqs.: strong non-linearity
and far from equilibrium statistical fluctuations. The most striking
quantitative feature of the fully developed turbulence, i.e.  the
highly non-Gaussian statistics and the existence of anomalous scaling
laws for the velocity moments, is reproduced in shell models, in good
quantitative agreement with the experimental results\cite{BJPV98,B03}.
One has:
\begin{equation}
\label{4.31}
<|u_n|^p> \sim k_n^{-\zeta(p)} \, ,
\end{equation}
where $\zeta(p) \neq {p \over 2} \zeta(2)$, which implies
that the velocity PDF's at different scales cannot be rescaled
by any  change of variables.\\
Let us now examine the numerical results concerning the response functions
in the shell model\cite{BDLV02}. 
Figure \ref{fig:4.2} shows the normalized diagonal  responses,
 $R_{nn}(t)$,  for $k_n$ in the inertial range, $n\in[7,14]$.
It is evident that the characteristic response time $\tau_R(k_n)$
decreases as $k_n$ increases.\\
In Figure \ref{fig:4.3}, one sees  a  clear difference 
between the response and the correlation function: this
is an additional  indication that the inertial-range statistics
are very far from  Gaussian.
Although the $R_{nn}(t)$ do not coincide with the (normalized)
 $C_{nn}(t)/C_{nn}(0)$, the times $\tau_C(k_n)$ and $\tau_R(k_n)$ 
are not very different $ \tau_C(k_n) \propto  \tau_R(k_n)$,
in agreement with the numerical result  
obtained by Kraichnan with the DIA at moderate Reynolds numbers\cite{Kr66}.
In addition, one has a  scaling behavior
$ \tau_R(k_n) \propto  \tau_C(k_n) \sim k_n^{-\gamma}\,$,
where $\gamma$ is close to the value $2/3$ of the 
Kolmogorov scaling.

\subsection{Beyond the  linear FDR and climate}

In the case of finite perturbations, the FDR (\ref{3.6})
is typically non-linear in the perturbation $\Delta{\bf x}_0$, and
thus there are no  simple relations analogous to (\ref{3.8}).
Nevertheless Eq. (\ref{3.6}) guarantees the existence of a link
between equilibrium properties of the system and the response to
finite perturbations.  This fact has a straightforward consequence for
systems with a single characteristic time, \textit{e.g.} a low dimensional
system: a generic correlation function 
 gives information about the relaxation time of finite size
perturbations, even when the invariant measure $\rho$ is not known
\cite{BLMV03}.
We saw that in systems with many different characteristic times, 
such as the fully
developed turbulence, one has a more complicated scenario: different
correlation functions show different behaviors.\\
For a finite perturbation $\delta {\bf x}(0)$ such that
$A({\bf x}(0)+\delta {\bf x}(0))- A({\bf x}(0))=\delta A(0)$, 
instead of (\ref{3.10}) one has: 
\begin{equation} 
\label{4.32}
\overline{\delta A(t)}= 
\langle A \left( {\bf x}(t) \right)
F \left( {\bf x}(0), \delta {\bf x}(0) \right) 
\rangle \, ,
\end{equation}
therefore  the relaxation properties depend
explicitly on the initial perturbation $\delta {\bf x}(0)$
and not only on the value of $\delta A(0)$.
In the following we will discuss the relevance of the 
amplitude of the perturbation, which may play a major role in
determining the response,  because different amplitudes may
trigger different response mechanisms with different time scales
\cite{BLMV03,BDLV02}.

\subsubsection{Response to finite perturbations}

\begin{figure}[htbp]
\includegraphics[angle=-0,width=15cm,clip=true]{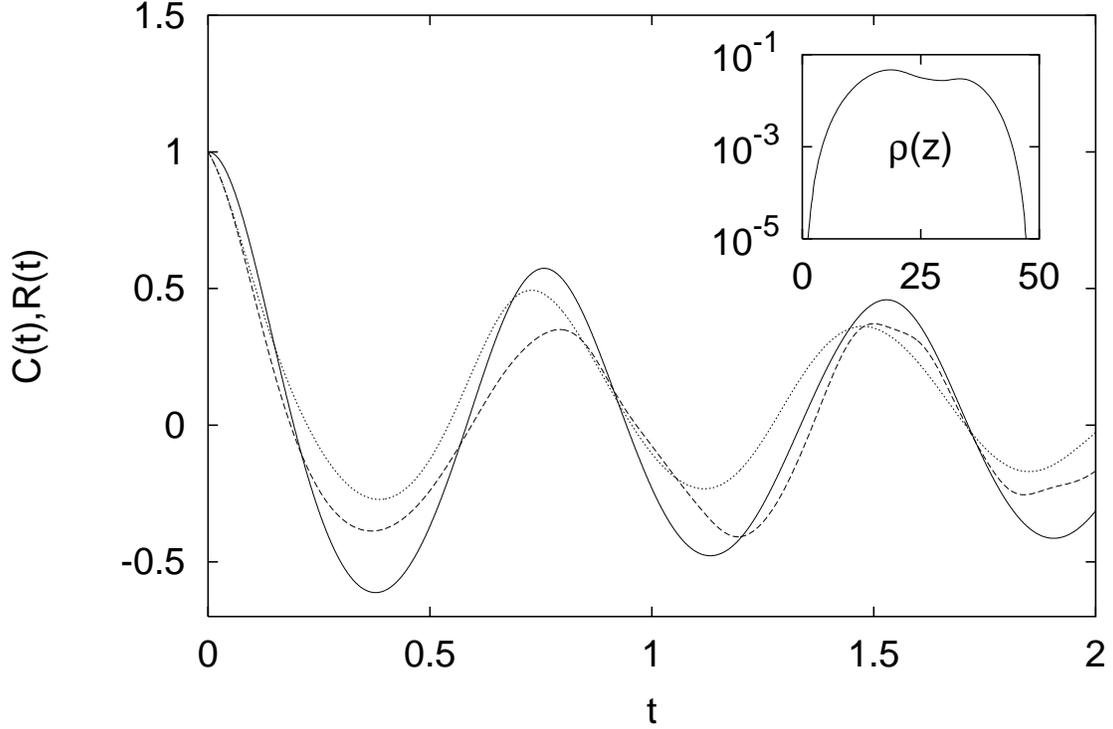}
\caption{ Normalized correlation function of the $z$ variable of
Lorenz model (solid line) compared with the mean response to different
perturbations of the same variable.  $\delta z_0 = 10^{-2} \sigma$
(dashed line), $\delta z_0 = \sigma$ (dotted line), with $\sigma^2 =
\langle z^2 \rangle - {\langle z \rangle}^2 = 75.17$.  In the inset
the invariant probability $\rho(z)$ versus $z$. Being a 1D projection,
$\rho(z)$ results regular, even if the full 3D distribution is
fractal.
\label{fig:4.4}}
\end{figure}

Consider two cases with a single characteristic time: a low
dimensional deterministic chaotic system, known as the Lorenz model,
and a nonlinear Langevin equation.  We begin with the Lorenz model
\cite{L63}:
\begin{eqnarray}
{dx \over dt} & = & \sigma (y-x) \nonumber \\
{dy \over dt} & = & -xz +rx -y \\
\label{4.33}
{dz \over dt} & = & xy -b z \nonumber
\end{eqnarray}
with standard parameters for chaotic behavior: $b=8/3$, $\sigma=10$ and 
$r=28$.
The correlation function of the variable $z$,
shown in Figure~\ref{fig:4.4}, qualitatively reproduces the behavior
of the response to different sizes of the perturbation of the $z$ variable, 
ranging from infinitesimal ones up to those of the size of the attractor.
The agreement  between the response function and the normalized
correlation function is just qualitative and it does not depend too
much on the  amplitude of the perturbation, 
because the invariant distribution is
not Gaussian (see inset of Fig.~\ref{fig:4.4}) and  the general correlation
(\ref{3.13}) should be used.

\begin{figure}[htbp]
\includegraphics[angle=-0,width=15cm,clip=true]{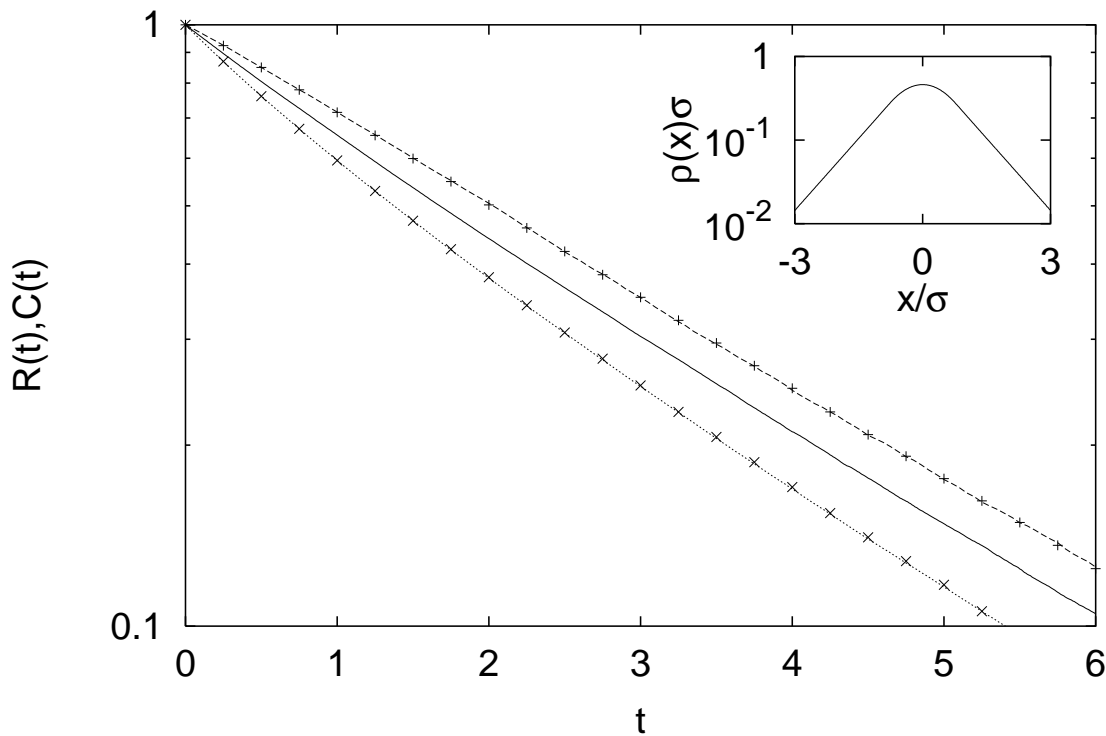}
\caption{
Response of the stochastic differential equation
(\ref{4.34}) with  $D=1$ and $U(x)$ as in (\ref{4.36}), 
to different perturbations: 
large $\delta x_0 = 2.3 \sigma$ ($+$) and 
infinitesimal $\delta x_0 = 7.6 \times 10^{-3} \sigma $ ($\times$). 
In both cases the mean response is exactly predicted by the correlation
function $<x(t)f(x(0))>$, with $f(x)$ given by (\ref{3.6},\ref{3.7}),
dashed line for $\delta x_0 = 2.3 \sigma$ 
and dotted line for $\delta x_0 = 7.6 \times 10^{-3}$.  
The correlation function $\langle x(t)x(0) \rangle / \sigma^2$
(solid line) is shown for comparison.
In the inset the invariant probability distribution 
$\rho(x)\sigma$ versus $x/\sigma$ with 
$\sigma^2 = \langle x^2 \rangle - {\langle x \rangle}^2 = 1.15$.
\label{fig:4.5}}
\end{figure}

To better illustrate this point, we consider
a system whose invariant probability distribution is known.
In this case, we can quantitatively compare the differences 
between the responses to infinitesimal and finite perturbations.
Let us consider the stochastic process $x(t)$ described by  
\begin{equation}
\label{4.34}
{dx \over dt} = - {dU(x) \over dx} + \sqrt{2D} \eta(t)
\end{equation}
where $\eta(t)$ is a white noise.
The invariant probability distribution is \cite{G90}:
\begin{equation}
\label{4.35}
\rho(x) = {\mathcal N} e^{-U(x)/D}
\end{equation} 
where ${\mathcal N}$ is fixed by normalization, and
\begin{equation}
U = \left\{ 
\begin{array}{ll}
{1 \over 2} x^2 &, |x| < 1 \\
|x| -{1 \over 2} &, |x| > 1 .
\end{array} 
\right.
\label{4.36}
\end{equation}
The resulting pdf, shown in the inset of Fig.~\ref{fig:4.5},
has a Gaussian core, with exponential tails.
Figure~\ref{fig:4.5} also shows the response function for
an infinitesimal and for a finite size perturbation.
For both perturbations, the response function measured from 
the perturbed trajectories is exactly predicted by the statistics
of the unperturbed system according to (\ref{3.13}),
while the Gaussian correlation 
$C(t) = \langle x(t)x(0) \rangle / \sigma^2$ is only
an approximation  of the response. 
By construction, the pdf of this system has larger tails than 
in the Gaussian case, thus large fluctuations have slower decay than small ones.
On the contrary in the Gaussian case, i.e. $U(x)={1 \over 2}x^2$ for any $x$,
the  response is an  exponential function 
and does not depend on the amplitude of the initial perturbation 
$\delta x(0)$.

We have already seen how in systems with many characteristic times,
different correlation functions behave differently and a variety of
time scales emerges, which correspond to the different decay times of
the correlation functions.  In addition, at variance with systems with
a single time scale, the amplitude of the perturbation plays a major
role in determining the response, because different amplitudes affect
features characterized by different time scales.  To illustrate these
ideas, we refer to the shell model (\ref{4.30}) and
consider the total energy $E(t) = {1 \over 2} \sum_{n=1}^N |u_n(t)|^2$
which is conserved in the inviscid, unforced limit \cite{BJPV98}.  We
study the response to perturbations of $E$ with different amplitude,
of the following form:
\begin{equation}
|\delta u_n^{(i)}(0)| = \left\{ \begin{array}{ll}
 0 & \, , \, 1 \le n \le i-1 \\
 \sqrt{\langle |u_n|^2 \rangle} & \, , \, i \le n \le N 
 \end{array} \right . \, .
\label{4.37}
\end{equation}
The value of $i = 1,..,N$ determines the largest spatial scale $\ell$
of the perturbation, $\ell \sim k_i^{-1}$.
This corresponds to a set of initial perturbations of the energy 
\begin{equation}
\delta E^{(i)}(0)  = {1 \over 2} \sum_{n=i}^N  
\langle|u_n|^2 \rangle .
\label{4.38}
\end{equation}
Such a kind of perturbation is motivated by the fact that, in the
unperturbed system, the energy is distributed among the shells
according to the Kolmogorov scaling $ \langle |u_n|^2 \rangle \sim
k_n^{-2/3}$, and the smaller scales give smaller contributions to the
energy $E(t)$.  Thus it is natural to assume that a small perturbation
of the energy affects mainly the small scales.

For each perturbation $\delta E^{(i)}(0)$, 
the average energy response
\begin{equation}
R_E(t)=
{ \overline{\delta E^{(i)}(t)}  \over \delta E^{(i)}(0)}
\label{4.39}
\end{equation}
reveals a behavior similar to  the correlation function of 
the largest perturbed shell $u_i(t)$, as shown in Fig.~{\ref{fig:4.6}.
An additional  measure of the relaxation time is provided by 
the halving times $T_h$ of the mean response, when  
$\overline{\delta E^{(i)}(T_h) } = {1 \over 2}  \delta E^{(i)}(0)$.
The dependence of the halving times
on the amplitude of the initial perturbation
 reflects the Kolmogorov scaling for  characteristic times.
Noting that  
$\tau_C(k_n) \sim \tau_R(k_n) \sim  k_n^{-2/3}
 \sim u_n^2 \sim \delta E^{(i)}(0)$ one has:
\begin{equation}
T_h \sim \delta E^{(i)}(0) \, \, .
\label{4.40}
\end{equation}
The above results  show that 
the response 
of a system with many characteristic times to a finite size perturbation
may depend on the amplitude of the perturbation.  
Thanks to the applicability of the FDR, it is possible to 
establish a link between relaxation times 
of different perturbations and characteristic times of the system. 

\begin{figure}[htbp]
\includegraphics[angle=-0,width=12cm,clip=true]{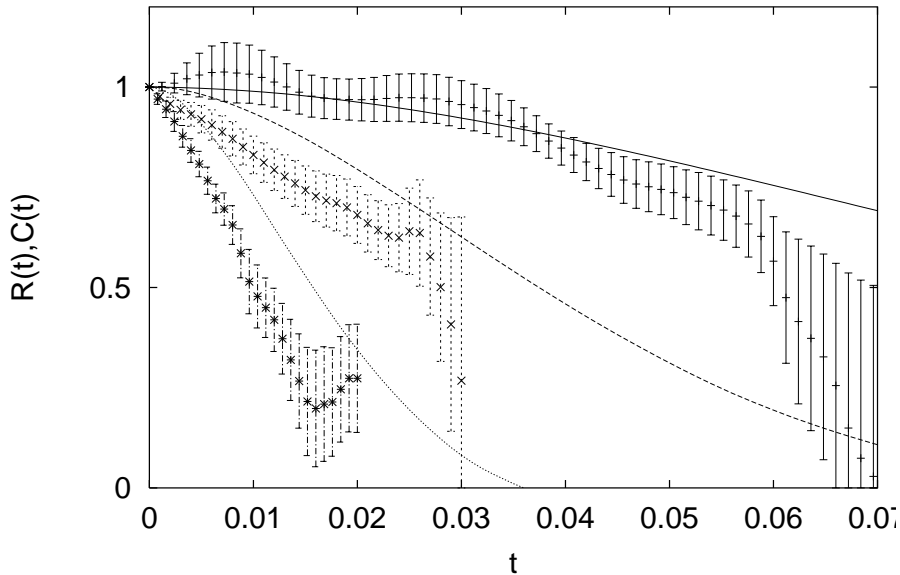}
\caption{
Mean response $R_E(t)$ 
of the total energy $E(t) = {1 \over 2} \sum |u_n|^2$
of the shell model (\ref{4.30})
to different amplitude perturbations:
$\delta E^{(12)}(0) =5.5 \times 10^{-3}$ ($+$),
$\delta E^{(14)}(0) =1.7 \times 10^{-3}$ ($\times$),
$\delta E^{(16)}(0) =4.5 \times 10^{-4} $ ($*$);
and the (normalized) 
correlation functions of the corresponding largest perturbed shell:
shell $n = 12$ (solid line), 
shell $n = 14$ (dashed line),
shell $n = 16$ (dotted line).
\label{fig:4.6}}
\end{figure}

\subsubsection{FDR and climate}

One of the key issues in  climate modeling is
the study of  the response to perturbations of the external forces,  or  control parameters.
In such a  context, the importance of the FDR is rather obvious: 
if it holds there is  the possibility, at least in principle, 
to understand the behavior of the system  under 
perturbations (e.g. a volcanic eruption, 
or change of the $C \, O_2$ concentration)  in terms of
the knowledge obtained from its past  time history\cite{L75,L78}.

To describe the climate system, let us consider the symbolic evolution equation:
\begin{equation}
\label{4.41}
{{d{\bf X}} \over {dt}}= {\bf Q}({\bf X})+ {\bf f}(t)
\end{equation}
where ${\bf X}$ is the state vector for the system, ${\bf Q}({\bf X})$
represents complicated dynamical processes and ${\bf f}(t)$ the external
influences.
The average effect of an infinitesimal
 perturbation $\delta {\bf f}(t)$ can be written in 
terms of the response matrix $R_{\alpha,\beta}(t)$; if 
 $\delta {\bf f}(t)=0$ for $t<0$ one has:
\begin{equation}
\label{4.42}
\overline{\delta X_{\alpha}(t) }=\sum_{\beta} \int_0^t 
 R_{\alpha,\beta}(t-t')  \delta  f_{\beta}(t') dt' \, .
\end{equation}
The goal  is to write $R_{\alpha,\beta}(t)$ in terms
of correlation functions of the unperturbed system.

Under many points of view, the questions concerning the FDR in climate systems are rather similar
to those posed in fluids dynamics, because one deals with  non Hamiltonian 
and non linear systems, whose invariant measures are non-Gaussian.
On the other hand, in climate studies, one hardly faces  equations obtained from first principles, and typically
it is necessary to work with  rather crude models, or  
with experimental time series~\cite{D99}.
Moreover, in climate problems and in geophysics,
the interest in infinitesimal perturbations is rather academic,
while the interesting problem is the behavior of the relaxation of large
fluctuations,  due to sudden changes of the parameters.\\\\
Starting from the seminal works of Leith \cite{L75,L78},
who proposed to employ the FDR to study the
response of climatic systems to changes in the external
forcing, many authors applied this relation to
different geophysical problems, ranging from simplified models
\cite{B80}, to general circulation models \cite{NBH93,CVS04}
and to the covariance of satellite radiance spectra \cite{KGC99}.
In most of these attempts,
the FDR has been used in its
Gaussian version, eq. (\ref{3.14}),  which has been acritically considered
a reasonable approximation, without investigating its limits
of applicability. For a recent application to climate models with fast and slow
variables see \cite{LV07}.

The results previously discussed show that a FDR
 holds also for non infinitesimal perturbations.
This generalization of the usual linear response 
theory has an obvious interest for  climate applications.
Although the FDR in its simplest form  (\ref{3.14}) 
does not hold, one has, at least, a qualitative agreement
between responses and correlations.
In particular, from features of the unperturbed dynamics, one infers the presence of different response times
associated with different variables.
This description is qualitatively satisfactory. But a precise
quantitative estimate faces  difficult technical problems such as:\\
a) the FDR involves the detailed form of the
invariant probability distribution (which is usually
unknown)\\
b) non infinitesimal perturbations  
 require good statistics to resolve the rare events.

\subsection{Granular materials}
\label{subsec:granular}

Granular systems~\cite{JN92,JNB96,JNB96b} constitute an example of
non-equilibrium physics, pervading our everyday life: sand, sugar,
coffee, pills, seeds, powders, etc., are all instances of the
``granular state of matter''. For the behavior of granular systems,
the environment temperature plays only a marginal role. At the same
time these systems are made of relatively many particles. This
suggests that statistical mechanical methods, such as kinetic theory,
hydrodynamics, thermodynamics and linear response can be adapted to
granular materials.

A starting point of the statistical approach is to identify a
distribution that is left invariant by the dynamics, as is the case of
the microcanonical distribution in Hamiltonian systems, and the conditions under which
 this distribution can be reached, starting from a generic
initial condition.  Unfortunately, statistically steady states of
granular matter are usually given by the balance of non-conservative
forces, such as external tapping and inelastic collisions or friction
with the boundaries, and the corresponding invariant measures are not
known.  It is therefore natural to ask whether it is possible to speak
at all of a granular temperature. This question has different answers,
depending on the degree of diluteness of the granular system under
examination. The most straightforward definition comes from the case
of dilute, strongly vibrated granular systems, which reach a
non-equilibrium stationary state: by analogy with molecular gases, a
``granular temperature'' $T_g$ can be defined in terms of the average
local kinetic energy per particle. This approach can a priori be
extended to denser, liquid-like, strongly vibrated
systems~\cite{DMGBLN03}. However, if the density is further
increased, or the external energy injection strongly reduced, the
assumption of ergodicity becomes less and less obvious: reaching a
stationary state may become experimentally or numerically very
difficult. Dense granular systems exhibit aging~\cite{NC99,BL00} and
memory~\cite{JTMJ00,BL01}.  Analogies with other aging systems have
also led to the definition of dynamic temperatures as quantifying the
violation of the Hamiltonian version of the fluctuation-dissipation
theorem~\cite{BCL02}. In this review, we do not touch on non-ergodic
statistical physics, hence we do not discuss glassy granular systems
and we focus only on systems that rapidly forget their initial
conditions. This class includes the so-called ``granular gases'', i.e.
granular systems which are fluidized by different means, for instance
strongly shaking the container, or otherwise letting a low-density fluid (e.g. air) pass
around the grains at high velocity.

In Appendix~\ref{app:granular} the reader can find a brief review of
the models used for the granular gases. Such models include the
following ingredients: a gas of particles, a collision rule that
dissipates kinetic energy, and a dynamical rule governing the dynamics between
collisions. This dynamics depends upon the external forcing. If the
system is driven from the boundaries (shaking the container) and
gravity can be neglected, for example in microgravity experiments, the
bulk dynamics is made of straight paths plus inelastic
collisions. If the container is large, the external energy may take a
long time to reach the interior of the gas: this motivates the study
of pure cooling models~\cite{D00}. They are considered as the building
blocks of the hydrodynamics of boundary driven granular
gases~\cite{BDKS98}. For this reason, several authors have considered
the application of the theory of linear response to perturbations of
these systems, as a valid alternative to kinetic theory, for the
derivation of transport coefficients~\cite{BP05}. Here, the analysis
is more difficult because there is no steady state, but
only a scaling behavior towards a state with all the particles
at rest, which undermines basic assumptions of kinetic theory, such as
the separation of scales and the fast equilibration of microscopic
modes~\cite{K99,G99}. On the other hand, experiments with uniform
drivings, i.e. setups where the injection of energy reaches all
grains, show a very rich phenomenology and are much easier to model:
the gas reaches a statistically steady state and the formalism
introduced in Sec. 3 can be applied.

\subsubsection{Cooling granular gases}

Most of the work devoted to linear response in cooling granular
gases, concerns the calculation of transport coefficients for small
perturbations of the homogeneous cooling
state, under different assumptions~\cite{DG01,GM04,BM04,BP05}. A well established
approach~\cite{DB02,DBB06} considers the Boltzmann equation for a
granular gas as the starting point, see Appendix~\ref{app:granular},
Eq.~\eqref{granuboltz} with $\Gamma=\gamma=0$, which describes the
evolution of a single particle velocity distribution $f(\bx,\bv,t)$, in
a gas with particles interaction given by inelastic
collisions, obeying the rule of Eq.~\eqref{collision}. The effect of
collisions is included in the collision integral,
Eq.~\eqref{granucolint}.

Following the standard kinetic theory procedure, it is possible to
derive balance equations for the usual fields:
\begin{eqnarray}
D_t n+n \nabla \cdot {\bf u}&=&0\\
D_t u_i + (mn)^{-1}\nabla_jP_{ij}&=&0\\
D_t T_g+\frac{2}{dn}(P_{ij}\nabla_j u_i+\nabla \cdot \mathbf{q})+T_g \zeta&=&0,
\end{eqnarray}
where $n$, ${\bf u}$ and $T_g$ are the density, velocity and
temperature fields respectively, and $D_t=\partial_t+\mathbf{u}\cdot
\nabla$. The pressure tensor $P_{ij}$, the heat flux $\mathbf{q}$ and
the cooling coefficient $\zeta$ depend upon the particular (time
dependent) form of the distribution function $f(\bx,\bv,t)$. A
homogeneous scaling solution
$f_0(\bx,\bv,t)=nv_0^{-d}(t)f_0^*(\bv/v_0(t))$ is supposed to exist:
it has no $\mathbf{x}$-dependence (homogeneity) and the
time-dependence is all included in the thermal velocity
$v_0(t)=\sqrt{2T_g(t)/m}$~\cite{BMC96}. Such a solution must satisfy
an equation for $T_g(t)$, which results in the so-called Haff
law~\cite{H83} for the homogeneous granular cooling. At large times,
in the homogeneous case, the granular temperature decays as $T_g(t)
\sim t^{-2}$.

It is possible to express a generic perturbation of the scaling state
$f_0$ in terms of the gradients of the macroscopic fields
$n(\bx,t),\bu(\bx,t),T_g(\bx,t)$. Substituting this form in the
Boltzmann equation and retaining only terms of first order in these
gradients results in expressions for the
pressure tensor and the heat flux, which are linear in the field gradients~\cite{DB02}:
\begin{eqnarray}
P_{ij}&=&p(\bx,t)\delta_{ij}-\eta(\nabla_j u_i+\nabla_i u_j-\frac{2}{d}\delta_{ij}\nabla \cdot \mathbf{u})\\
\mathbf{q}&=&-\kappa \nabla T_g-\mu \nabla n,
\end{eqnarray}
where $\mu$ is a new coefficient peculiar to granular
gases, which vanishes in the elastic limit, and the following
``Green-Kubo'' relations hold, giving the transport coefficients:
\begin{eqnarray}
\eta(t)&=&\frac{2 n m \ell v_0(t)}{d^2+d-2}\int_0^{s(t)} ds'\sum_{ij}\langle D_{ij}^*(s')\Phi^*_{2,ij}\rangle\text{e}^{-\frac{s'\zeta^*}{2}}\\
\kappa(t)&=&\frac{n \ell v_0(t)}{d}\int_0^{s(t)} ds'\langle \mathbf{S}^*(s')\cdot \pmb{\Phi}^*_3\rangle \text{e}^{\frac{s'\zeta^*}{2}}\\
\mu(t)&=&\frac{2T_g(t)\kappa(t)}{n}+\frac{m \ell v_0^3(t)}{d}\int_0^{s(t)} ds' \langle\mathbf{S}^*(s')\cdot(\pmb{\Phi}_1^*-\pmb{\Phi}_3^*)\rangle.
\end{eqnarray}
The transport coefficients depend upon the time $t$, which is related
to the time $s$ through the transformation $ds=dt v_0(t)/\ell$, with
$\ell$ the mean free path, i.e. $s$ is proportional to the total
number of collisions occurred since the initial time. In the above
expressions the transport coefficients are obtained as time integrals
of correlation functions calculated with the unperturbed distribution $f_0$
(the average $\langle ... \rangle$ refers to this distribution). These are
correlations between the currents at time $s'$
\begin{eqnarray}
D_{ij}^*&=&  V_i^*V_j^*-\frac{1}{d}V^{*2}\delta_{ij}\\
\mathbf{S}^*&=&\left(V^{*2}-\frac{d+2}{2}\right ) \mathbf{V}^*,
\end{eqnarray}
and new conjugated quantities at time $0$:
\begin{eqnarray}
\pmb{\Phi}_1^*&=&\frac{1}{2}\left ( \mathbf{V}^*+\frac{1}{2}\frac{\partial}{\partial \mathbf{V}^*}\ln f_0^* \right )\\
\Phi_{2,ij}^*&=&\frac{1}{2}\left ( \frac{1}{d}\delta_{ij} \mathbf{V}^* \cdot \frac{\partial}{\partial \mathbf{V}^*}-V_i^*\frac{\partial}{\partial V_j^*} \right ) \ln f_0^*\\
\pmb{\Phi}_3^*&=&\frac{1}{2}\left[\frac{\partial}{\partial \mathbf{V}^*}\ln f_0^*-\mathbf{V}^*\left (d+\mathbf{V}^*\cdot\frac{\partial}{\partial \mathbf{V}^*}\ln f_0^* \right ) \right].
\end{eqnarray}
The rescaled peculiar velocity $\mathbf{V}^*=(\bv-\bu(\bx,t))/v_0(t)$
and the rescaled cooling coefficient $\zeta^*=\zeta \ell/v_0$ have
been used in the above expressions.  In the elastic limit $r \to 1$,
the unperturbed state $f_0$ becomes the Maxwell distribution, so that
$\pmb{\Phi}_1^* \to 0$, $\Phi_{2,ij}^* \to D_{ij}^*$, $\pmb{\Phi}_3^*
\to \mathbf{S}^*(\mathbf{V})$, and therefore
\begin{eqnarray}
\mu \to \mu_0 &=& 0\\
\eta \to \eta_0&=&\frac{2 n m \ell v_0(t)}{d^2+d-2}\int_0^s ds'\sum_{ij}\langle D_{ij}^*(s')D_{ij}^*(0)\rangle\\
\kappa \to \kappa_0&=&\frac{n \ell v_0(t)}{d}\int_0^s ds'\langle \mathbf{S}^*(s')\cdot \mathbf{S}^* \rangle \,\,.
\end{eqnarray}
The difference occurring in the granular case ($r <1$), with respect
to the elastic case, are the following: the averages $\langle
... \rangle$ are based on an invariant measure $f_0$ which is not
Maxwellian; the conjugated currents at time $0$ are different from the
fluxes $\mathbf{S}^*$ and $D_{ij}^*$, as expected from the
non-Gaussianity of the unperturbed state and from the considerations of
section~\ref{sec:chaos}. The time integration is replaced by an
integration over the average collision number $s$, and contains an
additional time dependent factor. This term arises from the change of
the temperature over the duration of the integral.

\subsubsection{Driven granular gases}

The effect of a uniform driving acting upon all the particles of
the granular gas is modeled assimilating the driving to a thermal bath,
 on the macroscopic scale of grains (see
Appendix~\ref{app:granular} for details). All the variants of this
thermal bath model guarantee the achievement of a statistically
stationary state after the time needed to forget the  state in
which the system has been initially prepared. The study of
the FDR in these models is recent and has been
carried out by means of numerical
simulations~\cite{PBL02,BLP04,BBDLMP05,PBV07}. Some of the numerical
results has received an analytical interpretation in~\cite{G04}, and
experimental verifications have been attempted in~\cite{MPEU05}. Other idealized
models of inelastic energy exchange~\cite{SL04,SBL06,LL06} have also been
proposed. Some of the numerical results and their interpretation are reviewed here, in
view of relation~\eqref{3.9}. We start with
monodisperse weakly correlated systems, we continue with the bidisperse case,
and we conclude with an analysis of the effect of strong correlations
in dense systems. 

{\em Monodisperse gases with weak correlations}

A simple Fluctuation-Response experiment involving only one particle of a statistically
stationary gas made of identical particles, consists in applying an
instantaneous perturbation $\delta v_0(0)$ to the $x$-component of the
velocity of that particle at a time $0$ (assumed to be in the
stationary regime), and observing its relaxation averaged over many
different realizations:
\begin{equation} \label{risp_dv}
R(t)=\frac{\overline{\delta v_x(t)}}{\delta v_x(0)}.
\end{equation}
If the invariant measure describing the statistically stationary state
satisfies the following condition (the independence of the degrees of freedom)
\begin{equation} \label{eq:indep}
\rho(\{ \mathbf v_i, x_i\})=n^N\prod_{i=1}^N \prod_{\alpha=1}^d p_v(v_i^\alpha),
\end{equation}
where $v_i^{\alpha}$ is the $\alpha$ velocity component of the $i$-th particle,
with $n=N/V$, and if the single particle velocity pdf $p_v(v)$ is a Gaussian, then
Eq.~\eqref{3.9} predicts 
\begin{equation} \label{einst}
R(t)=C_1(t)=\frac{\langle v_x(t)v_x(0) \rangle}{\langle v_x^2 \rangle}
\end{equation}
for the response function.
If one of these assumptions does not apply, then relation~\eqref{einst}
is no more expected to hold.

\begin{figure}[ht]
\centerline {
\includegraphics[clip=true,width=10cm,keepaspectratio]{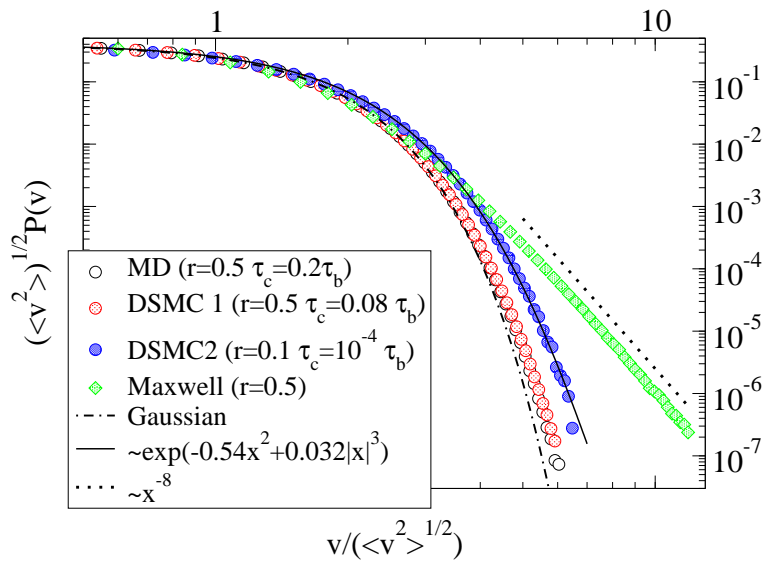}}
\caption{\label{fig:pure_dilute_pdf} Dilute cases, velocity pdf in different cases. }
\end{figure}

The above recipe has been applied~\cite{PBL02,BLP04,PBV07} to three
 monodisperse variants of the general model of driven granular gases
 of Appendix~\ref{app:granular}, Eq.~\eqref{gen_drive}, characterized by:
\begin{itemize}

\item i) very dilute conditions in $2D$, with volume fraction
 $\phi=n\frac{\sigma^2}{4} < 0.1$ (being $\sigma$ the diameter if the
 hard disks) and mean free time between collisions $\tau_c$ not much
 smaller than the viscosity time $\tau_b=1/\gamma$;

\item ii) same as model i) but with the additional assumption that
 positions of  particles are irrelevant and the probability of a
 collision is proportional to the relative velocity (Direct Simulation
 Monte Carlo, DSMC); 

\item iii)  same as model ii),
 with a constant collision probability (inelastic Maxwell model)
 and a deterministic driving. 

\end{itemize}
The last two variants are discussed in details in
 Appendix~\ref{app:granular}. All these models have the common
 property~\eqref{eq:indep} and non-Gaussian velocity pdfs $p_v(v)$. In
 particular model iii) displays $p_v(v)$ with high energy tails of the
 form $v^{-b}$ with $b= 4$ in $d=1$~\cite{BMP02,BMP02b} and $b>4$ in $d=2$ (a
 good estimate for not too high inelasticity is $b \simeq
 4/(1-r)$~\cite{EB02}). The extreme simplification of the dynamics of
 model iii) allows us to obtain a direct analytical computation of
 time correlations and responses (see Appendix~\ref{app:granular}). In
 Figure~\ref{fig:pure_dilute_pdf}, some examples of the velocity pdfs
 for these different models are shown. In Molecular Dynamics (MD)
 simulations of inelastic hard disks, even if strongly inelastic, but
 still dilute ($\phi < 0.1$), the velocity pdf is not far from a
 Gaussian, as in a DSMC with similar choices of the parameters
 ($\tau_c \sim 0.1 \tau_b$). Increasing $n$ in the DSMC leads to
 stationary regimes very far from thermal equilibrium, with $T_g \ll
 T_b$ and larger tails of the velocity pdf $p_v(v)$. A convenient fit
 of these velocity pdfs is the following
\begin{equation} \label{eq:fit}
p_v(v)=c_0 \exp(-c_1 v^2+c_2 |v^3|-c_3 v^4)
\end{equation}
where $c_0$ is the  normalization constant. In most of
observed cases $|c_3 v_{max}^4| \ll |c_2 v_{max}^3|$, with $v_{max}$
the largest value of $v$ in the histogram. Therefore, in
practice, we have a rather good fit even dropping the quartic term. Using
relation~\eqref{3.9} with the assumption of uncorrelated velocities, a
prediction for the response is obtained, in the form
\begin{eqnarray} \label{eq:respsingle}
R(t)=-2c_1\langle v_x(t)v_x(0) \rangle+3c_2\langle v_x(t)|v_x(0)|v_x(0)\rangle\\=-2 c_1 \langle v_x^2 \rangle C_1(t)+ 3 c_2 \langle |v_x(0)^3| \rangle C_2(t),
\end{eqnarray}
which also defines $C_2(t)$. 

\begin{figure}[ht]
\centerline {\includegraphics[clip=true,width=10cm,keepaspectratio]{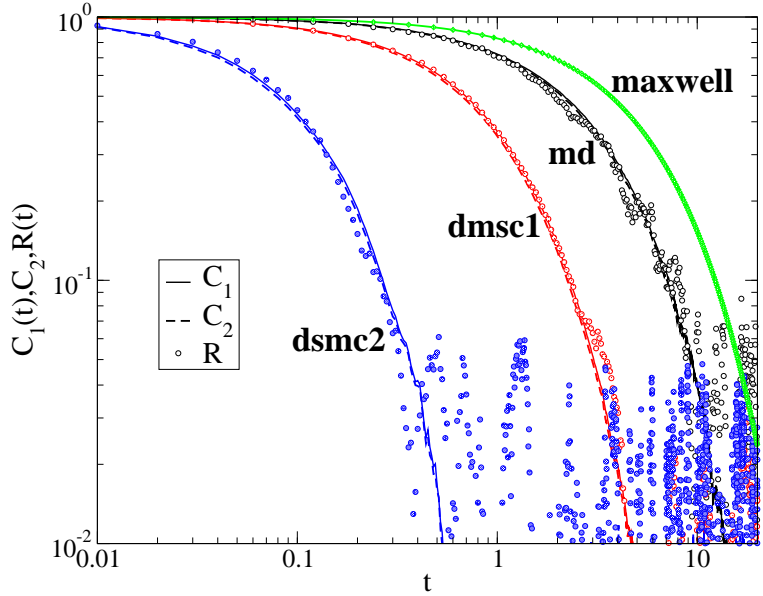}}
\caption{\label{fig:pure_dilute_inst} Dilute cases; experiment with
instantaneous displacement of the velocity of a tracer.  }
\end{figure}

Typical results of Fluctuation-Response numerical experiments on these three models are
shown in Figure~\ref{fig:pure_dilute_inst}, together with the
normalized velocity correlations $C_1(t)$ and $C_2(t)$. The general
picture that emerges from these numerical experiments is the
following:
\begin{itemize}
\item 
different correlations practically display the same behavior (we do not show $C_3(t)=\langle v(t)v^3(0)
\rangle/\langle v(0)^4 \rangle$, but the result is similar)
\item 
a very good agreement between $R(t)$ and $C_1(t)$ is observed, which amounts
to a verification, within the limits of numerical precision, of the
Einstein relation.
\end{itemize}
The observation that the self-correlations $C_k(t)$ are almost identical, at least for
$k=1,2,3$, is very robust: indeed with
large statistics one can appreciate very small differences only at late
times. Anyway, the data on the response function is usually quite
noisy, and, for the practical purpose of the linear combination
involved in the response, these small differences are negligible: the
Einstein relation is always satisfied, with $T_g$ as a proportionality
factor. In conclusion, the coincidence between velocity correlations
of different orders makes the non-Gaussianity of the velocity pdf
irrelevant for the linear response of the system.

It is interesting to note that an almost identical situation is
encountered in a gas of non-interacting particles, whose velocities
obey a Langevin equation with a non-quadratic potential:
\begin{equation}
\frac{d v(t)}{dt}=-\gamma\frac{d U(v)}{dv}+\sqrt{2 \gamma}\eta(t),
\end{equation}
with $U(v)=c_1 v^2-c_2 v|v|^2+c_3 v^4$ (with positive $c_1$, $c_2$ and
$c_3$). Numerical inspection, clearly indicates that
$C_1(t)$, $C_2(t)$ and $C_3(t)$ are practically indistinguishable.

A simple condition produces the observed behavior. In fact,
the time correlation of $v(t)$ with a generic function of the initial
velocity $f[v(0)]$ can be written as
\begin{eqnarray} \label{eq:angelo}
\langle v(t) f[v(0)] \rangle=\int dv_t \int dv_0 p_v(v_0) \mathcal{P}_t(v_t|v_0) v_t f[v_0]=\\\int dv_0 p_v(v_0) f[v_0] \langle v_t | v_0 \rangle,
\end{eqnarray}
where $\mathcal{P}_t(v_t|v_0)$ is the conditional probability of
observing $v(t)=v_t$ if $v(0)=v_0$ (time translation invariance is
assumed) and $\langle v_t | v_0 \rangle=\int dv_t
\mathcal{P}_t(v_t|v_0)v_t$ is the average of $v(t)$ conditioned to
$v(0)=v_0$. 

If, for some reason, the dependence on time and on $v(0)$ can be
factorized, i.e. $\langle v_t|v_0 \rangle=g(t)q(v_0)$, with $g$ and
$q$ two generic functions, then the time behavior results independent
of the choice of the function $f(v)$, i.e. of the order of the
correlation. This happens in model (iii), where in spite of the
non-Gaussian shape of the velocity pdf, the equivalence between $R(t)$
and $C_1(t)$, as well as any other correlation $C_f(t)=\langle
v(t)f[v(0)]\rangle/\langle v(0)f[v(0)]\rangle=
R(t)=\exp\left(-\frac{r(r+1)}{4}t\right )$, with any generic function
$f$ of the initial velocity value, is exact for the case $d=2$
and~\cite{BBDLMP05} for $d=1$, see Appendix~\ref{app:granular}.

Let us discuss further experiments, with
different kinds of perturbations, which  involve more than one
particle. A classical procedure, to avoid the
heating of the system when a global perturbation is applied, is the
following~\cite{CJM79}: once a steady-state has been reached, the system is
perturbed impulsively at time $0$ by a  force
applied (non-uniformly) on every particle. The response is then
monitored in time. The force acting on particle $i$ is
\begin{equation} \label{forcing}
\mathbf{F}(\mathbf{r}_i,t)=\zeta_i \boldsymbol{\xi}(\mathbf{r}_i,t)
\end{equation}
with the properties $\boldsymbol{\nabla} \times \boldsymbol{\xi} \neq
0$, $\boldsymbol{\nabla} \cdot \boldsymbol{\xi} = 0$, where $\zeta_i$
is a particle dependent variable with randomly assigned $\pm 1$
values. A simple case is realized by a transverse perturbation ${\bm
\xi}(\mathbf{r},t)= (0, \xi \cos (k_x x) \delta(t)
)$, where $k_x$ is compatible with the periodic
boundary conditions, i.e. $k_x=2 \pi n_k /L_x$ with $n_k$ integer and
$L_x$ the linear horizontal box size.  The staggered response function
(i.e. the current induced at $t$ by the perturbation at time $0$), and the
conjugate correlation,
\begin{eqnarray} 
R(t) & = & \frac{1}{\xi}\langle \sum_i \zeta_i \dot{y}_i(t)
\cos(k_x x_i(t)) \rangle \ ,\nonumber \\ 
C(t) & = &\langle \sum_i
\dot{y}_i(t) \dot{y}_i(0) \cos \{k_x[x_i(t)-x_i(0)] \} \rangle
\nonumber
\end{eqnarray}
are related, {\em at equilibrium}, by the Kubo relation $R(t,t_0) =
\frac{\beta}{2} C(t,t_0)$, $T_b=1/\beta$ being the bath
temperature. It is possible to derive the same
relation  from Eq.~\eqref{3.9} with the
assumption of uncorrelated variables and Gaussian velocities, 
and replacing $T_b$ with $T_g$ of the perturbed
species. In Figure~\ref{fig:shear}, the results of such numerical experiment are
shown in a parametrized form. The previous scenario is recovered: the
Kubo relation is satisfied with $T_g$ replacing $T_b$.

\begin{figure}[ht]
\center{\includegraphics[clip=true,width=10cm,keepaspectratio]{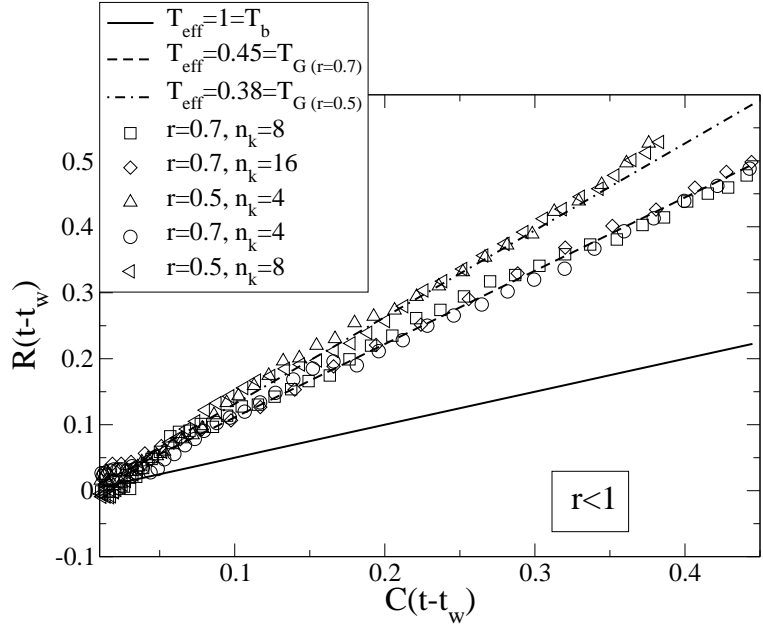}}
\caption{\label{fig:shear} Parametric plot of $R(t-t_w)$ vs. $C(t-t_w)$
for the numerical experiment with recipe $II$ (impulsive shear
perturbation) with $r<1$, with heating bath, and for different choices
of the wave number $n_k$ of the perturbation. $T_b=1$ and $\tau_b=10$,
$N=500$, $\tau_c=1$, $\Xi=0.01$, $n_k=8$, with averages over $10000$
realizations, using $t_w=100$.}
\end{figure}

{\em The binary mixture and the granular thermometer}

In the binary mixture case, one analyzes a system made of two
different components with $N_1$ and $N_2$ particles, with masses $m_1$
and $m_2$ respectively, coefficients of restitution $\alpha_{11}$ and
$\alpha_{22}$ for collisions among particles of the same species and
$\alpha_{12}$ for collisions among particles of different
species. Simulations as well as experiments and analytical
calculations have shown that, in this case, energy equipartition is
broken, i.e. $T_1 \neq T_2 \neq T_b$~\cite{PMP02,MP02,MP02b,BT02,FM02}.  At
the level of Boltzmann kinetic equation, the temperature ratio of a
binary granular mixture subject to stochastic driving of the form
given above has been obtained in~\cite{BT02} for the case
$\gamma_{s_i}=0$ and in~\cite{PMP02} for $\gamma_{s_i} \neq 0$. In the
case $\gamma_{s_i} \neq 0$ a bath temperature can still be defined as
$T_b=\Gamma_{s_i}\gamma_{s_i}$. Note that in general $\Gamma_{s_i}$
and $\gamma_{s_i}$ depend upon $m_i$ and the correct elastic limit
(i.e. equipartition) is recovered if and only if $T_b$ does not depend
on $m_i$. In~\cite{PMP02} it has been shown that a model with
$\Gamma_{s_i} \propto \sqrt{m_{s_i}}$ and $\gamma_{s_i} \propto
m_{s_i}$ fairly reproduces experimental results for the temperature
ratio $T_1/T_2$ measured in a gas of grains in a box vertically
vibrated. It is also known that equipartition is not recovered even in
the so-called tracer limit~\cite{MP99}, i.e. in the case $N_2=1$ and
$N_1 \gg 1$. For binary mixtures we discuss recent
results~\cite{BLP04} obtained for both molecular dynamics and
DSMC. The above Fluctuation-Response experiment with a small
perturbation introduced on a particle can be performed integrating in
time, measuring the mobility of a tracer particle and comparing it
with its self-diffusion coefficient: this is useful to reduce noise in
the measurement of the response function.  The mobility of the tracer
particle can be measured by applying a small constant drag force
$f_x$, in the $x$ direction for instance to a given particle, labeled
$0$, for times $t > 0$ (we assume the system to be isotropic). The
linearity of the response is always checked by changing the amplitude
of the perturbation.  The perturbed particle will reach at large times
a constant velocity $\mu f_x$, related to the instantaneous response
of Eq.~\eqref{risp_dv} by
\begin{equation}
\mu=\frac{1}{m_0}\int_0^t R(t') dt'.
\end{equation}
The mean-square displacement averaged over many unperturbed dynamics
$B(t) = \langle |{\bm r_0}(t) - {\bm r_0}(0)|^2 \rangle$
asymptotically grows linearly in time, $ B(t) \sim 4 D t$ (in
dimension $d=2$) with a self-diffusion coefficient given by
$D=\int_0^\infty \langle v(t) v(0) \rangle$. Integrating the
relation~\eqref{einst}, one obtains a prediction for the mobility in
the uncorrelated Gaussian case:
\begin{equation} \label{bidisp}
\mu=\frac{1}{T_s} D,
\end{equation}
where $T_s$ is the granular temperature of the species that the
perturbed particle  belongs to. 

The general result for a binary mixture~\cite{BLP04} is analogous to the one for
monodisperse gases: even when the velocity pdf is non Gaussian, the
closeness between correlations of different order makes
relation~\eqref{bidisp} true. By  using as test particle a
particle with index $1$ ($2$) for species $1$ ($2$), one obtains the two responses
$\chi_1=\overline{x_1(t)-x_1(0)} \simeq \langle \delta v_1 \rangle_{\infty}t$ and
$\chi_2=\overline{x_2(t)-x_2(0)} \simeq \langle \delta v_2 \rangle_{\infty}t$, and thus the mobilities $\mu_1$ and
$\mu_2$. Two independent Einstein relations ($\mu_i= D_i /T_i $) are
verified, by plotting $\chi_i$ vs. $B_i$. In
figure~\ref{fig:hom_kicks} we show, as an example, the verification of the
validity of Green-Kubo relations using DSMC in spatially homogeneous
regime. All the experiments, performed varying the restitution
coefficients and the masses of the two components, and with different
models and algorithms (homogeneous and non-homogeneous, DSMC and MD)
showed identical results, i.e the linearity of response-perturbation
relation with effective temperature equal to the granular temperature
of the perturbed species. The same  is true for other kinds of
perturbation, such as the current-shear perturbation experiment
discussed above.

In figure~\ref{fig:tracer} an even more striking result is portrayed:
the mobility-diffusion parametric graph is shown in the case of a
single tracer with different properties with respect to a bulk gas
($N_1=500$, $N_2=1$). In this case the tracer does not perturb
significantly the bulk. However the temperature of the tracer is quite
different from the bath temperature as well as from the gas
temperature~\cite{MP99}. Again, the effective temperature of the
tracer corresponds to {\it its} temperature and not to the temperature
of the bath or of the bulk. That is to say that a non-perturbing
thermometer, used to measure temperature of a granular gas through
Fluctuation-Response relations, would measure its own temperature and
not the bulk temperature. This is a consequence of the lack of
thermalization, or equipartition, due to energy dissipating collisions
which characterize stationary states of granular systems.

\begin{figure}[htb]
\centerline{ \includegraphics[clip=true,width=6cm,angle=0]{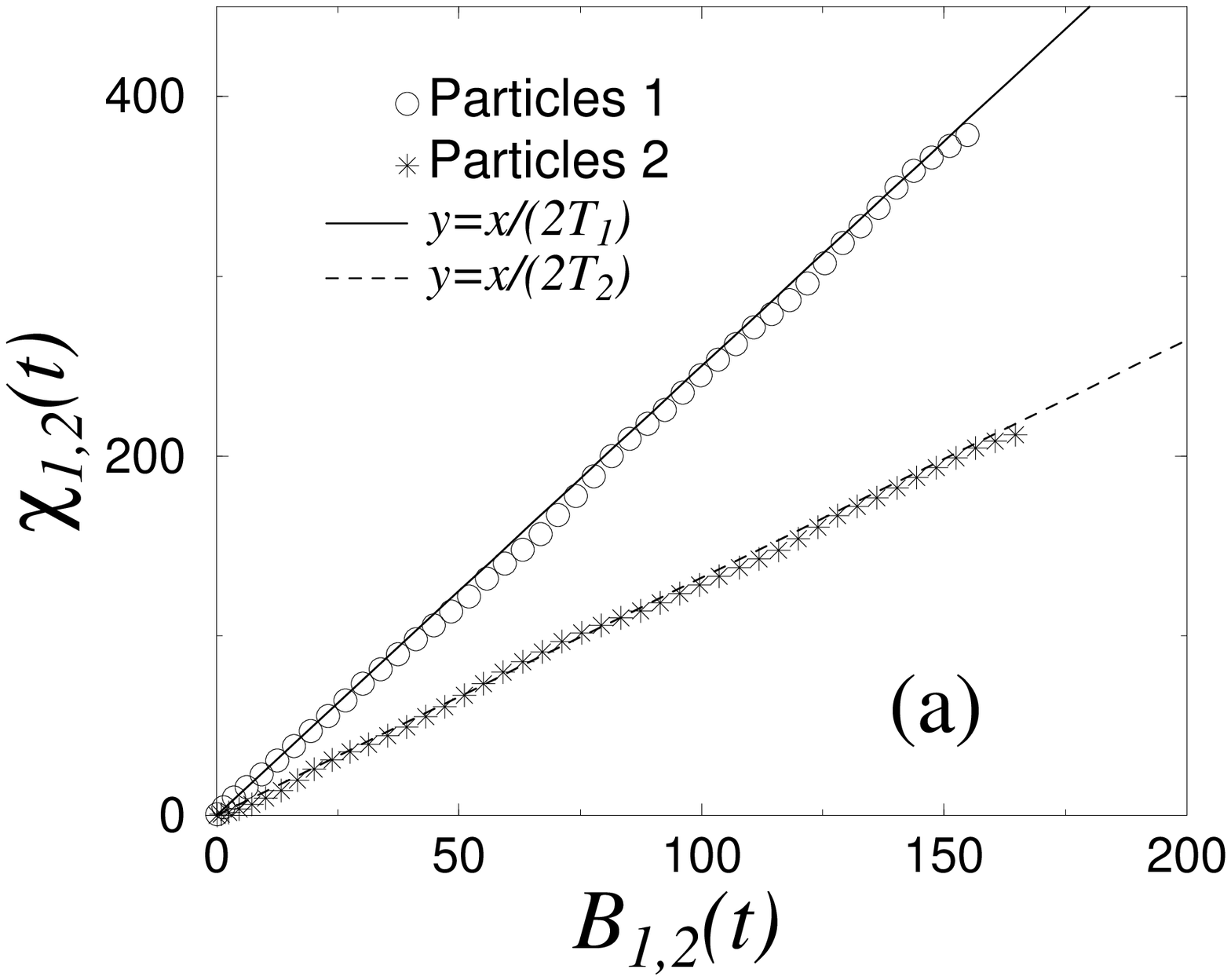}
        \includegraphics[clip=true,width=6cm,angle=0]{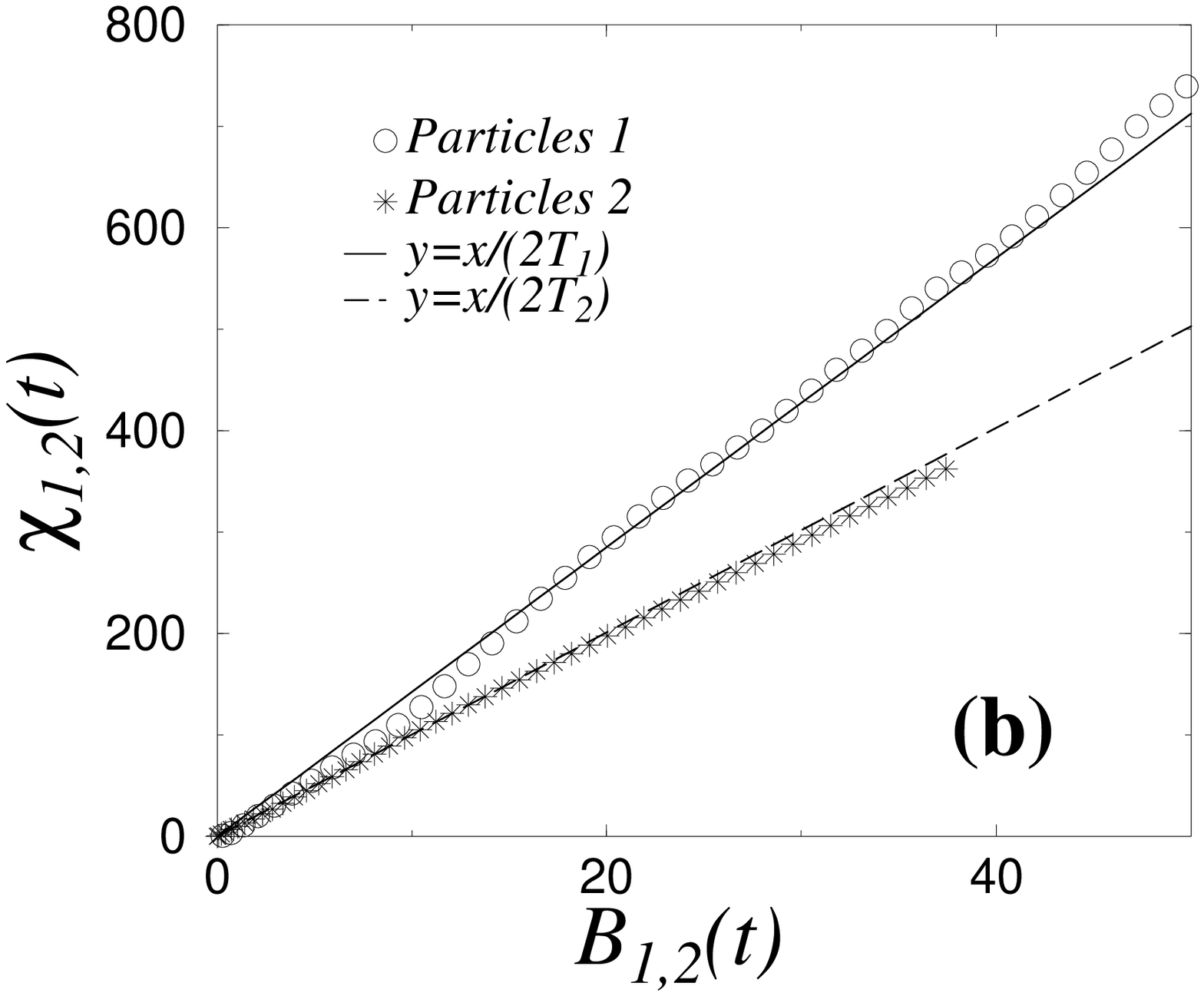} }
\caption{Binary mixture, homogeneous DSMC: mobility vs. mean-square
displacement; left: $\alpha_{11}=0.3$, $\alpha_{12}=0.5$,
$\alpha_{22}=0.7$, $m_2=3m_1$, $T_1\approx 0.2$, $T_2\approx 0.38$;
right: $\alpha_{11}=\alpha_{12}=\alpha_{22}=0.9$, $m_2=5m_1$,
$T_1\approx 0.035$, $T_2\approx 0.05$.  Symbols are numerical data,
lines have slope $1/(2T_1)$ and $1/(2T_2)$.  }
\label{fig:hom_kicks}
\end{figure}    

\begin{figure}[tb]
\centerline{ \includegraphics[clip=true,width=10cm,angle=0]{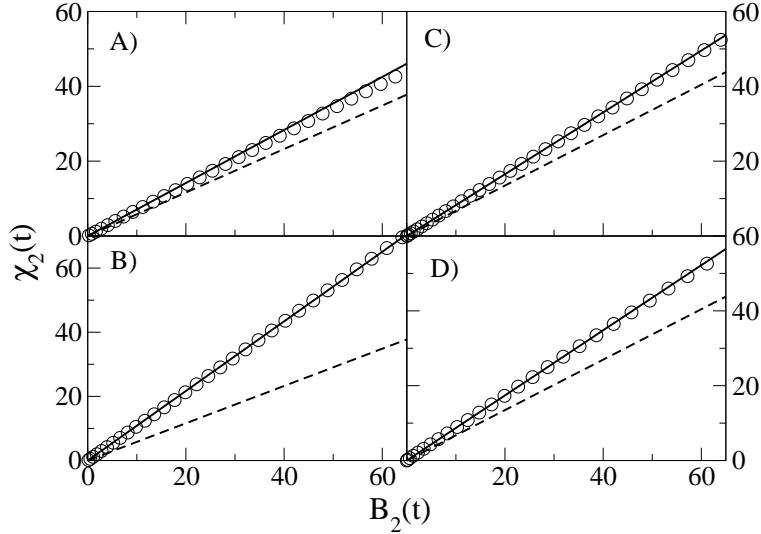} }
\caption{Binary mixture, homogeneous DSMC: mobility Vs. diffusion of a single particle of mass
$m_{tracer}$ in contact with $N=500$ particles of mass $m$, immersed
in a heat bath (i.e. random kicks plus viscosity).  We use the
following conventions: $\alpha_{tracer}=\alpha_{12}$ and
$\alpha=\alpha_{11}$. Only in case {\bf A)} the tracer is also in contact
with the external driving heat bath.  {\bf A)} $m_{tracer}=m$,
$\alpha=0.9$, $\alpha_{tracer}=0.4$, $T_g=0.86$, $T_g^{tracer}=0.70$;
{\bf B)} $m_{tracer}=m$, $\alpha=0.9$, $\alpha_{tracer}=0.4$,
$T_g=0.86$, $T_g^{tracer}=0.46$; {\bf C)} $m_{tracer}=7m$,
$\alpha=\alpha_{tracer}=0.7$, $T_g=0.74$, $T_g^{tracer}=0.60$; {\bf
D)} $m_{tracer}=4m$, $\alpha=\alpha_{tracer}=0.7$, $T_g=0.74$,
$T_g^{tracer}=0.57$.  The solid line has slope $T_g^{tracer}$, the
dashed line has slope $T_g$.  }
\label{fig:tracer}
\end{figure}

{\em Failure of the Einstein relation: correlated degrees of freedom}

The factorization of the invariant phase space measure,
Eq.~\eqref{eq:indep}, is no more obvious in model (i) when density
increases. Correlations between different degrees of freedom (d.o.f.),
that is positions and velocities of the same or of different
particles, appear also in homogeneously driven granular gases, as an
effect of the inelastic collisions that act similarly to an attractive
potential. Such a phenomenon has been discussed for this model of bath
in~\cite{PLMPV98,PLMV99,CDMP04,BTC07} and for other homogeneous thermostats
in~\cite{WM96,NETP99,TPNE01}. In~\cite{PLMPV98,PLMV99} it was also
discussed the interplay between local density and local granular
temperature, which in some very dissipative cases present strong
fluctuations correlated to each other. These correlations indicate a
breakdown of the factorization of the invariant measure, in particular
the velocity and the position distributions  of a given 
particle do not factorize. In general, even when relation~\eqref{eq:indep} does not hold,
one can define (and compute) the marginal probability density function
of the component $x$ of the velocity of one particle $i$, as the
marginalized of $\rho(\{ \bv_i, \bx_i\})$:
\begin{equation}
p_v^{(i)}(v_x)=\int \prod_{k=1}^{N} d\bx_i \prod_{k=1,k\neq i}^N d\bv_k dv_i^y\rho(\{ \bv, \bx\}).
\end{equation}
However not always does this function play a direct role in the
response function. For example, perturbing the $x$ component of the
velocity of the $i$-th particle and measuring the response of the same
component, one obtains
\begin{equation} \label{eq:respmulti}
R(t) = - \Biggl \langle v_i^x(t) \left.  \frac{\partial \ln \rho(\{
 \bv, \bx\})} {\partial v_i^x} \right|_{t=0}\Biggr \rangle \ \neq - \Biggl
 \langle v_i^x(t) \left.  \frac{\partial \ln p_v^{(i)}(v_i^x)}
 {\partial v_i^x} \right|_{t=0} \Biggr \rangle \, .
\end{equation}

\begin{figure}
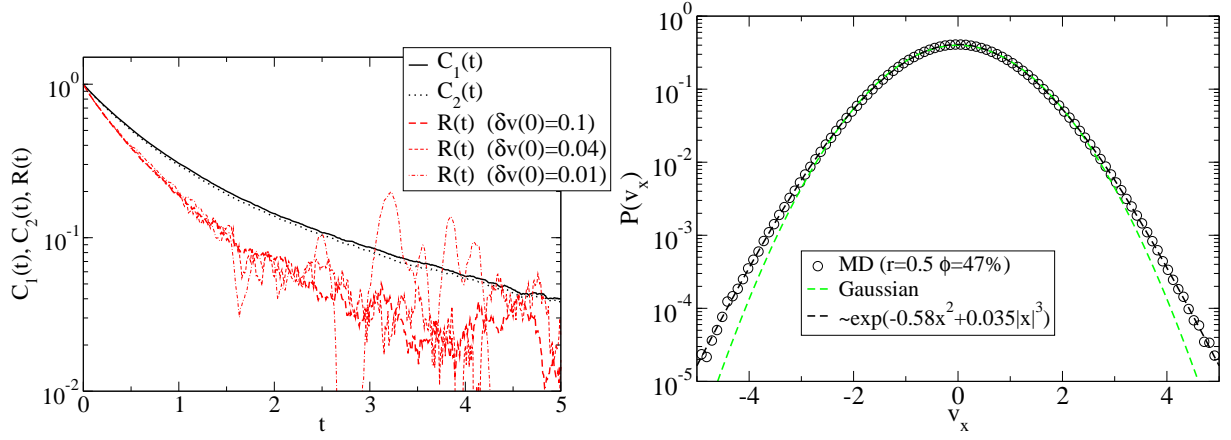

\includegraphics[width=8cm,clip=true]{denso_manyv0.eps}
\includegraphics[width=8cm,clip=true]{denso_pdf.eps}
\caption{Left: correlation functions and response. Right: velocity pdf. Both plots refer to the case of a dense granular gas \label{fig:denso}}
\end{figure}

This is exactly what happens in model (i) (here we discuss only the
monodisperse case~\cite{PBV07}) when density is increased. In
Figure~\ref{fig:denso}, left frame, the correlation functions $C_1(t)$
and $C_2(t)$ are shown, together with the response function measured
with different values of the perturbation $\delta v(0)$. The very good
agreement between different response functions guarantees that the
system is indeed linearly perturbed. At the same time, the different
correlation functions $C_k(t)$ are very close, reproducing the
phenomenology already observed in the previous dilute cases, with the
difference that the time dependence is not exponential but slower,
closer to a stretched exponential $\sim \exp(-(t/\tau)^\alpha)$ with
$\alpha<1$. Finally, looking at the velocity pdf of the gas, the
previously proposed exponential of a cubic expression,
Eq.~\eqref{eq:fit} with a negligible $c_3$ coefficient, is found to
perfectly fit the numerical results. Therefore, if the correlations
among the different d.o.f. are neglected, using
equation~\eqref{eq:respsingle} and the proportionality of the
functions $C_n(t)$, a verification of the Einstein formula $R(t)
\equiv C_1(t)$ is still expected. The results displayed in
Figure~\ref{fig:denso}, left frame, demonstrate that this is not the
case: the hypothesis of weak correlations among different d.o.f. is
not appropriate and the correct response function is given by Eq.~\eqref{eq:respmulti}.

\begin{center}
\begin{figure}
\includegraphics[width=12cm,clip=true]{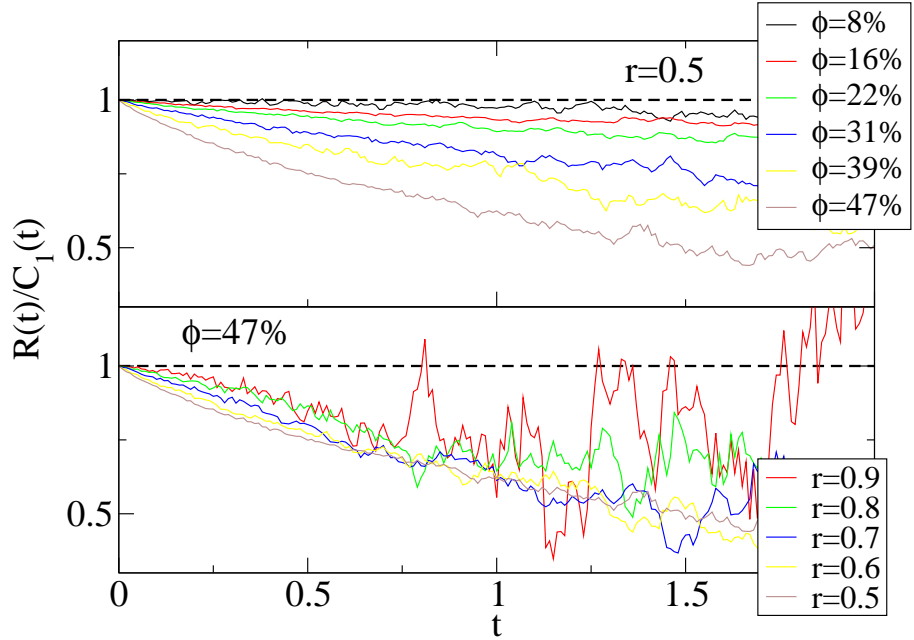}
\caption{Ratio between response and auto-correlation for different densities and restitution coefficients \label{fig:molti}}
\end{figure}
\end{center}

The degree of violation of the Einstein formula increases with the
volume fraction $\phi$ and the inelasticity $1-r$, as shown in
Figure~\ref{fig:molti}, where the ratio $R(t)/C_1(t)$ is reported
as a function of time. This observation is consistent with the above
argument: correlations among different d.o.f. increase when the
probability of repeated contacts (the so-called ``ring collisions'')
is enhanced, and this happens when the excluded volume is reduced as
well as when the post-collisional relative velocity is reduced. 

{\em A linear model with two correlated variables}

It is interesting to discuss a simple model with only two d.o.f. with
Gaussian marginalized pdfs. The two variables, $x,y$, obey the
following system of coupled Langevin equations:

\begin{eqnarray} \label{eq:lange}
\frac{dx(t)}{dt}&=&m_{11}x(t)+m_{12}v(t)+\sigma_{11} \eta_1(t)+\sigma_{12}\eta_2(t)\\
\frac{dv(t)}{dt}&=&m_{21}x(t)+m_{22}v(t)+\sigma_{21} \eta_1(t)+\sigma_{22}\eta_2(t)
\end{eqnarray}
When the matrices $\hat{m}$ and $\hat{\sigma}$ are diagonal, the two
variables are independent. Otherwise, they are correlated. Provided
that the matrix $\hat{m}$ is negative definite and $\det \hat{\sigma} \neq 0$, the pdf of the two variables
relaxes toward an invariant joint pdf, given by a bivariate Gaussian
function. Instead of discussing the general case, we suppose that the invariant joint pdf is expressed by
\begin{equation} \label{eq:joint}
\rho(x,v) \propto \exp(-\frac{x^2}{2}-\frac{v^2}{2}+\frac{xv}{2}).
\end{equation}

The marginal pdf of each variable is then a
Gaussian. Neglecting the correlation among $x$ and $v$, the response
of $v$ to a perturbation, would again be expected to be equal to
$C_1(t)=\langle v(t) v(0) \rangle/\langle v^2 \rangle$. On the
contrary, the correct response is given through the full
formula~\eqref{3.9}, applied to the joint pdf~\eqref{eq:joint}. The result is
\begin{equation} 
R(t)=\langle v(t)v(0) \rangle-\frac{1}{2} \langle v(t) x(0) \rangle.
\end{equation}

The difference between the Einstein formula and the correct response
is shown in Figure~\ref{fig:lange} for one choice of the matrix $\hat{m}$.
\begin{center}
\begin{figure}
\includegraphics[width=12cm,clip=true]{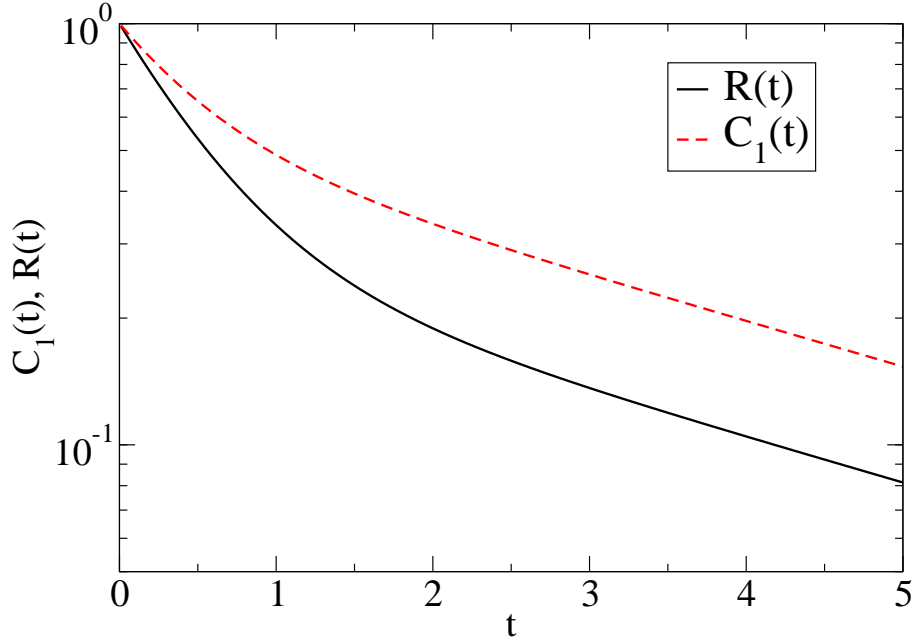}
\caption{Response $R(t)$ and velocity correlation $C_1(t)$ in the simple Langevin model with correlated variables discussed in Eq.~\eqref{eq:lange}, with parameters $m_{11}=-1.1$, $m_{12}=m_{21}=0.8$, $m_{22}=-1$. \label{fig:lange}}
\end{figure}
\end{center}

{\em The role of spatial inhomogeneity}

Inspired by a work about violations of the Einstein
relation in a non equilibrium model~\cite{ss06}, it is natural to conjecture an
effective spatial dependence of the pdf of the velocity component for
a particle at position $\bx$, at time $t$ of the form
\begin{equation} \label{eq:pdflocal}
p_v(v,\bx,t) \sim \exp\left\{-\frac{[v-u(\bx,t)]^2}{2T_g}\right\},
\end{equation}
where $u(\bx,t)$ is a local velocity average, defined on a small cell of
diameter $L_{box}$ centered in the particle. Such a hypothesis is
motivated by the fact that, at high density or inelasticities,
spatially structured velocity fluctuations appear in the system for
some time, even in the presence of external
noise~\cite{NETP99,TPNE01}. Following relation~\eqref{3.9}, with the
ansatz~\eqref{eq:pdflocal}, a formula for the response function follows:
\begin{equation} \label{eq:risplocal}
R(t)=C_s=\frac{1}{T_g}\langle v(t) \{v(0)-u[\bx(0)]\}\rangle.
\end{equation}
Figure~\ref{fig:seif} shows that relation~\eqref{eq:risplocal} is
fairly verified. Note however that the proposed
form~\eqref{eq:pdflocal} cannot be completely exact, a spatial
dependence of $T_g$ should also be included.

\begin{center}
\begin{figure}
\includegraphics[width=12cm,clip=true]{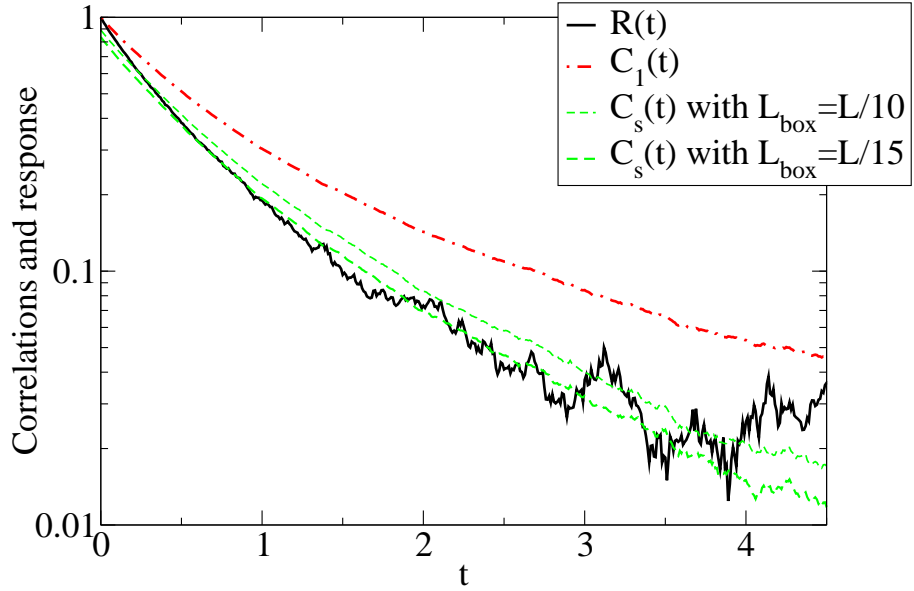}
\caption{Response $R(t)$ and different correlation functions for the
same MD simulation discussed in Figure~\ref{fig:denso}. The normalized
velocity self-correlations $C_1(t)$, as well as the correlation
$C_s(t)$ defined in Eq.~\eqref{eq:risplocal}, for different values of
the coarse graining radius $L_{box}$ are reported.\label{fig:seif}}
\end{figure}
\end{center}

\subsection{Further applications}

\subsubsection{Nano-systems}
\label{nanosys}
Since the early 60's, technology follows a trend towards miniaturization, 
which has led to the construction of devices of the size
of a few tens of nanometers, or smaller. In some circumstances, these 
devices are built by assembling single molecule after single molecule. 
While technology at the $\mu m$ scale is known to be more sophisticated,
but not substantially different from the macroscopic world technology, 
it is universally believed that nanotechnology has to face physical features 
which are quite different from those of the macroscopic world, especially when 
it comes to devices whose parts are made of an order $O(1)$ up to $O(10^3)$
atoms or molecules. In itself, this does not look particularly surprising.
Indeed, macroscopic bodies owe their properties to the very large number 
of their microscopic constituents, and to the fact that these constituents
interact with each other, in very precise fashions. This allows the onset of 
the state of local thermodynamic equilibrium (when the driving forces are not 
too large) and of its consequences, like the validity of the
hydrodynamic laws. The fact that the microscopic constituents of small systems,
like nano-devices, obey the same dynamical laws as those of the constituents 
of a macroscopic body is not sufficient for the small systems to behave as 
macroscopic objects.
Landau and Lifshitz express this concept as follows \cite{LandauFisStat}:

{\em ``At first sight, one may conclude that the growth of the number of 
particles makes infinitely complicated and intricate the properties of the
mechanical system at hand, so much that it becomes impossible to discover
any guiding law in the behaviour of the resulting macroscopic body.
But this is not the case and, as we will see below, new and very special 
laws emerge for a very large number of particles.

These are the so-called {\em statistical} laws which are due precisely to the 
exceedingly large number of particles of which the body is made. These laws are 
in no way reducible to mechanical laws, their specific characteristic is that 
they make no sense for mechanical systems of a small number of degrees of freedom.
Therefore, the presence of many degrees of freedom implies substantially different
laws, even though the motion of a system with many degrees of freedom follows the 
same mechanical laws of a system of a few degrees of freedom.''}

In particular, the hydrodynamic laws exclude that macroscopic observables may
spontaneously fluctuate: this is considered possible, but practically irrelevant, 
in many-particles systems. Differently, fluctuations are not negligible in
small systems or in small parts of macroscopic systems, where they might 
be directly observable \cite{bustamante,RITOreviwe}.

Nano-technological devices are then expected to behave differently from 
thermodynamic systems. Among them, we find, for instance, quantum dots, 
carbon nano-tubes and molecular machines, whose sizes range from a few to 
$O(100)$ nanometers. The question is: which are the laws that describe the 
behaviour of such small systems? This question is particularly important 
not only for its nano-technological implications, see e.g.\ Ref.\cite{Ferrari}, 
but also for understanding from a fundamental point of view the behaviour 
of microscopic systems in general and, in particular, biological systems.
Indeed, bacteria and cells are quite larger than the few tens of $n m$
which qualify nanosystems, but their behaviour is determined by events 
that occur at the nanoscale, like the transport of matter through the cell's 
membrane.

At the moment there is no clear answer to the above question. However, a widely 
shared view is that the mechanical laws, which are overshadowed
by the statistical laws in macroscopic systems, should play an important role
in nanosystems, hence it appears promising  to explore the nano-world from a 
dynamical systems point of view. In particular, the fluctuation relations 
which are the object of Section 5 are expected to be applicable to nano-phenomena. 
Indeed they have been tested and verified to various extents \cite{bustamante,RITOreviwe}, as 
discussed  in Section 5. The study of granular systems, in turn,
may provide insights on the above questions, since they would be mesoscopic 
objects, were their grains of the size of atoms or molecules.

\subsubsection{Fluctuations and response in biological systems}

Some applications of linear response theory to biological
systems have been  recently discussed by Kaneko and
co-workers~\cite{SIYK03,KF06}. The idea is
that fluctuations, always present in living organisms, are relevant
for the response of average properties to changes in the
environment. Environmental conditions can be seen as parameters determining the
distribution of fluctuations of measurable quantities: for example, the
concentration of a protein in a cell, as well as its fluctuations, are
influenced by many parameters, such as temperature, pH value and so
on. When an external perturbation is switched on, the parameters
change and the distributions of biological variables respond in some
way. A general relation for this response can be derived: in the
following, for simplicity but without loss of generality, we assume that the distribution of
fluctuations can be approximated by a Gaussian, and we discuss the
case of a scalar external parameter $a$ and a unique scalar observable
$x$. Upon a change $\Delta a$ of the parameter, the
distribution of $x$ changes from $\rho_a(x)$ to $\rho_{a+\Delta
a}(x)$, which can be written as
\begin{equation}
\rho_{a+\Delta a}(x)=\rho_a(x)\e^{\ln \rho_{a+\Delta a}(x)-\ln
\rho_a(x)}=\rho_a(x)\e^{\epsilon(a,\Delta a,x)}.
\end{equation}
The Gaussian assumption allows us to neglect the powers of $x$ higher
than $x^2$ in the expansion of $\epsilon(a,\Delta a,x)=\ln
\rho_{a+\Delta a}(x)-\ln \rho_a(x)$, obtaining
\begin{equation}
\rho_{a+\Delta a}(x)=\rho_a(x)N\e^{\epsilon^{(1)}(a,\Delta a)(x-\langle x \rangle_a)+\frac{1}{2}\epsilon^{(2)}(a,\Delta a)(x-\langle x \rangle_a)^2},
\end{equation}
where $\epsilon^{(n)}=(\partial^n \epsilon(a,\Delta a,x)/\partial
x^n)|_{x=\langle x \rangle_a}$ and $N=e^{\epsilon^{(0)}}$. Using this
 formula to calculate $\langle x \rangle_{a+\Delta a}$,  the
average of $x$ in the perturbed system, one obtains:
\begin{equation} \label{kaneko1}
\langle x \rangle_{a+\Delta a}-\langle x \rangle_{a}=\epsilon^{(1)}(a,\Delta a)\sigma^2_a \simeq b(a)\sigma_a^2 \Delta a
\end{equation}
where  the first order in the parameter change has
been considered, introducing $b(a)=\partial \epsilon^{(1)}(a,\Delta
a)/\partial \Delta a|_{\Delta a=0}$, and $\sigma_a^2=\langle
(x-\langle x\rangle_a)^2 \rangle$. Equation~\eqref{kaneko1} is a particular case of FDR, and
in fact can be derived from the generalized relation~\eqref{3.9}: it does not require thermal equilibrium, but only the
quasi-Gaussian assumption on $\rho(x)$. It must be noted, however,
that in relation~\eqref{kaneko1} the coefficient $b(a)$ can in
principle be related to $\sigma_a^2$. 

An experimental proof of relation~\eqref{kaneko1} has been obtained
studying the phenotype response to the exertion of an artificial
selection pressure over the genotype of a population of E. Coli
bacteria. The fluorescence intensity of the Green Fluorescent Protein
(GFP) has been used as a marker of the phenotype state $x$. The
external force has been identified as the selection pressure, providing
a constant rate of variation of the genotype, taken as the parameter
$a$.  The distribution of fluctuations of the fluorescence intensity
$\rho_a(x)$ has been measured in successive generations during the
application of the external force.  To  apply the
above relation, the DNA sequence of genes in the cell is regarded as a
parameter influencing the variables of the cell system, i.e. the
phenotype. The measured distribution $\rho_a(x)$ provides the average
$\langle x \rangle_a$ as well as its variance $\sigma^2_a$. The
proposed relation \eqref{kaneko1}, with a coefficient $b$ not
dependent on $a$, has been validated by these experimental data.

The above experiment probes the relevance of phenotype fluctuations
among clones, i.e. organisms with identical genes: these fluctuations
are always possible because living cells are of finite size and at
finite temperature. In nature, however, genetic diversity is the norm:
the parameter $a$ is not single valued in a population, but is
distributed around a mean. The variance of phenotype fluctuations is
then also induced by genotype variability. It is this variance that
was considered by Sir Ronald Fisher in his celebrated fundamental
theorem of natural selection~\cite{F99}, which asserts that evolution
speed and genetic variance are proportional. Even if the phenotipic
variance of the clones, $V_{ip}$, is not equivalent to that of
phenotypes in a genetically diverse population, $V_g$, they have been
shown to be related in~\cite{KF06}. There, an inequality has been
derived, stating that
\begin{equation} \label{variances}
V_g \leq V_{ip},
\end{equation}
only assuming that the phenotype distribution is single-peaked and not
too broad, which is equivalent to require a robustness of the mean
phenotype to genotype fluctuations. The equality sign in
Eq.~\eqref{variances} holds in the limit of high evolution speed, just
before the appearance of the error catastrophe, where the phenotype
distribution becomes too broad and the single-peak assumption fails.

Concluding, we can say that Fisher's theorem is somehow a biological
version of a general form of FDR.  Let us stress the
formal analogy with  statistical mechanics, whose small
fluctuations of macroscopic variables can actually  be detected
from the response functions.  Analogously, in biology the
fluctuations of the genotype variables, which are the corresponding
"microscopic" quantities, can be understood from the phenotype
features.

\section{Fluctuation relations}
\label{sec:fr}
While the theory of fluctuations around equilibrium states can be considered quite 
satisfactory and complete, the same cannot be said of the nonequilibrium theory.
In this context, the 1993 paper by Evans {\it et al.} \cite{ECM}, 
on the fluctuations of the entropy production rate, has been 
a pioneering attempt towards a unified theory of a wide range of nonequilibrium phenomena 
\cite{GGVienna}. The relation proposed and tested in \cite{ECM} is nowadays merely 
called {\em Fluctuation Relation} (FR). It constitutes one of the few general 
exact results, obtained on purely dynamical grounds, for systems almost arbitrarily 
far from equilibrium, and close to equilibrium it is consistent with the Green-Kubo 
and Onsager relations. A considerable number of works have been devoted to it, 
and various authors have derived it in different ways 
(e.g.\ \cite{GC,review,G07,CG07}). 

The FR of Ref.\cite{ECM} reads
\begin{equation}
\frac{{\rm Prob}_\zt(A)}{{\rm Prob}_\zt(-A)} = e^{\zt A}
\label{firstFR}
\end{equation}
where $A$ and $-A$ are averages of the normalized power dissipated in a driven system, in 
a long time $\zt$, and ${\rm Prob}_\zt(\pm A)$ is the steady state probability of 
observing values close to $\pm A$. The relation 
was derived for the following isoenergetic model of a 2-dimensional shearing fluid:
\begin{equation}
\left\{  
\begin{array}{l}
 \dfrac{d}{d t}{\bf q}_i = \dfrac{{\bf p}_i}{m} + \gamma \, y_i {\bf \hat{x}}  \\ 
 \\
 \dfrac{d}{d t}{\bf p}_i =  {\bf F}_i ({\bf q}) + \gamma  p_i^{(y)} {\bf \hat{x}} -\alpha_{th} {\bf p}_i
\end{array} 
\right.
\label{SLLODeqs}
\end{equation}
where $\zG=({\bf q}_1,{\bf p}_1,..., {\bf q}_N,{\bf p}_N)$ represents one point in phase space, 
$\gamma$ is the shear rate in the $y$ direction, ${\bf \hat{x}}$ is the unit vector in the $x$-direction,
and the friction term $\za_{th}$ (also called ``thermostat'') takes the form
\begin{equation}
\alpha_{th} (\zG) = - \frac{\zg}{\sum_{i=1}^N {\bf p}_i^2} ~
\sum_{i=1}^N p^{(x)}_i p^{(y)}_i \label{IKsllalpha}
\end{equation}
\noindent
determined by Gauss' principle of least constraint (cf.\ Subsection \ref{MDsec}). 
In this popular molecular dynamics model, the phase space contraction 
rate, defined by
\be
\zL = - \sum_k \sum_i \frac{\partial \dot{\bf q}^{(k)}_i}{\partial {\bf q}^{(k)}_i} + 
\frac{\partial \dot{\bf p}^{(k)}_i}{\partial {\bf p}^{(k)}_i} ~,
\ee
is proportional to the ``thermostat'' $\za_{th}$, hence to a quantity $\Omega$, which is the irreversible entropy
production, or the energy dissipation rate, divided by $\sum {\bf p}_i^2$.

Evans and Searles focused on the fluctuation properties 
of $\zW$ \cite{review,earlierpapers,generalized,SE2000,StephenPRE,stephennew},
which differs from $\zL$ in most cases.
They obtained ``transient'' FRs, which hold under 
the sole condition of reversibility of the microscopic dynamics. The term 
transient means that such relations concern the statistics of evolving ensembles 
of trajectories, rather than that of the steady states. These authors argued 
that, in the long $\zt$ limit, the same relations describe the steady state
fluctuations. We call \WFRs these relations, which will be described in 
subsection 5.3.2.

Gallavotti and Cohen provided the mathematical setting for the
results of Ref.\cite{ECM}, introducing the Chaotic Hypothesis
\cite{GC,GG-MPEJ,GGrevisited}:

\vskip 5pt \noi {\bf Chaotic Hypothesis:} {\it A reversible many-particle 
system in a stationary state can be regarded as a transitive Anosov system (cf.\ 
Appendix C) for the purpose of computing its macroscopic properties.}

\vskip 5pt \noi The result is a steady state FR for the fluctuations
of $\zL$, which we call $\zL$-FR, and we describe in subsection
5.3.1. Such a strong assumption as the Chaotic Hypothesis raised the
question of which models of physical interest are ``Anosov-like'',
since almost none of them is actually Anosov (lack of uniform
hyperbolicity being usual). The idea was that the Anosov property,
analogously to the Ergodic property, could hold ``in practice''.  This
view faces some difficulties which are evidenced and tackled, for
example, in Refs.\cite{ESR,BGGZ}.

Other relations which concern the statistics of nonequilibrium systems, and have 
generated much interest, are the Jarzynski and Crooks relations; they belong 
to the class of transient FRs, and relate free energy differences between 
equilibrium states to non-equilibrium processes \cite{CJ}, as described in 
subsection 5.3.3. 

The derivation of the FR's for deterministic systems motivated stochastic approaches, beginning with 
Kurchan's work \cite{Kurchan}. Among these studies, Ref.\cite{othertheory} has to be 
singled out since it solved, in a special case, a difficulty of systems 
with infinite state spaces, which are quite common. The works by Bodineau and Derrida 
\cite{BD}, and by Jona-Lasinio and coworkers \cite{BDSJL} also treat fluctuations 
in stochastic systems, but develop different theories, hence will be discussed 
separately in Section 7. 

We do not attempt an exhaustive review of this subject: we focus
mainly on the papers which initiated the various research
lines. Because entropy production is one of the quantities which are
most investigated in this context, we begin recalling basic facts of
Irreversible Thermodynamics. After that, we describe the models in
which the FRs were first considered.

\subsection{Irreversible entropy production}
\label{sec:models}
The main object of the FR's is the irreversible entropy production, for systems
in {\em Local Thermodynamic Equilibrium} (LTE), or the energy dissipation rate, more in general.

Irreversible Thermodynamics can be applied to systems that satisfy LTE \cite{DEGM}.
Physically, LTE is due
to the local exchange of momentum and energy in many particle collisions,
leading to a local Maxwell velocity distribution and local thermodynamics.
This allows one to
assume that the specific (i.e., per unit mass) local entropy $s$ at a
position ${\bf r}=(r_1,r_2,r_3)$ at time $t$ is a
function of the local specific energy $u$, of the local mass density
$n$, and of the local mass fractions $\cc=\{c_i\}_{i=1}^k$ of the $k$ constituents
out of which the system consists, as it is in equilibrium. In other words, LTE implies 
\begin{equation}
s({\bf r},t) = s[u({\bf r},t), n({\bf r},t), \cc({\bf r},t)] ~,
\label{local-s}
\end{equation}
and 
\begin{equation}
T d s = d u + p d v - \sum^k_{i=1} \mu_i d c_i
\label{gibbs-rel}
\end{equation}
where $p$ is the pressure, $v=1/n$ is the specific volume\footnote{In other words,
$v$ is the volume per unit mass.}, and $\mu_i$ the
chemical potential of component $i$. In particular, LTE assumes that
Eq.(\ref{gibbs-rel})
remains valid for a mass element followed along its center of mass motion,
even if the total system is not in equilibrium, so that one can write:
\be
T D_t s = D_t u + p D_t v - \sum^k_{i=1}\mu_i D_t c_i
\label{time-s}
\ee
where, denoting with $\mathbf{v}(\mathbf{x},t)$ the macroscopic velocity,
$$
D_t = \partial_t + {\bf v} \cdot \nabla_{\bf r}
$$ 
is the (barycentric) substantial time derivative \cite{DEGM}. 
It turns out that LTE holds for a wide variety of transport phenomena.

The hydrodynamical equations are based on LTE, as well as on the local 
conservation laws for $u$, $v$ and the $c_i$'s. Assuming local linear 
laws for their fluxes, one obtains, for instance, Fick's law for tracer 
diffusion, with local flux ${\bf J}$:
\be
{\bf J} ({\bf r},t) = - D~\nabla_{\bf r}~n({\bf r},t)
\label{fick}
\ee
where $\nabla_{\bf r} n$ is the local density gradient, and $D$ the diffusion 
coefficient.\footnote{More in general, $D$ may be a tensor.} For viscosity, the local flux is given by
Newton's law, and is expressed by
\be
J_{ij}({\bf r},t) = -\eta
\frac{\partial u_j}{\partial r_i}({\bf r},t)
\label{newton}
\ee
where $J_{ij}$ is the
$ij$-component of the local pressure tensor, giving the flow of the
$j$-component of the local momentum of a fluid in the $r_i$-direction, sheared
by a linear local velocity gradient $\partial u_j / \partial r_i$, and $\eta$ is 
the viscosity coefficient. Similarly, one obtains all the other linear laws of
Irreversible Thermodynamics. It is worth pointing out that states described by 
nonlinear laws, for the local quantities mentioned here, may be necessary in the
presence of strong external forces and gradients, but this does not mean that
LTE is broken: as long as one can properly speak of local thermodynamic quantities,
LTE must hold.

Equation (\ref{time-s}) and the hydrodynamic equations
lead then to the entropy balance equation for the
local variation of the entropy density:
\begin{equation}
n({\bf r},t) D_t s ({\bf r},t) = - \mbox{ div } {\bf J}_s({\bf r},t)
+ \sigma_s({\bf r},t) ~,
\label{local-balance}
\end{equation}
where ${\bf J}_s$ is the local entropy flux, and $\sigma_s \geq 0$ is the local
irreversible entropy production. This is a quantity related to the dissipated 
energy, hence should not be confused for the mere entropy variation, which 
takes place even in dissipationless phenomena, and does not need LTE. Indeed, 
in the linear regime,
$\zs_s$ is usually written as $\sum_i X_i J_i$, where $X_i$ and $J_i$ are
conjugate thermodynamic forces and fluxes, respectively.

\subsection{Ergodic hypothesis: the Khinchin's approach}

As in the study of equilibrium phenomena, it is desirable to develop a microscopic 
interpretation of the macroscopic theory outlined above, in order to extend our 
understanding beyond the realm of the previous subsection. To connect the theory 
of irreversible thermodynamics with a microscopic description, in analogy with 
equilibrium statistical mechanics, models are needed to assess the various 
hypothesis and to derive relations of practical interest. 
Indeed, while the equilibrium theory is solidly based on the Gibbs ensembles 
and on the Ergodic Hypothesis, the nonequilibrium counterparts of these ensembles 
and hypotheses are not identified yet, although various proposals have been 
made in the literature. Before we continue, it is therefore useful to 
recall how the microscopic description of equilibrium phenomena is
based on the Ergodic Hypothesis, following
Khinchin's approach \cite{Khinchin}. This will suggest possible avenues of
investigation, for the case of nonequilibrium phenomena.

Khinchin looked at the ergodic problem from a physical
perspective, avoiding the physically scarcely relevant concept of metrical 
transitivity, implied by the Birkhoff theorem. 
His general idea stems from the following facts:
\begin{itemize}
\item[a)] statistical mechanics concerns systems with a large
   number of degrees of freedom;
\item[b)] the physical observables are not generic
   (in mathematical sense) functions;
\item[c)] it is physically acceptable that ensemble averages do not coincide with
   time averages, for sets of initial conditions $\zG$ of measure tending
to zero when $N \to \infty$.
\end{itemize}
Khinchin considers systems with a separable Hamiltonian, i.e. systems whose Hamiltonian
can be written as
$$
H=\sum_{n=1}^N H_n({\bf q}_n,{\bf p}_n) 
$$
and a special class of observables (called \textit{sum functions}) of
the form
$$
f(\zG)=\sum_{n=1}^Nf_n({\bf q}_n, {\bf p}_n) 
$$
where $f_n=O(1)$. The pressure, the kinetic energy and the
single-particle distribution function are examples of sum functions.

Denote by $\langle \, \, \rangle$  the microcanonical
ensemble average, and let $\overline{f}({\zG})$ be
the time average of the observable $f$,
along a trajectory starting at ${\zG}$. Under quite general
hypotheses, and without invoking any metrical transitivity,
Khinchin showed that the following holds:
$$
\textrm{Prob} \left( { {|\overline{f} - \langle f \rangle|} \over
 |\langle f \rangle|} \ge K_1 N^{-1/4} \right) \le K_2 N^{-1/4}
$$
where $K_1$ and $K_2$ are $O(1)$. This means that ensemble averages of sum 
functions differ more than a given tolerance from time averages, only along
trajectories whose initial conditions have vanishing measure in the 
$N \to \infty$ limit.  
Mazur and van der Linden \cite{MVL1963} extended this result to systems of
particles interacting through short range potentials.

This shows that the details of the phase space dynamics (i.e.\ of the microscopic 
dynamics) are practically irrelevant for a notion of ergodicity that describes
physical systems and, indeed, they must be irrelevant for the thermodynamic 
behaviour to be as common as it is. In Khinchin's theory, the good statistical 
properties that are required are mainly explained as consequences of the fact that 
$N$ is very large.

\subsection{Molecular dynamics models} \label{MDsec}
The models developed in nonequilibrium molecular dynamics (NEMD) have been adopted,
in the development of microscopic theories of irreversible thermodynamics, because 
they are sufficiently simple to be analyzed in detail, and because of 
their success in describing various properties of nonequilibrium fluids and 
solids \cite{EM90,WH91}. It is known that results produced by simulations
of these systems are not reliable if quantum mechanical effects are important, 
if the inter-atomic forces are too complicated or insufficiently known, if the 
number of particles needs to be too large, or the simulations time too long.
But NEMD models are otherwise appropriate in computing quantities such as transport 
coefficients, and are a valid alternative to experiments. Furthermore,
the FRs were first conceived and investigated in one such model. The range 
of properties of a given physical system which they can describe, however, 
is not clearly delineated, and care should be used if the property under 
investigation is directly affected by the fictitious forces which characterize
NEMD. The simplest NEMD models are derived from 
Gauss' principle of least constraint \cite{lanczos,EM90}:

\vskip 5pt
\noindent
{\bf Gauss Principle (1829): }{\it 
Consider $N$ point particles of mass $m_1, ..., m_N$, subjected to
frictionless bilateral constraints ${\bf C}_i$ and to external
forces ${\bf F}_i$. Among all motions allowed by the constraints, 
the {\bf natural} one minimizes the ``curvature'', defined by
$$
\mathcal{C} = \sum_{i=1}^N
m_i \left( \ddot{\bf q}_i - \frac{{\bf F}_i}{m_i} \right)^2
= \sum_{i=1}^N \frac{1}{m_i} {\bf C}_i^2 ~.
$$
}
\vskip 5pt

\noindent
In the case of holonomic constraints, the equations of motion are Hamiltonian 
while non-holonomic constraints lead to non-Hamiltonian equations. The isokinetic 
($IK$) constraint, which fixes the kinetic energy 
$K = \sum_i {\bf p}_i^2 / 2m$, and the isoenergetic ($IE$) 
constraint, which fixes the internal energy $H_0 = K + \zF^{int}$, 
$\zF^{int}$ being the interaction potential, yield:
\begin{equation}
\left\{ 
 \begin{array}{l}
 \dfrac{d}{d t}{\bf q}_i = \dfrac{{\bf p}_i}{m} \\ \\
 \dfrac{d}{d t}{\bf p}_i =  {\bf F}_i^{int} ({\bf q}) + {\bf F}_i^{ext} ({\bf q}) 
  - \alpha_{th} (\zG) {\bf p}_i
\end{array} 
\right.
\quad i = 1, \ldots , N  
\label{thermosystem}
\end{equation}
 \noindent
where ${\bf F}_i^{int} ({\bf q})$ and ${\bf F}_i^{ext} ({\bf q})$ respectively 
denote the interactions among the particles and the external force driving the 
system. In particular, for ${\bf F}_i^{ext} ({\bf q}) = c_i  {\bf F}_e$, where
$c_i$ is a charge coupling particle $i$ to the external field ${\bf F}_e$, and
${\bf J} = \sum_{i=1}^N  c_i ({d}{\bf q}_i/dt)$, 
\begin{eqnarray}
\label{aIK}
&&\za_{th}(\zG) = \alpha_{IK} (\zG) = \frac {1}{2K} \left( {\bf J} \cdot  
{\bf F}_e + \sum_{i=1}^N \frac{\bf p_i}{m}
\cdot {\bf F}_i^{int} \right)  \quad \mbox{preserves } K ~, \\
&&\za_{th}(\zG) = \alpha_{IE} (\zG) = \frac {1}{2K} {\bf J} \cdot {\bf F}_e  
\qquad \qquad \qquad \quad \qquad \mbox{preserves } H_0 ~.
\label{aIE}
\end{eqnarray}
In the case of the SLLOD model, defined by Eqs.(\ref{SLLODeqs}), one has:
\begin{equation}
\alpha_{IK} (\zG) = \frac{1}{2K}~\sum_{i=1}^N \left( {\bf F}_i^{int} \cdot {\bf p}_i 
- \gamma p^{(x)}_i p^{(y)}_i \right) ~, \quad
\alpha_{IE} (\zG) = -\frac{\zg}{2K}~\sum_{i=1}^N p^{(x)}_i p^{(y)}_i \,\,.
\label{IKsllodalpha}
\end{equation}
\noindent

In the above examples,  
$\za_{IE}$ is the energy dissipation rate divided by the kinetic energy which, in 
local equilibrium, becomes the 
entropy production rate. Because $\zL = -\mbox{div}~(d \zG/d t)$
is in turn proportional to $\za_{IE}$, $\zL$ can be identified with the entropy 
production rate. However, real systems in nonequilibrium steady 
states can hardly be isoenergetic, and this identification appears to be accidental 
and of too limited validity. 

Depending on the physical property to be described, other constraints
can be used; e.g.\ isobaric, isochoric, isoenthalpic, constant stress
constraints.  The deterministic term $-\alpha_{th} {\bf p}$ is meant
to add or remove energy from the system, to balance the action of the
driving forces, so that a steady state can be
reached~\cite{EM90}~\footnote{This form of deterministic
``thermostat'' is computationally efficient, and makes the equations
of motion time reversal invariant~\cite{EM90}.}.  This constraint has
no immediate physical meaning; it merely serves the purpose of
replacing the many (practically impossible to treat) degrees of
freedom of a real thermostat, without appreciably perturbing the
properties of interest of the system under investigation. For
quantities not affected by how energy is removed from the system, its
form is irrelevant and various mechanical constraints will do
\cite{EM90,WH91,SEC98}.  The introduction of one or another such
constraint in the equations of motion is analogous to the choice of
one numerical algorithm or another, for the solution of a given
mathematical problem: one must choose the constraint or the algorithm
which alters as little as possible the quantity to be computed. The
introduction of hyperviscosities in the modellization of turbulence is
another analogous procedure.

Constant internal
energy $H_0$ or constant kinetic energy $K$, which prevent the system from
indefinitely ``heating up'', are popular constraints. The resulting systems 
are not Hamiltonian, but a Hamiltonian structure is not to 
be expected in systems in nonequilibrium steady states, if the thermostat degrees 
of freedom are not included \cite{RUPhysToday}. Indeed, a complete $N$-particle
model of a system and of its thermostat consists of Hamiltonian equations, which
may be written as:
\begin{equation}
\dfrac{d}{d t}{\zG} = \dfrac{d}{d t} \left( \begin{array}{c}
{\zG}_s \\
{\zG}_r \end{array} \right) = G(\zG) = \left( \begin{array}{c}
G_s(\zG_s,\zG_r) \\
G_r(\zG_s,\zG_r) \end{array} \right) ~, 
\label{NSNR}
\end{equation}
where $\zG_s = \{ {\bf q}_i,{\bf p}_i \}$, $i=1,...,N_s$, 
$\zG_r = \{ {\bf q}_i,{\bf p}_i \}$, $i=N_s+1,...,N$, and 
the subscript $s$ refers to the $N_s$ particles of the thermostatted system, 
while the subscript $r$ refers to the $N_r=N-N_s$ particles of the 
reservoir.
Moreover, if the time reversed evolution 
is allowed in phase space, it is also allowed in the projected space.
As the reservoirs remove energy from a driven system, on average,
the projected dynamics must be dissipative. \footnote{Systems of non-interacting particles
behave differently: usually projection leaves the dynamics Hamiltonian.}
Something similar happens in NEMD models, hence their non-Hamiltonian nature is not 
a hindrance, by itself. However, NEMD models are not obtained through the ideal 
projection procedure, and have been developed for specific practical purposes. 
Therefore, they must be used with care \cite{CR98,JR06}. 

Another popular deterministic model is the {\em  Nos\'e-Hoover thermostat}
\cite{EM90}, defined by:
\begin{equation}
\frac{d {\bf q}_i}{d {t}} = \frac{{\bf p}_i}{m} ~; \quad 
\frac{d {\bf p}_i}{d {t}} = {\bf F}_i^{int} + {\bf F}_i^{ext} - \xi {\bf p}_i ~; 
\quad \frac{d \xi}{d{t}} = \frac{1}{\theta} \left( 
\frac{K({\bf p})}{K_0} - 1 \right) ~; 
\label{NHeq}
\end{equation}
where $K_0$ is the chosen average of the kinetic energy 
$K({\bf p})$, and $\theta$ is a relaxation time. Normally, particles interact,
the dissipation is not too high and the dynamics are sufficiently ergodic that
the initial value of $\xi$ does not matter.  In the small $\theta$ limit, 
Nos\'e-Hoover approximates Gaussian $IK$ dynamics, but is  
more realistic and, in equilibrium, it generates canonical distributions 
as appropriate for macroscopic isothermal systems. 

The models illustrated above, like many others developed in NEMD, contain 
artificial forces, which are not present in Nature. The fact that they accurately 
describe certain properties of real systems, nonetheless, suggests that a form of 
equivalence of ensembles may be envisaged in nonequilibrium statistical mechanics. 
Several works have been devoted to clarify this issue. The 
first of them, as far as we know, are due to Evans and Morriss 
\cite{EMequiv,EM90}. For the equivalence of various thermostatted responses, 
see Refs.\cite{LBC,ESA,SEC98,HAHDG,HP04}, while \cite{HAHDG,HP04} show
that the phase space dimensionality loss, due to dissipation,
is a bulk phenomenon even when the thermostat acts only on the boundaries
\cite{AK}. Other aspects of the equivalence of deterministic thermostats are 
discussed in \cite{BTV,MP99}. In practice, one finds that response
to external drivings is not too sensitive to the choice of the thermostatting
mechanisms and that, in some cases, the equivalence of behaviors reaches rather subtle dynamical
properties \cite{GNS,GRS,CR98}.

\subsection{Deterministic systems: transient and steady state relations}
\label{sec:GC}
In this Section, we discuss three sets of fluctuation relations, part of which concerns
steady states, while the others describe properties of transient states arbitrarily
far form equilibrium. The Gallavotti-Cohen and the steady state Evans-Searles approaches
lead to steady state relations. The transient Evans-Searles, the Jarzynski and the Crooks
approaches lead to transient relations.
\subsubsection{The Gallavotti-Cohen approach}
\label{GCapproach}
The idea proposed by Gallavotti and Cohen~\cite{GC}, is that dissipative, 
reversible, transitive Anosov maps (cf.\ Appendix C) are idealizations of nonequilibrium
particle systems, hence that properties like the $\zL$-FR are enjoyed by physical 
systems as well. That the system evolves with discrete or continuous time, was 
thought to be a side issue in \cite{GC}, as apparently confirmed by Gentile's derivation for 
continuous time \cite{Gentile}. The proof of the $\zL$-FR for such maps followed 
these steps. First of all, it is known  \cite{SinaiPUP} that one Anosov map 
$S$\footnote{We denote by $S^k X$ the evolution that $S$ implies from the initial condition 
$X(0)=X$, i.e. $X(1)=S X$, $X(2)=S^2 X$, etc.}
admits a special kind of partition of its phase space $\mathcal{M}$, which is 
called Markovian. This is a subdivision of $\mathcal{M}$ in cells whose
interiors are disjoint from each other, and whose boundaries are
invariant sets constructed using the stable and unstable manifolds.
Consequently, the interior of a cell is mapped by $S$
in the interior of other cells, and not across two cells, which 
would include a piece of their boundary. Furthermore, arbitrarily fine 
partitions can be constructed, exploiting the time reversibility of the dynamics. 
Gallavotti and Cohen further assume that the dynamics is transitive, i.e.\ that
a typical trajectory explores all regions of $\mathcal{M}$, as finely as one wishes.
It is this structure that guarantees that probability weights of the kind conjectured 
in \cite{ECM}, cf.\ Eq.(\ref{sugge}), from which the \LFR follows, 
can be assigned to the cells of a finite Markov partition.

More precisely, let $\zL(X) = - \log J(X)$, where $J$ is the Jacobian determinant
of $S$,\footnote{If the point $X$ has $d$ coordinates, $X_i$, $i=1,...,d$, we can 
write $X_i(k+1)=f_i(X(k))$, where $f_i$ is a suitable function determined by $S$. 
Then $J(X)$ is the absolute value of the determinant of the matrix 
$\left( \partial f_i/\partial X_j \right)_X$.}
and consider the probability (or frequency of observation along one steady state 
trajectory) of the  dimensionless phase space contraction rate, obtained 
along a trajectory segment $w_{X,\zt}$, of origin $X \in \mathcal{M}$ and 
duration $\zt$. This quantity is defined by 
\begin{equation}
e_\zt(X) = \frac{1}{\zt \langle \zL \rangle} \sum_{k=-\zt/2}^{\zt/2-1} \zL(S^k X) 
\label{p}
\end{equation}
where $\langle . \rangle$ is the phase space average, with respect to the steady state
probability distribution.
Let $J^u$ be the Jacobian determinant of $S$ restricted to the unstable manifold $V^+$,
i.e.\ the product of the asymptotic factors of separation of nearby points, along the
directions in which distances asymptotically grow at an exponential rate.
{\em If the system is Anosov}, the probability that $e_\zt(X)$ falls in 
$B_{p,\ze}=(p-\ze,p+\ze)$ coincides, in the limit of fine Markov partitions 
and long $\zt$'s, with the sum of the weights 
\be
w_{X,\zt}=\prod_{k=-\zt/2}^{\zt/2-1} \frac{1}{J^u(S^k X)}
\ee
of the cells containing 
the points $X$ such that $e_\zt(X) \in B_{p,\ze}$. Then, if $\pi_{\tau}(B_{p,\ze})$ 
is the corresponding probability, one can write
\begin{equation}
\pi_\zt(e_\zt(X) \in B_{p,\ze}) \approx \frac{1}{M} \sum_{X, e_\zt(X)\in B_{p,\ze}} w_{X,\zt}
\label{Piofp}
\end{equation}
where $M$ is a normalization constant.
{\em If the support of the physical measure is} $\mathcal{M}$, which is the case 
if the dissipation is not exceedingly high \cite{ECSB}, time reversibility guarantees 
that the support of $\pi_{\tau}$ is an interval $[-p^*,p^*]$, $p^*>0$, and one can consider 
the ratio
\begin{equation}
\frac{\pi_\zt(B_{p,\ze})}{\pi_\zt(B_{-p,\ze})} \approx
\frac{\sum_{X,e_\zt(X) \in B_{p,\ze}} w_{X,\zt}}
{\sum_{X,e_\zt(X)\in B_{-p,\ze}} w_{X,\zt}} ~,
\label{pminusp}
\end{equation}
where each $X$ in the numerator has a counterpart in the denominator.
Denoting by $I$ the involution which replaces the initial condition of one
trajectory with the initial condition of the reversed 
trajectory, time reversibility\footnote{Time reversibility means: $I S^n = S^{-n} I$, where 
$I^2=$Identity.
For instance, $I$ is the reversal of momenta for Eqs.(\ref{thermosystem}), but
is more complicated for the shearing system of Ref.\cite{ECM}.}  
yields:
\begin{equation}
\zL(X)=-\zL(IX) ~, \quad w_{IX,\zt} = w_{X,\zt}^{-1} \quad \mbox{and~~~ }
\frac{w_{X,\zt}}{w_{IX,\zt}} = \exp(\zt \langle \zL \rangle p)
\end{equation}
if $e_\zt(X)=p$.
Taking small $\ze$ in $B_{p,\ze}$, the division of each term in the numerator
of (\ref{pminusp}) by its counterpart in the denominator approximately equals 
$e^{\zt \langle \zL \rangle p}$, which then equals the ratio in 
(\ref{pminusp}). In the limit of small $\ze$, infinitely fine Markov partition
and large $\zt$ one obtains:

\vskip 5pt\noindent
Theorem (Gallavotti-Cohen). {\em Let $(\mathcal{M},S)$ be dissipative
{\em (i.e.\ $\langle \zL \rangle > 0$)}, reversible 
and assume that the Chaotic Hypothesis holds. Then, 
\begin{equation}
\frac{\pi_{\tau}(B_{p,\ze})}{\pi_{\tau}(B_{-p,\ze})} = e^{\zt \langle \zL \rangle p} ~.
\label{largedev}
\end{equation}
with an error in the argument of the exponential which can be
estimated to be $p,\zt$ independent.
}
\vskip 5pt

\noindent
If the \LFR (hence the Chaotic Hypothesis on which it is based) holds,
the function $C(p;\zt,\ze)=(1/\zt
\langle \zL \rangle) \log \left[ \pi_{\tau}(B_{p,\ze})/\pi_{\tau}(B_{-p,\ze}) \right]$,
tends to a straight line of slope 1 for growing $\zt$, apart from small errors.
If $\zL$ can be identified with a physical observable, the \LFR
is a statement on the physics of nonequilibrium systems. 

Under the assumption that $\zL$ coincides with the entropy
production rate, the \LFR can be used to obtain the Green-Kubo
relations and the Onsager reciprocal relations, in the limit of small external 
drivings \cite{GG96}. This way, the \LFR appears to be an extension of such 
relations to nonequilibrium systems. Gallavotti assumes that the (Gaussian 
isokinetic, continuous time) system
is driven by the $\ell$ fields $F=(F_1, F_2,...,F_\ell)$, that the phase space 
contraction rate vanishes when all drivings vanish, and that
\begin{equation}
\zL(X) = \sum_{i=1}^\ell F_i J_i^0(X) + O(F^2) ~.
\end{equation}
This relation defines the linear ``currents'' $J_i^0$, which are proportional
to the forces $F_i$.
Then, the fast decay of the $\zL$-autocorrelation function, leads to this
expansion for the large deviation functional
\begin{equation}
\zeta(p) = - \lim_{\zt \to \infty} \frac{1}{\zt} \log \pi_\zt(p) = 
\frac{\langle \zL \rangle^2}{2 C_2} (p-1)^2 + O((p-1)^3 F^3)
\end{equation}
where 
$$
C_2 = \int_{-\infty}^\infty \langle \zL(S^t X) \zL(X)
\rangle_{\mbox{\small c}}~dt
$$ 
and $\langle . \rangle_{\mbox{\small c}}$ denotes the cumulant. Thus, the
$\zL$-FR, Eq.(\ref{largedev}), leads to
$\langle \zL \rangle = C_2/2 + O(F^3)$. Now, let the full (nonlinear) ``currents'' 
be defined by $J_i(X)=\partial_{F_i} \zL(X)$, and the transport coefficients be 
$L_{ij} = \partial_{F_j} \langle J_i \rangle |_{F=0}$. The derivatives 
with respect to the parameters $F$ require a property of differentiability
of SRB measures, which has been proven by Ruelle \cite{RueDiff}. 
Assuming this property, the validity 
of the \LFR and time reversibility, one can write
\begin{equation}
\langle \zL \rangle = \frac{1}{2} \sum_{i,j=1}^\ell (\partial_{F_j} 
\langle J_i \rangle + \partial_{F_i} \langle J_j \rangle) |_{F=0} F_i F _j
= \frac{1}{2} \sum_{i,j=1}^\ell ( L_{ij} + L_{ji} ) F_i F _j
\ee
to second order in the forces. Then, equating this with $C_2/2$ and considering 
$(L_{ij}+L_{ji})/2$ with $i=j$, one recovers the Green-Kubo relations.
To obtain the symmetry $L_{ij}=L_{ji}$,
Gallavotti extends the \LFR  to consider the joint distribution
of $\zL$ and its derivatives. He introduces the dimensionless current $q$, 
averaged over a long time $\zt$, through the relation
\begin{equation}
\frac{1}{\zt} \int_{-\zt/2}^{\zt/2} F_j \partial_{F_j} \zL(S^t X) d t
= F_j \langle \partial_{F_j} \zL \rangle q(X)
\end{equation}
and considers the joint distribution $\pi_\zt(p,q)$, with corresponding large
deviation functional 
$\zeta(p,q) = -  \lim_{\zt \to \infty} \frac{1}{\zt} \log \pi_\zt(p,q)$. The
result is a relation similar to the $\zL$-FR:
\begin{equation}
\lim_{\zt \to \infty} \frac{1}{\zt \langle \zL \rangle p} \log
\frac{\pi_\zt(p,q)}{\pi_\zt(-p,-q)} = 1 ~.
\label{jointlarge}
\end{equation}
This makes the difference $(\zeta(p,q) - \zeta(-p,-q))$ independent of $q$,
which leads to the desired result, $L_{ij}=L_{ji}$, in the limit of small
$F$. This work was refined
in \cite{GR97}; for related results, based on orbital measures, see 
Refs.\cite{RC98,LRladek}. 

Assuming that the currents and transport coefficients, here defined in terms 
of the phase space contraction rate, do represent physical quantities, these 
results show that the FRs are consistent with Irreversible Thermodynamics,
close to equilibrium. Hence they may be considered in the construction of 
a comprehensive nonequilibrium theory. However, some difficulties affect the 
present approach; the first being that $\zL$, which is directly related to the
thermostatting term $\za_{th}$, only in very special cases results proportional 
to the energy dissipation rate divided by the kinetic energy, $\zW$. 

Because global fluctuations are not observable in macroscopic systems,
local fluctuation relations have been devised, see e.g.
Refs.\cite{RTV00,CMlocal,GGlocal}, and in \cite{GRS} also a local version
the $\zL$-FR was tested numerically. In particular, the local $\zL$-FR of Ref.\cite{GGlocal}
concerns an
infinite chain of weakly interacting chaotic maps.  Let $V_0$ be a finite
region of the chain centered at the origin, $T_0 > 0$ be a time interval, and
define
\begin{equation} \langle \zL \rangle = \lim_{V_0,T_0 \rightarrow
\infty} \frac{1}{|V_0| T_0} \sum_{j=0}^{T_0-1} \zL_{V_0} (S^j X) ~,
\quad p = \frac{1}{\langle \zL \rangle |V|} \sum_{j=-T_0/2}^{T_0/2} 
\zL_{V_0} (S^j X) ~,
\label{etaplus}
\end{equation}
where $V = V_0 \times T_0$, $\zL_{V_0}(X)$ is the contribution to 
$\zL$ given by $V_0$, and $| E |$ denotes the volume of the set $E$.
Then, one obtains:
\begin{equation}
\pi_V(p) = e^{\zeta(p)|V| + O(|\partial V|)} ~, \quad
\mbox{with } \frac{\zeta(p) - \zeta(-p)}{p \langle \zL \rangle } = 1 \quad \mbox{and }
|p| < p^* ~,
\label{pilocal}
\end{equation}
where $|\partial V|$ is the size of the boundary of $V$, $p^* \ge 1$ and
$\zeta$ is analytic in $p$. The contribution of the boundary term $|\partial
V|$ should decrease with growing $V$, leading to the $\zL$-FR in
the limit of large (compared to microscopic scales) volume $V_0$ and long
times $T$.

The problem of local fluctuations, naturally leads to the possibility of
extending Onsager-Machlup theory to nonequilibrium systems. This has been done by
Gallavotti \cite{GG-OM1,GG-OM2}, under the assumption that the entropy production
rate is proportional to $\zL$. 

The fact that $\zL \ne \zW$ in general, and the identification of the systems of 
physical interest which verify the Chaotic Hypothesis, hence the $\zL$-FR, pose
an interesting question. For 
instance, Ref.\cite{GC} assumed that the \LFR could apply to systems which
have singular $\zL$, since
the \LFR had been proposed and verified in one such system \cite{ECM}. 
Later, however, the steady state \LFR was found to be hard, 
if not impossible, to verify in non-isoenergetic systems with singular $\zL$, close 
to equilibrium \cite{SE2000,DK,romans}.
In \cite{ESR} various scenarios are proposed to explain these facts. 
One of them concerns Gaussian isokinetic systems, whose $\zL$ is the sum 
of a dissipative term and a conservative term, and may be singular because of the 
interaction potentials (cf.\ Eqs.(\ref{aIK},\ref{IKsllodalpha})). In that case, the
dissipative term $\zW$ obeys the FR, while the conservative term does not, but its 
averages over long time intervals should become negligible with respect to 
the averages of $\zW$ as the length of the intervals grow \cite{romans,ESR}. Thus, in the long time 
limit, the \LFR should hold as a consequence of the validity of the \WFR \hskip -5pt, while 
the convergence times of the \LFR would diverge when equilibrium is approached, because $\zW$ 
vanishes as the square of the driving forces. Moreover, for reasons 
of symmetry, the range $[-p^*,p^*]$ of validity of the $\zL$-FR must shrink to $\{ 0 \}$ 
when the driving fields vanish, so that the \LFR may hold but only for trivial fluctuations. 

Trying to understand these facts, Refs.\cite{ESR,DJE03,BGGZ} concluded that in some cases
$\zL$ describes heat fluxes, not entropy productions, hence that in those cases the \LFR 
has to be modified, to mimic the heat FR of Van Zon and Cohen for stochastic systems 
\cite{vzc}. In particular, Ref.\cite{BGGZ} considers 
systems of the following form
\begin{equation}
\dfrac{d}{d t}{\q}_i = \p_i \qquad \dfrac{d}{d t}{\p}_i = E - \partial_{\q_i} \zF -\za \p_i
\qquad \zL = \zL^{(0)} - \zb \dfrac{d}{d t}{V}
\end{equation}
where $V$ is related to the interaction potential,\footnote{For instance, in the Gaussian
IK case of Eqs.(\ref{aIK},\ref{IKsllodalpha}), one has
$$
\frac{d}{d t} V = \frac {1}{2K}  \sum_{i=1}^N \frac{\bf p_i}{m}
\cdot {\bf F}_i^{int} 
$$} 
assuming that they have an equilibrium ($E=0$) 
distribution with exponentially decaying tails, while $\zL^{(0)}$ has Gaussian tails. 
It is then assumed that the tails have the same properties
when $E \ne 0$. Then, the average of $\zL$ in a time $\zt$ takes the form
\begin{equation}
\hskip -60pt
\overline{\zL}_{0,\zt}(X) = \frac{1}{\zt} \int_0^\zt \zL(S^t X) d t = \overline{\zL^{(0)}}_{0,\zt}(X) + 
\frac{\zb}{\zt} \left[ V(S^\zt X) - V(X) \right] \,\,.
\end{equation}
For chaotic systems, and large $\zt$, it is now reasonable to assume that
$\zL^{(0)}$, $V_f(X)=V(S^\zt X)$ and  $V_i(X)=V(X)$ 
are independently distributed. This, together with the exponential tails of 
$V$, leads to 
\be
\lim_{\zt \to \infty} \frac{1}{\zt} \log M(p^*) = \lim_{\zt \to \infty} \frac{1}{\zt} \log 
\int_{-p^* \langle \zL \rangle}^{p^* \langle \zL \rangle} d \zL^{(0)}
e^{\zt \tilde{\zeta}_0(\zL^{(0)}) - \zt | \zL - \zL^{(0)} |}
\ee
where 
\be
M(p^*) = \int_{-p^* \langle \zL \rangle}^{p^* \langle \zL \rangle}
d \zL^{(0)} \int_0^\infty d V_i \int_0^\infty d V_f
e^{\zt \tilde{\zeta}_0(\zL^{(0)}) -\zb(V_i+V_f)} \zd[\zt(\zL-\zL^{(0)})+
\zb(V_i - V_f)] 
\ee
and $\tilde{\zeta}_0(\zL^{(0)})$ is the rate (Cramer) function of $\zL^{(0)}$. Then,
one obtains
\bea
&&\hskip -20pt\tilde{\zeta}(\zL) = ~~~~~
\hskip -20pt\max_{\zL^{(0)} \in [-p^* \langle \zL \rangle,p^* \langle \zL \rangle]} 
\left[ \tilde{\zeta}_0(\zL^{(0)}) - | \zL - \zL^{(0)} | \right] =  \nonumber \\
&& \hskip 130pt = \left\{ \begin{array}{lll} \tilde{\zeta}_0(\zL_-) - \zL_- + \zL ~, & & \zL < \zL_- \\
\tilde{\zeta}_0(\zL) ~, & & \zL_- \le \zL \le \zL_+ \\
\tilde{\zeta}_0(\zL_+) + \zL_+ - \zL ~,& ~ & \zL > \zL_+ 
\end{array}
\right.
\eea
where $\tilde{\zeta}_0'(\zL_{\pm}) = \mp 1$. If the FR 
holds for $\zL^{(0)}$, with $| \zL^{(0)} | \le p^* \langle \zL \rangle$,
the fact that $\tilde{\zeta}_0( \langle \zL \rangle) = 0$ leads to
\begin{equation}
\tilde{\zeta}(\zL) - \tilde{\zeta}(-\zL) = 
\left\{ \begin{array}{lll} \zL~, & &  | \zL| < \langle \zL \rangle \\
\tilde{\zeta}_0(\zL) + \zL~, & & \langle \zL \rangle \le |\zL| \le \zL_+ \\
\tilde{\zeta}_0(\zL_+) + \zL_+~, & &  | \zL | > \zL_+  ~.
\end{array}
\right.
\label{Anmap}
\end{equation}
A relation similar to the heat FR of Van Zon and Cohen of \cite{vzc} is thus obtained 
for $\zL$. The authors of \cite{BGGZ}
conclude that the $\zL^{(0)}$-FR holds with $|p| \le p^*$, 
if $\zL^{(0)}$ is bounded or decays faster than exponential. They adopt Gentile's 
approach for Anosov flows, which reduces the flow to a 
Poincar\'e map \cite{Gentile}, and assume that the Chaotic Hypothesis apply
to a Poincar\'e map which avoids the singularities of $\zL$. This may be
done taking a level section $V=\bar{V}$, for a certain value $\bar{V}$.
Then, the volume contraction rate of the map, $\zL^{(0)}$, is bounded, the 
terms $(V_f - V_i)$ vanish, because $V_i=V_f=\bar{V}$, and the Chaotic Hypothesis
yields the $\zL^{(0)}$-FR. Here, the connection with the Van Zon - Cohen theory is 
made replacing the white noise with chaos due to uniform hyperbolicity.

\subsubsection{The Evans-Searles approach}
Because $\zL$ has no obvious physical meaning and the \LFR 
is based on strong assumptions which are not strictly enjoyed by systems of physical 
interest, one natural question comes to the fore: can one derive the steady state relation for 
the dissipated power divided by the kinetic energy, the $\zW$-FR, under 
more physical assumptions? In \cite{ESR2}, the Evans-Searles approach, first 
proposed in 1994 \cite{ES94}, has been polished to give an affirmative answer to
this question. 

Let us begin with the transient $\zW$-FR. Let ${\cal M}$ be the phase space, 
$S^\tau: {\cal M} \rightarrow {\cal M}$,
a reversible evolution with time reversal involution operation $I$. Take a
probability measure $d \mu(\zG) = f(\zG) d \zG$ on ${\cal M}$,
and let the observable $\phi : {\cal M} \rightarrow \zR$ be odd with 
respect to time reversal (i.e.\ $\phi(I \zG) = -\phi(\zG)$). Introduce 
\begin{equation}
\Ft(\zG) = \frac{1}{\tau} 
\int_{t_0}^{t_0+\tau} \phi(S^{s} \zG) d s = \frac{1}{\tau} \phi_{t_0,t_0+\zt}(\zG)
\label{phitau}
\end{equation}
and the {\em Dissipation Function} 
\begin{equation}
\Wt(\zG) = \frac{1}{\zt}
\ln \frac{f(S^{t_0}\zG)}{f(S^{t_0+\zt} \zG)} +
\overline{\zL}_{t_0,t_0+\zt}(\zG) 
\label{omegat}
\end{equation} 
for time even probability density $f$ (i.e.\ $f(I \zG)=f(\zG)$).
The derivation that follows holds in full generality, but $\zW$ equals 
the dissipated power only if $f$ is the equilibrium probability density 
for the given system, or is properly related to that. The existence of 
the logarithmic term of Eq.(\ref{omegat})
has been called {\em ergodic consistency} \cite{review}. For $\zd > 0$,
let $A^+_\zd=(A-\zd,A+\zd)$ and $A^-_\zd=(-A-\zd,-A+\zd)$, and let 
$E(\phi \in (a,b))$ be the set of points $\zG$ such that $\phi(\zG) \in (a,b)$. 
Then, $E(\Wz \in A^-_\zd) = I S^\zt E(\Wz \in A^+_\zd)$, and
the transformation $\zG = I S^\zt X$ has Jacobian
\begin{equation}
\left| 
\frac{d \zG}{d X} \right|
=\exp\left( - \int_0^\zt \zL(S^s X) d s \right) = e^{-\zL_{0,\zt}(X)} ~,
\label{ccordtransf}
\end{equation}
which leads to
\be
\frac{ \int_{E(\Wz \in A^+_\zd)} f (\zG) d \zG }{
\int_{E(\Wz \in A^-_\zd)} f(X) d X } 
=\frac{ \int_{E(\Wz \in A^+_\zd)} f (\zG) d \zG }{
\int_{E(\Wz \in A^+_\zd)} \exp\left[-\zW_{0,\zt}(X) \right] f(X) d X }   
\label{transW}
\ee
i.e.\ to the {\em transient} $\zW$-FR:
\be
\frac{\mu(E(\Wz \in A^+_\zd))}{\mu(E(\Wz \in A^-_\zd))} 
= e^{[A+\ze(\zd,A,\zt)]\zt} ~, \quad | \ze | \le \zd
\label{ESFR}
\ee
which concerns the non-invariant probability measure $\mu$ of 
density $f$. Time reversibility is essentially the only ingredient 
of the above derivation. 

To obtain the steady state $\zW$-FR, consider the ratio  
\begin{equation}
\frac{\mu(E(\Ft \in A^+_\zd)) }{\mu(E(\Ft \in A^-_\zd)) } =
\frac{ \int_{E(\Ft \in A^+_\zd)} f (\zG) d \zG }{
\int_{E(\Ft \in A^-_\zd)} f (\zG) d \zG }
\label{PpoverPp1}
\end{equation}
and take $t=2t_0+\zt$. Then 
\begin{equation}
E(\Ft \in A^-_\zd) = I S^{t} E(\Ft \in A^+_\zd)
\label{timeinversionp}
\end{equation} 
so that the transformation $\zG=IS^t W$, and the same algebra as
above, yield
\begin{equation}
\frac{\mu(E(\Ft \in A^+_\zd)) }{\mu(E(\Ft \in A^-_\zd)) } = \left\langle \exp \left( 
- \overline{\zW}_{0,t_0} t_0 \right) 
\right\rangle_{\overline{\phi}_{t_0,t_0+\zt} \in A^+_\zd}^{-1} \,\,.
\label{phiratio}
\end{equation}  
Move now the evolution from sets to measures, using
\begin{equation}
\mu(E) = \mu_{t_0}(S^{t_0} E) = \int_{S^{t_0}E} f_{t_0}(W) d W
\end{equation}
where $E$ is a subset of ${\cal M}$, $\mu_{t_0}$ is the evolved measure up to
time $t_0$, and $f_{t_0}$ its density.  Some algebra yields
\bea
\hskip -20pt
\frac{\mu_{ {t_0}}(E(\overline{\phi}_{0,\zt} \in A^+_\zd))}
{\mu_{ {t_0}}(E(\overline{\phi}_{0,\zt} \in A^-_\zd))} &=&
\frac{\mu_{ {t_0}}(S^{t_0}E(\Ft \in A^+_\zd))}
{\mu_{ {t_0}}(S^{t_0}E(\Ft \in A^-_\zd))} \\
&=&\frac{\mu(E(\Ft \in A^+_\zd))}{\mu(E(\Ft \in A^-_\zd))} =
\frac{1}{\langle \exp \left(
-\overline{\Omega}_{0,t} t \right) \rangle_{\Ft \in A^+_\zd}} \,\,.
\label{ESSFT}
\eea 
We call $\phi$-FR this relation. For $\Ft=\Wt$, taking the logarithm and dividing by
$\zt$, it produces:
\bea
&&\frac{1}{\zt} \ln \frac{\mu_{t_0}(E(\overline{\zW}_{0,\zt} \in A^+_\zd))}
{\mu_{t_0}(E(\overline{\zW}_{0,\zt} \in A^-_\zd))} = \nonumber \\
&&\hskip 60pt A + \ze(\zd,t_0,A,\zt) 
- \frac{1}{\zt} \ln 
\left\langle e^{- t_0(\overline{\zW}_{0,t_0} + \overline{\zW}_{t_0+\zt,2t_0+\zt})} \right\rangle_{\Wt \in A^+_\zd} \,\,.
\label{SSESFT}
\eea
If $\mu_{t_0}$ tends to a steady state $\mu_\infty$ when $t_0 \to \infty$, the
above should change from a statement on the ensemble $f_{t_0}$, to a statement
concerning also the statistics generated by a single typical trajectory. A result
like this
would be the {\em steady state} $\zW$-FR. This requires further assumptions, 
because $t_0$ tends to infinity before $\zt$ does and, in principle, the growth 
of $t_0$ could make the conditional average in (\ref{SSESFT}) diverge. 
Nevertheless, the decay of the auto-correlation of $\zW$ suffices.
Indeed, the conservation of probability yields
\begin{equation}
\left\langle e^{-s\overline{\zW}_{0,s}} \right\rangle = 1 ~, \quad \mbox{for every } s \in \zR ~,
\label{NEPI}
\end{equation}
and if the $\zW$-autocorrelation time vanishes, one can write:
\bea
&&\hskip -10pt\left\langle e^{-t_0 \overline{\zW}_{0,t_0}} e^{-t_0 \overline{\zW}_{t_0+\zt,2t_0+\zt}} 
\right\rangle_{\Wt\in A^+_\zd} = \nonumber \\
&&\hskip 60pt \left\langle e^{-t_0(\overline{\zW}_{0,t_0} + \overline{\zW}_{t_0+\zt,2t_0+\zt})} \right\rangle =
\left\langle e^{-t_0 \overline{\zW}_{0,t_0}} 
\right\rangle \left\langle e^{- t_0\overline\zW_{t_0+\zt,2t_0+\zt}} 
\right\rangle
\label{condave-full}
\eea
and
\begin{equation}
1 = \left\langle e^{-s\overline{\zW}_{0,s} -(t-s)\overline{\zW}_{s,t} } \right\rangle = 
\left\langle e^{-\zW_{0,s}} \right\rangle \left\langle e^{-\zW_{s,t}} \right\rangle \Longrightarrow
\left\langle e^{-(t-s)\overline{\zW}_{s,t}} \right\rangle =1 \quad \mbox{for all~} s, t ~.
\end{equation}
Hence, the logarithmic correction term in (\ref{SSESFT}) identically 
vanishes for all $t_0,\zt$, and the \WFR is verified even at short $\zt$'s. 
Of course, this idealized situation does not need to be realized. Nevertheless,
tests performed on molecular dynamics systems indicate that there exists a 
constant $K$, such that~\cite{ESR2,ESRP} 
\begin{equation}
0 < \frac{1}{K} \le \left\langle e^{-t_0(\overline{\zW}_{0,t_0} + \overline{\zW}_{t_0+\zt,2t_0+\zt})} 
\right\rangle_{\Wt \in A^+_\zd} \le K ~.
\label{aveBDD}
\end{equation}
As a matter of fact, the decorrelation (or Maxwell time), $t_M$,  
expresses a physical property of the system, thus it does not 
depend on $t_0$ or $\zt$, and depends only mildly on the  
external field [usually, $t_M(F_e) = t_M(0) + O (F_e^2)$]. Its 
order of magnitude is that of the mean free times. If these 
scenarios are realized, Eq.(\ref{aveBDD}) follows and the 
logarithmic correction term of Eq.(\ref{SSESFT}) vanishes 
as $1/\zt$, with a characteristic scale of order $O(t_M)$. 

It results that the steady state $\zW$-FR can be obtained basically only 
from time reversibility and from the decay of the $\zW$-autocorrelation, and
one can write
\be
\frac{1}{\zt} \ln \frac{\mu_\infty (E(\overline{\zW}_{0,\zt} \in A^+_\zd))}
{\mu_\infty (E(\overline{\zW}_{0,\zt} \in A^-_\zd))} = A + \mbox{small correction}
\label{TSSESFT}
\ee where the small correction can be made arbitrarily small taking
sufficiently large $\zt$ and sufficiently small $\zd$.  This decay of
correlations could be relaxed or replaced by other assumptions, but is
needed for the convergence to a steady state, and is verified in most
systems of physical interest. Furthermore, this approach explains why
the relevant convergence times are functions of material properties of
the systems, and do not diverge in the equilibrium limit. The
decorrelation times of $\zW$ are indeed material properties.

Various other relations can now be obtained. For instance, for
any odd $\phi$, any $\zd>0$, any $t_{0}$ and any $\zt$, one has
\begin{equation} 
\langle
\exp \left( -t \overline{\Omega}_{0,t} \right) \rangle_{\Ft \in (-\zd,\zd)}
= 1 \label{inter} 
\end{equation}
which, in the $\zd\to\infty$ limit, leads
to the well known normalization property - the so-called Nonequilibrium Partition Identity
(\ref{NEPI}).
Given a second
density $\tilde{f}$ and the corresponding $\tilde{\zW}$, one has
\begin{equation} 
\left\langle \exp \left( -t\overline{\tilde{\Omega}}_{0,t} \right)
\right\rangle_{\Ft \in A^+_\zd} \approx \frac{\mu_{t_0}(E(\overline{\phi}_{0,\zt}
\in A^-_\zd))} {\mu_{t_0}(E(\overline{\phi}_{0,\zt} \in
A^+_\zd))} \approx \langle \exp \left( -t \overline{\Omega}_{0,t} \right) 
\rangle_{\Ft \in A^+_\zd} ~, 
\end{equation}
for large $t_{0}$, all $\zt$ and all allowed pairs $A$ and $-A$, if $d \mu = f d \zG$ and
$\d \tilde{\mu} = \tilde{f} d \zG$ converge to the same steady state. Then, as far as the
FRs are concerned, the different dissipation functions are
equivalent. The Dissipation Relation
\be
\langle \phi \rangle_t = \int_0^t d s \langle \zW(0) \phi(s) \rangle ~,
\label{theDR}
\ee
where $\langle . \rangle_t$ the average with respect to $\mu_t$,
is another direct consequence of the approach followed in this section
\cite{ESdissir}.

The theory outlined above is consistent with Irreversible Thermodynamics, since 
it leads to the Green-Kubo relations, close to equilibrium \cite{ESR}; hence it 
can be considered in the construction of a more general theory, which encompasses
far from equilibrium phenomena.
Differently from Ref.\cite{GG96}, which deals with $\zL$ and with asymptotic times, 
Ref.\cite{ESR} deals with $\zW$ and stresses the role of the physical time scales. 
This is illustrated on a Nos\'e-Hoover thermostatted system [Eqs.(\ref{NHeq})], whose
equilibrium state is given by the extended canonical density
\begin{equation}
f_c(\zG,\xi) =
\frac{e^{-\zb \left(H_0+Q \xi^2 /2 \right)}}{\int d \xi ~ d \zG~~
e^{-\zb \left(H_0+Q \xi^2 /2 \right)}} \qquad \mbox{with } ~~~
Q = 2 K_0 \theta = \frac{g \theta}{\beta}
\end{equation}
where $H_0$ is the internal energy, $K_0$ si the reference kinetic energy corresponding to the inverse
temperature $\beta$, and $g$ depends on the number of degrees of freedom \cite{EM90}.
In the most common case, in which 
${\bf F}_i^{ext}=c_i {\bf F}_e$ and ${\bf J} = \sum_{i=1}^N c_i( d{\bf q}_i/d t )$,
one has
$$
\xi = -\frac{1}{2K} \left[ \frac{d H_0}{d t} + {\bf J} \cdot {\bf F}_e V \right]
$$
and
$$
\zL = - \mbox{div} \left( \frac{d \zG}{d t} \right) - {\partial}_\xi \left( \frac{d \xi}{d t} \right) 
= d N \xi ~,
$$
where $d$ is the spatial dimension,
which shows that $\zL$ and its fluctuations are not directly related to the
dissipation rate $\zW = \beta {\bf J} \cdot {\bf F}_e V$, while its average is, because
$\langle H_0 \rangle = 0$. Integrating $f_c(x,\xi)$ yields
\begin{equation}
f_c(\xi) = \int d \zG f_c(\zG,\xi) = \sqrt{\frac{\zb Q}{2 \pi}} ~ \exp \left[- \zb Q \xi^2 /2 \right] ~.
\end{equation}
Therefore, the distribution of $\overline{\xi}_{0,t}$ is also Gaussian
in equilibrium, with variance proportional to that of
$\overline{h}_{0,t}=(1/t) \int_0^t \dot{H}_0 d \zt$, and near
equilibrium it can be assumed to remain Gaussian, around its mean, for
large $t$, as a consequence of the Central Limit Theorem (CLT). To use
the $\zL$-FR together with the CLT, the values $A$ and $-A$ of the
random variable $\overline{\xi}_{0,t}$ must be a small number of standard
deviations away from $\langle \zW \rangle$.  However,
as explained in Refs.\cite{romans,ESR,BGGZ} and briefly recalled in
Section \ref{GCapproach}, the distribution of $d N\overline{\xi}_{0,t}$ verifies the $\zL$-FR only for times $t$
sufficiently large that the standard deviation of the random variable
$\overline{h}_{0,t}$ is negligible with respect to that of  $\overline{\Omega}_{0,t}$, $\zs_{\overline{\Omega}_{0,t}}$
say.

In \cite{SE2000} it was proven that the variance of the average current obeys
\be
t \zs_{\overline{J}_{0,t}}^2(F_e) = \frac{2 L(F_e) k_{_B} T}{V} + O\left( \frac{F_e^2}{t N} \right)
\label{sigmaJ}
\ee
hence that of $\overline{\zW}_{0,t}$ obeys
\be
t \zs_{\overline{\Omega}_{0,t}}^2(F_e) = \left( \frac{F_e V}{2 K_0} \right)^2 \left[\frac{2 L(F_e) k_{_B} T}{V} + O\left( \frac{F_e^2}{t N} \right)
\right] \,\,,
\label{sigmaO}
\ee
where
$$
L(F_e) = \zb V \int_0^\infty d t \langle (J(t)-\langle J \rangle)
(J(0)-\langle J \rangle) \rangle
$$
and $L(0) = \lim_{F_e \to 0} L(F_e)$ is the corresponding linear transport coefficient. 
Equation (\ref{sigmaJ}) shows that the standard deviation of the average current
decreases when $t$ grows at fixed $F_e$, while it tends to a positive constant, 
when $F_e$ decreases at fixed $t$. Consequently, the standard deviation of $\overline{\Omega}_{0,t}$ tends to zero when $F_e \to 0$ at fixed $t$. 
Differently, the variance of $\overline{h}_{0,t}$ tends to a constant when $F_e$ tends
to zero at fixed $t$. Therefore, the smaller $F_e$ the longer the time $t$ needed
for $\zs_{\overline{\Omega}_{0,t}}$ to dominate over the variance of $\overline{h}_{0,t}$, and this time grows 
without bounds when $F_e$ tends to zero. Let us assume for simplicity, that the 
variations of these standard deviations are monotonic when either $F_e$ or $t$
varies and the other variable is fixed. Then, given $F_e$ and $A$, there is $t_\zs(F_e,A)$ 
such that the standard deviation of $\overline{\zL}_{0,t}$ is sufficiently large that  
$A$ and $-A$ are within a few standard deviations from the mean only if
$t < t_\zs(F_e,A)$. At the same time, let $t_\zd(F_e,A)$ be sufficiently large that
the steady state $\zL$-FR applies to the values $A$ and $-A$,
with accuracy $\zd$. To derive the Green-Kubo relations from the $\zL$-FR, one would need
\begin{equation}
t_\zd(F_e,A) < t < t_\zs(F_e,A)
\label{inequ}
\end{equation}
which is not satisfied if $t_\zd(F_e,A)$ grows too fast with decreasing $F_e$. This
suggests that the derivation of the Green-Kubo relation from the $\zL$-FR is 
problematic, for systems like the Nos\'e-Hoover systems or the Gaussian isokinetic 
systems with singular interaction potentials. In Ref.\cite{ESR} it is further shown
that the validity of Eq.(\ref{inequ}) is necessary but not sufficient to obtain the result. 

Differently, there is no problem in obtaining the Green-Kubo relation from the \WFR, 
because the corresponding convergence times do not diverge when $F_e$ tends to zero (as seen 
above, they become of the order of the Maxwell times), and because $\zW$ is the observable 
of interest (the energy dissipation rate divided by the kinetic energy). 
For times $t$ of the order of the Maxwell time, one may then write 
\cite{ESR}:
\begin{equation}
\langle \zW \rangle = \frac{t}{2} \zs_{\overline{\zW}_{0,t}}^2 \quad \mbox{or } ~~
L(0) = \lim_{F_e \to 0} \frac{\langle J \rangle}{F_e} = \zb V 
\int_0^\infty d s ~ \langle J(0) J(s) \rangle
\label{GKES}
\end{equation}
where the first equality is due to the validity of the  $\zW$-FR and the second to \eqref{sigmaO}.\footnote{ Equating Eqs.(33,34) in Ref.\cite{ESR}, one obtains
$\beta V F_e = \langle J \rangle/t \sigma^2_{\overline{J}_{0,t}}$. Then, the fact that $\Omega = \beta V F_e J$ implies
$\beta V F_e = 2 \langle \Omega \rangle / t \sigma_{\overline{J}_{0,t}} \beta V F_e$, from which Eq.(\ref{GKES}) follows.}

\subsubsection{The Jarzynski and the Crooks relations}
A transient relation for the fluctuations of non-dissipative 
systems has been obtained by Jarzynski 
\cite{CJ}.\footnote{The non-invariant ensemble, which characterizes the
``transient'' nature of the Jarzynski equality, is the initial canonical ensemble.}
To introduce this relation, the author considers a Hamiltonian finite particles system, 
in equilibrium with a heat bath at
temperature $T$,  which is driven away  from equilibrium, by performing some work $W$ on
it. If  $\zl$ is  a parameter that controls the Hamiltonian $H_\zl$ say, the work is performed varying $\zl$. 
Let $\zl=\zl(t)$ be a given protocol of duration $\tau$ for a variation of $\zl$ from a value $\zl(0)=A$ to another value $\zl(\tau)=B$. Suppose that the same protocol 
is repeated many times, always in the  same manner, and always from the equilibrium corresponding to $\zl=A$,
to build the statistics  
of the work done when $\zl$ evolves from $A$ to $B$. 
Let $\rho$ be the PDF of the externally performed work and $\langle
e^{-\zb W} \rangle_{A \to B}$ the average over all works done in
varying $\zl$ from $A$ to $B$. These works are not the
thermodynamic work done on the system, if the process is not
performed quasi-statically, in the presence of local equilibrium
\cite{CohMauz}, but is nevertheless a measurable quantity. The
Jarzynski equality states that \cite{CJ}:
\begin{equation}
\left\langle e^{-\zb W} \right\rangle_{A \to B} = 
\int d W ~ \rho(W) e^{- \zb W} = e^{-\zb \left[ F(B) - F(A) \right]}
\label{JR}
\end{equation} where $\zb = 1 / k_{_B} T$, and $\left[ F(B) - F(A) \right]$ is the free
energy difference between the initial equilibrium, with $\zl=A$
and the equilibrium state corresponding to $\zl=B$. This state exists and is unique, hence the value of the state
variable $F$ at $\zl=B$ is well defined, but it does not need to be
the actual state of the system at time $\zt$. Indeed, the work $W$ is
 performed on the system in the time interval $[0,\zt]$, while
relaxation to the equilibrium corresponding to $\zl=B$ may require
longer times, depending on the explicit form of the protocol
$\zl(t)$. Note that the process always begins in the same equilibrium state
corresponding to $\zl=A$, but in different microscopic states. In general, at time $\zt$, the system will
not even be in local equilibrium: Eq.(\ref{JR}) is supposed to hold
for all possible protocols, hence even arbitrarily far from
equilibrium (large $\frac{d}{d t}{\zl}$). Therefore, the presence of
the equilibrium quantities $F(A)$ and $F(B)$ may look puzzling. The puzzle is solved observing that
the externally measured work does not need to coincide with the
thermodynamic work (which would not differ from experiment to
experiment, if performed quasi-statically, cf. Section
\ref{ensevsth}). From an operational point of view, it does not matter
whether the system is in local equilibrium or not: the externally
applied forces can be measured, the resulting motions can be recorded,
hence the externally performed works are defined and can be
quantified. The difference of this process from a standard
thermodynamic one is that the external work cannot be equated to the
internal thermodynamic work, if the process is not performed
quasi-statically.

The limitation of Jarzynski's derivation, that the dynamics be Hamiltonian, has been 
overcome by Evans in terms of NEMD models \cite{EvansJR}. Therefore, the Jarzynsky 
equality, similarly to the transient $\zW$-FR, results of quite wide applicability,
since it does not rest on particular conditions on the microscopic dynamics. It is also
consistent  with the second  law of thermodynamics, since it yields
\begin{equation}
\left\langle \zb W \right\rangle_{A \to B} \ge \zb \left[ F(B) - F(A) \right]
\end{equation}
because $\ln \left\langle \Phi \right\rangle \ge \left\langle \ln \Phi
\right\rangle$ for positive observables $\Phi$, although $\left\langle
W \right\rangle_{A \to B}$ is not a thermodynamic quantity in general.

Similarly, computing the ratio of the probability that the work done in  the 
forward transformation is $W$, to the probability that it is $-W$ in the $B$
to $A$ transformation, with reversed protocol $-\frac{d}{d t}{\zl}$,
produces the Crooks relation \cite{GK}:
\begin{equation}
\frac{{\rm Prob}_{A \to B} (W = a)}{{\rm Prob}_{B \to A} (W = -a)} = 
e^{-\zb \left[ F(B) - F(A) \right]} ~ e^a
\label{CrRel}
\end{equation}
which  leads to the Jarzynsky equality, by  a simple
integration:
\begin{eqnarray}
&&\left\langle e^{-\zb W} \right\rangle_{A \to B} = \nonumber \\ 
&&\hskip -15pt\int {\rm Prob}_{A \to B} (W = a) e^{-a} d a = e^{-\zb \left[ F(B) - F(A) \right]}
\int {\rm Prob}_{B \to A} (W = -a) d a \nonumber \\
&=& e^{-\zb \left[ F(B) - F(A) \right]} \,\,.
\end{eqnarray}
A relation exists between these results and the $\zW$-FR. In the first place,
the transient \WFR may be applied to the protocols of the Jarzynski equality
and of the Crooks relation, \cite{DJE03}. Then, let $f_A$ and $f_B$ be the
canonical distributions at same inverse temperature $\zb$, for the Hamiltonians
$H_A$ and $H_B$.  The
corresponding Helmholtz free energies are $F_i = - k_{_B} T \ln \int d\mathbf{q}d\mathbf{p} 
\exp[-H_i/k_{_B} T]$ for $i=A,B$. For simplicity, let $\zl$ go from $A$ to $B$
in a time $\zt$, with rate $\frac{d}{d t}{\zl} = 1/\zt$, and from $B$ to $A$ with rate
$\frac{d}{d t}{\zl} = -1/\zt$.  Correspondingly, the microscopic thermostatted
evolution equations are
\begin{eqnarray}
&&\dfrac{d}{d t}{\q} = \frac{\partial H_{\zl}(\q,\p)}{\partial \p} ~, \quad
\dfrac{d}{d t}{\p} = - \frac{\partial H_{\zl}(\q,\p)}{\partial \q} - \epsilon_i \za(x) \p ~, 
\\
&&\dfrac{d}{d t}{\zl} = \pm \frac{1}{\zt} ~, \quad \mbox{with } \mbox{$+$ ~ for } 
A \to B \quad \mbox{and } \mbox{$-$ ~ for } B \to A
\end{eqnarray}
where $\epsilon_i = 1$ for $i=1,...N_w$ and $\epsilon_i = 0$ for $i>N_w$, so that the 
deterministic thermostat acts only on the $N_w$ particles of the walls of 
the system, to fix their kinetic temperature and mimic a heat bath 
at temperature  $T$. Then, the work performed by the external
forces is  given  by  $\zb W  =  \zb  [  H_B -  H_A]  -
\zL_{0,\zt}$.  

If $A=B$, the \WFR applies directly, and
the Jarzynski  equality is  an immediate consequence  of the \WFR  because the
\WFR implies
\begin{equation}
\left\langle e^{-\zb W} \right\rangle = \int P(W) e^{-\zb W} d W = \int d W P(-W) = 1 \,\,.
\end{equation}

The transient $\zW$-FR, the Jarzynski equality and the Crooks relation 
do not have the same range of applicability, the Crooks relation being the most
general. It is remarkable how all of them connect equilibrium to nonequilibrium 
properties of physical systems: the transient $\zW$-FR deals with the energy
dissipation associated with the application of a dissipative field to an ensemble 
of systems initially in equilibrium; the Jarzynski and Crooks relations relate the
work done to the free energy differences. The point of view of NEMD is important 
in this context, because finite Hamiltonian systems do not afford nonequilibrium 
steady states and also because, in general, external drivings do not need to preserve 
the Hamiltonian nature of the dynamics. 
Therefore, the agreement, where appropriate, with the NEMD approach strengthens 
the results. The interest of all these relations is bound to grow with our 
understanding of microscopic systems, particularly in biophysics and nanotechnology.

The connection between equilibrium and nonequilibrium properties made by the $\zW$-FR, 
the Jarzynski equality and the Crooks relation is, in a sense, opposite to that made 
by the Fluctuation Response formula: while the relations treated in this subsection 
obtain information about equilibrium states from nonequilibrium dynamics, the Fluctuation 
Response formula uses equilibrium fluctuations to gain information on nonequilibrium states.

\subsubsection{Ensemble vs thermodynamic relations}
\label{ensevsth}

There is a substantial difference between the transient relations considered in this 
Section (see also, e.g.\ \cite{VanDB}), and the Fluctuation Response formula. The Fluctuation Response formula concerns
thermodynamic properties of macroscopic objects, hence it holds for the behaviour
of a single system. Differently, as previously observed, the transient relations 
concern the statistics of ensembles of objects. Nevertheless, the transient
relations are sometimes confused for thermodynamic statements. This is due to the fact 
that the transient relations either have forms similar to the second law of thermodynamics, 
or can be used to obtain relations formally like the second law. Also, the statistical 
nature of the results concerning a collection of particle systems, is often confused for 
the statistics concerning the many particles of a single system. Obviously, this is not 
appropriate, and the importance of the transient relations must be found precisely in 
their unavoidable ensemble nature, not in their thermodynamic content. Indeed, the 
differences between the two kinds of relations 
have practical consequences, which can be understood as follows:

\begin{itemize}

\item[\bf 1.] Thermodynamic statements do not refer to ensembles of
systems, they concern single systems. If one has a collection of systems
in the same thermodynamic state, they all behave in the same way, hence
their average behaviour is, {\em a fortiori}, bound to be that of a single 
member of the collection. For this reason, one may resort to a
microscopic statistical description, in which each single system is in
a different microscopic state, in order to recover the unique common
thermodynamic behaviour. This macroscopic behaviour emerges when the
time and space scales are well separated from the microscopic scales, and 
sufficiently mixing processes take place. In that case, the
differences among the ensemble members are not observable: they cannot
be expressed in terms of thermodynamic quantities, such as the thermodynamic 
work $W$ done on a system. Equivalently, one may say that
the differences vanish typically as $1/\sqrt{N}$, where $N$ is the number of
particles of each system in the collection. Thus, sufficiently large $N$
implies practically no differences in the behaviours of the collection
members.

\item[\bf 2.] If work is done in a reversible fashion on a macroscopic
system, one can draw the corresponding path of thermodynamic states in
the, e.g.\ pressure-volume plane. Integrating along this path, one
obtains a given value for the work done on the system, which is
operationally accessible, by measuring the work done by the external
equipment. In this case, the system and the working apparatus are
globally in close contact all the time, and the energy given to the
system, under a fixed protocol, is always the same. Therefore, the
ensemble average and the single works are equal, 
and equal the internal and the external works.

\item[\bf 3.] If work is not done in a reversible fashion, one obtains different
values for $W$, even if the same protocol is repeated. The energy
that a single system may acquire from the experimental apparatus cannot
be predicted, and only averages over many repetitions of the
protocol may be inferred. The transient relations 
serve to this purpose. It is interesting to observe that this is
the case even for large $N$. Indeed, given a large collection 
of systems, the different values 
of the works performed on them (not to be confused with fluctuations in the 
evolution of a single system) are not due to the size of $N$, but to the fact that the process
is not quasi-static, and corresponds to no path in the space of thermodynamic
states. The reason is that the energy exchanged 
between system and driving apparatus depends on the microscopic initial state,
not just on the driving protocol. Every time the protocol is repeated, the microstate
is different, hence a different value for $W$ is obtained. This value is unpredictable, because 
the initial microstate is unknown, moreover different initial ensembles lead to
different statistics.
\end{itemize}
Therefore, the content of the ensemble relations in far from equilibrium
processes is not, e.g.\ the impossible quantification of the dissipative works,
but the statistical description of large collections of systems.
Indeed, they describe intriguing properties of the chosen initial ensembles.
The fact that only in certain limits are the transient relations equivalent to 
thermodynamic statements is not a difficulty, because they have 
the advantage of being valid even when the thermodynamic description
fails (e.g.\ in the absence of local equilibrium, or for systems of a small
number of particles).

\section{Fluctuation relations in stochastic systems and applications}

\subsection{Stochastic systems: bounded and unbounded state spaces}
\label{sec:vzc}

The derivations of the FR for stochastic systems are much easier than
those of the $\zL$-FR, because, as noted in \cite{Kurchanlast}, the
difficulties connected with singular invariant phase space
distributions are avoided, and one can deal with smooth probability
distributions. \footnote{Note that for the same reason the derivation
of the $\zW$-FR given in Section 5, which deals with the evolution of probability
densities, is technically much easier, and of wider
applicability than that of the $\zL$-FR.} The first of the stochastic
FRs was produced by Kurchan, who obtained a modified detailed balance
property for Langevin processes of finite systems, and a FR for the
entropy production, under a few assumptions, like {\em e.g.} the
boundedness of the potential \cite{Kurchan}.  Lebowitz and Spohn
\cite{LS} extended this result to generic Markov processes with a
finite number of states.  Additionally, under the assumption that
local detailed balance is attained, they showed that the Gibbs entropy
production is related to the action functional that satisfies the
FR. This means that for Markov processes the Gibbs entropy production
plays the role of the phase space contraction rate in the
$\Lambda$-FR.  Maes obtained a large deviation principle for discrete
space-time Gibbs measures \cite{CM99}. He showed that using the Gibbs
formalism to describe the steady state of time discrete lattice systems
leads to a FR for a kind of Gibbs entropy production. These results
can be seen as a generalization of the $\Lambda$-FR and of its
stochastic versions, as stochastic dynamics and thermostatted systems
satisfying the chaotic hypothesis are particular cases of space-time
Gibbs measures.

Farago pointed out that singularities may cause difficulties in the
conventional use of the FR, for stochastic systems \cite{Farago}.
Wang, {\em et al.}, reported the experiment of a colloidal particle
that is dragged through water by means a moving optical trap
\cite{WSMSE}, confirming an integrated version of the transient
FR. This experiment may be modeled through a stochastic Langevin
evolution describing a Brownian particle, dragged in a liquid by a
moving harmonic potential, a problem studied by various authors
\cite{Farago,mazonka,narayan,Maesrecent}. In particular, analyzing the
results of \cite{WSMSE}, Van Zon and Cohen \cite{vzc} observed that,
in the presence of a potential term, the work done on the system in a
time $\tau$, $W_\tau$, consists of two parts: the heat $Q_\tau$
corresponding to the energy {\em dissipated} in the liquid, and the
potential energy difference of the Brownian particle $\Delta U_\tau$:
\begin{equation} \label{eq:e-balance}
W_\tau = Q_\tau + \Delta U_\tau \ .
\end{equation}
A Brownian particle in a fluid,  subjected to a  harmonic potential
that moves with a constant velocity  $v^*$, is described by
the overdamped Langevin process
\begin{equation} \label {eq:brownian}
\deriv{x(t)}{t} = -[x(t) - x^*(t)] + \zeta(t) \ ,
\end{equation}
where $x(t)$ is the position of the  particle at time $t$, $x^*(t) = v^*t$ is the
prescribed position of the minimum of the  potential and $\zeta(t)$ is a white noise term
representing the fluctuating force the fluid exerts on the particle. The units
used are such that $k_BT = 1$.
The total work is given by
\begin{equation} \label{eq:vzc-work}
W_\tau = -v^* \int_0^\tau [x(t) - x^*(t)] dt \ .
\end{equation}
In a comoving frame Eq.~(\ref{eq:brownian}) reduces to a standard
Ornstein-Uhlenbeck process and thus, the stationary probability
distribution and Green's function are Gaussian in the particle's
position.  Since the total work is linear in the particle's position,
$W_\tau$ is Gaussian as well.  Because of this and of
Eq.~(\ref{eq:vzc-work}), in the case of transient fluctuations, the
variance of $W_\tau$ equals $2 \langle W_\zt \rangle$. Since $W_\tau$
is Gaussian, the total work satisfies the transient FR. In the
$\tau\rightarrow\infty$ limit, the variance of $W_\tau$ remains twice
its mean, hence the total work satisfies the steady state FR.

Van Zon and Cohen pointed out  that the quantity measured in the experiment of
\cite{WSMSE} was  precisely the total work  and thus, in  agreement with their
Langevin treatment.  However, they also clarified that, while the  PDF of the 
total work is Gaussian, at
equilibrium  the PDF  of the  potential energy  is exponential: 
$P(\Delta  U) \sim \exp(-\mbox{const.}~ \Delta U)$, and is  expected to remain such
in the  steady states. Therefore, while the small fluctuations
of heat  are expected  to coincide  with those of  the total  work, since the
contribution of the potential energy is only $\mathcal{O}(1)$, 
large heat fluctuations are more likely to be  due to a
large fluctuation of the potential energy.

Consider  the  harmonic potential $U(t) \equiv \frac{1}{2}|x(t) - x^*(t)|^2$
in Eq.~(\ref{eq:e-balance}). The heat $Q_\tau$ is nonlinear in
the particle's position, hence its PDF needs not be Gaussian.
Compute the Fourier transform of this PDF:
\begin{equation}
\hat{P}_\tau(q) \equiv \int_{-\infty}^\infty dQ_\tau e^{iqQ_\tau}
{\rm Prob}_\tau(Q_\tau) \ .
\end{equation}
Writing ${\rm Prob}_\tau(Q_\tau)$ in terms of the joint distribution of the work
$W_\tau$ and of the positions $x(0), x(\tau)$, one obtains
\begin{equation} \label{eq:vzc-fourier}
\hat{P}_\tau(q) =
\frac{\exp\left\{w\left(i-q\right)\left(\tau -
      \frac{2q^2(1-e^{-\tau})^2}{1 + (1-e^{-2\tau})q^2}\right)\right\}}{[1 +
  (1-e^{-2\tau})q^2]^{3/2}} \ ,
\end{equation}
where $w=\langle W_\zt \rangle / \zt$ is  the rate of work  done in the system, related  to the stationary
state average. Anti-transforming $\hat{P}_\tau(q)$, one  considers the  heat
fluctuation function
\begin{equation} \label{eq:Q-fluc}
f_\tau(p) =
\frac{1}{w\tau}\ln\left[\frac{{\rm Prob}_\tau(pw\tau)}{{\rm Prob}_\tau(-pw\tau)}\right] \ ,
\end{equation}
where  $p=Q_\tau/\langle Q_\tau\rangle$ and  $\langle Q_\tau\rangle  = \langle
W_\tau\rangle - \langle \Delta U_\tau\rangle  = w\tau$, since
$\langle\Delta U_\tau\rangle = 0$,  in the steady state.
To   obtain an asymptotic    analytical   expression   of
Eq.~(\ref{eq:Q-fluc}), consider the quantity
\begin{equation} \label{eq:e}
e(\lambda) \equiv \lim_{\tau\rightarrow\infty}-\frac{1}{w\tau}\langle
e^{-\lambda Q_\tau} \rangle \ ,
\end{equation}
for large $\tau$, and
\begin{equation} \label{eq:Pq}
{\rm Prob}_\tau(Q_\tau) \sim e^{-w\tau\hat{e}(Q_\tau/w\tau)} \ ,
\end{equation}
where  $\hat{e}(p)= \max_{\{\lambda\}} \ [e(\lambda) -  \lambda p]$  is  the Legendre
transform  of $e(\lambda)$.   
Analytically continuing  $\hat{P}_\tau$  to imaginary
arguments,  one   can    write   $\langle    e^{-\lambda    Q_\tau} \rangle   =
\hat{P}_\tau(i\lambda)$, i.e.\
\begin{equation} \label{eq:e-sol}
\langle e^{-\lambda Q_\tau}\rangle =
\frac{\exp\left[-w\lambda(1-\lambda)\left\{\tau+\frac{2\lambda^2(1-e^{-\tau})^2}{1-(1-e^{-2\tau})\lambda^2}\right\}\right]}{[1-(1-e^{-2\tau})\lambda^2]^{3/2}} ~,
\end{equation}
which  is  singular  for  $\lambda=\pm
(1-e^{-2\tau})^{-1/2}$. Using Eqs.~(\ref{eq:e}) and (\ref{eq:e-sol}) and 
taking the $\tau\rightarrow\infty$ limit, the singularities
move to $\pm 1$, yielding
\begin{equation} \label{eq:e-small}
e(\lambda) = \lambda(1-\lambda) \qquad \textrm{for} \ |\lambda|<1 \ .
\end{equation}
This implies that the conventional steady state FR, $\lim_{\tau\rightarrow\infty}f_\tau(p) = p$,
holds as in Ref.\cite{LS}.
Outside the interval $(-1,1)$, the integral
in  Eq.~(\ref{eq:Pq})  diverges due  to  the exponential  tails  of  the PDF  of
$Q_\tau$: $e(\lambda)=-\infty$ for $|\lambda|>1$.
Finally, taking these values in the Legendre transform $\hat{e}(p)$, and 
substituting in  Eq.~(\ref{eq:Pq}) one obtains
\be
\lim_{\tau\rightarrow\infty} f_\tau(p) = 
\left\{
\begin{array}{ll}
p & \textrm{~~~~for~~} \ 0 \le p < 1\\
p-(p-1)^2/4 & \textrm{~~~~for~~} \ 1 \le p < 3\\
2 & \textrm{~~~~for~~} \ p \ge 3\\
\end{array}
\right.
\label{eq:VZCFT}
\ee
and the values for negative $p$ are obtained from $f_\tau(-p)=-f_\tau(p)$.
The fluctuations of heat that are smaller than $\langle Q_\tau\rangle$ satisfy
the conventional FR, as those of $W_\zt$; larger heat fluctuations behave
differently, in particular, the probability ratio $f_\zt$ is much larger than
that predicted by the conventional FR. A similar phenomenon can occur in
deterministic systems \cite{ESR,BGGZ}, and the logarithmic term in the
definition of $\zW$, Eq.(\ref{omegat}), represents the corresponding boundary
terms.

The validity of the extended FR of Van Zon and Cohen has been
addressed in other stochastic systems. For instance, Ref.\cite{PRV06}, where a
granular system is also considered, studies the extended FR, and shows   
that the boundary terms may be important in many common situations. One 
interesting feature of the models 
considered in~\cite{PRV06}, is that the granular system behaves very similarly to 
a simple Markov chain with unbounded state space, as discussed in Subsection 5.4.1.
Somewhat different results are reported in Ref.\cite{germans}, which shows that 
the extended FR does not hold in the partially asymmetric zero-range process 
with open boundaries. Hence the range of validity of the Van Zon - Cohen
scenario is still to be fully understood. 

Various other studies have dealt with the problem of a Brownian particle and
Langevin process in general, like those of 
Refs.\cite{PO98,HS01,seifert05,seifert06,Kurchanlast,V06,imparato,lastimpa}. 
In particular, Ref.\cite{V06} points out that exponential
distributions of the boundary terms may not always have the same effect on
FRs, hence the problem of their relevance in FR remains open.

\begin{center}
\begin{tabular}{| c | c | c | c |} \hline \hline
Relation & State & Dynamics & Observable \\
\hline
(\ref{firstFR}) & NESS & Gaussian thermostatted & DP \\
\hline
(\ref{largedev}) & NESS & Reversible transitive Anosov & $\zL$ \\
\hline
(\ref{pilocal}) & NESS & Reversible transitive Anosov & Local-$\zL$ \\
\hline
(\ref{Anmap}) & NESS & Reversible flow with transitive & Local-$\zL$ \\
 & & \raisebox{6pt}{Anosov Poincar\'e map} &  \\
\hline
(\ref{transW}) & SMe-NE & Reversible and dissipative & $\zW$ \\
\hline
(\ref{phiratio}) & SMe-NE & Reversible and dissipative & $\phi$ \\
\hline
(\ref{NEPI}) & SMe-NE & Reversible and dissipative & $\zW$ \\
\hline
(\ref{theDR}) & SMe-NE & Reversible and dissipative & $\zW$ \\
\hline
(\ref{TSSESFT}) & NESS & Reversible and dissipative & $\zW$  \\
 & & \raisebox{6pt}{$\zW$-correlation decay with} & \\
 & & \raisebox{12pt}{respect to initial measure} & \\
\hline
(\ref{JR}) & C-NE & {Hamiltonian dynamics} & $F$ \\
\hline
(\ref{CrRel}) & C-NE & Hamiltonian dynamics & $F$ \\
& & \raisebox{6pt}{or Markov process} & \\
\hline
(\ref{eq:VZCFT}) & NESS & Markovian model of & $HF$ \\
 & & \raisebox{6pt}{colloidal particle in optical trap} & \\
\hline \hline
\end{tabular} \vskip -4pt
\end{center}
{\small {\bf Table 1.} Synoptic table of fluctuation relations. The
first column gives the equation number.  The second column gives the
details of the state for which the relation holds: NESS means
nonequilibrium steady state, C-NE means experiments starting in
canonical equilibrium and continuing under nonequilibrium dynamics,
SMe-NE means initial smooth ensemble followed by nonequilibrium
dynamics. The third column specifies the dynamics for which the
relation has been obtained.  In the fourth column, the observables for
which the relation holds are given: DP means dissipated power, $\zL$
means phase space contraction rate, $\zW$ means dissipation function,
$\phi$ is any odd observable, $F$ is the Helmholtz free energy. Local
means integrated over a given region of the relevant space, HF means
heat flux. }

\subsubsection{Markov chains with an infinite number of states}

\label{markovchain}
The problem of singularities in unbounded spaces can be understood in
detail using a simple model~\cite{PVTW06,PRV06}. Consider a Markov
chain with $N$ (possibly infinite) states with invariant probability
measure $\bmu=\{\mu_{\sigma} \}$ and transition rates $K_{ab}$ which denote
the conditional probability of going from state $a$ to state $b$. We
are interested in the fluctuations of the following two ``action
functionals'':
\begin{subequations} \label{eq:functionals}
\begin{align}
W(\tau)&=\sum_{i=0}^\tau \log\frac{K_{\sigma_i \sigma_{i+1}}}{K_{\sigma_{i+1}
   \sigma_i}}\\
\oW(\tau)&=\sum_{i=0}^\tau \log\frac{K_{\sigma_i  \sigma_{i+1}}}{K_{\sigma_{i+1}  \sigma_i}}+\log\frac{\mu_{\sigma_0}}{\mu_{\sigma_{\tau+1}}}\\
\end{align}
\end{subequations}
where $\sigma_i \in \{1, 2, ... N\}$ is the state of the system at
time $i$. Both functionals have been defined by Lebowitz and Spohn,
in~\cite{LS99}, who focused on the properties of $W$, neglecting
the importance of the difference $B=\oW-W$. Both $W$ and $\oW$
associate a real number to any finite trajectory (any realization of
the Markov chain). In a stationary state which verifies the detailed
balance condition $\mu_a K_{ab}=\mu_b K_{ba}$, one has $\oW(t)
\equiv 0$, while $W(t) \ne 0$ (except when $K_{ab}=K_{ba}$). For $t$
large enough, almost all the trajectories yield $\lim_{s \to \infty}
W(s)/s = \lim_{s \to \infty} \oW(s)/s=\langle W(t)/t \rangle=\langle
\oW(t)/t \rangle$, where (assuming an ergodic and stationary system)
$\langle \rangle$ indicates an average over many independent segments
from a single very long trajectory. Let $S=-\sum_{i=1}^N \nu_i \log
\nu_i$ be the Gibbs entropy of the system at time $t$, where
$\nu_i(t)$ is the probability to be in the state $i$ at time $t$; then
\begin{equation}
S(t+1)-S(t)=\Sigma(t)-A(t)
\end{equation}
where $\Sigma(t)$ is always non-negative, $A(t)$ is a linear function
with respect to $\bnu(t)=\{\nu_i(t)\}$, and $\langle W(t)
\rangle=\langle \oW(t) \rangle \equiv \int_0^t \dd t'
A(t')$. In~\cite{LS99} this has been shown for continuous time Markov
processes, but the proof is valid also in the discrete time case (see
for example~\cite{GA04}). This leads to consider $W(t)$ and $\oW(t)$
equivalent for the contribution of a single trajectory to the total
entropy flux. In a stationary state $A(t)=\Sigma(t)$ and therefore the
flux is equivalent to the production. For large enough $t$ one has:
\begin{enumerate}

\item at
 equilibrium (i.e. when there is detailed balance) $\langle W(t)
 \rangle=\langle \oW(t)\rangle=0$; 

\item out of equilibrium $\langle W(t)
 \rangle>0$ and $\langle \oW(t)\rangle>0$.
\end{enumerate}

Consider the fluctuations of the functional $W$.  We first define an
extended probability vector $\bp(t,W)$, whose component $p_i(t,W)$ is
the probability of finding the system at time $t$ in the state $i$,
with value $W$ for the Lebowitz and Spohn functional. This means
that $\sum_i p_i(t,W)=f(t,W)$ is the probability density function at
time $t$ for the functional $W$, while $\int \dd W p_i(t,W)=\nu_i(t)$
is the probability of finding the system in state $i$ at time t and
$\int \dd W \sum_i p_i(t,W)=1$.  The evolution of $\bp(t,W)$ is given
by the following equation:
\begin{equation}
p_i(t+1,W)=\sum_j K_{ji}p_j(t,W-\Delta w_{ji})
\end{equation}
where $K$ is the previously defined transition matrix and $\Delta
w_{ij}$ is the variation of $W$ due to a jump from the state $i$ to
the state $j$. This reads $\Delta w_{ij}=\ln \frac{K_{ij}}{K_{ji}}$.
Then define the function $\bpt(t,\lambda)=\int \dd W
\exp(-\lambda W) \bp(t,W)$ and obtain, for its evolution
\begin{equation}
\bpt(t+1,\lambda)=A(\lambda) \bpt(t,\lambda)
\end{equation}
$A(\lambda)$ being the matrix defined by
\begin{equation}
A_{ij}(\lambda)=K_{ji}^{1-\lambda}K_{ij}^\lambda,
\end{equation}
and, therefore,
\begin{equation}
\bpt(t,\lambda)=A(\lambda)^t\bpt(0,\lambda)
\end{equation}
with $\pt_i(0,\lambda)=\int \dd W \exp(-\lambda W) \nu_i(0)\delta(W)=\nu_i(0)$.

The characteristic function of the distribution of $W$ is obtained
summing over all the states:
\begin{equation}
\ftilde(t,\lambda)=\sum_i \pt_i(t,\lambda)=\sum_i 
\sum_j [A(\lambda)^t]_{ij}\nu_j(0)=\sum_j \nu_j(0) \sum_i [A(\lambda)^t]_{ij}.
\end{equation}
The distribution of $W$ at time $t$ is recovered by inverting the
Laplace transform:
\begin{equation}
f(t,W)=\int_{-i \infty}^{+i \infty} \dd\lambda \exp(\lambda W) \ftilde(t,\lambda).
\end{equation}
Furthermore, the knowledge of the characteristic function suffices to compute the 
moments or the cumulants of $W$: 
\begin{subequations} \label{cumulants}
\begin{align} 
\langle W_t^n \rangle&=(-1)^n\left . \frac{d^n}{d\lambda^n}
\ftilde(t,\lambda) \right\lvert_{\lambda=0}\\
\langle W_t^n \rangle_c&=(-1)^n\left .  \frac{d^n}{d\lambda^n}
\log \ftilde(t,\lambda) \right\lvert_{\lambda=0}.
\end{align}
\end{subequations}

At large times, the evolution operator $A(\lambda)^t$ is dominated by
the largest eigenvalue $y_1(\lambda)$ of $A(\lambda)$. Defining
$y_1(\lambda)=\exp(\zeta(\lambda))$ it follows that
\begin{subequations} \label{largetimes}
\begin{align}
\bpt(t,\lambda) &\sim \exp(\zeta(\lambda) t) \left[\sum_j x^{(1)}_j(\lambda) \nu_j(0)\right]
\mathbf{x}^{(1)}(\lambda)\\
\tilde{f}(t,\lambda)=\sum_i \pt_i(t,\lambda) &\sim \exp(\zeta(\lambda)t) \left[\sum_j x_j^{(1)}(\lambda) \nu_j(0)\right] \sum_i x^{(1)}_i(\lambda) ,
\end{align}
\end{subequations}
where $\mathbf{x}^{(1)}(\lambda)$ is the eigenvector of $A(\lambda)$
associated to the largest eigenvalue $y_1(\lambda)$.  One expects,
from the above large times behavior, an analogous large time behavior
for the density distribution of $W$, i.e.
\begin{multline} \label{saddle}
f(t,W)=\int_{-i \infty}^{+i \infty} \dd\lambda \exp(\lambda W) \sum_j \nu_j(0) \sum_i [A(\lambda)^t]_{ij} \\
\sim \exp\left[t \max_\lambda\left(\lambda \frac{W}{t}+\zeta(\lambda)\right)\right]=\exp[t \pi(W/t)]
\end{multline}
where we have introduced $\pi(w)=\lim_{t \to \infty} \frac{1}{t}\log
f(t,wt)$, the large deviation function associated with $f(t,W)$, which
is obtained as
a Legendre transform of $\zeta(\lambda)$, i.e. $\pi(w)=\lambda^*
w+\zeta(\lambda^*)$ with
$\frac{d}{d\lambda}\zeta(\lambda)\vert_{\lambda=\lambda^*}=-w$, under the validity of the last chain of equalities.

The previously defined Fluctuation Relation appears at this stage. In
fact, it is evident that $A(\lambda)=A^T(1-\lambda)$, being $A^T$ the
transpose of $A$. This implies that $\zeta(\lambda)=\zeta(1-\lambda)$
which suffices to get $\pi(w)=\pi(-w)+w$, i.e. $f(t,W)=f(t,-W)\exp(W)$
at large times. The validity of the expansion in~\eqref{saddle} is
crucial for the validity of this relation, but it is not guaranteed
when the integrand presents non-analyticities in the $\lambda$ complex
plane. Such a catastrophe can happen, for example, when the number of
states $N$ becomes infinite and the initial (final) probability
$\mu_{j(i)}(0)$ has some unbounded form (see below). The physical
meaning of such a catastrophe is that the large fluctuations in the
initial and final state cannot be neglected, because they contribute
to the tails (i.e. the large deviations) of $f(t,W)$ at all times. Note
that the breaking of analyticity of $\ftilde(t,\lambda)$ is associated
with large fluctuations of invariant probabilities. Many studies,
including numerical simulations~\cite{PVTW06}, analytical
calculations~\cite{F02,F04,V06} and heuristic arguments~\cite{ZC03}, show that the initial and final configurations
modify the standard picture when the distribution of the boundary term
$B=\ln \mu_{\sigma(0)} -\ln \mu_{\sigma(t)}$ has exponential, or
higher, tails in the stationary state. This may occur {\em even if
$\mu_i$ has Gaussian tails}.

Let us see what happens to the functional $\oW$ which differs from $W$
only for the non time-extensive term $B$, letting the prime denote
quantities which involve this functional.  For example, we denote
the extended measure vector by $\obp(t,\oW)$ and the probability density
function at time $t$ by $\of(t,\oW)$, the characteristic function
by $\oft(t,\lambda)$, etc.  The increment of the functional at each jump
reads $\Delta w'_{ij}=\Delta
w_{ij}\log\left(\frac{\mu_i}{\mu_j}\right)$, recalling that $\mu_i$ is
the invariant measure for the state $i$. The evolution matrix
therefore reads
\begin{equation}
\oA_{ij}(\lambda)=K_{ji}^{1-\lambda}K_{ij}^\lambda \mu_j^{-\lambda} \mu_i^\lambda,
\end{equation}
and the evolution equation reads
\begin{equation}
\obpt(t+1,\lambda)=\oA(\lambda) \obpt(t,\lambda).
\end{equation}
The fundamental difference between $\oW$ and the original $W$ appears
now. Since $\obpt(0,\lambda)=\bnu(0)$, if one takes
$\bnu(0)\equiv\bmu$ (the system is in the stationary regime from the beginning),
it happens that
\begin{equation}
\oft(1,\lambda)=\sum_i \opt_i(1,\lambda)=\sum_i \sum_j \oA_{ij}(\lambda) \mu_j=\sum_i \sum_j K_{ji}^{1-\lambda}K_{ij}^\lambda \mu_j^{1-\lambda} \mu_i^\lambda,
\end{equation}
i.e. $\oft(1,\lambda)=\oft(1,1-\lambda)$. By recursion, one realizes
that this is the case for all times $t$, i.e.\
\begin{equation}
\oft(t,\lambda)=\oft(t,1-\lambda),
\end{equation}
in general, which leads immediately to a finite-time symmetry relation
\begin{equation}
\of(t,\oW)=\of(t,-\oW)\exp(\oW),
\end{equation}
valid for any $t$.

The functional $\oW$ contains a term that cancels the effects of the
fluctuations of the steady state measure, leading to a conservation of
the symmetry $\lambda \to 1-\lambda$ all along the evolution. Such a
conservation prevents  surprises at large times also in presence of
large fluctuations of the initial measure.

As an example of the above discussion~\cite{PRV06}, consider a Markov chain with
$N+2$ states, labeled $A$, $B$ and $C_i$ with $i \in \{1, 2,
... N\}$ and with a transition matrix $K$ defined by
\begin{equation}
\begin{pmatrix}
p_{AA}&   p_{AB}&  (1-p_{AA}-p_{AB})k_1& (1-p_{AA}-p_{AB})k_2&
\dotsb& (1-p_{AA}-p_{AB})k_N\\
p_{BA}&    0&     (1-p_{BA})k_1& (1-p_{BA})k_2&   \dotsb& (1-p_{BA})k_N\\
p_{CA}&       1-p_{CA}&     0& 0&              \dotsb& 0\\
p_{CA}&       1-p_{CA}&     0& 0&              \dotsb& 0\\
\vdots& \vdots&  \\
\end{pmatrix}
\end{equation}
with $\sum_{i=1}^N k_i=1$, under the constraint that, for every 
jump of positive transition probability, the reversed jump is 
also possible. The invariant probability $\boldsymbol{\mu}$ for 
$N=1$ is given by
\begin{subequations}
\begin{align}
\mu_A&=\frac{p_{BA}+p_{CA}-p_{BA}p_{CA}}{\mathcal{N}}\\
\mu_B&=\frac{1-p_{AA}+p_{CA}(p_{AB}+p_{AA}-1)}{\mathcal{N}}\\
\mu_C&=\frac{1-p_{AA}-p_{AB}p_{BA}}{\mathcal{N}}
\end{align}
\end{subequations}
with $\mathcal{N}=2+p_{AA}(-2+p_{CA})+p_{AB}p_{CA}-p_{BA}(-1 + p_{AB} + p_{CA})$.

It can be easily seen that the invariant measure for $N>1$ is closely related to
the case $N=1$ with a decomposition of the measure of state
$C_1$ into the measures of $C_i$ proportional to the values $k_i$:
\begin{equation}
\mu_{C_i}=\mu_C k_i
\end{equation}
In numerical simulations of this model~\cite{PRV06}, the functionals $W(\tau)$ and
$\oW(\tau)$ can be measured along independent non-overlapping segments
of duration $\tau$, extracted from a unique trajectory after the
stationary regime has been achieved. Then, the following relation can be verified:
\begin{equation}
  G_\tau(X)=\log F(\tau,X)-\log F(\tau,-X)=X \label{as_es}
\end{equation}
where $F(t,X)$ is the probability density function of finding one of
the two functionals $W$ or $\oW$ after a time $t$ equal to $X$
(i.e. $F(t,X)$ corresponds to $f(t,W)$ or to $f'(t,\oW)$ depending on
the cases). Some typical results are shown in
figures~\ref{fig:map_pdf}, which show the graph of $G_\tau$ vs $x$ and
the pdfs of $W$ and $\oW$. The phenomenology can be divided in $3$
main cases, depending on the choice of the transition rates. Whenever
detailed balance is satisfied (here when $A$ is disconnected from all
$C_i$'s, i.e. when $p_{AB}=1-p_{AA}$ and $p_{CA}=0$),
$\oW(\tau)$ is identically zero and does not fluctuate (its pdf is a
delta in zero), as expected, and therefore $W(\tau)$ coincides with the opposite of
the boundary term $B=\log\frac{\mu_{\sigma_1}}{\mu_{\sigma_\tau}}$:
they both have symmetric fluctuations around zero with exponential
tails, and $G_\tau \equiv 0$. When detailed balance is violated
(i.e. when $A$ is connected to all $C_i$'s) both $W(\tau)$ and
$\oW(\tau)$ fluctuate around a nonzero (positive) value which, for
$\tau$ large enough, is the same for the two functions. In general the
fluctuations of the boundary term
$B=\log\frac{\mu_{\sigma_1}}{\mu_{\sigma_\tau}}$ have a pdf with
exponential tails and the pdf of $\oW$ is almost perfectly
Gaussian. This is not true for the pdf of $W$.  At small times
($\tau=100$) the pdf of $W$ resembles the pdf of $B$, evidencing that
at this time $W$ is completely dominated by $B$. At larger times the
pdf of $W$ and $B$ start to deviate, in particular that of $W$ tends
to that of $\oW$ in the bulk, with evidently different tails
of exponential form. The tails of the pdf of $W$ are always
dominated by the fluctuations of $B$. The symmetry relation for the
functional $W$ is not verified in any of these simulations, both at
small and large times. The asymmetry $G_\tau(W)$ has a slope
near $1$ only for small values of $W \ll \langle W \rangle$, then
deviates and saturates to a constant value in good agreement with the
value $2\langle W \rangle$ predicted by van Zon and
Cohen~\cite{ZC03}. Note that the observations at large times are
perfectly compatible with those at small times, i.e. if both ordinates
and abscissas are divided by $\tau$, the curves are similar (in the
first two cases they overlap very well). Therefore the ``reduction''
of the violation at large times is only apparent. On the other hand
the symmetry relation for $\oW$ is {\em always} satisfied, at all
times and for all the choices of the parameters. Finally, when the
choice of the rates $k_i$ is such that the invariant measure on states
$C_i$ is an exponential with a very high slope, the
numerical results show that the fluctuations of $W$ are much closer to
those of $\oW$ already at small times. This is reflected on the good
agreement with the FR of both functionals $W$ and
$\oW$. One can still be doubtful about this verification,
because of the limited range of values of $W$ available, which can
possibly hide a failure at larger values. The distribution of the
boundary term, in fact, still has exponential tails (invisible at our
resolution) and  it can be argued that these tails (being
Gaussian those of $W$) will dominate at very large values.

\begin{figure}[htbp]
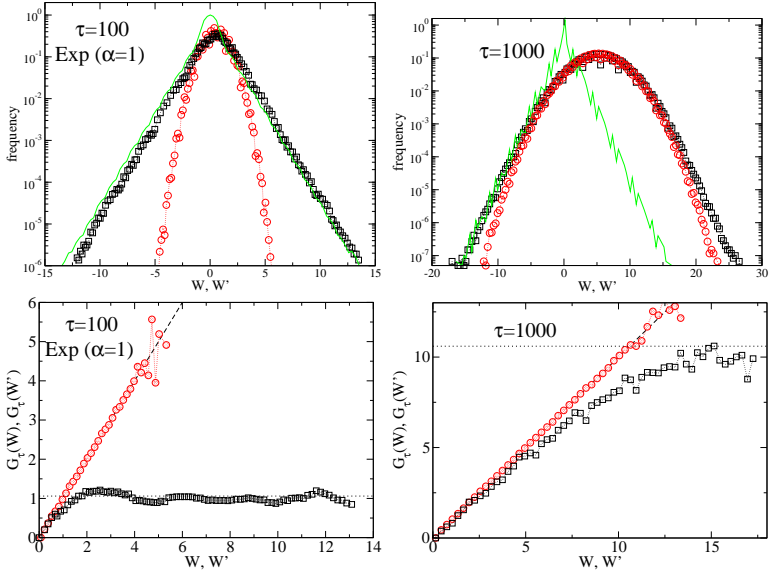

  \includegraphics[width=5cm,clip=true]{pdf_exp1_t100.eps}
  \includegraphics[width=5cm,clip=true]{pdf_exp1_t1000.eps}\\
  \includegraphics[width=5cm,clip=true]{plot_exp1_t100.eps}
  \includegraphics[width=5cm,clip=true]{plot_exp1_t1000.eps}
\caption{Study of the boundary terms for the Fluctuation Relations in
the Markov chain discussed in the text.  Top graphs: probabilities of
observing $W$ (black squares), $\oW$ (red circles) and $B=\oW-W$
(green lines). Bottom graphs: $G_\tau(W)$ vs. $W$ (black squares) and
$G_\tau(\oW)$ vs. $\oW$ (red circles). The dashed line has slope
$1$. The dotted horizontal line marks the van Zon and Cohen prediction
for the FR violation, $G_\tau(W)=2\langle W \rangle$ for large
$W$~\cite{ZC03}.  The left column is at time $\tau=100$ and the
right column at time $\tau=1000$. In the simulation we have used
$N=50$, $p_{AA}=0.2$, $p_{AB}=0.3$, $p_{BA}=0.3$ and $p_{CA}=0.5$ and
$k_i \propto exp(-\alpha i)$ with the value of $\alpha$ given in the
plot. In all cases $\langle W(\tau) \rangle=\langle \oW(\tau)
\rangle=0.0053\tau$, i.e. $0.53$ for $\tau=100$ and $5.3$ for
$\tau=1000$.  \label{fig:map_pdf}}.
\end{figure}

\subsection{Numerical works}
An exhaustive account of the numerical works devoted to the study of FRs
is practically impossible here, because of space limitations. Therefore, we briefly 
mention a selected group of works, and we concentrate only on two of them.

The steady state FR for nonequilibrium systems was introduced and numerically verified 
by Evans {\it et al.}~\cite{ECM}, in a model in which $\Lambda=\Omega$. 
After that, some numerical tests have been devoted to the $\zL$-FR, in 
order to understand the extent to which the 
Chaotic Hypothesis and the $\Lambda$-FR describe physical systems, see e.g.\ 
\cite{BGG,GRS,LRB,BR01,SE2000,RM03,romans,DK,GZG,TG,stephennew}. The result is that the 
$\Lambda$-FR is easily verified when $\Lambda=\zW$, while difficulties emerge 
when this is not the case. The largest group of tests concerns quantities 
related in various fashions to the entropy production or dissipated energy
rates, both in transient and steady states, see e.g.\ 
Refs.\cite{BGG,LLP1,RS,GRS,LRB,BR01,earlierpapers,generalized,SE2000,HeatFlow,chatelier}.
It has been found that FRs for these physical observables are ubiquitously verified, in
models of nonequilibrium fluids.

\subsubsection{The nonequilibrium FPU chain}
Lepri {\it et al.} \cite{FPU} modified the Fermi-Pasta-Ulam model, imposing 
nonequilibrium boundary conditions by means of two Nos\'e-Hoover thermostats at 
different temperatures \cite{LLP1}. This work constitutes a numerical investigation 
of a steady state FR, meant to test the physical applicability of 
the relation in a model substantially different from the original one, i.e.\ in a 
solid rather than in a fluid. The test of \cite{LLP1} concerned the fluctuations 
of the heat flux, which verified the FR, supporting the idea 
that such relations do have a quite wide range of applicability.

The model studied in Ref.\cite{LLP1} consists of $N$ anharmonic oscillators 
of mass $m$ and positions $x_j$, $j=1,..., N$, whose first and last oscillators
($j=1,N$) interact with the left ($\ell$) and the right ($r$) thermostats,
respectively, and with still walls, while all oscillators interact with their
nearest neighbours through the potential:
\be
V(q)=\frac{q^2}{2} + \beta\frac{q^4}{4} ~,
\label{potential}
\ee
where $q_j=x_j-ja$ is the displacement of the $j$-th oscillator from its equilibrium 
position, and $a$ is the equilibrium distance between two nearest neighbours. The 
equations of motion take the form
\bea
&&m\frac{d^2}{d t^2}{q}_1 = F(q_1 - q_0) - F(q_2-q_1) - \zeta_\ell \frac{d{q}_1}{dt} ~,
\nonumber \\
&&m\frac{d^2}{d t^2}{q}_i = F(q_i - q_{i-1}) - F(q_{i+1}-q_i) ~, \quad
\mbox{for } i = 2,..., N-1 \label{eqsmot} \\
&&m\frac{d^2}{d t^2}{q}_N = F(q_N - q_{N-1}) - F(q_{N+1}-q_N) - \zeta_r \frac{d{q}_N}{dt}, \nonumber
\eea
where $q_0 = q_{N+1} = 0$, $F = - d V/ d z$, 
\be
\frac{d}{d t}{\zeta}_r = \frac{1}{\theta_r^2} \left( \frac{\dot{q}_N^2}{T_r} -1 \right)
~, \quad \quad
\frac{d}{d t}{\zeta}_\ell = \frac{1}{\theta_\ell^2} \left( \frac{\dot{q}_1^2}{\ts} -1
\right) ~,
\label{NHthermo}
\ee
and $\theta_{r},\theta_{l}$ are relaxation times which determine the efficiency of the
thermostats.
In the case that $T_r \ne T_\ell$, Eqs.(\ref{eqsmot},\ref{NHthermo}) define the 
nonequilibrium FPU-$\zb$ model, which is time reversal invariant, with time reversal 
operation defined by $I(q,p,\zeta) = (q,-p,-\zeta)$. The system is also dissipative, and
the time average of the phase space contraction rate,
$(\overline{\zeta}_r + \overline{\zeta}_\ell)$, is positive.

The authors argue that the heat flux $J$ can be identified, in this case,
with the flux of potential energy. Then, they
find that the average fluxes at the left, $J_L$, and right, $J_R$, ends 
of the chain are given by
\begin{equation} \label{eq:llp-1}
J_{L,R} = -\langle\zeta_{L,R}\rangle T_{L,R} \ ,
\end{equation}
where  $\langle\zeta_{L,R}\rangle$  is the  mean  value  of the  corresponding
effective momentum  of the  thermostat.  By subtracting  from one  another the
fluxes in  Eq.~(\ref{eq:llp-1}) and using the fact $J_L=-J_R=J$, on average,
in the steady state, one obtains
\begin{equation} \label{eq:llp-2}
\langle \zL \rangle = \langle\zeta_L\rangle + \langle\zeta_R\rangle = 
\langle J \rangle \left(\frac{1}{T_R} -
\frac{1}{T_L}\right)  \,\,.
\end{equation}
This relation is analogous to the one for the global entropy production 
as obtained from linear response, but while it holds for the averages, 
the instantaneous values and the fluctuations of $\zL$ are not simply 
proportional to those of $J$. In Ref.\cite{LLP1}, the steady state FR for $J$ 
is tested and found to hold, even though the model does not enjoy normal energy 
transport, in the large N limit. This is not a contradiction in itself,
although the validity of the FR, close to equilibrium, implies the validity of 
the Green-Kubo relations (\ref{GKES}). The reason is that the nonequilibrium 
FPU chain at finite $N$, enjoys normal transport. The anomalous behaviour is 
reflected only in the divergence of the heat conductivity with growing $N$.

\subsubsection{Granular materials}

Some difficulties are encountered in experimental or numerical studies
of the FR. An example from heated granular gases is discussed
here. The theoretical analysis correctly predicts the asymptotic
(large deviations, $t \to \infty$) behavior of the fluctuations of the
work done on the gas, but measurements done at finite time lead to
contrasting conclusions~\cite{VPBTW05,VPBTW06}.

As already discussed in Sect.~\ref{subsec:granular} and
detailed in the Appendix~\ref{app:granular}, a standard way of
modeling a granular gas is to consider a set of $N$ hard-spheres
undergoing inelastic collisions, in which a fraction $(1-r^2)$ of the
relative kinetic energy is dissipated. The restitution
coefficient $r$ lies between $0$ and $1$, where $1$ corresponds to elastic collisions. The
inelastic collision rule~\eqref{collision} breaks microscopic
reversibility, i.e. for any given trajectory in phase space, the
time-reversed one is not always physically realizable. One can write an
energy balance equation for the total kinetic energy
$E(t)=\sum_i\frac{1}{2}\bv_i^2$, which varies according to
\begin{equation} 
\label{energybalance}
\Delta E=E(t)-E(0)=W(t)-D(t)
\end{equation}
where $W(t)$ is the energy injected by an external driving mechanism,
while $D(t)\geq 0$ is the energy irreversibly dissipated by the
inelastic collisions. The average variation rate of $D$ can be estimated as
the collision rate times the energy dissipated through a collision,
\begin{equation}
  \frac{\dd}{\dd t}\langle D\rangle=-\frac{1-r^2}{4\ell}\langle|\bv_{12}\cdot\bsigma|^3\rangle
\end{equation}
where $\ell$ is the mean free path, $\bsigma$ is the unit vector
joining the centers of the colliding particles and $\bv_{12}$ is their
relative velocity. The mean kinetic energy per particle provides a
typical energy scale, also termed {\it granular temperature}, and it
is defined as
\begin{equation}
  T_g=\langle\bv_i^2\rangle/d=\beta_g^{-1} \,\,,
\end{equation}
where $d$ is the spatial dimension.
A typical heating mechanism is provided by Eq.~\eqref{gen_drive}. We
discuss here the monodisperse case which achieves a statistically
stationary state, thanks to the injection of energy by means of independent
random forces acting on each individual particle.
This heating mechanism is easier to be handled mathematically and
leads to a uniform stationary state. The work provided by the external
source reads
\begin{equation}
W(t)=\sum_i\int_0^t\dd\tau \;{\pmb \eta}_i(\tau)\cdot {\bv}_i(\tau)
\end{equation}
Given that $\langle W\rangle/t=2d\Gamma $ (where $\Gamma$ is a
parameter of the model, see Appendix A), the typical energy scale is
set by Eq.~\eqref{tgeq}. Following a procedure similar to that of
Section~\ref{markovchain}, we compute the large deviation
function for the injected work $W$.

We begin introducing a phase space density $\rho(\bP,W,t)$ that counts
the number of systems in state $\bP$ (the $N$-velocities phase space
point) which, in the time window $[0,t]$, accumulated a total work
$W(t)=W$. A generalized Liouville equation can be written for $\rho$,
in which the Liouville operator ${\mathcal L}_W$ can be split into a
$W$ conserving part, ${\mathcal L}_\text{coll}$, and a part accounting for
changes in $W$ under the effect of the external injection mechanism
${\mathcal L}_\text{inj}$:
\begin{equation}
  \partial_t\rho={\mathcal L}_{W}(\bP,W) \rho={\mathcal
    L}_\text{inj}(\bP,W)\rho+{\mathcal L}_\text{coll}(\bP) \rho
\end{equation}
It is convenient to introduce the Laplace transform of $\rho$, 
\begin{equation}\hat{\rho}(\bP,\lambda,t)=\int\dd W\text{e}^{-\lambda
  W}\rho(\bP,W,t),
\end{equation} 
and rewrite the Liouville equation in terms of
$\hat{\rho}$,
\begin{equation}
  \partial_t\hat{\rho}={\mathcal{L}_{W}}(\bP,
  \lambda)\hat{\rho}={\hcL}_\text{inj}(\bP,
  \lambda)\hat{\rho}+{\hcL}_\text{coll}(\bP) \hat{\rho}
\end{equation}
The largest eigenvalue $\mu(\lambda)$ of ${\mathcal L}_W(\lambda)$ 
governs the asymptotic behavior of $\hat{\rho}$,
\begin{equation}\label{notLiou}
  \hat{\rho}(\bP,\lambda,t)\simeq C(\lambda) \text{e}^{\mu(\lambda)t}\tilde{\rho}(\bP,\lambda)
\end{equation}
where $\tilde{\rho}(\bP,\lambda)$ is the (right) eigenvector of
${\hcL_{W}}(\bP,\lambda)$, associated with $\mu(\lambda)$, and
$C(\lambda)$ is the projection of the initial state on this
eigenvector. It is then possible to project on the single-particle
distribution $f^{(1)}(\bv,\lambda,t)=\int d\bP_{N-1}
\hat{\rho}(\bP,\lambda,t)$. We arrive at
\begin{equation}\label{lambdaBoltzmann}
   \partial_t f^{(1)}(\bv,\lambda,t)=\Gamma\Delta_{\bv}f^{(1)}+2\lambda \Gamma
    \bv\cdot \bv f^{(1)}+\Gamma(\lambda^2 v^2-d\lambda)f^{(1)}+ \text{Coll.}
\end{equation}
where in the rhs of (\ref{lambdaBoltzmann})  ``Coll'' is a shorthand
for the collision operator which, resorting to the
molecular chaos hypothesis, reads
\begin{equation}
  \text{ Coll.}=\frac{1}{\ell}\int_{\bv_{12} \cdot \bsigma >0}
  \dd \bv_2\,\dd\bsigma \,(\bv_{12}\cdot \bsigma)\,\left(\frac{1}{r^2}
    f^{(1)}(\bv_1^{*},\lambda)f^{(1)}(\bv_2^{*},\lambda)-
    f^{(1)}(\bv_1,\lambda)f^{(1)}(\bv_2,\lambda)\right),
\end{equation}
where $\bv_1^{*}$ and $\bv_2^{*}$ are the pre-collisional velocities,
defined in the Appendix~\ref{app:granular}.  The steady velocity
pdf equation (Boltzmann equation) is recovered at $\lambda=0$.  The
process encoded in~\eqref{lambdaBoltzmann} can be read off as the
original granular gas dynamics in which additional particles are
created when $\lambda>0$ or annihilated when $\lambda<0$, at a
velocity dependent rate $\lambda v^2$.  The largest eigenvalue of ${\mathcal
L}_W(\lambda)$, $\mu(\lambda)$ is then interpreted as the population
growth rate. This remark was numerically exploited, for somewhat
different systems in~\cite{GKP06}. The splitting of $f^{(2)}$ as a
product of independent one particle distributions is indeed a
kind of molecular chaos hypothesis, for the system with particle
non-conserving fictitious dynamics. The Boltzmann equation toolbox
offers many ways to deduce an expression for $\mu(\lambda)$. The
simplest approximation is to project $\tilde{f}^{(1)}(\bv,\lambda)$
onto a Gaussian with a $\lambda$-dependent variance, denoted as
$T(\lambda)$, as explained in~\cite{VPBTW06}.

\begin{figure}[t] 
\begin{minipage}[t]{.7\linewidth}
\includegraphics[clip=true,width=1 \textwidth]{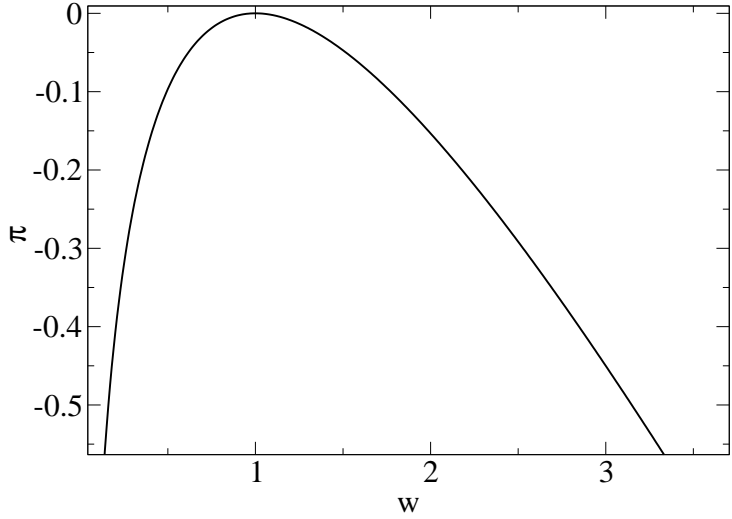} 
\caption{Plot of the Cramer function for the injected power $\pi(w)=\lim_{t \to \infty} \frac{1}{t}\log
f(t,wt)$, where $f(t,W)$ is the pdf of $W$ at time $t$.}
\end{minipage} 
\end{figure}

Working in terms of rescaled quantities, with the
granular temperature precisely given by
\begin{equation} \label{tgeq} T_g=\left( \frac{2 d \ell \Gamma \sqrt{\pi}}{(1-r^2)
\Omega_d} \right)^2\,\,,  \end{equation} with
$\Omega_d=2\pi^{d/2}/\Gamma(d/2)$, and performing the following
substitutions
\begin{equation}\label{scales}
  \mu \to \mu\frac{T_g}{d\Gamma N}\,\,, \qquad \qquad
\lambda \to \lambda T_g 
\end{equation}
one has:
\begin{equation}
\mu(\lambda)=-\lambda+\frac{1}{2}\frac{T(\lambda)}{T_g}\lambda^2
\end{equation} 
for the dimensionless quantities, where $\sqrt{T(\lambda)}$ sets the typical velocity scale
for trajectories characterized by $\lambda$. At large values of $\lambda$,
corresponding to values of $W$ small with respect to $\langle
W\rangle$, we expect that $T(\lambda)\ll T_g$ and indeed 
$\frac{T(\lambda)}{T_g}\simeq \frac{2}{\lambda}$ as
$\lambda\to+\infty$~\cite{VPBTW05,VPBTW06}. 
This can be further refined to
obtain the behavior of $\mu(\lambda)$ at $\lambda\to+\infty$,
\begin{equation} 
  \mu(\lambda) \simeq -\lambda^{1/4} \,\,. 
\end{equation} 
Besides, $\mu(\lambda)$ possesses a cut in the $\lambda$-plane at
$\lambda_c=-3/2^{8/3}$, such that, as $\lambda\to\lambda_c^+$,
\begin{equation} 
  \mu(\lambda)=3/
  2^{2/3}-3^{3/2}2^{1/6}\sqrt{\lambda-\lambda_c}+{\mathcal
    O}(\lambda-\lambda_c) 
\end{equation} 
The presence of this cut is responsible for the exponential decay of
the pdf of $W$ at large values of $W$. However, the non-analytic
behavior at $\lambda\to+\infty$ leads to a non-analytic behavior as
$w=W/t\to 0^+$,
\begin{equation}
  \pi(w\to 0^+)\sim-w^{-1/3}\,\,,\qquad 
\pi(w\to\langle w\rangle) \simeq -(w-1)^2/2\,\,,\qquad
\pi(w\to\infty)\sim -w\,\,.
\end{equation}

\begin{figure}[t] 
\begin{minipage}[t]{.46\linewidth}
\includegraphics[clip=true,width=1 \textwidth]{gc_tau.eps}
\caption{\label{fig:gc} Plot of $\pi(w,t)-\pi(-w,t)$ for $t=1,2$ and
  $3$ mean free times (mft). The inset shows the probability density
  function of $w(t)$ for the same times.} 
\end{minipage}
\hfill 
\begin{minipage}[t]{.46\linewidth}
\includegraphics[clip=true,width=1 \textwidth]{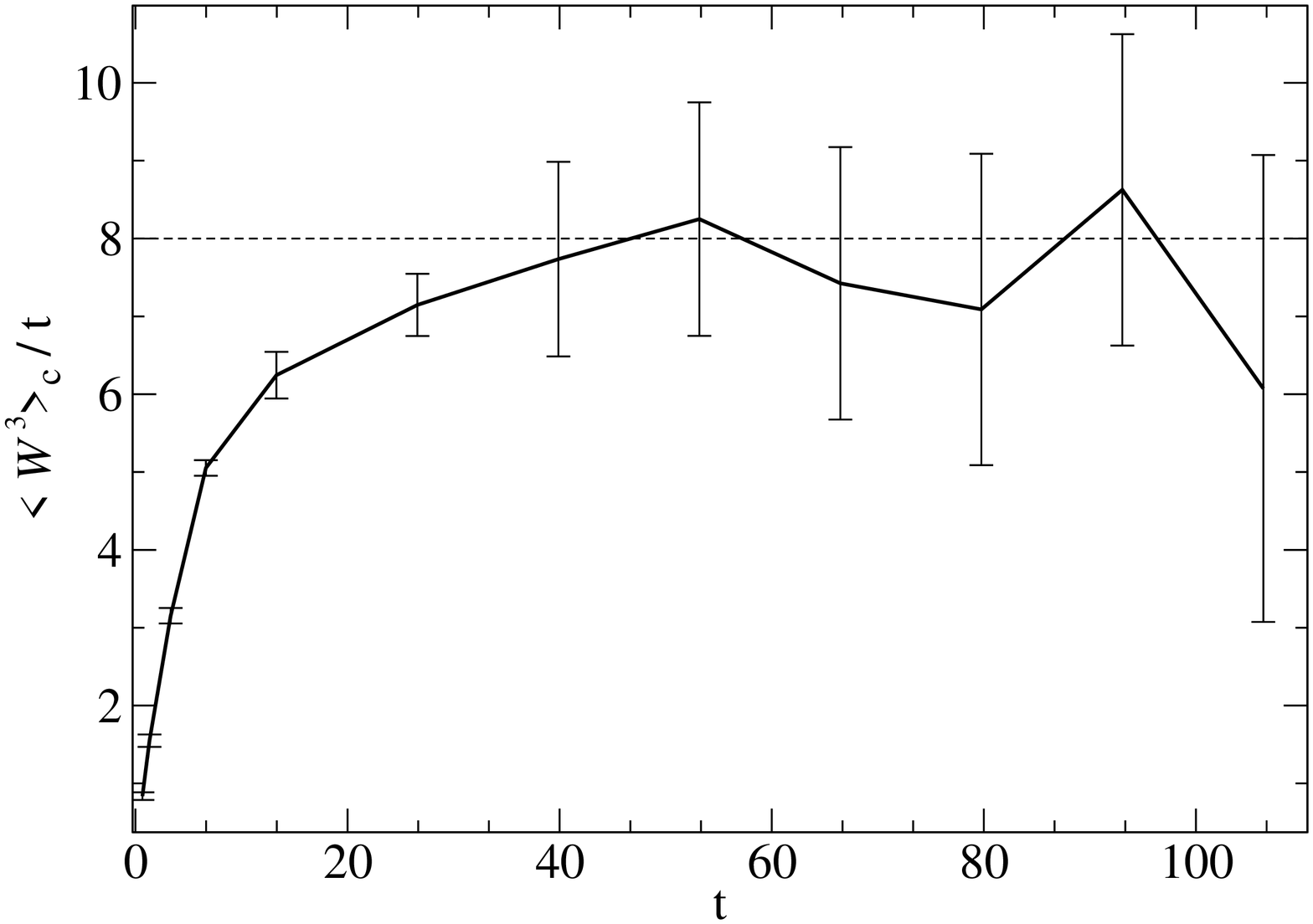} 
\caption{\label{fig:cumulant}Plot
of the third cumulant of $W$ divided by time. The asymptotic regime is reached after fifty mft.} 
\end{minipage} 
\end{figure}

Simulations of the granular gas heated with the stochastic
thermostat~\cite{VPBTW05,VPBTW06} confirm the above results and thus
the validity of these approximations. The total work $W(t)$ provided
by the random forces over $[0,t]$ is of course, on average, a positive
quantity, but at finite times there exist phase space trajectories
which yield a negative $W$.  However, one of the key hypotheses
underlying the Fluctuation Relation is that the dynamics features at
least a weak form of time reversibility, for which trajectories (or
ensembles of trajectories) possess time reversed counterparts (however
unlikely). In our granular gas, dissipative collisions prevent time
reversed trajectories to be physically acceptable trajectories at
all. It should therefore be no surprise that no specific fluctuation
relation holds in our case. This is confirmed by the explicit
calculations presented above, that show no specific
symmetry property of the injected power large deviation function.  In
spite of this, it may be instructive to plot $w$ vs. $\frac{1}{t}\ln
\frac{P(w t,t)}{P(-wt,t)}$ at possibly large times, since this is
exactly $\pi(w)-\pi(-w)$, being $\pi(w,t)=\frac{1}{t}\ln P(w t,t)$, in
the infinite time limit.  This is shown in figure \ref{fig:gc}: one
actually observes a straight line with slope $\beta_g$, while the analytic results suggest the absence of
such a straight line, even for values of $w$ not too far from its
average. In the first place, a numerical simulation is always carried
out for finite times, and thus one measures $\pi(w,t)$ rather than its
$t\to\infty$ limit, $\pi(w)$. Though $\pi(w,t)$ deviates from a
quadratic form, which would correspond to Gaussian $P(W,t)$, this is
no proof that the asymptotic regime has been reached. For that matter
it is instructive to investigate the behavior of the third cumulant
$\langle W^3\rangle/t$ as a function of time, presented in figure
\ref{fig:cumulant}. The conclusion is that the third cumulant reaches
its asymptotic value, consistent with the analytic expression, at
times a few tens as large as those for which negative values of
$w=W/t$ can be measured, as shown in figure \ref{fig:gc}. A similar
plot of the fourth cumulant would signal that the asymptotic regime
has not been reached over the chosen time window. What one actually
observes is simply the remnant of a short time quadratic behavior for
$\pi(w)$. This is also consistent with the quadratic approximation for
$\mu(\lambda)$, which would indeed imply $\pi(w)-\pi(-w)=\beta_g w$,
after restoring the appropriate physical scales as in (\ref{scales}).

\subsection{Experimental works}
Experimental tests of FRs are usually rather delicate, because large fluctuations are not 
directly observable in macroscopic systems. Therefore, one must either rely on the validity 
of local versions of the FRs, or consider small systems, like micro-biological systems 
or nano-technological devices. Some experimental works on the FRs, 
and on similar relations are the following:
\begin{itemize}
\item
Ciliberto and Laroche's experiment \cite{ciliberto-1}, for the
temperature fluctuations in a fluid undergoing Rayleigh-B\'enard convection.
A linear law was observed, but with a different (perhaps not asymptotic) slope.
\item
Wang {\it et al.} experiments, for the 
transient and steady state fluctuations of work done on a colloidal particle, 
dragged through water by means of an optical trap, \cite{othertheory,CRWSSE}.
These experiments concerned nanoscale systems and motivated Ref.\cite{othertheory}
\item
Feitosa and Menon's experiment on a mechanically driven inelastic granular gas
in  a  fluidized  steady  state  \cite{FM04}.  They claimed a verification
of a local version of FR, but their analysis has been criticized (cf.\ Section 6.3.1).
\item
Garnier and Ciliberto's experiment on the fluctuations of the dissipated power
of an  electric dipole  \cite{exptss}. They considered the work and heat 
fluctuations, both related with the fluctuations of the power dissipated  
by the resistor, and partly verified the Van  Zon - Cohen scenario.
\item
Shang {\it et al.} experiment on the fluctuations of a local
entropy  production  in  turbulent  thermal  convection \cite{shang}. 
They considered a cylindrical cell filled with water of aspect ratio one,
and verified the steady stat FR.
\item
Ciliberto {\it et al.} experiment
on turbulent flows, verified the local version of the FR,
proposed in \cite{GNS,GRS}
\item
Douarche {\it et al.} experiment on the
steady state and transient work fluctuations of a damped harmonic oscillator,
kept out of equilibrium by an external force \cite{ciliberto-2}. The 
transient FR was confirmed for any averaging time of the work. The steady-state 
version was observed to converge to the asymptotic behaviour. 
\item
Tietz {\em et al.} experiment on
the entropy production  in a single two-level system,  a defect center
in natural  IIa-type diamond \cite{tietz06}. 
\end{itemize}
It must be noted that almost all of these experiments afford verifications of the
FRs only to a certain degree, because various parameters of the relevant theories 
are not accessible or controllable in them. Some authors even argue that a direct 
verification of the (steady state) FRs may be practically impossible \cite{ZampReview}. 
This opinion is likely too extreme, but the difficulty, so far, remains.
Therefore, the observation of any approximately linear arrangement of data, with 
any slope, is often taken as sufficient evidence for the validity of the steady 
state relation, even if obtained
for relatively short averaging times. Easier is the case of the transient relations.
Because these relations are algebraic ``identities'', which hold under almost 
no conditions, one may find odd that they are tested all, but there are at least
two reasons for doing it. The first is to verify the relevance of the models for 
which these relations have been derived, to the physical reality. Indeed, those
models have been devised and proved to be successful for certain purposes which,
in general, are not those of the FRs, as recalled in Subsection 5.2. The second reason 
is that one can still rely on a  few quantitative theoretical results, for the
interpretation of certain nonequilibrium phenomena (protein stretching experiments, 
etc.). 

Despite the
difficulty in gathering the necessary statistics, the transient relations offer
some of the few parameter-free quantitative predictions, which may be used to infer
equilibrium properties of certain systems, from nonequilibrium experiments.
The Jarzynski and Crooks relations, in particular, have been tested in a good number of
experiments including molecular and biophysical experiments. The reason is
that this equality can be used to estimate the equilibrium free energy
out of measurements of dissipated work in nonequilibrium
processes. This is particularly useful in systems for which no other
method to estimate the free energy exists. 
Douarche {\it et al.} experimentally confirmed both in
\cite{ciliberto-3}.  An experiment for the colloidal particle in a
time-dependent non-harmonic potential was reported by Blickle {\it et
al.}~\cite{blickle}. The Jarzynski equality was
first confirmed in 2002 by Liphardt {\it et al.} in measurements of
the dissipated work in the folding-unfolding process of a single molecule
of RNA \cite{liphardt}.  Collin {\it et al.} experimentally confirmed the
Crooks fluctuation relation by pulling an RNA hairpin
\cite{ritort-1}. As discussed in detail in 6.3.2, using optical
tweezers they measured the fluctuations of the dissipative work during
the unfolding and refolding of a small RNA molecule.  Their
measurements confirmed the Crooks relation near and far from
equilibrium.  They also estimated the free energy from the
nonequilibrium measurements.  The nonequilibrium steady-state equality
of Hatano and Sasa \cite{HS01} has been verified by Trepagnier {\it et
al.} \cite{trepagnier}.  A detailed discussion of these experiments can be found
in the review papers \cite{ritort} and \cite{bustamante}.

\subsubsection{Experiments with granular gases}

Recently an experiment has been performed by Menon and
Feitosa~\cite{FM04,PVBTW05} using a granular gas shaken in a container
at high frequency.  The setup consisted of a $2D$ box containing $N$
identical glass beads, vibrated at frequency $f$ and amplitude
$A$. The authors observed the kinetic energy variations $\Delta E_\tau$,
over time windows of duration $\tau$, in a central sub-region of the
system characterized by an almost homogeneous temperature and density.
They subdivided  this variation into two contributions:
\begin{equation} \label{balance}
\Delta E_\tau=W_\tau-D_\tau,
\end{equation}
where $D_\tau$ is the energy dissipated in inelastic collisions and
$W_\tau$ is the energy flux through the boundaries, due to the kinetic
energy transported by incoming and outgoing particles. The authors of
the experiment have conjectured that $W_\tau$, being a measure of
injected power in the sub-system, can be related to the entropy flow
or the entropy produced by the thermostat constituted by the rest of
the gas (which is equal to the internal entropy production in the
steady state). They have measured its pdf $f(W_\tau)$ and found that
\begin{equation} \label{gc_exp}
\ln\frac{f(W_\tau)}{f(-W_\tau)}=\beta W_\tau
\end{equation}
with $\beta\neq 1/T_g$. By lack of a reasonable explanation for the
value of $\beta$, the authors have concluded to have experimentally
verified the FR with an ``effective temperature'' $T_{eff}=1/\beta$,
suggesting its use as a possible non-equilibrium generalization of the
usual granular temperature. The same results have been found in MD
simulations of inelastic hard disks with a similar setup, see
Figure~\ref{fig:pdf}.
There is a  major objection to this reasoning: in the limit of
zero inelasticity (i.e. in the case of ideal elastic grains) the pdf
$f(W_\tau)$ becomes symmetric with respect to the average injected
power through the borders of the central region, which is $0$, so that
$\beta \to 0$ and the effective temperature would diverge instead of
coinciding with the equilibrium temperature. There is also a 
subtler problem in the interpretation of this experiment: from
Equation~\eqref{balance}, it can be argued that the large deviation
function of $f(W_\tau)$  equals the large deviation function of
$f(D_\tau)$. Recall that  $f(W_\tau) \sim
exp(\tau \pi(W_\tau/\tau))$ when $\tau \to \infty$, and that $\pi(q)$ is the large deviation
function of $f(W_\tau)$, with $q=W_\tau/\tau$. Since $D_\tau$ is {\em
always positive}, it follows that the relation~\eqref{gc_exp} cannot
hold for $\tau \to \infty$. 

\begin{figure}[htbp]
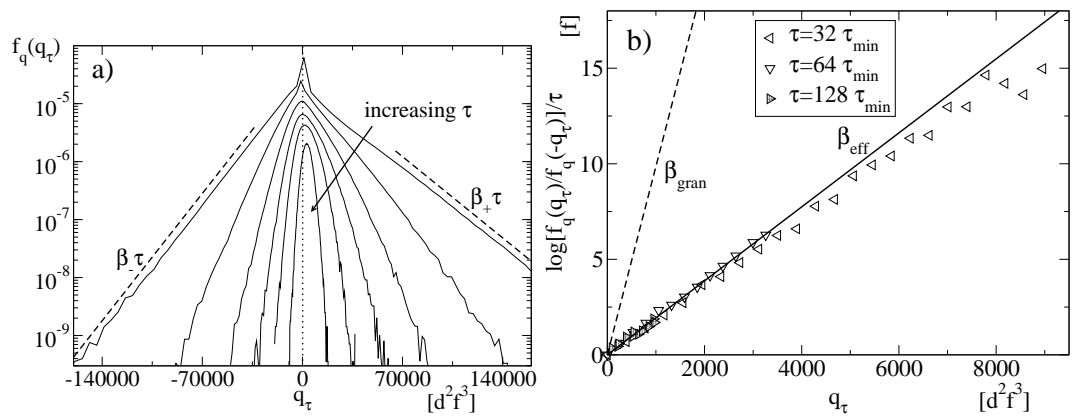

\includegraphics[width=7cm,clip=true]{pdf.eps}
\includegraphics[width=7cm,clip=true]{gc.eps}
\caption{{\bf a)} pdfs of injected power $f_q(q_\tau)$ from MD
  simulations for different values of $\tau=(1,2,4,8,16,32)\times
  \tau_{min}$ with $\tau_{min}=0.015 \tau_{box}$. Here $N=270$ and
  $\alpha=0.9$. The distributions are shifted vertically for
  clarity.  The dashed lines put in evidence the exponential tails of
  the pdf at $\tau=\tau_{min}$.  {\bf b) } plot of
  $(1/\tau)\log(f_q(q_\tau)/f_q(-q_\tau))$ vs. $q_\tau$ from MD
  simulations (same parameters as above) at large values of $\tau$. The
  solid curve is a linear fit (with slope $\beta_{eff}$) of the data
  at $\tau=128 \tau_{min}$. The dashed line has a slope
  $\beta_{gran}=1/T_{gran}$\label{fig:pdf}}
\end{figure}

A different explanation has been proposed in~\cite{PVBTW05}, which
does not involve the FR but rather imputes the observed results to
effects of finite time and poor statistics due to the limits of the
measurement procedure.  It appears that the injected power measured in
the experiment can be written as
\begin{equation} \label{qu}
W_\tau=\frac{1}{2}\left(\sum_{i=1}^{n_+} 
v_{i+}^2-\sum_{i=1}^{n_-} v_{i-}^2\right),
\end{equation}
where $n_-$ (resp. $n_+$) is the number of particles leaving
(resp. entering) the sub-region during the time $\tau$ (with frequency
$\omega$), and $v_{i-}^2$ ($v_{i+}^2$) are the squared moduli of their
velocities. In order to analyze the statistics of $W_{\tau}$ we take
$n_-$ and $n_+$ as Poisson-distributed random variables with average
$\omega \tau$, neglecting correlations among particles entering or
leaving successively the central region. As supported by the direct
observation of the simulation~\cite{PVBTW05}, the velocities ${\bf v}_{i+}$ and ${\bf
v}_{i-}$ are assumed to originate from two distinct populations with
different temperatures $T_+$ and $T_-$ respectively.  Indeed, compared
with the population entering the central region, the particles that
leave it have suffered on average more inelastic collisions, so that
$T_-<T_+$.  Finally Gaussian velocity pdfs~\footnote{while non-Gaussian tails
are quite common in granular gases, it has been
checked that they play a negligible role here} are assumed. This allows one
to calculate the generating function of $f(W_\tau)$, obtaining
$g(z)=\exp(\tau\mu(z))$ for any $\tau$, with
\begin{equation} \label{mu}
\mu(z)=\omega\left(-2+(1-T_+z)^{-D/2}+(1+T_-z)^{-D/2}\right),
\end{equation}
which also automatically coincides with the large deviation function
of $g(z)$. We recall that the two large deviation functions $\mu(z)$
and $\pi(q)$ are related by a Legendre transform. A FR with a slope
$\beta$ implies that $\mu(z)=\mu(\beta-z)$ and this relation is
not fulfilled by Eq.~\eqref{mu}. It has, however, been
observed~\cite{AFMP00} that $\pi(q)-\pi(-q)
\approx 2\pi'(0)q+o(q^3)$ at small values of $q$. This means that the
linear relation~\eqref{gc_exp} with $\beta_{eff}=2\pi'(0)$ can be
observed if the range of $W_\tau$ is small, which is usually the case,
since in experiments and simulations the statistics of negative events is very
poor. From equation~\eqref{mu} a formula for the slope $\beta$ follows: 
\begin{equation}
\beta_{eff}=2\frac{\gamma^{\delta}-1}{\gamma+\gamma^{\delta}}\frac{1}{T_-}
\label{eq:beta}
\end{equation}
with $\gamma=T_+/T_-$ and $\delta=2/(2+d)$. When $\gamma=1$ (i.e. when
the gas is elastic) $\beta_{eff}=0$. This formula is in good agreement
with the slope observed in the simulations and in the experiment.

The fact that expression~\eqref{eq:beta} is in good
quantitative agreement with the experiment is consistent with the
observation of negative events, i.e. not too large $\tau$. We can thus argue that the experiment and
the simulations have been performed over times large enough to test the
validity of formula~\eqref{eq:beta}, but not so large that
correlations may be disregarded.

\subsubsection{Applications of the Jarzynski relation to biophysical systems}

The second principle of thermodynamics implies that the  free
energy difference $F(B)-F(A)$ between two equilibrium states $A$ and
$B$ is smaller or equal than the work $W_\tau$ done by the external force over
the system during the $A \to B$ transformation taking a time $\tau$:
\begin{equation}
W_\tau \ge \Delta F=F(B)-F(A).
\end{equation}
The equal sign applies only to infinitely slow transformations,
i.e. $\tau \to \infty$. For any finite $\tau$, even averaging over
many realizations, say $M$, of the same experiment $A \to B$, the
measured $\langle W_\tau \rangle=\frac{1}{M}\sum_{i=1}^M W_{\tau,i}$ is bound by
the second principle to be larger than $\Delta F$.

The Jarzynski equality~\eqref{JR} allows to measure
$\Delta F$ by means of non-adiabatic transformations $A \to B$,
provided that the average $\langle W_\tau \rangle$ is replaced by a
weighted average $\langle W_\tau \rangle_J$, i.e. the equal sign is
always true in $\Delta F=\langle W_\tau \rangle_J$ with
\begin{equation} \label{free}
\langle W_\tau \rangle_J =-\beta^{-1} \ln \langle \exp(-\beta W_\tau) \rangle \approx -\beta^{-1} \ln \left[ \frac{1}{M}\sum_{i=1}^M \exp(-\beta W_{\tau,i})\right].
\end{equation} 

As noticed in Section 5.4.3, this argument reverses the link between
equilibrium and non-equilibrium phenomena implied by the classical
FDR. For instance, Green-Kubo formulas give a practical recipe to
calculate transport coefficients from correlation functions in the
unperturbed system, i.e. they extract non-equilibrium properties from
equilibrium ones. On the contrary, in the case of Eq.~\eqref{free}, an
equilibrium property, $\Delta F$, is obtained by means of
non-equilibrium measurements. This is
particularly useful in biology, where the reconstruction of the
free-energy landscape of (large) complex molecules, such as proteins is
crucial for determining folding pathways, the topology of the folded
state, and the biological utility of the protein. The speed of a fair
realization of reversibility along a reaction coordinate is determined
by the relaxation time of the protein, which can be very long:
therefore a naive estimate of $\langle W_\tau \rangle$ would require an
exceedingly large $\tau$. Before the Jarzynski relation, the only
known correction to the rough estimate $\langle W_\tau \rangle$
was~\cite{H91}
\begin{equation} \label{freefdr}
\Delta F \approx \Delta F_{FDR}= \langle W_\tau\rangle -\beta \sigma^2/2, 
\end{equation}
with $\sigma^2$ the variance of $W_\tau$, which is valid only near
the reversible path, i.e. still for large $\tau$. This correction can
be obtained from the FDR for Hamiltonian systems. A posteriori, from
the Jarzynski relation, Eq.~\eqref{freefdr} is easily obtained,
assuming that the distribution of $W_\tau$ is Gaussian.

The alternative provided by Eq.~\eqref{free}, suggested in~\cite{HS01b}
and experimentally tested in~\cite{liphardt}  for the first time, consists in
performing $M$ realizations of a pulling experiment on a single
bio-molecule: the external force, provided by the cantilever of an
Atomic Force Microscopy or by moving the laser trap of an optical
tweezer (this was done in the real experiment), is applied to a
terminal of the molecule, through a bead that plays the role of a
handle, guiding the transformation from a folded $A$ to an unfolded
$B$ state in a time $\tau$. Measures in numerical and real experiments
have shown that the weighted average~\eqref{free} can be effectively
used to have a better estimate of $\Delta F$ with respect to $\Delta
F_{FDR}$ Eq.~\eqref{freefdr}, for almost the whole range of
displacements between $A$ and $B$. Actually the applicability of
Eq.~\eqref{free} is not trivial, since a careful evaluation of all
possible errors is required: reducing $\tau$ allows to increase the
number of experiments $M$, but at the same time makes wider and wider
the distribution of $W_\tau$ and, therefore, more and more uncertain
the estimate of the exponential average $\langle \exp(-\beta W_\tau)
\rangle$ through a finite sum. However it has been
shown~\cite{HJ01,R07} that a range of $\tau$ exists where the estimate
is reliable. Moreover, this procedure has the advantage of providing a
direct estimate of the statistical error of the measure, which is not
possible with a small number of measurements and large $\tau$. On the
other side, when applying a similar procedure to large systems, the
difficulty arises that sampling negative $W_\tau$ requires the number of
independent experiments $M$ to grow exponentially with the system
size~\cite{LG05}.

\section{Large deviations and FR}
\label{sec:ld}

In Sections 2, 3 and 4 we discussed the properties of the mean responses,
 and their link with correlation functions.  Sections 5 and 6 have been
 devoted to a more detailed statistical description of the probability distribution of time averages on 
 certain intervals.  Now, we want to treat at a more advanced level the
 statistical description of the time history of the system, as
 described by its observables.

\subsection{Onsager-Machlup approach to fluctuations
of thermodynamic variables}

As far as we know the first attempt in this direction was made by
Onsager and Machlup, about half a century ago~\cite{OM53,MO53}.
In  two influential papers they studied the probability of a given
succession of nonequilibrium states, of a spontaneously fluctuating
thermodynamic system, under the assumption that the $N$ variables
$\{\alpha\}$ determining the state are Gaussian stochastic processes.
Basically, Onsager and Machlup assume that the evolution of
$\{\alpha\}$ is determined by a set of linear Langevin equations:
\begin{equation}
\label{l.1}
 \sum_{j=1}^N R_{i,j} { {d \alpha_j} \over {dt}}=
-\sum_{j=1}^N s_{i,j}\alpha_j + \sqrt{2c}\eta_i \,\, ,
\end{equation}
where $\{\eta\}$ is a white noise vector, i.e. a Gaussian process with
 $<\eta_k(t)>=0$ and $<\eta_k(t)\eta_j(t')>=\delta(t-t')\delta_{k,j}$.
 The symmetric matrices $\mathbf{R}=\{R_{i,j}\}$ and
 $\mathbf{s}=\{s_{i,j}\}$ determine the {\it thermodynamic forces}
 $\{X\}$ and the entropy $S(\{\alpha\})$:
\begin{equation}
\label{l.2}
X_i=
{ {\partial S} \over {\partial \alpha_i} }=
-\sum_{j=1}^N s_{i,j}\alpha_j \,\, .
\end{equation}
In this  approximation, one can express the entropy as
\begin{equation}
\label{l.3}
S(\{\alpha\})=S_0-
{1 \over 2}  \sum_{i,j}s_{i,j} \alpha_j \alpha_i \,\, ,
\end{equation}
whereas the dissipation function is
\begin{equation}
\label{l.4}
\phi=
{1 \over 2}  \sum_{i,j}R_{i,j} { {d \alpha_i} \over {dt}}
{ {d \alpha_j} \over {dt}}  \,\, ,
\end{equation}
and the corresponding entropy functional  is
\begin{equation}
\label{l.5}
\psi=
{1 \over 2}  \sum_{i,j}L_{i,j}X_i X_j  \,\, ,
\end{equation}
where the matrix $\mathbf{L}$ is defined by $\mathbf{L}{\bf
X}=d\pmb{\alpha}/dt$ i.e.  $\mathbf{L}=\mathbf{R}^{-1}$.\footnote{ As
stressed by many authors, e.g. (\cite{OM53}), a thermodynamic
description of a system requires an answer to the general question ``{\it
how do you know you have taken enough variables, for it to be
Markovian?}''  For instance a noisy $RC$
circuit is Markovian in the variables $(q,{\dot q})$ but not for $q$
(or ${\dot q}$) alone.}  Because of the Onsager reciprocal relations,
the matrices $\mathbf{L} , \mathbf{R}$ and $\mathbf{s}$ are symmetric
(see section~\ref{sec:reciprocal}), and the Boltzmann principle
implies that the invariant probability distribution is given by
\begin{equation} P_{inv}(\{\alpha\}) \propto \exp \left[ S(\{\alpha\})\right] \propto \exp \Bigl[
 - {1 \over 2} \sum_{i,j}s_{i,j} \alpha_j \alpha_i \Bigr] \,\, ,
 \label{multi} \end{equation}
 where we took $k_B=1$. The value of $c$
 in~\eqref{l.1} is determined in such a way that
 Eq.~\eqref{multi} holds.  Above  we have  used the
 original notation of the paper by Onsager and Machlup. In 
modern notation one would have
\begin{equation}
\label{new}
{d\alpha_k \over dt}=- \sum_{j=1}^N B_{j,k}\alpha_j+
\sqrt{2c}\tilde{\eta}_k
\,\,\,\,\, k=1,...,N \,\, ,
\end{equation}
where now $\{ \tilde{\eta} \}$ is a Gaussian process with
 $<\tilde{\eta}_k(t)>=0$ and
 $<\tilde{\eta}_k(t)\tilde{\eta}_j(t')>=C_{k,j} \delta(t-t')$.
The link with the original notation, is given by
$ \mathbf{B}=\mathbf{R}^{-1}\mathbf{s}$ and
$\mathbf{C}=\mathbf{R}^{-2}$.\\
Let, for the sake of simplicity, $N=1$:
\begin{equation}
\label{l.6}
{d\alpha \over dt}=- \gamma \alpha+ \sqrt{2c}\eta \,\, ,
\end{equation}
with $\gamma=s/R$.
The conditional probability density for $\alpha(t_1)=\alpha_1$,
given   $\alpha(t_0)=\alpha_0$, is
\begin{equation}
\label{l.7}
P_1(\alpha_1, t_1|\alpha_0, t_0)=
{ 1 \over {\sqrt{2 \pi C (1- a(\tau)^2)}} } \times
\exp \Bigr(-{ {(\alpha_1-a(\tau)\alpha_0)^2} \over {2C[1-a(\tau)^2]}
}\Bigl)
\,\,\ ,
\end{equation}
where $t_1=t_0+\tau$ and
$$a(\tau)=e^{-\gamma \tau} \,\,\,\,,\,\,\,\,  C={c \over \gamma}
={1  \over s} \,\,\,. $$
Using the Markovian nature of the process, it is easy to
determine  the probability density of a discrete trajectory,
$\alpha(t_1)=\alpha_1, \alpha(t_2)=\alpha_2, ...
\alpha(t_n)=\alpha_n...\alpha(t_M)=\alpha_M$, where $t_n=t_0+n\tau$
and  $\alpha(t_0)=\alpha_0$. One gets
\begin{equation}
\label{l.8}
P_M(\alpha(t_1)=\alpha_1, ..., \alpha(t_M)=\alpha_M|\alpha_0, t_0)=
\prod_{n=1}^M
P_1(\alpha_n, t_n|\alpha_{n-1}, t_{n-1}) \,\, .
\end{equation}

In the limit of small $\tau$, one obtains the rather transparent
expression:
 \begin{equation}
\label{l.9}
P_1(\alpha_n, t_n= t_{n-1}+\tau |\alpha_{n-1}, t_{n-1})=
{ 1 \over {\sqrt{ 4\pi D \tau }} } \times
\exp \left[-{\ell( \alpha_n, \alpha_{n-1}) \tau }\right]
\,\,\ ,
\end{equation}
where
\begin{equation}
\label{l.10}
\ell( \alpha_n, \alpha_{n-1})={1 \over {4D}}
{\Bigr[ { {(\alpha_n-\alpha_{n-1})} \over {\tau}} + \gamma \alpha_n \Bigr]
^2 },
\,\,\  \mbox{with} \,\,   D=C\gamma={1 \over R} \,\, ,
\end{equation}
which leads to
\begin{equation}
\label{l.11}
P_M(\alpha(t_1)=\alpha_1, ..., \alpha(t_M)=\alpha_M|\alpha_0, t_0)=
\Bigr({ 1 \over {4 \pi D \tau} }\Bigl)^{M/2} \times
\exp \left[-\sum_{n=1}^M{\ell( \alpha_n, \alpha_{n-1}) \tau }\right] \,\,.
\end{equation}

The conditional  probability distribution
$P_1(\alpha_M, t_M|\alpha_0, t_0)$ can be obtained from \eqref{l.11},
integrating over
$\alpha_{1}, \alpha_{2},...,\alpha_{M-1}$.
Performing the limit $\tau \to 0$ and $M \to \infty$
in such a way that $\tau M =constant=t-t_0$, one obtains
\begin{equation}
\label{l.12}
P_1(\alpha(t), t|\alpha(t_0), t_0)=
\int_{\alpha(t_0)}^{\alpha(t)}{\cal D}{\alpha(t')}
\exp  \left\{ -\int_{t_0}^t dt' \mathcal{L}[\alpha(t'), \dot{\alpha}(t')]\right\}
\end{equation}
where the Lagrangian function is
\begin{equation}
\label{l.13}
 \mathcal{L}[\alpha, \dot{\alpha}]={1 \over 4 D}
\Bigr( \dot{\alpha}+\gamma \alpha \Bigl)^2
\,\,\,\,\,\, ,
\end{equation}
and the functional integral is defined in the
usual way, i.e.
\begin{equation}
\label{l.14}
\int_{\alpha(t_0)}^{\alpha(t)}{\cal D}{\alpha(t')} X_t[\{\alpha(t')\}]
=\lim_{M \to \infty}
\Bigr({1 \over {4 \pi D \tau}}\Bigl)^{M/2}
\int d\alpha_1 \int,...,\int d\alpha_M
X_t[\{\alpha(t_1),...,\{\alpha(t_M)]
\end{equation}
for any functional $X_t[\{\alpha(t')\}]$.

The probability density of a path $\{ \alpha(t)\}$ is
\begin{equation}
\label{l.15}
P(\{ \alpha(t') \}, t_0<t'<t) \sim
\exp  \left\{ -\int_{t_0}^t dt' \mathcal{L}[\alpha(t'), \dot{\alpha}(t')]\right\}.
\end{equation}
Noting that
\begin{equation}
\label{l.17}
 \mathcal{L}[\alpha, \dot{\alpha}]=
{R \over 4}\Bigl(\dot \alpha + \gamma \alpha \Bigr)^2=
{1 \over 2}\Bigl( \phi  +  \psi - {dS \over dt} \Bigr)
\end{equation}
one can recast~\eqref{l.15} into
\begin{equation}
P(\{ \alpha(t') \}, t_0<t'<t) \sim
\exp \left\{ - \frac{1}{2}\int_{t_0}^t dt' (\phi+\psi)+\frac{1}{2}[S(\alpha_t)-S(\alpha_0)]\right\}.
\end{equation}
This result plays a role analogous to the Boltzmann principle in
nonequilibrium statistical mechanics because it expresses the
probability of a state in terms of its entropy.  Here the probability
of a sequence of states is expressed in terms of the entropy and of
the dissipation function.

The most probable value of $\alpha(t)$ is obtained minimizing
$\int_{t_0}^t dt' \mathcal{L}[\alpha(t'), \dot{\alpha}(t')]$:
\begin{equation}
\label{l.16}
{d\alpha \over dt}=- \gamma \alpha \,\,\,
\mbox{with} \,\,\, \alpha(t_0)=\alpha_0 \,\,\,
\end{equation}
i.e. $ \alpha(t)= \alpha(0)e^{-\gamma(t-t_0)}$.
In other words, the most probable  value of $\alpha(t)$ coincides with the
conditional average $<\alpha(t)|\alpha(t_0)>$.

Analogously, the most probable path starting from $\alpha_0$
and ending in $\alpha_t$ is given by the minimum of $\int_{t_0}^t dt'
\mathcal{L}[\alpha(t'), \dot{\alpha}(t')]$ with the constraints
$\alpha(t_0)=\alpha_0$ and $\alpha(t)=\alpha_t$. The minimum condition
corresponds to
\begin{equation}
\label{l.18}
\Bigl[\int_{t_0}^t (\phi + \psi) dt'  \Bigr]_{min}=
2 \Bigl(S(\alpha_t) - S(\alpha_0) \Bigr).
\end{equation}

The extension to $N\ge 2$,
does not present particular difficulties.
For the case described by (\ref{new}),
 the symmetry of the matrix $\mathbf{C}$ allows us to work
in the diagonal representation of the white noise $\tilde{\eta}_k$, 
i.e. to deal with $N$ independent Markov processes. Therefore it suffices
 to replace the $\mathcal{L}$ in \eqref{l.13} for $N=1$ with
\begin{equation}
\label{l.19}
\mathcal{L}(\pmb{\alpha}, \dot{\pmb{\alpha}})={1 \over 4c}
\sum_{k,k'}\Bigl(\dot{\alpha}_k +\sum_jB_{k,j} \alpha_j \Bigr)
A_{k,k'}  \Bigl(\dot{\alpha}_{k'} +\sum_jB_{k',j} \alpha_j \Bigr)
\end{equation}
where  $\mathbf{A}={\mathbf{C}}^{-1}$.
In addition one has
\begin{equation}
\label{pippo}
\mathcal{L}(\pmb{\alpha}, \dot{\pmb{\alpha}})=
{1 \over 2}\Bigl( \phi(\pmb{\alpha}, \dot{\pmb{\alpha}})
  +  \psi({\bf X}) - { dS(\pmb{\alpha}) \over dt} \Bigr) \,\, ,
\end{equation}
and all the previous results hold for $N>1$.\\

\subsubsection{The non Gaussian case}

The generalization of the above results to the nonlinear case, i.e.
\begin{equation}
\label{nl}
{d\alpha_k \over dt}= f_k(\pmb{\alpha})+ \sqrt{2 c}{\eta}_k
\,\,\,\,\, k=1,...,N \,\, ,
\end{equation}
where $\{ \eta \}$ is a Gaussian process with
$\langle \eta_k \rangle=0$ and
$\langle \eta_k(t) \eta_{k'}(t') \rangle=\delta_{k,k'}\delta(t-t')$,
is rather simple, at least at a formal level.
Instead of (\ref{l.19}), one obtains
\begin{equation}
\mathcal{L}(\pmb{\alpha}, \dot{\pmb{\alpha}})={1 \over 4 c}
\sum_{k,k'}\Bigl(\dot{\alpha}_k -f_k \Bigr)
A_{k,k'}  \Bigl(\dot{\alpha}_{k'} -f_{k'} \Bigr) \,\,\,  .
\end{equation}
A simple way  to derive this result is the following:
repeat the argument to obtain~\eqref{l.11}
for small $\tau$, and  perform the path integral.
Let us sketch the basic idea, for the one variable case:
$$
{d\alpha \over dt}=f(\alpha) +\sqrt{2c}\eta \, .
$$
Denote $\alpha_n=\alpha(t_0+n\tau)$ and write the above Langevin equation
in its  discrete time approximation:
$$
 \alpha_n -\alpha_{n-1}=f(\alpha_n)\tau + \sqrt{2c\tau}w_n
$$
where $\{ w_n \}$ are Gaussian random variables with
$\langle w_n \rangle =0$ and   $\langle w_n w_{n'}  \rangle
=\delta_{n,n'}$.
Since the pdf of $w_n$ is  Gaussian, the generalization of (\ref{l.11}) merely requires that
$\ell$ in (\ref{l.11}) be replaced by
$$
\ell( \alpha_n, \alpha_{n-1})={1 \over 4c }\Bigl[
{(\alpha_n -\alpha_{n-1}) \over \tau} -f(\alpha_n) \Bigr]^2 \, .
$$
The above procedure is  heuristic and the result is  not completely exact.
 Nevertheless it yields the leading term for small values of $c$.
We do not discuss here this delicate point, the interested reader can
see~\cite{GGG80,FW98}.

The theory here outlined, for homogeneous systems, can be reformulated for spatially
extended systems, as illustrated in the next subsections.

\subsection{Large deviations in extended nonequilibrium systems}
Various attempts have been made to extend Onsager-Machlup theory
\cite{OM53} to the large fluctuations of physical systems in
nonequilibrium steady states~\cite{ELS}. Among them, the work by Jona-Lasinio and
collaborators~\cite{bsgj01,BDSJL}, is particularly relevant to
our discussion. Such approach generalizes those of Derrida {\it et al.}
\cite{DLS}, and in addition it leads to an independent derivation of
the FR. The theory of \cite{bsgj01,BDSJL} begins from the
assumption that a hydrodynamic-like description of the system at hand
is possible hence, physically, it can be applied as far from
equilibrium as the validity of the local thermodynamic equilibrium
allows.\footnote{This, of course, includes a very wide range of
phenomena, well beyond the linear regime, which has quite wide
applicability by itself.}

The theory leads to the conclusion that the nonequilibrium entropy functional, which 
generalizes the Onsager-Machlup entropy to extended systems, is a non-local functional 
of the thermodynamic variables, hence that correlations are present over macroscopic 
scales. To illustrate these facts, consider stochastic models of interacting particles, 
whose number is locally conserved, in contact with
particles reservoirs. Assume that these systems admit 
the hydrodynamic description
\begin{equation}
\partial_{t}\varrho = \nabla \cdot
\left[\frac{1}{2} D\left(\varrho \right)\nabla \varrho \right] \equiv \mathcal{D}\left(\varrho
\right) ~, \quad \quad \varrho = \varrho(\x,t) ~,
\label{hydro}
\end{equation}
where $\varrho$ is the vector of macroscopic observables, $\x$ is the macroscopic 
space variable, $t$ is the macroscopic time, $D$ is the diffusion matrix. 
Equation (\ref{hydro}) is analogous, in the spatially extended case, to Eq.(7.7) without noise.

The mathematical theory is necessarily developed
for very idealized models, such as the simple exclusion or the zero range stochastic
processes.\footnote{These models are usually one-dimensional. Their 
hydrodynamic limit consists of a scaling of the microscopic space and time variables, 
$\zt$ and ${\bf r}$, with the macroscopic space and time variables given by $t=\zt/N^2$ 
and $\x={\bf r}/N$, and the number of particles per unit length $N$ tending to infinity.} 
Nevertheless, the assumptions under which the theory holds are thought to be valid much more generally 
than in these cases, and are the following:

\vskip 2pt \noindent {\bf Assumptions:} 1) {\it The mesoscopic evolution is
  given by a Markov process $X_t$, which represents the configuration of the
  system at time $t$. The nonequilibrium steady state is described
  by a probability measure ${\rm Prob}$ over the trajectories of $X_t$;}

\noindent 2) {\it the macroscopic description is given in terms of
  fields $\varrho$ which constitute the local thermodynamic variables, whose
  evolution is described by (\ref{hydro}), which has a unique stationary
  solution $\tilde{\varrho}$, under the given nonequilibrium boundary
  conditions;}

\noindent
3) {\it Denoting by $I$ the time inversion operator defined by $I X_t = X_{-t}$, 
the probability measure ${\rm Prob}^*$, describing the evolution of the time 
reversed process $X_t^*$, and ${\rm Prob}$ are related by
\begin{equation}
{\rm Prob}^*(X_t^*=\phi_t, t\in[t_1,t_2]) = {\rm Prob}(X_t=\phi_{-t},t\in[-t_2,-t_1]).
\label{Pst*}
\end{equation}
Moreover, if $L$ is the generator of $X_t$, the adjoint dynamics is generated
by the adjoint (with respect to the invariant measure $\mu$) operator $L^*$,
which admits the {\em adjoint} hydrodynamic description
\begin{equation}
  \partial_{t}\varrho =\mathcal{D}^{*}\left(\varrho \right),
\label{adjointH}
\end{equation}

\noindent
4) The measure ${\rm Prob}$ admits a large deviation principle
describing the fluctuations of $\varrho$, i.e.\ the probability for a large number $N$ of particles
 that the evolution of the random variable $\varrho_N$ deviates from the solution of (\ref{hydro}),
to follow a given path $\hat{\varrho}(t)$, in the interval $[t_i,t_f]$, goes like
\be
{\rm Prob} \left( \varrho_N \left(X_{N^2t} \right) \sim \hat{\varrho}(t), [t_i,t_f] \right)
\approx e^{-N^d \left[ \mathcal{S}(\hat{\varrho}(t_i)+J_{[t_i,t_f]}(\hat{\varrho}) \right]}
\ee
where $d$ is the spatial dimension, $\mathcal{S}(\hat{\varrho}(t_i))$ is the entropy cost to produce 
$\hat{\varrho}(t_i)$, and $J$ is the extra cost required to follow the given path (taking
$\mathcal{S}(\tilde{\varrho})=0$, $J$ vanishes at $\varrho=\tilde{\varrho}$).
}
\vskip 5pt \noi  
This machinery leads to the following results:
\begin{itemize}
\item
The Onsager-Machlup theory is generalized by the introduction of the
``adjoint hydrodynamic'' equation (\ref{adjointH}), for the spontaneous fluctuations
around nonequilibrium steady states. Then, assuming that  $\mathcal{D}$ can be decomposed as
\begin{equation}
\mathcal{D}(\varrho) = \frac{1}{2} \nabla \cdot \left( \chi(\varrho) \nabla
\frac{\delta \mathcal{S}}{\delta \varrho} \right) + \mathcal{A} ~
\label{decompose}
\end{equation}
where $\mathcal{A}$ is a vector field orthogonal to the thermodynamic force
$\delta \mathcal{S}/\delta \varrho$ (the functional derivative of the entropy with
respect to the state), a temporal asymmetry arises in the fluctuation-relaxation
paths. Indeed, a spontaneous fluctuation out of a nonequilibrium steady state follows
a trajectory which is the {\em time reversal} of the relaxation path, according to the 
adjoint hydrodynamics, i.e.\ it solves
\begin{equation}
\partial_t \varrho = - \mathcal{D}^*(\varrho) = - \mathcal{D}(\varrho) + 2 \mathcal{A} ~,
\label{Ddiff}
\end{equation}
which is not merely the time reversal of the hydrodynamic
equation. Being orthogonal to the thermodynamic force, the term that
breaks the time symmetry, $\mathcal{A}$, is not the one that
contributes to the entropy production. Note that Eq.~\eqref{adjointH}
describes the adjoint hydrodynamics, while Eq.~\eqref{Ddiff} refers to
the evolution of spontaneous fluctuations.

\item
A Hamilton-Jacobi type equation is given for the macroscopic entropy $\mathcal{S}(\rho)$.
Expressing $J$ as
\be
J_{[t_i,t_f]}(\hat{\varrho}) = \int_{t_i}^{t_f} dt ~ \mathcal{L}(\hat{\varrho}(t),\partial_t \hat{\varrho}(t))
\ee
and introducing the Hamiltonian as the Legendre transform of $\mathcal{L}({\varrho},\partial_t {\varrho})$,
$$
\mathcal{H}(\varrho,Y) = \sup_\xi \left\{ \int \xi Y d ~{\bf x} - \mathcal{L}(\varrho,\xi) \right\} ~,
$$
the Hamilton-Jacobi type equation associated with the relation between $\mathcal{S}$ and $J$ is 
\be
\mathcal{H} \left( \varrho , \frac{\delta \mathcal{S}}{\delta \varrho} \right) = 0 ~,
\ee
where $\mathcal{S}(\varrho) = \inf_{\hat{\varrho}} J_{[-\infty,0]}(\hat{\varrho})$ is computed over
all trajectories $\hat{\varrho}(t)$ connecting $\tilde{\varrho}$ to $\varrho$.
Then, the steady state satisfies $\mathcal{H}(\tilde{\varrho},0)=0$.
\item
An H-theorem is given for $\mathcal{S}$:
\be
\frac{d}{d t} \mathcal{S} \le 0 ~.
\ee
\item
A nonequilibrium fluctuation dissipation relation holds:
\be
\mathcal{D}(\varrho) + \mathcal{D}^*(\varrho) = \nabla \cdot \left( \chi(\varrho) \nabla
\frac{\delta \mathcal{S}}{\delta \varrho} \right) ~.
\ee
\end{itemize}
These results have been later generalized to the case of systems 
subjected to external bulk forces, like electric fields \cite{BDSJL}.

\subsubsection{Temporal asymmetries in deterministic systems}
The question arises as to which aspects of the large deviation theory described above may
be verified experimentally, or numerically tested in models other than the lattice gases 
for which it has been rigorously established. For instance, one usually assumes that
the stochastic description is a reduced representation of some microscopic reversible dynamics, 
and should be recovered from that in some limit. However, various difficulties are to be faced, 
in order to answer that question.  For instance, the dynamics of the Lorentz gas are proven equivalent
to Markov processes, obtained by coarse graining with Markov partitions the relevant phase 
space \cite{BSC}. On the other hand, the kind of stochastic process that one obtains depends 
on the particular form of graining, and concerns the phase space instead of the real space. 
Furthermore, it is not always possible to point out without ambiguities which quantities of 
the stochastic systems correspond to given quantities of the deterministic systems. Therefore, 
the identification of the deterministic dynamics which are consistent with a given stochastic 
process or, conversely, the identification of the stochastic processes which preserve the 
observable properties of given microscopic dynamics, is a rather difficult task, in general.

This, on the one hand, indicates that a direct comparison between the predictions of 
Refs.\cite{bsgj01,BDSJL,BD} and the fluctuations of deterministic particle systems 
may be rather problematic, if not impossible, although desired \cite{BD}. At the same time, 
the behaviour of a macroscopic system cannot depend on such subtle and subjective issues like 
the form of the partitioning of the phase space; thus a test on deterministic systems of some 
aspect of the stochastic theory should be possible, in principle at least. Therefore, in 
Refs.\cite{GR04,GRV-jona,PSR} the prediction of the temporal asymmetries of fluctuations
is tested, because the differences between hydrodynamics and adjoint hydrodynamics are 
largely responsible for the other theoretical predictions, and because such asymmetries 
look amenable to direct test. It would be interesting to test, on deterministic systems, 
other features of the stochastic theory.

The origin of the temporal asymmetry can be easily understood as follows. 
Consider the macroscopic deterministic dynamics described by
\begin{equation}
\dfrac{d}{d t}{\varrho} = {\cal D}(\varrho) ~,
\label{xdotb}
\end{equation}
on ${\cal M} \subset \zR^n$, where ${\cal D}$ is a vector field with a
globally attracting fixed point $\tilde{\varrho} \in {\cal M}$. The $n$
components of $\varrho$ may represent the values taken by a scalar
thermodynamic observable on the $n$ different sites of a spatially discrete
system. Let the local mesoscopic dynamics be a perturbation of Eq.(\ref{xdotb}),
with a Gaussian
noise of covariance $\left\langle \xi _{i}\left( t\right) \xi _{j}\left(
t'\right) \right\rangle = K_{ij}\delta \left( t-t'\right)$ and mean
$\left\langle \xi(t) \right\rangle = 0$, where $\mathbf{K}$ is a symmetric,
positive definite matrix:\footnote{Note that here we have adopted the convention 
of Ref.\cite{bsgj01} for $\mathbf{K}$, which misses a factor 2 with respect to more standard 
notation used in Section 7.1.}
\begin{equation}
\dfrac{d}{d t}{\varrho} = {\cal D}(\varrho) + \xi ~.
\label{stoc-pertu}
\end{equation}
This allows different evolutions between one initial state
\(\varrho_{i}=\varrho \left( t_{i}\right)\) and one later state
\(\varrho_{f}=\varrho \left( t_{f}\right)\). 
The different paths connecting \( \varrho_{i} \) to \( \varrho _{f} \) occur
with different probabilities and Eq.(\ref{xdotb}) can be
obtained from the maximization of probabilities of the form \( P \propto \exp
( -F_{\left[ t_{i},t_{f}\right] }(\varrho) ) \), which depend on 
$\varrho_{i}$, $\varrho _{f}$ and on the path connecting them:
\begin{equation}
F_{\left[ t_{i},t_{f}\right] }\left( \varrho \right) =
\frac{1}{2}\int _{t_{i}}^{t_{f}}\left\langle
\dfrac{d}{d t}{\varrho}-\mathcal{D}\, ,\, \dfrac{d}{d t}{\varrho}-\mathcal{D}\right\rangle \,
\mathrm{d}t~,
\label{scalpr}
\end{equation}
where the scalar product is defined by $\langle x , y \rangle = x^T \mathbf{K}^{-1} y$,
and the superscript $T$ indicates transposition.
Therefore, Eq.(\ref{stoc-pertu}) is analogous to Eq.(7.7), and 
Eq.(\ref{scalpr}) corresponds to Eq.(7.22), with $\mathbf{K}^{-1}$ in place of $\mathbf{A}/2c$.
Now, decompose the vector field \( \mathcal{D} \) as
\begin{equation}
\mathcal{D}\left( \varrho \right) =-\frac{1}{2}{\bf K}\nabla_\varrho V
\left( \varrho \right)
+\mathcal{A}\left( \varrho \right) ,\qquad \textrm{with } ~ \left\langle
{\bf K}\nabla_\varrho V\, ,\, \mathcal{A}\right\rangle = 0 ~,
\label{decdissipative}
\end{equation}
where $\nabla_\varrho$ indicates differentiation with respect to the components of $\varrho$,
and let \( \tilde{\varrho} \) be a minimum of \( V \), with $V(\tilde{\varrho})=0$, 
separating dissipative contributions to $\mathcal{D}$ from non-dissipative ones,
as common in diffusion processes described by finite dimensional Langevin equations \cite{FW98}. 
Substituting (\ref{decdissipative}) in (\ref{scalpr}), one obtains 
\be
2 F_{\left[ t_{i},t_{f}\right] }\left( \varrho \right) = 
\int _{t_{i}}^{t_{f}}\left\langle \dfrac{d}{d t}{\varrho }+\frac{1}{2}{\bf K}
\nabla_\varrho V-\mathcal{A}\, ,\, \dfrac{d}{d t}{\varrho }+\frac{1}{2}{\bf K}
\nabla_\varrho V-\mathcal{A}\right\rangle \, \mathrm{d}t \label{firstall} 
\ee
which is minimized by the relaxational path converging to \( \tilde{\varrho} \) and
obeying
\be
\dfrac{d}{d t}{\varrho} = -\frac{1}{2}{\bf K}\nabla_\varrho V
\left( \varrho \right)
+\mathcal{A}\left( \varrho \right) ~,
\ee
i.e.\ obeying (\ref{xdotb}). However, the integrand of (\ref{firstall}) is a quadratic form
and has a second minimizing path, which is revealed by an integration by parts. This yields:
\begin{eqnarray}
2 F_{\left[ t_{i},t_{f}\right] }\left( \varrho \right) &=& \int _{t_{i}}^{t_{f}}\left\langle \dfrac{d}{d t}{\varrho }-
\frac{1}{2}{\bf K}\nabla_\varrho V-\mathcal{A}\, ,\, \dfrac{d}{d t}{\varrho }-\frac{1}{2}
{\bf K}\nabla_\varrho V-\mathcal{A}\right\rangle \, \mathrm{d}t + \nonumber \\
&&+ 2\LARGE[ V\left( \varrho _{f}\right) -V\left( \varrho _{i}\right) \LARGE] 
\label{jt1t2}
\end{eqnarray}
whose second term has no variation, once the initial and final states are fixed. Therefore,
the second possible kind of evolution for $\varrho$ corresponds to the fluctuations 
away from \( \tilde{\varrho} \), which obey 
\begin{equation}
\label{invhydr}
\dfrac{d}{d t}{\varrho} = - \mathcal{D}^{*}\left( \varrho \right) =
\frac{1}{2}{\bf K}\nabla_\varrho V+\mathcal{A}\left( \varrho \right) =  -\mathcal{D}
+ 2 \mathcal{A} ~.
\end{equation}
The qualitative properties of the deterministic dynamics do not depend on
$\mathcal{A}$, as the time derivative of the Lyapunov function $V$ does not depend 
on $\mathcal{A}$. Curiously, the asymmetry of the fluctuation paths depends only 
on this non dissipative term.

In the hydrodynamic limit, \( \mathcal{D} \) becomes the elliptic differential operator of 
Eqs.(\ref{hydro},\ref{decompose}), where \( {\bf K} \) turns into $\nabla \cdot \chi(\varrho) \nabla$,
$\nabla_\varrho$ into the functional derivative $\delta /\delta \varrho$ and 
$V(\varrho)$ into $\mathcal{S}(\varrho)$.
In the theory developed by Onsager and Machlup, $\mathcal{A}=0$ because only
the small fluctuations were considered, and the linear response regime was assumed.

The above suggests that sufficiently chaotic systems of interacting particles should 
typically have asymmetric fluctuation paths. This has been indeed confirmed in 
\cite{GRV-jona,PSR} for the nonequilibrium FPU model, as well as for the shearing 
system of Eq.(\ref{SLLODeqs}). In Ref.\cite{GRV-jona} it was then explained that the 
current fluctuations of the many particles nonequilibrium Lorentz gas, studied in 
\cite{GR04}, had to be symmetric, as observed, because of the lack of interactions 
among those particles. It remains an open question to see which kinds of interactions 
lead to symmetric and which to asymmetric nonequilibrium fluctuation paths.

To properly deal with deterministic systems of finitely many particles, 
Refs.\cite{GR04,GRV-jona,PSR} introduced various notions of fluctuation-relaxation 
paths, together with various criteria to assess their asymmetry. In the following
subsection, we consider 
only one of these notions, for the local heat flux of the nonequilibrium FPU model.

\subsubsection{Heat flux fluctuation-relaxation paths in the FPU model}
An observable $X: {\cal M} \to \zR$, for a system made of a finite number 
of particles, looks very noisy, in general. 
Therefore, given an initial condition $\zG$ in the steady state, 
we search for the most likely path
that starts at $X_{t_i}=\overline{X}$, reaches a certain fluctuation value
${\cal T}(X)$ at time $\hat{t}$, and later returns to $\overline{X}=X_{t_f}$. 
By path we mean the trajectory segment \{$X_{\hat{t}+\tau},\ \tau \in
[-\tau_0,\tau_0]$\} where, for simplicity, $\xt$ denotes $X(S^t \zG)$,
and $\overline{X}$ is the corresponding time average of $X$.
The time $\zt_0$ is a positive constant, that needs to be identified
case by case, and represents the typical fluctuation time. This notion 
may look odd, since a part of a given fluctuation-relaxation path is sampled more than 
once (typically twice), but is appropriate to study the symmetry properties 
of paths, because it does not alter their symmetry, and does not neglect any of 
them. Furthermore, the results obtained so far are robust against variations 
of this definition \cite{PSR,GRV-jona}.

Given $n$ paths, they can be arranged to produce a two dimensional histogram, by 
partitioning the rectangle
$[-\tau_0,\tau_0]\times [\min_{\tau,s} X_\tau^\se,\ \max_{\tau,s} X_\tau^\se]$
with rectangular bins, and evaluating the frequency of visitation of each bin. 
The ``crest'' of the histogram represents the most likely path in the sample, 
the one which should become the unique (deterministic) path in the large system
limit. In \cite{GRV-jona}, a number $M$ of tests is performed, each producing 
a different histogram and a different crest, so that the maximum, minimum 
and average crests can be defined as follows. Let 
$t \in [-\tau_0,\tau_0]$ and $j=1,...,M$ be the index of the crest, then,
denoting by ${\cal C}^j$ the crest produced in the $j$-th test,
\be
{\cal C}_{av}(t)=\frac{1}{M} \sum_{j=1}^{M}{\cal C}^j(t) 
\label{avcrest}
\ee
is the average crest; 
\be
\underline{\cal C}(t)=\min_{j=1,\ldots,M}{\cal C}^j(t)
\label{mincrest}
\ee
is the minimum crest, and 
\be
\overline{\cal C}(t)=\max_{j=1,\ldots,M}{\cal C}^j(t)
\label{maxcrest}
\ee
is the maximum crest.

\noi
To assess the symmetry properties of a crest ${\cal C}$, discretize the
time interval $[0,\zt_0]$ in $b_\zt$ sub-intervals, and introduce
the {\em asymmetry coefficient} as
\be
\alpha_c({\cal C})=\frac{2}{b_\tau(\kT-\overline{X})}\left(
\sum_{p=\frac{b_\tau}{2}+1}^{b_\tau} {\cal C}(p) -
\sum_{p=1}^{\frac{b_\tau}{2}} {\cal C}(p) \right ) ~,
\label{crestasymm}
\ee
where, the normalizing factors are introduced only for convenience. A crest 
is called {\em  symmetric} if its asymmetry coefficient vanishes.
The asymmetry coefficients can take positive as well as negative values, 
and random oscillations of a fluctuation path around its symmetric relaxation 
image result in a vanishing asymmetry coefficient. With these definitions,
and $M=20$, the heat flux in a central portion of the chain, of the nonequilibrium 
FPU model, with $N=150$ oscillators, produces results like
$$
\alpha_c(\underline{\cal C}) = 0.170 ~, \quad \alpha_c({\cal C}_{av}) = 0.194 ~,
\quad \alpha_c(\overline{\cal C}) = 0.211
$$
for a large temperature difference between the ends of the chain ($T_\ell=200$ and $T_r=20$), 
and for a fluctuation value ${\cal T}$ three standard deviations higher than the mean. Here, 
the gap between the maximum and minimum crests asymmetries can be taken as the numerical 
uncertainty on the asymmetry of the crest. The result is that temporally asymmetric 
fluctuation-relaxation paths are found in time reversal invariant dynamics. 
\begin{figure}
\centering
\includegraphics[width=6.9cm,height=6.5cm]{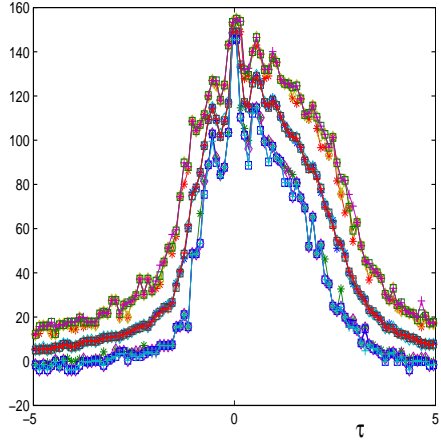}
\caption{\small Crest of the heat flux $f$ for an FPU chain with $N=66, \ts=70,\td=20$,
               $\overline{f}\simeq 16.2$, and $\mathcal{T}\simeq 148$. The highest line represents
               $\overline{\cal C}$, the intermediate line represents ${\cal C}_{av}$, and
               the lowest line represents $\underline{\cal C}$. Each line is
               the overlap of the four crests obtained with $M=39,59,79,99$.}
\label{20infnity}
\end{figure}
Figure \ref{20infnity} illustrates these facts for a chain with fewer oscillators, for
which a relatively large number of tests can be performed. In this case, it is interesting
to see that the three kinds of crests do not appreciably change when the number of tests
grows over $M=20$, which is an indication of the accuracy achieved in the construction of 
the crests. In Ref.\cite{PSR2}, a justification of the ubiquity of the temporally asymmetric 
fluctuation-relaxation paths, in nonequilibrium particle systems, is given in terms 
of correlation functions.

\subsection{The additivity principle}
The theory developed in Refs.\cite{bsgj01,BDSJL} has a non-local nature, as
easily understood in the works of Derrida and collaborators \cite{DLS,BD}, who 
independently worked out the explicit solutions of what can now be seen as specific 
examples of the theory of \cite{bsgj01,BDSJL}. Actually, explicitly computing 
the large deviation functional of the current in systems like the symmetric and asymmetric 
simple exclusion processes, Bodineau and Derrida realized that a certain {\em additivity 
principle} describes the corresponding steady states. 

The idea is the following: let a 
one dimensional system have length $L+L'$ and be in contact with two reservoirs of
particles at densities $\varrho_a$ and $\varrho_b$. Express the probability of the 
current integrated in a time $t$, $Q_t = j t$ say, as
\be
{\rm Prob}_{L+L'}(j,\varrho_a,\varrho_b) \sim e^{-t F_{L+L'}(j,\varrho_a,\varrho_b)} ~.
\ee
Then, the additivity principle relates $F_{L+L'}$ to the large deviation functionals of the 
subsystems of sizes $L$ and $L'$, by taking 
\be
{\rm Prob}_{L+L'}(j,\varrho_a,\varrho_b) \sim \max_\varrho \left\{ {\rm Prob}_L (j,\varrho_a,\varrho)
{\rm Prob}_{L'}(j,\varrho,\varrho_b) \right\} ~.
\ee

\begin{figure}[h]
\includegraphics[width=8cm,clip=true]{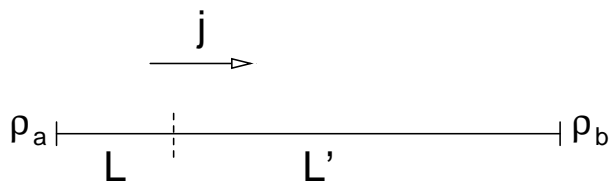}
\caption{Sketch of a one-dimensional system between two reservoirs
$\rho_a$ and $\rho_b$, which is virtually divided into two sub-systems
of length $L$ and $L'$ respectively. An average current $j$ flows
through the system (assuming for instance that $\rho_a>\rho_b$).}
\end{figure}

In other words, one assumes that the probability of transporting $j$
between the reservoirs at densities $\varrho_a$ and $\varrho_b$, is
the maximum over $\varrho$ of the probabilities of transporting the
same currents between two reservoirs at densities $\varrho_a$ and
$\varrho$, times the probabilities of transporting it between two
reservoirs at densities $\varrho$ and $\varrho_b$.  Then, in the large
system limit, in which boundary contributions become negligible, the
relevant large deviation functional takes the form
\be
F_{L+L'}(j,\varrho_a,\varrho_b) = \max_\varrho \left\{ F_L (j,\varrho_a,\varrho) + F_{L'} (j,\varrho,\varrho_b) \right\} \,\,.
\ee
In the case of
nonequilibrium boundary conditions (i.e.\ $\varrho_a \ne \varrho_b$), this results in
a non-local relation for the steady state density profile, as shown in the next subsection
for an exactly solvable model. Introducing the scaling hypothesis
\be
F_L(j,\varrho_a,\varrho_b) \simeq \frac 1 L G(Lj,\varrho_a,\varrho_b)
\ee
and observing that the optimal profile connecting $\varrho_a$ to $\varrho_b$
is the same for both $j$ and $-j$, in the steady state, Bodineau and Derrida 
obtain also the validity of the FR for the current $j$, in the form
\be
G(-j,\varrho_a,\varrho_b) = G(j,\varrho_a,\varrho_b) - 2 j \int_{\varrho_b}^{\varrho_a} 
\frac{D(\varrho)}{\zs(\varrho)} ~ d \varrho
\ee
where $D(\varrho)$ and $\zs(\varrho)$, in the large $L$ limit, are defined as follows:
let $\varrho_a = \varrho + \Delta \varrho$ and $\varrho_b = \varrho$, with small $\Delta \varrho$,
then $\langle Q_t \rangle / t = D(\varrho) \Delta \varrho / L$ and $\langle Q_t^2 \rangle / t = \zs(\varrho)/L$.

\subsubsection{The symmetric simple exclusion process}
The symmetric simple exclusion process affords one of the few
nonequilibrium systems for which the large deviation function of the
density can be computed in an explicit way. This model consists of a
1-dimensional lattice with $L$ sites, which can be either empty or
occupied by one particle. At each time step, of length $\Delta t$
every particle can jump with equal probability $\Delta t$ to the
right, if the site on the right is empty, or to the left, if the site
on the left is empty. At the left boundary of the lattice (site 1),
particles are injected with probability $\za \zD t$, if 1 is empty,
and are removed with probability $\zg \zD t$, if 1 is
occupied. Similarly, at the right boundary (site L), particles are
injected with probability $\zd \zD t$ and removed with probability
$\zb \zD t$. This is equivalent to have a reservoir of particles at
density $\varrho_a = \za / (\za + \zg)$ on the left, and one of
density $\varrho_b = \zd / (\zd + \zb)$ on the right of the
lattice. Therefore, if $\zt_i=0,1$ is the random variable equal to 0
when site $i$ is empty, and equal to 1 when it is occupied, one
obtains \bea &&\frac{d \langle \zt_1 \rangle}{d t} = \za - (\za + \zg
+ 1)\langle \zt_1 \rangle + \langle \zt_2 \rangle \\ &&\frac{d \langle
\zt_i \rangle}{d t} = \langle \zt_{i-1} \rangle - 2 \langle \zt_i
\rangle + \langle \zt_{i+1} \rangle ~, \qquad 2 \le i \le L-1 \\
&&\frac{d \langle \zt_L \rangle}{d t} = \langle \zt_{L-1} \rangle -
(\zd + \zb + 1) \langle \zt_L \rangle + \zd \,\,.\eea The steady state,
which corresponds to vanishing derivatives, takes the form \be \langle
\zt_i \rangle = \frac{\varrho_a \left( L + \frac{1}{\zb + \zd} - i
\right) + \varrho_a \left( \frac{1}{\za + \zg} + i - 1 \right)}{L +
\frac{1}{\za + \zg} + \frac{1}{\zb + \zd} - 1} \ee and the current
takes the form \be J = \langle \zt_i - \zt_{i+1} \rangle =
\frac{\varrho_a - \varrho_b}{L + \frac{\varrho_a}{\za} +
\frac{\varrho_b}{\zd} -1} \ee which is proportional to the gradient
$(\varrho_a - \varrho_b)/L$, when $L \gg 1, \varrho_a/\za , \varrho_b
/ \zd$.  At equilibrium, with $\varrho_a = \varrho_b = \hat{\varrho}$,
and in the large $L$ limit, if the lattice has unit length, the large
deviation functional takes the form \be F(\varrho) = \int_0^1
B(\varrho(x),\hat{\varrho}) \d x ~, \quad \mbox{with ~~~}
B(\varrho,\hat{\varrho}) = (1-\varrho) \log
\frac{1-\varrho}{1-\hat{\varrho}} + \varrho \log
\frac{\varrho}{\hat{\varrho}}\,\,. \ee The functional is thus a local
function of $\varrho(x)$, and is convex. Let us note that
$B(\varrho,\hat{\varrho})$ is nothing but the Creamer's function
corresponding to the binomial distribution, see Appendix
B. Differently, in the nonequilibrium case, $\varrho_a \ne \varrho_b$,
one has \be F(\varrho) = \int_0^1 \left[ B(\varrho(x), f(x)) + \log
\frac{f'(x)}{\varrho_b-\varrho_a} \right] \ee where $f$ is the
monotone solution of \be \varrho = f + \frac{f(1-f)f''}{f'^2} ~, \quad
f(0)=\varrho_a ~, ~~~~ f(1) = \varrho_b \,\,. \ee This result, which was
first obtained in \cite{DLS-PRL} and, later, as a special case of the
theory of \cite{bsgj01}, shows that the value of $f$ at position $x$ depends on the profile
$\varrho(y)$ at all points $y$, hence that $F$ is non-local.

\section{Conclusions}
\label{sec:end}

In this report we have summarized the history and the state of the art of the studies
concerning the response of physical systems to external actions. This has evidenced
how the statistical mechanics community has gradually shifted its focus from the
investigation of equilibrium and near equilibrium phenomena, to strongly nonequilibrium
phenomena, which cannot be understood in terms of thermodynamic quantities. Similarly,
the interest has moved from the construction of a microscopic interpretation of the
macroscopic phenomena, for which the thermodynamic limit of finite systems is necessary, 
to the study of the properties of mesoscopic and microscopic objects, which can be
understood as systems of finitely many microscopic components. These objects are
presently under intense investigation, because they appear to be extremely interesting
from both fundamental and applied viewpoints. From the applied viewpoint, it is important
to understand their behaviour, because of the emerging nanotechnology, which impacts also 
on our ability to manipulate biological systems. From the fundamental point of view,
it is now obvious that they must be described by phenomenological equations which differ from those of the macroscopic world.
But the classification of the phenomena which are possible at the mesoscopic and microscopic
level is still to be done, and a consistent theory of such phenomena is still lacking.

Nevertheless, the body of knowledge acquired in the past century, in
the construction of the microscopic theory of the response of
macroscopic systems, appears to be an ideal springboard to decisively
leap into the largely unexplored realm of strongly nonequilibrium
systems, and of microscopic objects. Indeed, this knowledge has been
based on the characterization of the fluctuations which, although
macroscopically not observable, are evident if the thermodynamic limit
is not taken. This is why we believe that a review of the response
theory stretching up to present times, with emphasis on non-standard
issues (granular media, fluids, nano-systems and biological systems),
was needed and should be useful in advancing our understanding of the 
physical world.

\vspace{1cm}

{\bf Acknowledgments}  

We thank our colleagues with whom we collaborated on the issues here
discussed: A. Baldassarri, A. Barrat, G. Benettin, L. Biferale,
G. Boffetta, F. Cecconi, M. Cencini, E.G.D. Cohen, I. Daumont,
D.J. Evans, M. Falcioni, G. Gallavotti, A. Gamba, C. Giberti,
S. Isola, O.G. Jepps, G. Lacorata, S. Lepri, V. Loreto, J. Lloyd,
C. Mejia-Monasterio, G. Morriss, S. Musacchio, M. Niemeyer, C. Paneni,
D.J. Searles, E. Segre, T. T\'el, E. Trizac, F. van Wijland, C. Vernia, 
P. Visco and J. Vollmer.

In addition we are grateful to  K. Kaneko, M.H. Lee and F. Ritort
for correspondence and useful remarks.

\appendix 

\section{Appendix: Models of granular gases}

\label{app:granular}

The most common model of fluidized granular matter is the gas of
inelastic hard spheres in $d$ dimensions~(reviews of recent results
can be found in~\cite{PL01} and~\cite{PL03}), interacting 
through binary inelastic collisions. Initially the $N$ particles are placed in a
box of volume $V=L^d$, at random positions and with random 
velocities, typically taken from a Gaussian distribution. Since we
discuss monodisperse as well as bidisperse systems, we denote by $s_i$,
$\sigma_i$ and $m_i$ respectively the species, the diameter and the
mass of particle $i$. The use of $\sigma$ and $m$ without indexes will
denote the monodisperse case. When particles $i$ and $j$ collide, the
instantaneous change of their velocities is given by
\begin{equation} \label{collision}
\mathbf{v}_i'=\mathbf{v}_i-(1+r_{ij})\frac{m_j}{m_i+m_j}[(\mathbf{v}_i-\mathbf{v}_j)\cdot \hat{\mathbf{\sigma}}_{ij}]\hat{\mathbf{\sigma}}_{ij}
\end{equation}
where $r_{ij} \in [0,1]$ is the restitution coefficient, $m_i$ and $m_j$
are the masses of the particles and $\hat{\mathbf{\sigma}}_{ij}$ is
the unit vector joining the centers of the particles. When $r_{ij}=1$ the
gas is elastic and no energy is dissipated in the collisions. In all
other cases the collisions are dissipative. Different kinds of
experimental setups can be reproduced by this model, provided that a
suitable driving mechanism is added. A widely accepted model for the
investigation of strongly fluidized granular gases, reduces the effect
of the interaction between the grains and the fluidizing agent
(e.g. an air flow or random collisions with the vibrating plate) to
that of a ``thermal'' bath coupled to every particle. Between
collisions, the particles are subjected to a random force given by white noise, with the possible
addition of a viscous term. The equation of motion for a particle is
then
\begin{equation} \label{gen_drive}
m_i \frac{d {\bm v}_i}{dt} = {\bm F}^{col}_i + \pmb{\eta}_i - \gamma_{s_i} {\bm v}_i
\end{equation}
where ${\bm F}^{col}_i$ is the force due to inelastic collisions,
$\gamma_{s_i}$ is the viscosity coefficient of species $s_i$ and
$\pmb{\eta}_i$ is the Gaussian stochastic force, with $\langle
\eta_{i\alpha}(t) \eta_{j\beta}(t') \rangle = \Gamma_{s_i}^2
\delta_{ij}\delta_{\alpha\beta} \delta(t-t')$, where the Greek indices
refer to Cartesian coordinates.  The granular temperature of species
$s$ is given by its mean kinetic energy $T_g^s = m_s \langle
v^2\rangle_s/d$ where $\langle \rangle_s$ is an average restricted
only to particles of species $s$. This definition is important in view
of the lack of energy equipartition observed in granular mixtures.

From the generic equation of motion~\eqref{gen_drive}, many different
models can be obtained. When $\gamma_1=\gamma_2=\gamma$,
$\Gamma_1=\Gamma_2=\Gamma$, $r_{11}=r_{12}=r_{22}=r$, and all the particles are
identical, the gas is monodisperse. The ``cooling granular gas'' is
recovered when $\gamma=\Gamma=0$~\cite{BMC96}.  In such a case the particles follow
ballistic trajectories until they collide and lose energy. The kinetic energy of the particles in their
center of mass frame decreases with time, justifying the term
``cooling''. The homogeneous cooling is an idealized situation
obtained when the positions of the particles are disregarded~\cite{BMC96}. This
situation is considered the analogous of the perfect rarefied gas in
the realm of granular matter. For this model an equation for the
evolution of the single particle velocity pdf $f(\bv,t)$ can be written,
in the form~\cite{NE98}
\begin{equation} \label{granuboltz}
(\partial_t+\bv \cdot \nabla)f(\bv,t)=J[f,f](\bv,t),
\end{equation}
where, in the Boltzmann-Grad limit, the collision operator $J$, for hard spheres, takes the form:
\begin{equation} \label{granucolint}
J[f,f](\bv_1,t)=\frac{1}{\ell}\int_{\bv_{12} \cdot \bsigma >0}
  \dd \bv_2\,\dd\bsigma \,(\bv_{12}\cdot \bsigma)\,\left(\frac{1}{r^2}
    f(\bv_1^{*},t)f(\bv_2^{*},t)-
    f(\bv_1,t)f(\bv_2,t)\right) \,\,.
\end{equation}
 In this last expression, the
relative velocity $\bv_{12}=\bv_1-\bv_2$ has been used, as well as the
pre-collisional velocities $\bv_1^{*},\bv_2^{*}$ which can be
obtained from $\bv_1,\bv_2$ inverting the collision
rule~\eqref{collision} for particles $1$ and $2$. 

When $\Gamma \neq 0$ the effect of external random forces balances the
energy dissipated in collisions, and a statistically stationary state
is reached. Two possibilities have been considered in the literature:

\begin{itemize}

\item the heat bath with viscosity~\cite{PLMPV98,PLMV99,CDMP04,BTC07}, i.e. $\gamma \neq 0$: in this case
a ``bath temperature'' can be defined as $T_b=\Gamma^2/2\gamma$. This
corresponds to the temperature of a gas obeying
equation~(\ref{gen_drive}) with elastic collisions or without
collisions. The same temperature can be observed if the viscosity is
very high, i.e. when $\gamma \gg 1/\tau_c$ where $\tau_c$ is the mean
free time between collisions;  when $\gamma \ll 1/\tau_c$ and $r<1$ the
gas still reaches a stationary regime, but its granular temperature is
in general smaller than $T_b$ and, therefore, the system is out of
equilibrium~\cite{PCV05};

\item the heat bath without viscosity~\cite{WM96,NETP99,TPNE01},
i.e. $\gamma=0$; in this case to obtain the correct elastic limit one
must reduce the driving intensity $\Gamma \sim 1-r^2$ as $r \to 1$, so
that the average kinetic energy per particle and per degree of freedom
$T_g$ does not diverge.

\end{itemize}

These two models show a rich range
of phenomena, that are usually observed in driven granular gases. The
statistical properties of the stationary state obtained by those
models are different from those of an elastic gas in contact with a
thermal bath: the most important difference is that the velocity
distribution is non-Gaussian with enhanced high-energy tails~\cite{NE98}. The
other typical feature is that the inelastic collisions reduce the
relative velocity of particles and make re-collisions more likely:
this is a source of correlations which emerges in the form of density
clusters and velocity parallelization (vortices and other
structures~\cite{PLMV99,NETP99,TPNE01}). The hydrodynamic transport coefficients for these models
have been calculated from the Chapman-Enskog  approximated solution of the Boltzmann
equation in~\cite{GM02}, while an analytical study of the Einstein
relation can be found in~\cite{G04}.

Other models can be derived from the previous ones. A useful
approximation is the so-called Direct Simulation Monte Carlo, or
DSMC~\cite{B94,MS00,MG02}, which consists in disregarding the relative
position of particles in order to drastically reduce the probability
of a re-collision: this is usually considered to be equivalent to the
Molecular Chaos assumption in the $N \to \infty$ limit . Here, we
consider the uniform gas. This model is nonetheless useful to test
also spatial effects, see for example~\cite{PLMV99,MS00}. In this case
the collisions occur with a probability which is proportional to the
relative velocity of the particles, in order to be consistent with the
kernel of the collision integral of the Boltzmann
equation~\eqref{granucolint}. A further simplification is operated in
the so-called Inelastic Maxwell Models (see~\cite{BMP02,BMP02b} and
references therein). These models are gases of inelastic particles
without positions, where collisions occur with a constant probability,
i.e. not dependent upon the relative velocity. They are inspired by
the so-called Maxwell Molecules, an elastic kinetic model having the
same property due to the particular inter-particle (soft core)
potential. Inelastic Maxwell Models have a very simple Boltzmann
equation that can be analytically solved in the one dimensional
case~\cite{BMP02,BMP02b}. Other properties can be obtained in an analytical
form, for example it is possible to demonstrate a general relation
between the response $R(t)$ of a particle to the sudden perturbation
of its velocity at time $0$, with a very large set of correlation
functions, including the self-correlation~\cite{BBDLMP05,PBV07}:
\begin{equation}
R(t)=C_f(t)=\frac{\langle v(t)f[v(0)]\rangle}{\langle v(0)f[v(0)]\rangle}=
\exp\left(-\frac{r(r+1)}{4}t\right )
\end{equation}
 with any generic function $f$ of the initial velocity value.

\section{Appendix: Large Deviations in a nutshell}

For the sake of self-consistency,  we briefly discuss the
basic aspects of the Large Deviations theory.
Roughly speaking, one can say that the Large Deviations theory is
 a generalization of the two most important
limit theorems of probability theory, i.e. the law of large numbers,
and the central limit theorem~\cite{E99,V03}.\\
Consider a  simple example:
 a sequence of independent tosses of an unfair coin.
The possible outcomes are head ($+1$) or tail ($-1$).  Denote the possible
result  of the $n$-th toss by $x_n$, where  head  has probability
$\alpha$, and tail  has probability $1-\alpha$.
Let $Y_N$ be the mean value after $N$ tosses,
\begin{equation}
\label{LD.1}
Y_N= {1 \over N} \sum_{n=1}^N x_n.
\end{equation}
If $N \gg 1$, a straightforward application of the central limit theorem
gives
\begin{equation}
\label{LD.2}
Prob(Y<Y_N<Y+dY)\simeq P_N(Y) dY
\sim   e^ {-{{N(Y-\langle Y \rangle)^2} \over {2 \sigma^2}} } dY  \,\, ,
\end{equation}
where $\langle Y \rangle=2\alpha-1$, $\sigma^2=4\alpha(1-\alpha)$,
$P_N(Y)$ is the pdf of the random variable \eqref{LD.1}.
Since $Y_N$ lies between $-1$ and $1$,
it is easy to realize that~\eqref{LD.2} is accurate only for
small deviations of $Y_N$ from the mean (namely $|Y_N-\langle Y \rangle
|<O(1/\sqrt{N})$).\\
A natural way to introduce the large deviations and show their  deep
relation
with the concept of entropy is to perform a combinatorial computation.
The number of ways in which $K$ heads occur in $N$ tosses is $N!/[K!(N-K)!]$, therefore, the exact binomial distribution yields
\begin{equation}
\label{LD.3}
Prob(Y_N={2K \over N} -1)={{N!} \over {K!(N-K)!}} \alpha^{K}
(1-\alpha)^{N-K} \,\, .
\end{equation}
Using Stirling's approximation and writing $K=pN$ and $N-K=(1-p)N$
one obtains
\begin{equation}
\label{LD.4}
 P_N(Y=2p-1) \sim
 e^ {-N I(\alpha,p)}  \,\, ,
\end{equation}
where
\begin{equation}
\label{LD.5}
I(\alpha,p)=p \ln { p \over \alpha} +
           (1-p)\ln {{1-p} \over {1-\alpha}} \,\, .
\end{equation}
Note that $I(\alpha,p)$ is called ``relative entropy'' (or
Kullback-Leibler divergence), and $I(\alpha,p)=0$ for $ \alpha=p$,
while $I(\alpha,p)>0$ for $ \alpha \neq p$.  It is easy to repeat the
argument for the multinomial case, where $x_1, ..., x_N$ are
independent variables that take $m$ possible different values
$a_1,a_2, ...,a_m$ with probabilities $\pi_1, \pi_2 ,..., \pi_m$.  In
the limit $N \gg 1$, the probability of observing the frequencies
$f_1,f_2,...,f_m$ is
$$
Prob_N(\{ f_j \}\simeq \{ p_j \}) \sim e^{-N I(\{ p \}, \{ \pi \})}
$$
where
$$
I(\{ p \}, \{ \pi \})=\sum_{j=1}^m p_j \ln {p_j \over \pi_j} \,\,\, ,
$$
is called ``relative entropy'' of the probability $\{ p \}$, with respect to the
probability $\{ \pi \}$. Such a quantity measures the discrepancy between
 $\{ p \}$ and  $\{ \pi \}$ in the sense that
$I(\{ p \}, \{ \pi \})=0$ if and only if $\{ p \}= \{ \pi \}$,
and  $I(\{ p \}, \{ \pi \})>0$ if $\{ p \}\neq \{ \pi \}$.

From the above  computation one understands that it is
possible to go beyond the central limit theory, and to
estimate  the statistical features of extreme (or tail) events,
as the number of observations grows without bounds.
Writing $I(p,\alpha)$ in terms of $Y=2p-1$, Eq.~(\ref{LD.4}) becomes
\begin{equation}
\label{LD.6}
 P_N(Y) \sim
 e^ {-N S(Y)}  \,\, ,
\end{equation}
with
$$
S(Y)={{1+Y} \over 2 } \ln {{1+Y} \over {2 \alpha}}
+ {{1-Y} \over 2 } \ln {{1-Y} \over {2 (1- \alpha)}} \,\, .
$$ The function $S(Y)$ is called Cramer's function.  Of course, for $p$
close to $\alpha$, i.e. $Y\simeq \langle Y \rangle$, a Taylor
expansion reproduces the central limit theorem (\ref{LD.2}).

However Eq.~\eqref{LD.6} has a general validity, and can be derived in different
ways, which show how the shape of $S(Y)$ is related to the
behaviour of the moments of the variable $x$.  In particular it is possible to see
that $S(Y)$ can be expressed in terms of a Legendre transform:
\begin{equation}
\label{LD.7}
S(Y)=\sup_{q}\Bigl[qY-L(q)\Bigr] \,\, ,
\end{equation}
where $L(q)$ is the ``Cumulants Generating Function'' given by
\begin{equation}
\label{LD.8}
L(q)=\ln \langle e^{qx} \rangle  \,\, .
\end{equation}
Let us sketch the argument.
Consider the quantities $\langle e^{qNY_N}\rangle$ which can be written in two ways:
\begin{align}
\langle e^{qNY_N}\rangle&= \langle e^{qx}\rangle^N=e^{N L(q)}\\
\langle e^{qNY_N}\rangle&=\int  e^{qNY_N}  P_N(Y_N) dY_N \sim \int  e^{[qY- S(Y)]N} dY
\end{align}
whose identification leads to
\begin{equation}
\label{LD.9}
\int  e^{[qY- S(Y)]N} dY \sim
e^{N L(q)}
\end{equation}
and, in the limit of large $N$, using the steepest descent method, one has
\begin{equation}
\label{LD.10}
L(q)= \sup_{Y}\Bigl[qY-S(Y)\Bigr] \,\, ,
\end{equation}
which is the inverse of~(\eqref{LD.7}).
Since it is possible to show that $S(Y)$ is convex,
Equations (\ref{LD.7}) and (\ref{LD.10}) are fully equivalent.

For the more general and interesting case of dependent
variables, $L(q)$ is defined as
$$
L(q)=\lim_{N \to \infty} {1 \over N}
\ln \langle e^{q\sum_{n=1}^Nx_n} \rangle  \,\, ,
$$
and (\ref{LD.7}) is exact if $S(Y)$ is convex,
otherwise Eq.~(\ref{LD.7}) gives the convex envelop of the correct
$S(Y)$.\\
Let us note that the Cramer function must obey some constraints:\\
a) $S(Y)>0$ for $Y\neq \langle Y \rangle$;\\
b) $S(Y)=0$ for $Y= \langle Y \rangle$;\\
c) $S(Y)\simeq (Y -  \langle Y \rangle)^2/(2 \sigma^2)$,
  where $\sigma^2=\langle(x -  \langle x \rangle)^2 \rangle$,
  if $Y$ is close to $\langle Y \rangle$.\\
Of course a)  and b) are consequences of the law of large numbers,
and c) is  nothing but the central limit theorem. A mathematical introduction to the Large Deviation theory is given in~\cite{E99}.

\section{Appendix: Anosov systems}
Anosov systems play an important pedagogical role, in the theory of dynamical systems: their behaviour 
is highly chaotic and rich in structure. The celebrated 
Arnold cat map, defined by 
\be
\left( \begin{array}{c} x_{n+1} \\ y_{n+1} \end{array}  \right) = \left( \begin{array}{lr}
1~ & ~1 \\ 1~ & ~2 \end{array} \right) \left( \begin{array}{c} x_{n} \\ y_{n} \end{array}  \right)
~~ \mbox{mod } 1
\ee
on the unit square $\mathcal{M} = [0,1]\times[0,1]$,
belong to this class.\footnote{Because of the mod operation, the Arnold cat map is not a
linear map.} 

Although they can be understood in rather simple terms, they are 
quite  effective in expressing the concept of deterministic chaos. Anosov dynamics $S^t$ 
can be discrete (Anosov diffeomorphisms) or continuous (Anosov flows), and are defined on 
smooth manifolds $\mathcal{M}$ (the phase space). Anosov diffeomorphisms and flows are 
not completely equivalent, although from a flow one may always extract a map, observing
the continuous time dynamics at discrete instants of time. Here, we limit our attention
to properties which are similar in the two cases, presenting the general theory for
continuous time systems, since physical systems mostly have continuous time, and illustrating 
it on the Arnold cat map, which can be explicitly worked out in detail.

For simplicity, assume that the dynamics explores densely $\mathcal{M}$, i.e.\ that it is {\em transitive}.
Then, the main feature of an Anosov system is that $\mathcal{M}$ is hyperbolic, i.e.\ that for 
all $X \in \mathcal{M}$, 
the tangent space $T_X \mathcal{M}$ continuously splits in a stable, an unstable and a neutral 
linear space \cite{GuckHolm}, $E_X^s, E_X^u$ and $E_X^0$ respectively. 
The tangent space of a flow at a point $X \in \mathcal{M}$ can be interpreted as the space of all 
possible velocities at that point. Indeed, the evolution remains in $\mathcal{M}$, hence these 
velocities can only be tangent to $\mathcal{M}$ in that point, and constitute a linear space of 
dimension equal to the dimension of $\mathcal{M}$. The points in $T_X \mathcal{M}$ are then also 
called tangent vectors. More precisely, let $D S^t(X)$ be the Jacobian map of the transformation 
$S^t$ (which evolves any initial
condition $X$ into $X_t=S^t X$) evaluated at the point $X$. Clearly, $D S^t(X)$ depends on 
both $t$ and $X$. Then, for a flow, transitivity and hyperbolicity mean that:
\begin{description}
\item[a)]  $T_X \mathcal{M} = E_X^u \oplus  E_X^s \oplus E_X^0$ is continuous (or smooth) 
in $X$, i.e.\, in any coordinate chart, the 
components of a constant vector field are continuous (or smooth) 
functions of the coordinates of the point $X$. $E_X^0$ is a one-dimensional linear 
subspace in case of flows, while it reduces to $\{ 0 \}$ in case $S^t$ is a map.
\item[b)] The linear subspaces $E_X^s$ and $E_X^u$ are invariant with respect to the tangent 
(or Jacobian) map $D S^t$, \ie\ for all  $X \in \mathcal{M}$ $ \; D S^t(X) E_X^s = E_{S^t(X)}^s$ 
 and  $D S^t(X) E_X^u = E_{S^t(X)}^u$.
\item[c)] There exist $C > 0$ and $1> \theta > 0$, such 
that for all $t \in \zR$, we have $\parallel D S^t(X) \xi  \parallel 
\leq  C  \, \theta^t \! \parallel 
\xi \parallel$, for all $\xi \in E_X^s$ and
$\parallel D S^{-t}(X)  \eta  \parallel \leq  C  \, \theta^{t} \!  
\parallel \eta \parallel$, for all 
$\forall \eta \in E_X^u$.
\end{description}
The tangent space at $X$ may also be seen as the space which contains the infinitesimal perturbations 
$\zd X = X' - X$ of the initial point of a given trajectory, that produce nearby trajectories. Because 
$X$ and $X'$ lie in 
$\mathcal{M}$, because $\mathcal{M}$ is assumed to be smooth, hence to have a tangent in $X$, 
and because $\zd X$ is infinitesimal, $\zd X$ lies in this tangent space, of dimension equal to 
that of $\mathcal{M}$. The tangent vectors evolve according to the rule $DS^t(X) \zd X$, hence 
condition c) implies that they all grow or decrease at exponential rates, which are everywhere 
bounded away from zero, except for the tangent vectors directed along the neutral space. 

\noindent
Furthermore, transitive Anosov systems enjoy the following properties \cite{GuckHolm,Ruellebook}: 
\begin{enumerate}
\item the stable and unstable manifolds  
$$\hskip -14pt W_X^s = \{ X' \in \mathcal{M} : \lim_{t \to \infty} \| S^{t} X' - S^{t} X \| = 0 \}$$
$$W_X^u = \{ X' \in \mathcal{M} : \lim_{t \to \infty} \| S^{-t} X' - S^{-t} X \| = 0 \}$$ 
are globally defined and dense in $\mathcal{M}$, for every $X \in \mathcal{M}$ (transitivity);
\item periodic orbits are dense in $\mathcal{M}$;
\item there is an invariant measure $\mu$ which represents the statistics of the forward time 
evolution, and which has a density along the unstable directions.
Such a measure is usually called Sinai-Ruelle-Bowen 
(SRB) measure and satisfies
\begin{equation}
\lim_{T \rightarrow \infty} \frac{1}{T} \int_0^T \zF(S^t X) =
\int \zF(y) d \mu (y) = \langle \Phi \rangle
\nonumber
\end{equation}
for (Lebesgue) almost all $X \in \mathcal{M}$. For maps, the same holds,
if the time integral is replaced by a sum over time steps.

\end{enumerate}

\noindent
Thanks to property (2), $\mu$ can be represented in terms of orbital measures.  
If $P \, (T,T+ \epsilon ]$ is the set of the unstable periodic orbits of 
the system, with period $\tau \in (T,T + \epsilon]$, for a fixed $\epsilon > 0$ 
and for any $T>0$, the following holds \cite{PARRY}:\footnote{This theorem 
holds more generally for Axiom-A systems, which may have a finite number of
attracting sets.}

\vskip 5pt
\noi
{\bf Theorem:}{\em
Let $(\mathcal{M},S^t)$ be an
Anosov flow (with $S^t \in C^2$). Let $\zF \in C^1$, and let $j$ be an
unstable periodic orbit of period $\tau_j$. For any $X_j \in j$, denote by $J^u_j$ 
the Jacobian of the dynamics restricted to the unstable manifold, with initial condition $X_j$. 
Then, for all $ \epsilon > 0$,
\begin{equation}
\langle \zF \rangle = \lim_{T \rightarrow\infty}
\dfrac{\sum_{j \in P_{(T , T +\epsilon]}} (J^u_j)^{-1} 
\int_0^{\tau_j} \zF(S^t X_j) d t}{
\sum_{j \in P_{(T, T +\epsilon]}} (J^u_j)^{-1} {\tau_j}}
\label{parry}
\end{equation}
where $X_j$ is any point in the orbit $j$.
}

\vskip 2pt
\noi
A version of Eq.(\ref{parry}) is available for maps, and is formally
identical to (\ref{parry}): it suffices to replace the integrals by sums.
It  is interesting to note that $\langle \zF \rangle$ is obtained 
as a weighted average of orbital averages, with weights 
\be
(J^u_j)^{-1} \cdot {\tau_j}
\label{sugge}
\ee
which have 
the suggestive form ``inverse of instability $\times$ period'': i.e.\ one orbit 
contributes more than another to $\langle \zF \rangle$ if it has larger period,
but contributes less if it is more unstable. Furthermore, it has been observed that
the neighbourhoods of the periodic orbits of a given interval of periods afford, at
least in some systems, a hierarchical tessellation of the phase space, which is finer 
where $\mu$ is higher \cite{LNRM}.
This should yield a rather quick convergence of the limit in (\ref{parry}), because the 
low period orbits cover the parts of phase space which have higher probability, leaving 
to the long period ones the task to cover even the less probable parts. However, 
Eq.(\ref{parry}) has proved highly non-trivial to use for practical computations, 
and it has been tested and used only in low-dimensional systems, such as the Lorentz 
gas \cite{GaspBook,MR}.  This does not prevent the use of that formula for theoretical
calculations, hence various results of statistical mechanics interest have been obtained 
through it (see, e.g. \cite{RC98,DRsmoo,LRladek}). Furthermore, it was the use of 
such weights that led the authors of \cite{ECM} to propose the $\zW$-FR (which happened 
to coincide with the $\zL$-FR in their case). This is why the mathematical formulation 
of the \LFR given by Gallavotti and Cohen was based on the Chaotic Hypothesis,
i.e.\ on the hypothesis that systems of physical interest behave as if they
were transitive Anosov systems. 

All the above is easily illustrated by means of the Arnold cat map. Being a diffeomorphism, it
has no neutral direction, and its tangent space at any point $X \in \mathcal{M}$ is 
$T_X \mathcal{M} = \zR^2$.
The eigenvalues and the eigenvectors of the matrix defining the map are 
$\zl^{u,s}= \left( 3 \pm \sqrt{5} \right)/2$, and
\be
v^u = \left( \begin{array}{c} 1 \\ \frac{1}{2} \left( 1 + \sqrt{5} \right)  \end{array} \right) =
 \left( \begin{array}{c} 1 \\ \phi  \end{array} \right) ~, \qquad
v^s = \left( \begin{array}{c} 1 \\ \frac{1}{2} \left( 1 - \sqrt{5} \right)  \end{array} \right) = 
\left( \begin{array}{c} 1 \\ -\phi^{-1}  \end{array} \right)
\ee
where $\phi$ is the golden ratio. Therefore, points that lie close to $(0,0)$, in the straight line 
of slope $\phi$ passing through the point $(0,0)$ remain in that line and their distance grows by a 
factor $\phi$ at each iteration. When these points cross the boundary of the unit square,
they re-enter the square from the opposite side, and continue to move on a line of same slope, while
their distance continues to grow by the same factor, except for the effect produced by the mod operation.
Therefore, the infinite line of slope $\phi$, passing through $(0,0)$, cut and moved by the mod 1 operation,
constitutes the unstable manifold of the point $(0,0)$. Similarly, the line of slope $-\phi^{-1}$, passing
through $(0,0)$, cut and moved by the mod 1 operation, constitutes the stable manifold of $(0,0)$,
because its points are moved along that direction, while their distances are reduced by the factor $\phi^{-1}$
at every iteration of the map. Similarly, one finds that each point $X \in \mathcal{M}$ has stable and
unstable manifolds which are the straight lines of slopes $\phi$ and $-\phi^{-1}$, passing through $X$,
mod 1. The exponential growth and decrease of the sizes of the tangent vectors are precisely the factors
$\phi$ and $\phi^{-1}$, because the tangent vectors are linear combinations of $v^u$ and $v^s$.
The density of the unstable manifold of the point $(0,0)$, in the Arnold cat map, is easily understood
observing that its slope is irrational. Therefore, if one thinks of it as the trajectory
of a point particle, which moves at constant velocity, exits the square at given instants
of times, and re-enters the square form the opposite side, one realizes that it can never 
re-enter at a point which has been previously visited. In other words, this trajectory, i.e.\
the unstable manifold, wraps around densely exploring all of $\mathcal{M}$. The invariant 
SRB measure is the Lebesgue measure $d \mu = d x d y$, and the density of the periodic orbits
follows from the fact that the pieces of stable and unstable manifolds of a point intersect forming
a grid (called Markov partition), whose cells have invariant borders, inside each of which there are
periodic points. Because the manifolds are dense, hence can form arbitrarily fine grids, the
density of the periodic points follows. To find the periodic points
of period $n$, one must solve the equation
\be
\left( \begin{array}{c} x \\ y \end{array}  \right) = \left( \begin{array}{lr}
1~ & ~1 \\ 1~ & ~2 \end{array} \right)^n \left( \begin{array}{c} x \\ y \end{array}  \right)
~~ \mbox{mod } 1
\ee
For instance, $(0,0)$ is a fixed point, hence it is periodic of all periods.
All the periodic orbits are unstable, because of the ubiquitous hyperbolic structure.

Normally, the Anosov systems that are considered to illustrate the idea of chaos in physics 
are low dimensional, like the Arnold cat, because their dimensionality is not particularly 
important for their statistical 
properties. These properties are due to the uniform hyperbolicity, i.e. to the violent and 
ubiquitous trend of the trajectories to separate at an exponential rate, which produces an 
extremely high degree of mixing in the phase space. This makes possible to formally obtain
in low dimensional Anosov dynamics certain properties that are expected to hold in 
thermodynamic systems, like the symmetry of Onsager relations \cite{GR97}. Obviously, the
physical content of that symmetry in a system of one particle is totally different from the
content of that symmetry in a thermodynamic system, and is obtained for completely different 
reasons. Adding that hardly any system of physical interest is of the Anosov kind, 
 that the phase space dynamics is not the real space dynamics, and recalling the meaning 
of the statistical laws, as expressed e.g.\ in \cite{LandauFisStat}, cf.\ Sect.\ref{nanosys}, 
leads to the conclusion that some care must be used in describing physical phenomena by means of Anosov 
systems.

\section{Appendix: A pedagogical example}

It is very instructive, also from a pedagogical point of view, to work
out, for a system simple enough, the connection between some of the
results discussed in Sections 5, 6 and 7. 
We closely follow a recent interesting paper by Astumian~\cite{A06},
which considers the overdamped dynamics of a colloidal particle
immersed in a fluid with viscosity $\gamma$, subject to a conservative
force $F_c=-U'(x)$ and pulled by a non-conservative force
$F_{nc}(t)=g(t)F_{ext}$. The position $x$ of the particle obeys the
Langevin equation
\begin{equation} \label{astu1}
\gamma \frac{dx}{dt}=F(t)+\sqrt{2\gamma T}\eta(t),
\end{equation}
where $F(t)=F_c+F_{nc}(t)$, $\eta(t)$ is a Gaussian white noise. This
is also the equation that governs the process of
pulling a terminal of a macromolecule anchored to a  surface and
surrounded by water; this system has been studied in recent experiments~\cite{liphardt,R07}.

Following the discussion of Onsager-Machlup theory in Sec. 7.1,, in
particular the non-linear generalization of Par. 7.1.1, one can obtain
the probability density of a path starting at time
$t_0$ and ending at time $t$. Again it is sufficient to consider
discrete times $t_0+k\tau$ with $\tau$ arbitrarily small and
$k\in[0,n]$ with $n$ being the integer part of $(t-t_0)/\tau$. Since
the noise is Gaussian and delta-correlated, the sequence of variables
$\eta_k=\eta(t_0+k\tau)$ has the probability density 
\begin{equation}
P[(\eta_n,t|...|\eta_0,0)]\propto \exp\left(-\frac{1}{2} \sum_{k=0}^n \eta_k^2\tau \right)
\end{equation}
which, in the limit $\tau \to 0$, becomes
\begin{equation}
P[(\eta_n,t|...|\eta_0,0)]\propto \exp\left(- \frac{1}{2}\int_0^t ds \eta^2(s) \right).
\end{equation}
Equation~\eqref{astu1} tells us that
$\eta(t)=(\frac{dx}{dt}-F/\gamma)/\sqrt{2D}$, with $D=T/\gamma$, which
finally gives us
\begin{equation} \label{astu2}
P[\{\eta(t)\}] \propto \exp(-L),
\end{equation}
where
\begin{equation} \label{astu3}
L=\frac{1}{4D}\int_0^t ds\left(\frac{dx}{ds}-\frac{F}{\gamma}\right)^2
=\frac{\Delta U-w_{ext}[\{\eta(t)\}]}{2\gamma D}+\frac{1}{4D}\int_0^t ds \left[\left (\frac{dx}{ds} \right)^2+\frac{F^2}{\gamma^2} \right]
\end{equation}
is the thermodynamic action,
\begin{equation}
w_{ext}[\{\eta(t)\}]=\int_0^t ds \left(F_{nc}(s) \frac{dx}{ds}\right)
\end{equation}
is the work done by the external forcing $F_{nc}(t)$ and $\Delta
U=-\int_{x_0}^{x_t}F_c(x)dx$ is the difference of potential between
the final and the initial position. To find the most probable path
from $(x_0,0)$ to $(x_t,t)$, it is sufficient to minimize the
action~\eqref{astu3} while keeping fixed the endpoints.  Note that to
obtain, from Eq.~\eqref{astu2}, the correct probability density for
the trajectories $x(t)$, one has to calculate the Jacobian of the
transformation $\eta(t) \to x(t)$. Anyway, even
disregarding this problem, many interesting results discussed in this
review can already be reproduced.
\begin{itemize}

\item A difference (asymmetry) between the probability of a path
$\{\eta(t)\}$ and its time-reversed $\{I\eta(t)\}$,  follows
from the fact that the last integral in~\eqref{astu3} does not change
sign upon applying the time-reversal operator $I$. This corresponds,
in this example, to the asymmetry of paths discussed in 7.2.1. Of
course the richness of the recent results  in Sections 7.2 and 7.3
(such as the non-locality of the entropy functional), which concerns
spatially extended systems (fields) cannot be reproduced by this
example.

\item The ratio between the probability density of a path and that of
its time-reversed reads:
\begin{equation} \label{astu4}
\frac{P[\eta(t)]}{P[I\eta(t)]}=\exp\left(\frac{w_{ext}[\eta(t)]-\Delta U}{k_B T} \right),
\end{equation}
which, for large times, allows one to identify the work $w_{ext}$ done
by the external non-conservative force (divided by $k_B T$) as the
entropy produced during the time $t$. This is an example of the result
by Kurchan~\cite{Kurchan} and by Lebowitz and Spohn~\cite{LS99} on the FR for
stochastic systems. Even without referring to the FR,
Eq.~\eqref{astu4} is interesting because the r.h.s. does not depend
explicitly on time.

\item
By a simple calculation it is immediate to see that 
\begin{equation}
\frac{P(w_{ext}[\eta(t)]=W)}{P(w_{ext}[\eta(t)]=-W)}=\exp\left(\frac{W}{k_b T} \right)
\end{equation}
which is the FR for stochastic systems.

\item Finally, one has
\begin{equation}
\left \langle \exp \left( \frac{-W}{k_B T}\right) \right \rangle=
\int_{-\infty}^{\infty} dW \exp \left( \frac{-W}{k_B T}\right) P(W)=\int_{-\infty}^{\infty}dW P(-W)=1
\end{equation}
which is nothing but the Jarzynski relation, or the partition identity, in this steady state case.

\end{itemize}

\bibliographystyle{elsart-num}
\bibliography{fluct.bib}

\end{document}